\documentclass[pagesize=auto,bibliography=totocnumbered]{scrbook}

\usepackage{comment}
\usepackage{etoolbox}
\usepackage{fancyvrb}
\usepackage{graphicx}
\usepackage{subcaption}
\usepackage{hyperref}
\usepackage[numbers]{natbib}
\usepackage{bibunits}
\usepackage{xspace}

\usepackage{fontenc}
\usepackage[utf8]{inputenc}
\usepackage[spanish]{babel}

\addto\captionsspanish{}
\addto\captionsspanish{}

\usepackage{makeidx}

\hypersetup{
    colorlinks=true,
    linkcolor=blue,
    filecolor=blue,
    urlcolor=blue,
    }

\usepackage[nogroupskip,toc]{glossaries}

\usepackage{listings}
\usepackage[dvipsnames]{xcolor}

\usepackage{amssymb}
\usepackage{yfonts}

\decimalpoint

\usepackage{amsmath,amsthm}
\usepackage{tcolorbox}
\tcbuselibrary{theorems}

\newcommand\nextpage[1][]{
\ifdefined\HCode {
  \HCode{<mbp:pagebreak />}}
\else
  \newpage
\fi
}

\def\surl#1{{\small{\url{#1}}}}

\newcommand{\abs}[1]{\left|{#1}\right|}
\newcommand{\ket}[1]{\left|{#1}\right\rangle}

\newcommand{\braket}[2]{\left\langle{#1} \middle| {#2}\right\rangle}

\newcommand{\prob}[1]{\mathcal{P}\{{#1}\}}

\newcommand{\granZ}{\mathcal{Z}}

\newcommand{\HBoltzmann}{\mathcal{H}}

\def\dbar{{\mathchar'26\mkern-12mu d}}

\newcommand{\E}[1]{\left\langle{#1}\right\rangle}

\theoremstyle{plain} 
\newtheorem{teorema}{Teorema}[chapter]
\newtheorem{lema1}[teorema]{Teorema} 

\theoremstyle{plain} 
\newtheorem{definition1}{Definición}[chapter]

\newtcolorbox{definitionbox}{
    colback=blue!1, 
    colframe=blue!10, 
    fonttitle=\bfseries,
}

\newenvironment{definition}
{\begin{definitionbox}\begin{definition1}}
{\end{definition1}\end{definitionbox}}

\newenvironment{lema}
{\begin{definitionbox}\begin{lema1}}
{\end{lema1}\end{definitionbox}}

\newcounter{exercise}[section] 

\renewcommand{\theexercise}{\thesection.\arabic{exercise}}

\newenvironment{exercise}{%
    \refstepcounter{exercise}
    \par\medskip\noindent
    \textbf{\theexercise. }
}{\medskip} 

\ifdefined\HCode{\KOMAoptions{twoside=false}} \fi

\makeglossaries
\newglossaryentry{omega}{
    name={\ensuremath{\Omega}},
    sort=omega,
    description={Número de estados accesibles}
}

\newglossaryentry{mu}{
    name={\ensuremath{\mu}},
    sort=mu,
    description={Potencial químico}
}

\newglossaryentry{muB}{
    name={\ensuremath{\textswab{m}}},
    sort=muB,
    description={Momento dipolar magnético}
}

\newglossaryentry{t}{
    name={\ensuremath{T}},
    sort=t,
    description={Temperatura absoluta}
}

\newglossaryentry{mi}{
    name={\ensuremath{\bar{\mu}}},
    sort=mi,
    description={Potencial electroquímico}
}

\newglossaryentry{f}{
    name={\ensuremath{F}},
    sort=f,
    description={Energía libre de Helmholtz}
}

\newglossaryentry{g}{
    name={\ensuremath{G}},
    sort=g,
    description={Energía libre de Gibbs}
}

\newglossaryentry{kb}{
    name={\ensuremath{k_B}},
    sort=kb,
    description={Constante de Boltzmann}
}

\newglossaryentry{s}{
    name={\ensuremath{S}},
    sort=s,
    description={Entropía}
}

\newglossaryentry{configuration}{
    name={\ensuremath{\{n_i\}}},
    sort=configuración,
    description={Configuración}
}

\makeindex

\begin{document}



\thispagestyle{empty}

\vspace{3cm}
  \begin{center}
	\bfseries \Huge Física Estadística   
        ~\\
	\bfseries \LARGE Teoría, Ejemplos Trabajados, Modelos y Simulaciones \\   
        ~\\
        \bfseries \Large David Alejandro Miranda Mercado, Ph.D\\Carlos José Paéz González, Ph.D\par   

        \vspace{3cm}
    
    \end{center}
    
\par

\newpage

\pagenumbering{roman}

\tableofcontents

\printglossaries


\addchap{Prefacio}

La inspiración para escribir este libro surgió de la necesidad de nuestros estudiantes del curso Física Estadística por contar con un material de estudio moderno y en concordancia con el programa del curso. A su vez, la colaboración entre los profesores a cargo del curso ha sido una motivación para la construcción de este recurso didáctico, lo que se enmarca en un esfuerzo más amplio por construir y enriquecer a nuestra Escuela de Física.

A lo largo de los últimos años, nuestra experiencia enseñando el curso de Física Estadística ha permeado la creación de esta primera edición del libro. En ella, hemos incorporado elementos teóricos, ejemplos trabajados, simulaciones y análisis de investigaciones científicas actuales en el campo. Este último aspecto es particularmente relevante, ya que a menudo se omite en textos más tradicionales.

Con este libro, buscamos poner a disposición de nuestros estudiantes e interesados en la física estadística un material didáctico autocontenido y actualizado. En su contenido se incluyen ejemplos trabajados, simulaciones y análisis de investigaciones científicas recientes en las que la física estadística ha jugado un papel fundamental para la descripción de fenómenos naturales. \\ \\

{\raggedleft
David A. Miranda \\
Carlos J. Paez \\
2023\\
}


\addchap{Introducción}

La Física Estadística es un área de la física que se encarga del estudio del comportamiento de sistemas compuestos por numerosas partes, generalmente consideradas como partículas. Un método para visualizar esta composición consiste en imaginar un sistema macroscópico que es dividido en partes hasta llegar a una división tal que no puede seguir dividiéndose. En este punto, dicho sistema puede ser descrito por las porciones más pequeñas posibles, las cuales denominaremos partículas. La idea que la materia está formada por partículas indivisibles es antigua y se atribuye al filósofo griego Leucipo de Mileto, quien postuló la existencia de los átomos. En el marco de la Física Estadística, consideramos a la partícula como la porción más pequeña de un sistema, la cual puede considerarse indivisible bajo ciertas condiciones\footnote{Es importante tener en cuenta que un sistema puede considerarse compuesto por moléculas, las cuales, según la Física Estadística, pueden tratarse como partículas. Sin embargo, al introducir suficiente energía en una molécula, esta puede dividirse en átomos y, al suministrar suficiente energía a los átomos, estos se pueden dividir en protones, electrones y neutrones. El estudio de la división en partículas más pequeñas pertenece al ámbito de la física de altas energías, un campo de estudio tanto teórico como experimental.}. El comportamiento de estas partículas está determinado por las leyes de la física, en general, por la mecánica cuántica (aunque en algunos casos se puede recurrir a la física clásica, cuando se considera el límite clásico). Sin embargo, la descripción analítica de un sistema formado por muchas partículas es restrictivo (por no decir imposible), y es aquí donde los conceptos estadísticos juegan un papel fundamental para la descripción del sistema.

La descripción de las partículas en términos de las leyes de la física se conoce como una descripción microscópica del sistema. Este término tiene un origen histórico, cuando los tamaños más pequeños medibles eran de micrómetros. En la actualidad, sin embargo, existen técnicas experimentales que permiten alcanzar resoluciones mayores, logrando observar estructuras de tamaños nanométricos (por ejemplo, mediante la microscopía de fuerza atómica) o incluso menores (por ejemplo, mediante la microscopía electrónica de transmisión). A pesar de ello, seguiremos utilizando el término \lq\lq microscópico\rq\rq\, para referirnos a la descripción del sistema a la escala de sus partes constituyentes, independientemente de si estas tienen tamaños de micrómetros o menores.

En la descripción estadística de un sistema compuesto por partículas, se pueden adoptar dos enfoques: el análisis de la evolución temporal del sistema o el estudio de conjuntos estadísticos de sistemas. La concepción actual de la Física Estadística data de los tiempos en los que Boltzmann estudiaba el gas ideal clásico con un enfoque atomístico, época en la que ni siquiera estaba clara la existencia de los átomos y moléculas (finales del siglo XIX y principios del siglo XX). En su célebre libro, Boltzmann \cite{Boltzmann1995} propone seguir el enfoque de ensambles estadísticos de sistemas, que consiste en considerar muchos sistemas preparados en las mismas condiciones, pero cada preparación (sistema) completamente aislado del otro. Así, la forma en la que se agrupan los sistemas del ensamble da lugar a una descripción estadística, la cual debe ser independiente del agrupamiento. En otras palabras, la forma en que se organiza el ensamble no debe alterar la descripción física del sistema. Usualmente en Física Estadística se utilizan tres tipos de agrupaciones o ensambles. \textit{Primero}, un conjunto formado por el sistema aislado bajo la condición que su energía está en el rango comprendido entre $E$ y $E+\delta E$; a este ensamble se le conoce como microcanónico. \textit{Segundo}, un conjunto compuesto por sistemas en equilibrio termodinámico con un reservorio de calor a una temperatura constante $T$, con el que el sistema puede intercambiar energía; a este conjunto se le conoce como ensamble canónico. \textit{Tercero}, un conjunto compuesto por sistemas en equilibrio termodinámico con un reservorio de calor a una temperatura $T$ y potencial químico $\mu$ constantes, con el cual el sistema puede intercambiar energía y partículas; a este se le conoce como ensamble gran canónico. Es evidente que se pueden realizar otros tipos de agrupaciones estadísticas, dando lugar a otros tipos de ensamble.

En este libro se abordará el estudio de la Física Estadística bajo condiciones de equilibrio termodinámico, partiendo de la definición de ensambles estadísticos. La obra está dividida en cuatro capítulos: \textit{Primero}, una breve introducción a los métodos estadísticos. \textit{Segundo}, la descripción estadística de sistemas aislados, correspondiente a los ensambles microcanónicos. \textit{Tercero}, la descripción estadística de sistemas en contacto con un reservorio de calor a temperatura constante $T$, conocida como ensamble canónico. Y por último, la descripción de sistemas en contacto con un reservorio de calor a temperatura $T$ y con un potencial químico $\mu$ constantes; el ensamble gran canónico. En cada capítulo, se presentará una descripción teórica, ejemplos trabajados, enlaces a simulaciones (ver \href{https://github.com/davidalejandromiranda/StatisticalPhysics}{repositorio en GitHub}) y discusiones sobre reportes científicos recientes. Además, se han dispuesto algunos vídeos en un \href{https://www.youtube.com/playlist?list=PLQcmiXk5CJeZAbK1Iw0LnygynpWyNg7_7}{canal de YouTube}, con los cuales los lectores podrán complementar sus estudios. 


\chapter{Capítulo I. Introducción a los métodos estadísticos}

\pagenumbering{arabic}

La comprensión de la realidad implica la construcción de modelos teóricos, usualmente, en lenguaje matemático, que constituyen uno de los pilares fundamentales de la Física. Conceptos como partículas, ondas, partículas cuánticas, campos, entre otros, son utilizados en la construcción de dichos modelos, todo ello con el fin de comprender la realidad, para así hacer predicciones.  A pesar que en ciertas situaciones la certeza de las predicciones realizadas por la Física es muy alta, hay otras en que es imposible realizar predicciones con una certeza absoluta, como en sistemas altamente confinados, entre los cuales un electrón en un átomo de hidrógeno aislado del resto del universo es una situación de interés, entre otras, por la existencia de una descripción analítica.

En un átomo de hidrógeno, el electrón es descrito por un vector de estado $\ket{n, l, m_l, m_s}$, determinado por los números cuánticos principal ($n$), de momento angular ($l$), proyección en el eje z del momento angular ($m_l$) y espín ($m_s$). Supongamos que se quiere predecir el resultado de medir una cierta cantidad física $A$, en el sistema formado por el electrón en el estado $\ket{n, l, m_l, m_s}$, siendo $\hat{A}$ el observable asociado a dicha cantidad física. Según la mecánica cuántica, dado que el sistema está aislado, al realizar la medición esta corresponderá con uno de los posibles valores propios $a$ del observable $\hat{A}$, con vector propio $\ket{a}$, es decir, $\hat{A}\ket{n, l, m_l, m_s} = a\ket{a}$\footnote{Note que los valores propios del observable $\hat{A}$ forman un conjunto $\{ a \}$, cuyos elementos representan cada uno de los posibles resultados del proceso de medición de la cantidad física $A$.}. La mecánica cuántica predice que al medir la cantidad $A$ existe una probabilidad $|\braket{a}{n, l, m_l, m_s}|^2$ de obtener el valor $a$, donde dicha probabilidad es un número entre cero y uno, pero no necesariamente uno, es decir, es imposible predecir con precisión el valor de la cantidad $A$ cuando se realiza una medición.

La situación se vuelve más interesante cuando consideramos más átomos de hidrógeno, o más electrones en átomos con mayor número atómico. Si se piensa en un sistema macroscópicamente pequeño, digamos de unos pocos gramos, la cantidad de electrones involucrados es tan grande como el número de Avogadro ($\sim 6\times 10^{26} / mol$), haciendo restrictivo el uso de la mecánica cuántica de manera directa. Sin embargo, se pueden utilizar conceptos estadísticos, con los cuales se pueda describir el comportamiento macroscópico de las partes constituyentes, mientras que con la mecánica cuántica se describir cada parte. En este sentido, la estadística permite realizar la extrapolación de la descripción de muchas de las partes individuales para considerar el todo. Este es el propósito de estudiar física estadística, aunque no se limita a solo sistemas confinados; la física estadística permite describir todo tipo de sistemas cuyas partes constituyentes están gobernadas por ciertas reglas, descritas por la Física.

En este primer capitulo centraremos la atención en los conceptos estadísticos básicos, con el fin de ganar cierta familiaridad con su aplicación. Se estudiará los conceptos variable y proceso aleatorio, distribuciones estadísticas, momentos de una variable aleatoria, normalidad y pruebas de normalidad, así como el teorema del límite central. En el capítulo se incluyen varios casos de estudio, algunos de ellos acompañados de simulaciones disponibles en línea. La lectura cuidadosa del material aquí presentado es clave para la comprensión a profundidad de la física estadística.

\section{Variable aleatoria}
\index{Variable aleatoria}

Se suele llamar variable a una cierta cantidad que varía respecto a otra, por ejemplo, respecto al tiempo, o el espacio. Una variable muy conocida es la altura $y(\vec{x})$ de una pelota de baloncesto lanzada por un jugador, donde $\vec{x}$ es la distancia medida desde el lugar donde se realizó el lanzamiento, siendo $\vec{x}=\vec{x}(t)$ una cantidad que varía con el tiempo. Otro ejemplo de una variable, también mecánica, es la posición $\vec{R}$ de una mota de polvo en un cuarto por donde entra un rayo de luz que permite observar su movimiento aleatorio\footnote{El movimiento de la mota de polvo se conoce como movimiento browniano.}. Otros ejemplos de variables, de nuestro mayor agrado e interés, corresponde con las ondas de presión $P(t)$ generadas por nuestros audífonos cuando reproducimos la canción preferida\footnote{En la descripción de esta onda de presión, $P(t)$, se supone que se mide por nuestros oidos, por ello se omite el carácter vectorial de la onda y esta se describe en términos del desplazamiento del tímpano que es interpretado como sonido por nuestro cerebro.}, o cuando escuchamos una emisora; claro, en el caso de una mala señal de radio, domina un ruido de fondo que podemos nombrar $N(t)$.

\begin{figure}[!ht]
    \centering
    \includegraphics[width=\textwidth]{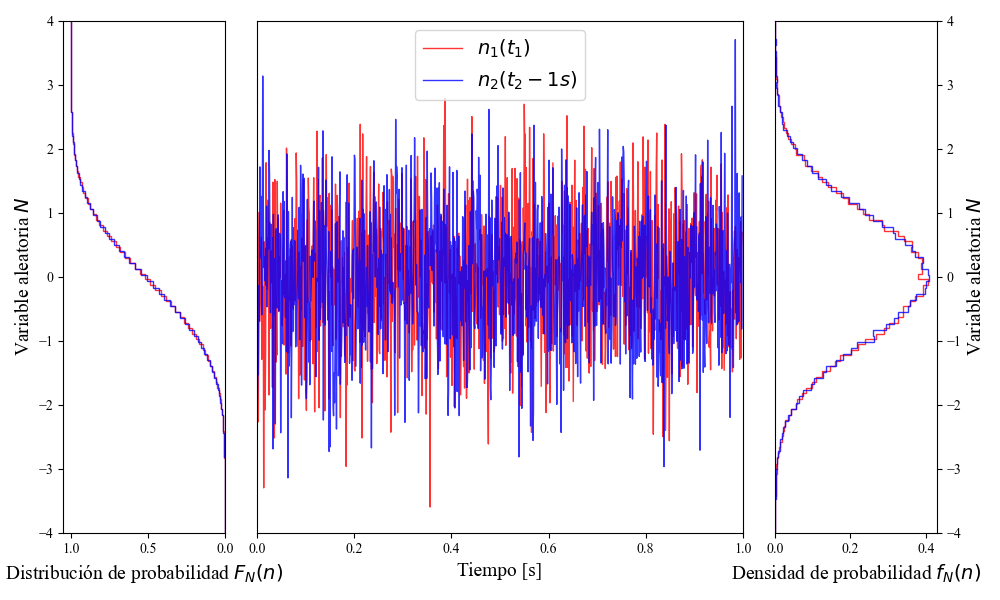}
    \caption{Dos muestras ({\color{red}$n_1$} y {\color{blue}$n_2$}) de la variable aletaoria $N$ (al centro), su distribución de probabilidad $F_N(n)$ (a la izquierda) y su densidad de probabilidad $f_N(n)$ (a la derecha). Cada muestra tiene $10^5$ observaciones.}
    \label{fig:ch1_variable-aleatoria-N}
\end{figure}

De los cuatro ejemplos de arriaba, dos pueden ser descritos de manera determinística ($y$, $P$), mientras que los otros dos es imposible describirlos de manera determinística ($\vec{R}$, $N$). Estas dos últimas variables se conocen como variables aleatorias; en ambos casos ($\vec{R}$, $N$) se nota que existe un fenómeno aleatorio asociado, el movimiento browniano para el caso de $\vec{R}$ y el ruido de fondo (por lo general ruido blanco), para $N(t)$. La existencia de un fenómeno aleatorio al cual se asocia la variable aleatoria es una de las características de esta, además, si se observa la variable aleatoria en diferentes ocasiones, cada observación será diferente a la anterior. Por ejemplo, supongamos que $0 \leq t_1 < 1s$ y $1s \leq t_2 < 2s$ son dos intervalos de tiempo; al medir la variable aleatoria $N$ en el primer intervalo se obtiene $n_1(t_1)$ y en el segundo intervalo $n_2(t_2)$, que corresponden con observaciones diferentes, sin embargo, tanto su densidad de probabilidad como su distribución de probabilidad acumulativa son similares, ver Figura~\ref{fig:ch1_variable-aleatoria-N}.

Nótese que hay que distinguir la variable aleatoria $N$ de las observaciones $n_1$ y $n_2$ de dicha variable aleatoria. A pesar que $n_1(t_1) \neq n_2(t_2 - 1s)$, las densidades de probabilidad (también la distribución de probabilidad) son similares. Esto nos quiere decir que las variables aleatoria se caracterizan por sus densidades de probabilidad (o por sus distribuciones de probabilidad), más no por los valores numéricos que se pueden obtener en cada observación, dado que los valores numéricos pueden cambiar de observación en observación, pero su densidad de probabilidad (y su distribución de probabilidad) será la misma. 

Tanto la distribución $F_X(x)$ como la densidad $f_X(x)$ de probabilidad pueden ser usadas en la descripción de una variable aleatoria $X$, donde $x$ es una cierta observación de esta. En la Figura~\ref{fig:ch1_variable-aleatoria-N} se puede observar que la distribución de probabilidad $F_X(x)$ (también conocida como distribución de probabilidad acumulativa) se obtiene al sumar la probabilidad que la variable aleatoria $X$ tome valores menores a una cierta observación $x$, es decir, $F_X(x) = \prob{X \leq x}$. Por esta razón, como los posibles valores de las observaciones cumplen la condición $|x| < 4$, la distribución de probabilidad para el extremo $x=4$ está dada por $F_X(4) = \prob{X \leq 4} = 1$, donde el valor de uno quiere decir que es 100\% probable que la variable aleatoria tome un valor $x \leq 4$. Si analizamos el otro extremo, $x=-4$, entonces, la distribución de probabilidad es $F_X(-4) = \prob{X \leq -4} = 0$, que nos indica la imposibilidad de una observación donde $x$ sea menor que $-4$. Por otra parte, la densidad de probabilidad nos permite observar las observaciones $x$ más probable, en este caso $x=0$. Es conveniente precisar los conceptos estudiados en definiciones formales.

\begin{definition}
    Una variable aleatoria $X$ corresponde con una cantidad numérica asociada a un proceso aleatorio, la cual está determinada por una función de distribución $F_X(x)$, donde $x$ corresponde a una cierta observación de la variable aleatoria $X$.
    \index{Variable aleatoria!definición}
\end{definition}

\begin{definition}
    La probabilidad $\prob{ }$ es una medida que cuantifica la posibilidad que ocurra un cierto evento específico y toma valores entre cero, cuando es improbable obtener el evento y uno, cuando se espera obtener, con toda seguridad, el evento. 
\end{definition}

\begin{definition}
    Dada una variable aleatoria $X$, la \textbf{fundición de distribución} $F_X(x)$ (también conocida como función de distribución acumulativa) se define como la probabilidad que $X$ sea menor o igual a $x$, lo cual se suele escribir como $F_X(x) = \prob{X\leq x}$.
    \index{Función de distribución, definición.}
\end{definition}

\begin{definition}
    Dada una variable aleatoria $X$, la probabilidad que $X$ tome valores entre $x$ y $x + dx$ está dada por $\prob{x \leq X < x+dx} = f_X(x) dx$, donde $f_X(x)$ se conoce como la densidad de probabilidad de $X$ y se relaciona con la función de distribución $F_X(x)$ por medio de la siguiente expresión,

    \begin{equation}
    \label{eq:dist-prob}
        F_X(x) = \int_{-\infty}^x f_X(y) dy
    \end{equation}
    \index{Densidad de probabilidad, definición.}
\end{definition}

Nótese que de la relación entre la distribución $F_X(x)$ y la densidad $f_X(x)$ de probabilidad, dada por la ecuación \eqref{eq:dist-prob}, se tiene que $dF_X(x)/dx = f_X(x)$, por lo tanto, si $F_X(x)$ no es diferenciable, $f_X(x)$ no existe, mientras que la distribución de probabilidad $F_X(x)$ existe. Esto quiere decir que la distribución de probabilidad $F_X(x)$ puede definirse de manera precisa incluso para casos discontinuos, mientras que la densidad de probabilidad $f_X(x)$ solo existe cuando la distribución de probabilidad es continua. 

Un caso simple corresponde con las distribuciones de probabilidad discretas, en cuyo caso no tiene sentido definir una densidad de probabilidad, más bien se definen valores de probabilidad asociados a una cierta observación; por ejemplo, al lanzar una moneda con dos posibles observaciones, $x=0$ o $x=1$, con igual probabilidad, se puede definir de manera correcta la distribución de probabilidad $F_X(x)$, pero no tiene sentido definir la densidad de probabilidad, en vez de ello se tiene la probabilidad que $x=0$, dada por $\prob{X=0}=1/2$ y la probabilidad que $x=1$, dada por $\prob{X=1}=1/2$. 

Otro detalle a tener en cuenta es que mientras que la distribución de probabilidad $F_X(x)$ es una cantidad adimensional que puede tomar valores entre cero y uno, la densidad de probabilidad $f_X(x)$ tiene unidades del inverso de las unidades de la variable aleatoria, tales que al calcular la probabilidad que $X$ tome valores entre $x$ y $x+dx$, $\prob{x \leq X < x+dx} = f_X(x) dx$, se obtiene una cantidad adimensional.

%
%
%
%

\newpage
\section{Momentos de una variable aleatoria}

Dada una cierta variable aleatoria $X$, existen cantidades llamadas momentos que permiten describir sus características. Para definir los momentos de una variable aleatoria es necesario comenzar con la operación valor esperado, denotada por $\E{ }$. A continuación, se presentan las definiciones correspondientes.

\begin{definition}
    Sea una cierta cantidad $f(x)$, donde $x$ es una observación de la variable aleatoria $X$. El \textbf{valor esperado} de $f(x)$, denotado como $\E{f(x)}$, está dado por la ecuación \eqref{eq:valor_esperado_discreto}, en el caso que $X$ sea una variable aleatoria discreta con probabilidad $\prob{X=x_n} = p_n$ y por la ecuación \eqref{eq:valor_esperado_continuo}, para una variable aleatoria $X$ continua con densidad de probabilidad $f_X(x) = \rho(x)$.
    \index{Variable aleatoria!valor esperado}

    \begin{equation}
        \label{eq:valor_esperado_discreto}
        \E{f(x)} = \sum_n f(x_n) p_n
    \end{equation}
    
    \begin{equation}
        \label{eq:valor_esperado_continuo}
        \E{f(x)} = \int_{-\infty}^\infty f(x) \rho(x) dx
    \end{equation}

\end{definition}

\begin{definition}
    El n-ésimo \textbf{momento} de una variable aleatoria $X$ se define como el valor esperado de la n-ésima potencia de dicha variable, dado por $\mu'_r = \E{X^n}$.
    \index{Variable aleatoria!momento}
\end{definition}

\begin{definition}
    Se define la \textbf{media} de una variable aleatoria $X$ como su primer momento y se denota como $\mu = \E{X}$. A la media también se le conoce como valor esperado de la variable aleatoria.
    \index{Variable aleatoria!media}
\end{definition}

\begin{definition}
    El n-ésimo \textbf{momento central} de una variable aleatoria $X$ se define como el valor esperado de la n-ésima potencia de dicha variable alrededor de la media y está dado por $\mu_r = \E{(X - \mu)^n}$.
    \index{Variable aleatoria!momentos centrales}
\end{definition}

\begin{definition}
    Se define la \textbf{desviación estándar} de una variable aleatoria $X$ como la raíz cuadrada de su segundo momento central y se denota como $\sigma = \sqrt{\E{(X - \mu)^2}}$. Al cuadrado de la desviación estándar, $\sigma^2=\E{X^2} - \mu^2$, se le conoce como \textbf{varianza} de la variable aleatoria.
    \index{Variable aleatoria!desviación estándar}\index{Variable aleatoria!varianza}
\end{definition}

Para complementar el estudio de los momentos de una variable aleatoria, se sugiere responder las preguntas de autoexplicación del ejemplo trabajado titulado: \href{https://colab.research.google.com/github/davidalejandromiranda/StatisticalPhysics/blob/main/notebooks/es_MomentoVariablesAleatorias.ipynb}{momentos de una variable aleatoria}.

\section{Probabilidad conjunta}
\index{Probabilidad conjunta}

Antes de mostrar algunos ejemplos de variables aleatorias con diferentes distribuciones estadísticas, es necesario determinar cómo se calcula la probabilidad cuando se tienen dos o más variables aleatorias. A esta probabilidad se le conoce como probabilidad conjunta y, en el caso más general, el resultado de observar una de las variables aleatorias que determinan la probabilidad conjunta podría afectar el resultado de la observación de otra de las variables. A pesar de ello, en física estadística el caso en que una observación no afecta otra es de gran relevancia y está determinado por la independencia estadística de las variables aleatorias.

\begin{definition}
    Dos variables aleatorias $X_1$ y $X_2$ son \textbf{estadísticamente independientes}, si la observación de una no afecta la observación de la otra.
\end{definition}

\begin{definition}
    Sean $X_1,\, X_2, \cdots,\, X_N$ variables aleatorias estadísticamente independientes. La probabilidad que al observar $X_i$ se obtenga $x_i$, para $i=1, 2, \cdots N$, donde $p_i = \prob{X_i=x_i}$, se conoce como probabilidad conjunta y está dada por,

    \begin{equation}
    \label{eq:ch1_prob_conjunta}
        \prob{X_1 = x_1, X_2 = x_2, \cdots, X_n = x_n} = p_1 p_2 \cdots p_N = \prod_{i=1}^N p_i
    \end{equation}
    \index{Independencia estadística}
\end{definition}

\section{Distribución binomial}
\index{Distribución binomial}
Antes de definir formalmente la distribución binomial, es conveniente estudiar un ejemplo simple. Sea una moneda cuyas caras están marcadas con los números $0$ y $1$, donde la probabilidad que al lanzarla se observe $c=0$ es igual a la probabilidad de observar $c=1$, es decir, $\prob{C=0} = \prob{C=1}=1/2$. Si $X$ es una variable aleatoria que representa el número de veces que se observa cara, ¿cuánto es la probabilidad $\prob{X=k; n} = W_n[k]$ que al lanzar $n$ veces la moneda se observe $k$ veces $c=0$? 

Hay dos maneras de realizar el experimento para responder la pregunta, una consiste en realizar $n$ medidas independientes con la misma moneda, mientras se lleva la cuenta de las veces que se observa $c=0$ y la otra consiste en tomar $n$ monedas iguales, lanzarlas, sin que interactúen unas con otras (aisladas entre si) y contar en cuantas monedas se observa $c=0$, esto se conoce como construir un ensamble (micro canónico). Se espera que ambos experimentos den resultados similares, a pesar que la metodología para realizarlos es diferente\footnote{Esta afirmación conduce al principio ergódico, el cual será estudiado en el capítulo 2.}; en ambos casos la probabilidad $W_n[k] = \Omega(k)/n$, donde $\Omega(k)$ es el número de posibilidades de obtener $k$ veces $c=0$ y,

$$n = \sum_k \Omega(k)$$

\begin{figure}[t]
    \centering
    \includegraphics[height=5cm]{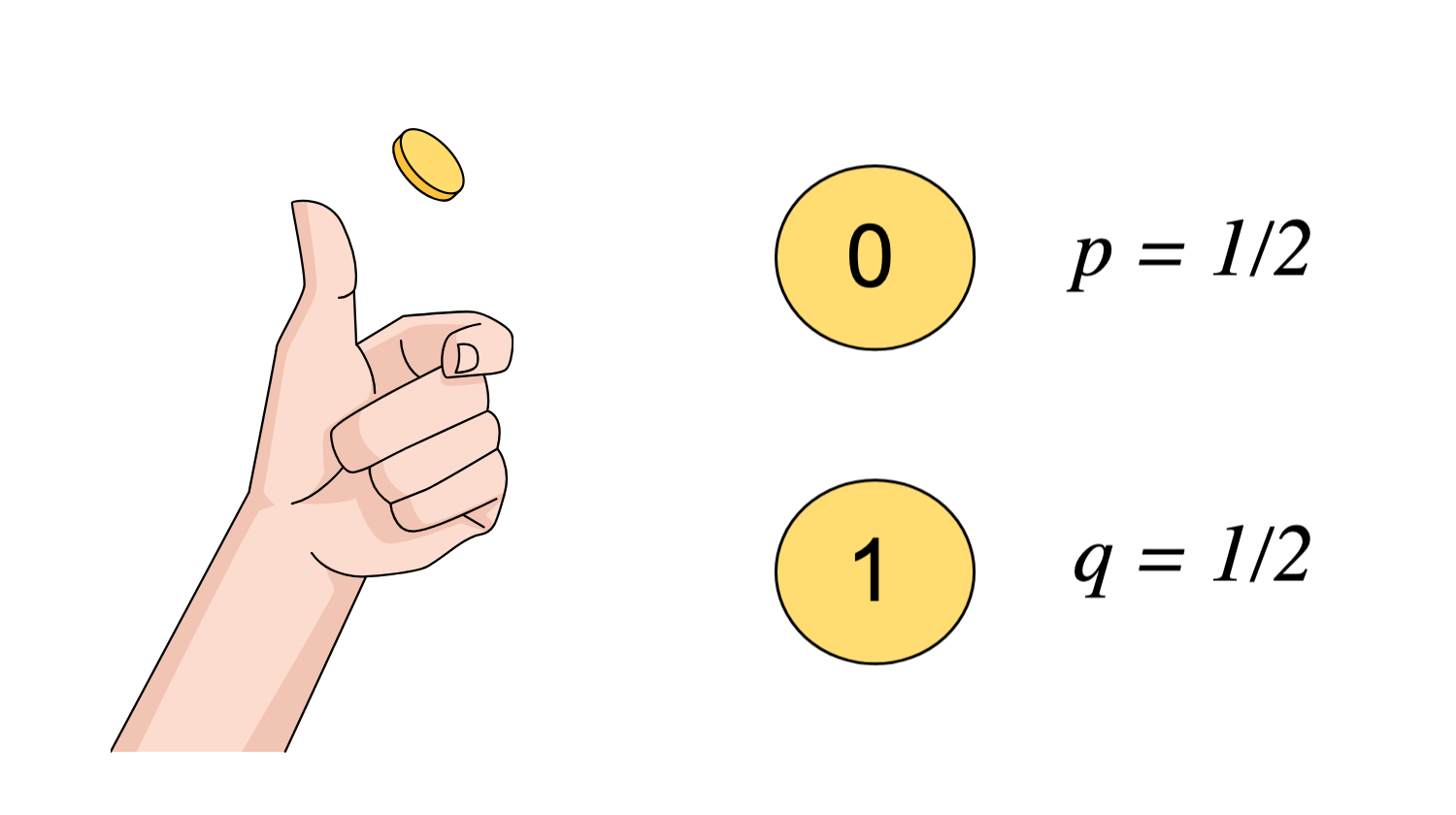}
    \caption{Representación gráfica del experimento de lanzar una moneda, donde la probabilidad de obtener $c=0$ es $p=\prob{C=0}=1/2$ y la de obtener $c=1$, $q=\prob{C=1}=1/2$.}
    \label{fig:ch1_lanzamiento_una_moneda}
\end{figure}

Con la finalidad de resolver la pregunta, analizaremos el caso de dos monedas, donde podemos hacer lanzamientos de la misma moneda o conformar un ensamble formado por dos monedas idénticas no interactuantes. En tal caso, la probabilidad de obtener $c=0$ en el primer lanzamiento es $p=\prob{C=0}=1/2$, que es exactamente igual a la probabilidad de obtener $c=1$, $q=\prob{c=1}=1/2$, ver Figura~\ref{fig:ch1_lanzamiento_una_moneda}.

\begin{figure}[ht]
    \centering
    \includegraphics[height=5cm]{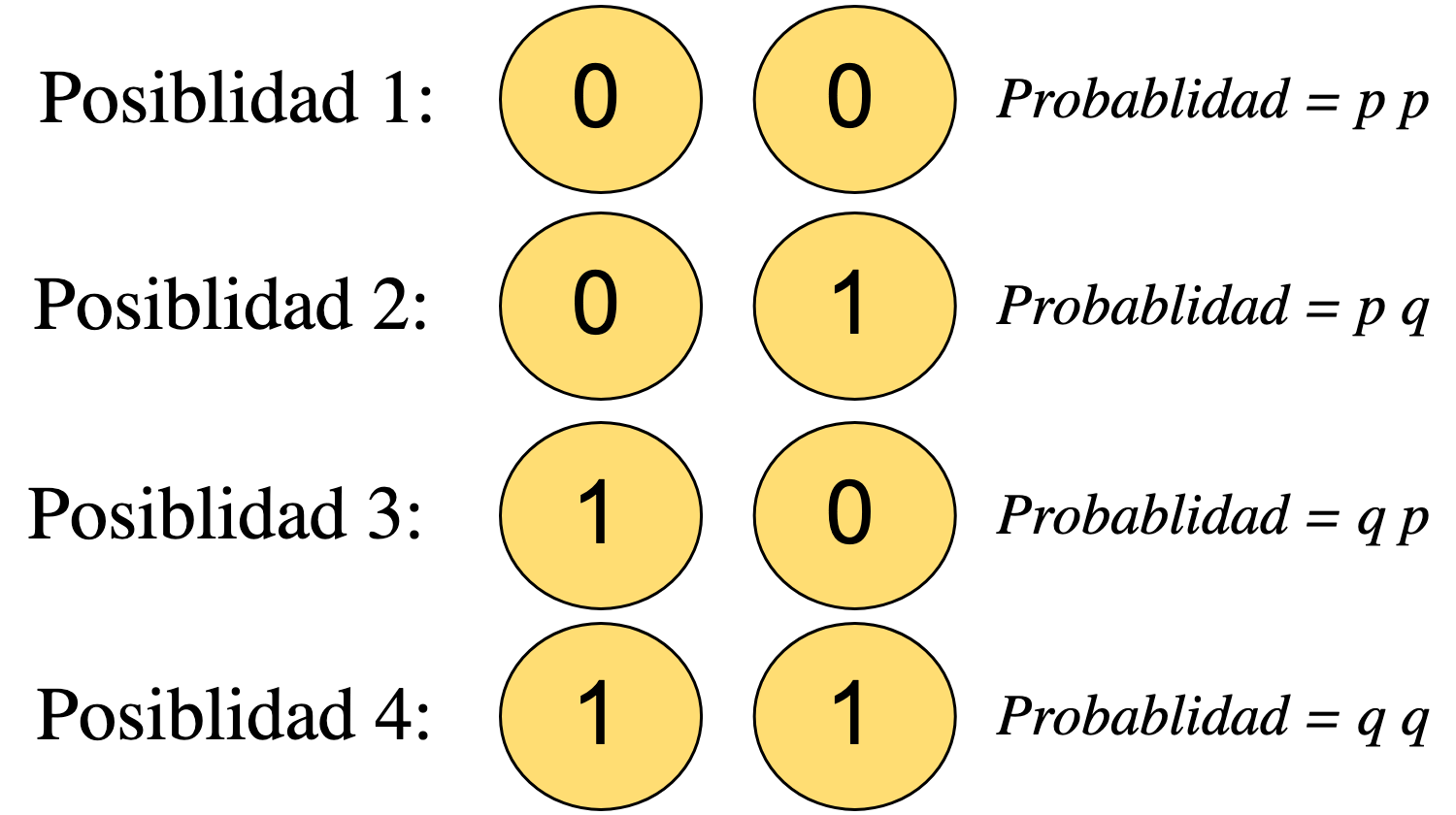}
    \caption{Posibles resultados de lanzar dos monedas, donde, para cada posibilidad la probabilidad está dada por el producto de las probabilidades individuales.}
    \label{fig:ch1_dos_monedas_probabilidades}
\end{figure}

En el experimento con dos observaciones, existen cuatro posibles resultados, como se observa en la Figura~\ref{fig:ch1_dos_monedas_probabilidades}. Entonces, la probabilidad de obtener una vez $c=0$, que corresponde con $k=1$, está dada por $W_2[1] = pq + qp = 2 pq = 1/2$ y la probabilidad de obtener dos veces $c=0$, por $W_2[2] = pp = p^2 = 1/2$.  Nótese que $pp = p^2$ indica la probabilidad conjunta que en el primero de una secuencia de dos lanzamientos (o la primera moneda, para el caso del ensamble) se observe $c=0$ y en el segundo lanzamiento (o segunda moneda, para el ensamble), se observe $c=0$; de igual manera, $pq$ corresponde con la probabilidad conjunta de obtener $c=0$ en el primer lanzamiento y $c=1$ en el segundo lanzamiento. La cantidad de posibilidades donde se obtiene $k$ veces $c=0$ la hemos denominado número de estados accesibles y se representa por $\Omega(1)=2$ (para $k=1$), $\Omega(2)=1$ para $k=2$ y $\Omega(0)=1$ para $k=0$. Esto implica que se puede escribir la probabilidad de obtener $k$ veces $c=0$ en términos del número de estados accesibles,

$$W_2[2] = \Omega(2) p^2 q^{2-2}=1$$
$$W_2[1] = \Omega(2)p^{1}q^{2-1}=2$$
$$W_2[0]=\Omega(0)p^{0}q^{2-0}=1$$

Para tres monedas se puede construir un ensamble como se muestra en la Tabla~\ref{tab:ch1_ensamble_tres_dados}, donde el número de posibles casos para cada valor $k$ están tabulados en la Tabla~\ref{tab:ch1_posibilidades_tres_dados}. Tomemos el caso $k=3$, para el cual hay tres posibilidades ($001$, $010$ y $100$), correspondientes a las todas las posibles combinaciones de elegir dos elementos de tres $\binom{3}{2}$, dada por,

$$\binom{3}{2} =\frac{3!}{2!(3-2)!} = 3$$

\begin{table}[!ht]
\centering
\caption{Observaciones del experimento consistente en lanzar tres monedas (moneda 1, 2 y 3) cuyas caras están marcadas con los números $0$ y $1$. Cada posible resultado se enumera en la primera columna y la probabilidad del mismo, en la última columna.}
\label{tab:ch1_ensamble_tres_dados}
\begin{tabular}{|c|ccc|l|}
\hline
Posibilidad & Moneda 1 & Moneda 2 & Moneda 3 & Probabilidad \\
\hline \hline
1 & 0 & 0 & 0 & $ppp=p^3$ \\ 
2 & 0 & 0 & 1 & $ppq=p^2q$ \\ 
3 & 0 & 1 & 0 & $pqp=p^2q$ \\ 
4 & 0 & 1 & 1 & $pqq=pq^2$ \\ 
5 & 1 & 0 & 0 & $qpp=p^2q$ \\ 
6 & 1 & 0 & 1 & $qpq=pq^2$ \\ 
7 & 1 & 1 & 0 & $qqp=pq^2$ \\ 
8 & 1 & 1 & 1 & $qqq=q^3$ \\ 
\hline
\end{tabular}
\end{table}

\begin{table}[ht]
\centering
\caption{Número de posibilidades de obtener $k$ veces el valor $x=0$ en un experimento con tres monedas donde cada uno de los dos posibles valores tiene la misma probabilidad.}
\label{tab:ch1_posibilidades_tres_dados}
\begin{tabular}{|c|c|c|}
\hline 
$k$ & $\Omega(k)$ & Probabilidad \\
\hline \hline
0 & 1 & $q^3$ \\
1 & 3 & $pq^2$ \\
2 & 3 & $p^2q$ \\
3 & 1 & $p^3$ \\
\hline
\end{tabular}
\label{tab:probabilidades}
\end{table}

Donde, la probabilidad de obtener dos observaciones con $c=0$ (es decir, $k=2$), está dada por,

$$W_3[2] = \Omega(2)p^2q^{3-2} = \binom{3}{2} p^2 q^{(3-2)} = \frac{3}{8}$$

Como era de esperarse, este número es exactamente igual a dividir el número de posibilidades para $k=2$, $\Omega(2) = 3$, sobre el total de posibilidades ($n$), que se obtiene al sumar todos los estas accesibles,

$$\sum_{k=0}^3\Omega(k) = 8$$

\begin{definition}
    Se conoce como \textbf{ensayo de Bernoulli} a un experimento aleatorio con dos posibles resultados, uno llamado éxito, con probabilidad $p$ y el otro, fracaso con probabilidad $q=1-p$.\index{Ensayo de Bernoulli}
\end{definition}

\begin{lema}
    La probabilidad que al observar $n$ ensayos de Bernoulli, con probabilidad de éxito $p$, se obtengan $k$ éxitos es una probabilidad binomial dada por,
    
    \begin{equation}
        \label{eq:ch1_probabilidad_binomial}
        W_n[k] = \binom{n}{k} p^k q^{n-k} =\frac{n!}{k!(n-k)!} p^k q^{n-k}
    \end{equation}
    \index{Probabilidad binomial}
\end{lema}

\begin{definition}
     Una variable aleatoria $X$ es binomial si su probabilidad está dada por una distribución binomial $W_n[k]$. Esto implica que una observación de $X$ corresponde con uno de $n$ posibles valores $x_k$, donde $k=1, 2, \cdots, n$; cada posible observación $x_k$ se comporta como un ensayo de Bernoulli y la probabilidad de observar $x_k$ está dada por $W_n[k]$.\index{Variable aleatoria!binomial}
\end{definition}

\subsection*{Ejemplo ilustrativo: potencia media para batería defectuosa}

\begin{figure}[ht]
    \centering
    \includegraphics[height=5cm]{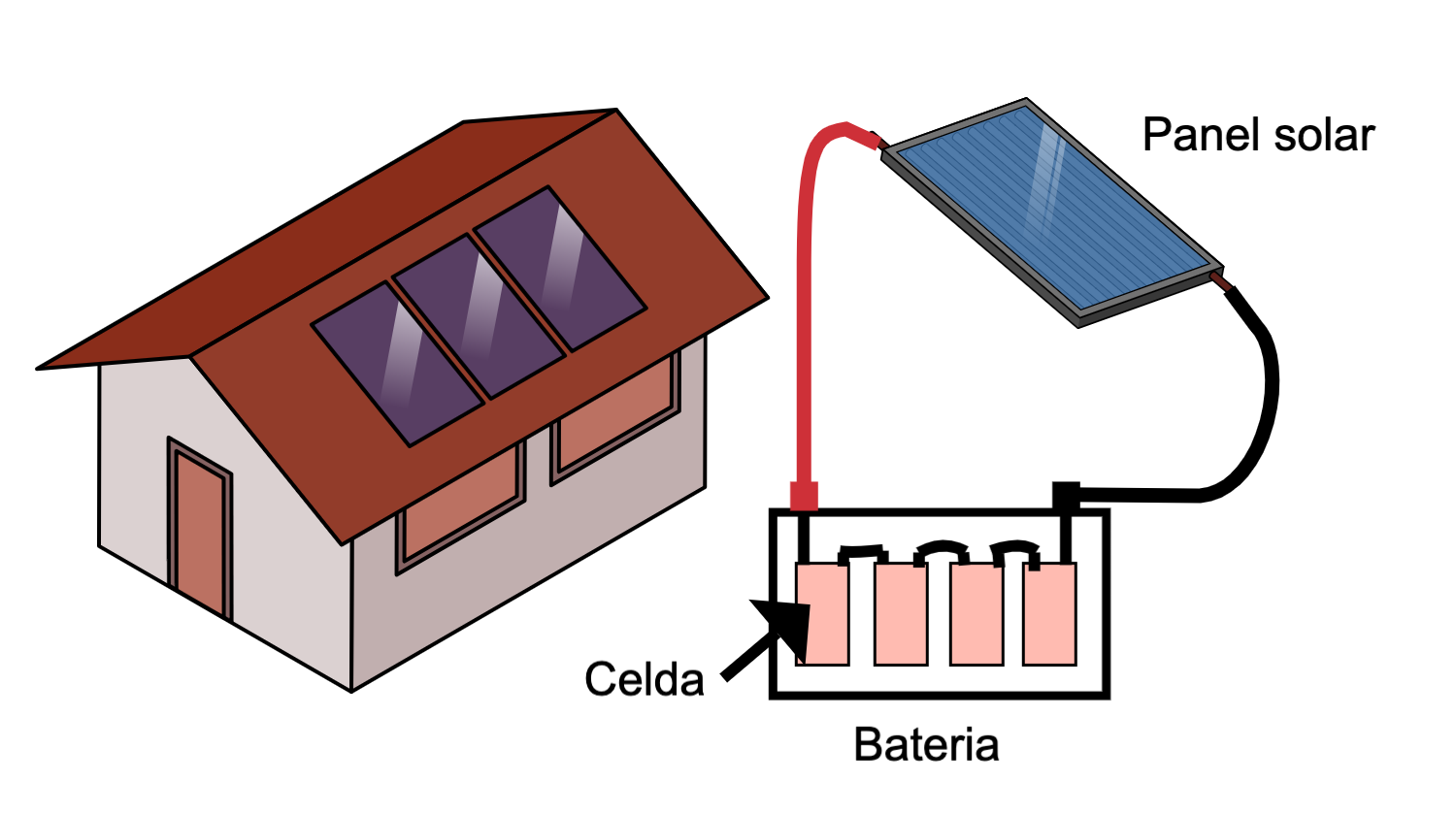}
    \caption{Representación simplificada de un sistema de energía solar (se omite el inversor) donde un panel solar transforma energía proveniende del sol en energía eléctrica que se almacena en una batería electroquímica formada por varias celdas (cuatro en la figura).}
    \label{fig:ch1_sistema_panel_solar}
\end{figure}

En una finca se utiliza energía solar para proveer la energía requerida. La energía se obtiene por paneles solares que colectan la energía del sol y la transforman en energía eléctrica que se almacena en una batería electroquímica con $n$ celdas, ver Figura~\ref{fig:ch1_sistema_panel_solar}. Cuando una celda funciona correctamente, esta contribuye con un voltaje $v_0$, pero si no funciona, su aporte es cero. De tal manera que si se tienen funcionando correctamente $k$ celdas, el voltaje de la batería está dado por $V=kv_0$. Si se tiene la probabilidad $p$ que una celda funcione correctamente y $q=1-p$, que no funcione, ¿cuál es el valor medio de la potencia consumida las luces de la finca, las cuales se pueden modelar con una resistencia $R$?

Como las celdas que conforman la batería cuando funcionan suministran una diferencia de potencial $v$, pero cuando no lo hacen su aporte es cero, entonces, se tienen dos posibles resultados al realizar la observación de la diferencia de potencial $v_i$ de la i-ésima celda. Esto quiere decir que el voltaje de la celda $i$ se puede modelar como un ensayo de Bernoulli con probabilidad $p$ de observar $v=v_0$ y $q=1-p$, de observar $v=0$. Por lo tanto, si hay $k$ celdas operativas, el voltaje total de la batería $V$ está dado por,

$$V = \sum_{i=1}^n v_i = kv_0$$

Por lo tanto, $V$ puede tomar valores $v\in\{0, v_0, \cdots, kv_0, \cdots, nv_0\}$, donde cada posible valor corresponde con un ensayo de Bernoulli con probabilidad $p$; esto quiere decir que $V$ es una variable aleatoria binomial.

La potencia consumida por una resistencia $R$ está dada por $P=V^2/R$ y su valor esperado se obtiene utilizando la ecuación \eqref{eq:valor_esperado_discreto}, donde $p_n$ está dado por la probabilidad para la distribución binomial, ecuación \eqref{eq:ch1_probabilidad_binomial},

$$\E{P} = \sum_{k=0}^n P_k W_n[k] = \frac{1}{R} \sum_{k=0}^n V_k^2 W_n[k] $$

Como $v_i$ solo puede tomar dos valores (cero o $v_0$), entonces, $V_k = k v$ y el valor esperado para la potencia consumida está dado por, 

$$\E{P} = \frac{v_0^2}{R} \sum_{k=0}^n k^2 W_n[k] = \frac{v_0^2}{R} \sum_{k=0}^n \frac{n!}{k!(n-k)!}p^kq^{n-k} = \frac{v_0^2}{R} \E{k^2}$$

Con el fin de calcular este valor esperado, es conveniente utilizar la siguiente identidad,

\begin{equation}
    \label{eq:ch1_identidad_dist_binomial}
    p\frac{\partial}{\partial p} \sum_{k=0}^n p^kq^{n-k} = \sum_{k=0}^n k p^kq^{n-k}
\end{equation}

Nótese que esta identidad permite expresar $\E{k^2}$ en términos de derivadas parciales de la probabilidad binomial $W_n[k]$ respecto a $p$,

$$\E{P} = \frac{v^2}{R} p\frac{\partial}{\partial p} \left( p\frac{\partial}{\partial p} \sum_{k=0}^n \frac{n!}{k!(n-k)!}p^kq^{n-k} \right) = \frac{v^2}{R} p\frac{\partial}{\partial p} \left( p\frac{\partial}{\partial p} \sum_{k=0}^n W_n[k] \right)$$

Por el teorema del binomio se sabe que,

$$\sum_k \frac{n!}{k!(n-k)!}p^kq^{n-k} = (p + q)^n,$$

Al reemplazar en la expresión para $\E{P}$, realizar las derivadas parciales y teniendo en cuenta que $p+q=1$, se encuentra la respuesta a la pregunta,

$$\E{P} = \frac{v_0^2}{R} pn (1- p + pn)$$

\section{Distribución uniforme}
\index{Distribución uniforme}

La distribución uniforme es muy útil en el estudio de sistemas físicos, de hecho, en el siguiente capítulo se utilizará para describir la probabilidad de cada estado de un sistema aislado cuya energía se encuentra en un rango entre $E$ y $E+dE$, es decir, supondrá que todos los posibles estados del sistema (en dicho rango de energía) tienen la misma probabilidad. 

\begin{definition}
    Sea una variable aleatoria uniforme discreta $X$, donde los posibles resultados de observar dicha variable están dados por $x_k$ para $k=1, 2, ..., N$. La probabilidad que al observar la variable aleatoria se obtenga $x_k$ está dada por,\index{Variable aleatoria!uniforme}
    
    $$\prob{X=x_k} = \frac{1}{N}$$
\end{definition}

\begin{definition}
    Sea una variable aleatoria uniforme continua $X$, donde los posibles resultados de observar dicha variable están dados por $a \leq x \leq b$, para $b>a$. La probabilidad que al observar la variable aleatoria se obtenga un valor entre $x$ y $x+dx$ está dada por,
    
    $$\prob{x \leq X < x + dx} = f_X(x)dx = \frac{1}{b-a}dx$$
\end{definition}

\newpage
\subsection*{Ejemplo ilustrativo: máxima probabilidad al lanzar dos dados}

Cuando se lanzan dos dados de seis caras, como suele ocurrir en juegos como el parqués, se tiene la intuición que al sumar los valores numéricos en las caras superiores, los resultados menos probables son el dos (doble uno) y el 12 (doble seis), mientras que el resultado más probable es el número siete. ¿Cómo se llega a la conclusión que la máxima probabilidad al sumar los valores numéricos de estas caras es precisamente el número siete?

Sean $X_1$ y $X_2$ dos variables aleatorias discretas uniforme que pueden tomar valores $x = \{1, 2, 3, \cdots, 6\}$. Cada variable aleatoria modela uno de los dados, por lo tanto, el valor esperado de la suma del resultado de lanzar los dados corresponde con $\E{X_1 + X_2} = \E{X_1} + \E{X_2} = 2 \E{X}$, donde,

$$\E{X} = \sum_{k=1}^6 k q = \frac{1}{6}\sum_{k=1}^6 k = \frac{21}{6} = 3.5$$

Por lo tanto, $\E{X_1 + X_2} = 2 \E{X} = 7$.

Con el objetivo de responder las otras dos partes de la pregunta, es decir, la probabilidad de obtener dos (doble uno) y 12 (doble seis), implica un análisis más detallado del problema, para lo cual, es conveniente definir el número de estados accesibles $\Omega$ y con este calcular las diferentes probabilidades.

\begin{definition}
    Un estado en que se puede encontrar un sistema se conoce como \textbf{estado accesible}.\index{Estado accesible}
\end{definition}

\begin{definition}
    El \textbf{número de estados accesibles} \gls{omega} se define como la cantidad de estados en que se puede observar a un cierto sistema.\index{Número de estados accesibles}
\end{definition}

De esta manera, cuando se lanzan dos dados de seis caras, donde cada cara tiene la misma probabilidad $\prob{X=x_k} = 1/6$, se obtienen 36 posibles resultados diferentes, como se muestra en la Tabla \ref{tab:ch1_dist_uniforme_estados_accesibles}. Al analizar el número de resultados que comparten el mismo valor en la columna de la suma, es posible determinar el número total de estados accesibles $\Omega$, ver Tabla \ref{tab:ch1_dist_uniforme_Omega}.

\begin{table}[ht]
\centering
\caption{Estados accesibles obtenidos al lanzar dos dados de seis caras, cada cara con la misma probabilidad, donde la columna suma corresponde con la adición de los valores en las caras superiores de cada uno de los dos dados (dado 1 y dado 2).}
\label{tab:ch1_dist_uniforme_estados_accesibles}
\begin{tabular}{|ccc|c|ccc|}
\cline{1-3} \cline{5-7}
Dado 1 & Dado 2 & Suma & & Dado 1 & Dado 2 & Suma \\
\cline{1-3} \cline{5-7} 
1 & 1 & 2 & & 4 & 1 & 5 \\
1 & 2 & 3 & & 4 & 2 & 6 \\
1 & 3 & 4 & & 4 & 3 & 7 \\
1 & 4 & 5 & & 4 & 4 & 8 \\
1 & 5 & 6 & & 4 & 5 & 9 \\
1 & 6 & 7 & & 4 & 6 & 10 \\
2 & 1 & 3 & & 5 & 1 & 6 \\
2 & 2 & 4 & & 5 & 2 & 7 \\
2 & 3 & 5 & & 5 & 3 & 8 \\
2 & 4 & 6 & & 5 & 4 & 9 \\
2 & 5 & 7 & & 5 & 5 & 10 \\
2 & 6 & 8 & & 5 & 6 & 11 \\
3 & 1 & 4 & & 6 & 1 & 7 \\
3 & 2 & 5 & & 6 & 2 & 8 \\
3 & 3 & 6 & & 6 & 3 & 9 \\
3 & 4 & 7 & & 6 & 4 & 10 \\
3 & 5 & 8 & & 6 & 5 & 11 \\
3 & 6 & 9 & & 6 & 6 & 12 \\
\cline{1-3} \cline{5-7}
\end{tabular}
\end{table}

\begin{table}[!ht]
\centering
\caption{Número de estados accesibles para cada valor de suma del resultado obtenido al lanzar dos dados de seis caras con igual probabilidad para cada una de ellas.}
\label{tab:ch1_dist_uniforme_Omega}
\begin{tabular}{|cc|c|cc|c|cc|}
\cline{1-2} \cline{4-5} \cline{7-8}
Suma & $\Omega$ & & Suma & $\Omega$ & & Suma & $\Omega$ \\
\cline{1-2} \cline{4-5} \cline{7-8}
1 & 0 & & 5 & 4 & & 9 & 4 \\
2 & 1 & & 6 & 5 & & 10 & 3 \\
3 & 2 & & 7 & 6 & & 11 & 2 \\
4 & 3 & & 8 & 5 & & 12 & 1 \\
\cline{1-2} \cline{4-5} \cline{7-8}
\end{tabular}
\end{table}

\begin{equation}
    \label{eq:ch1_dist_uniforme_Omega}
    p_n  = \frac{\Omega_n}{\sum_n{\Omega_n}}
\end{equation}

\begin{figure}[th]
    \begin{subfigure}{.5\textwidth}
    \centering
    \includegraphics[width=\linewidth]{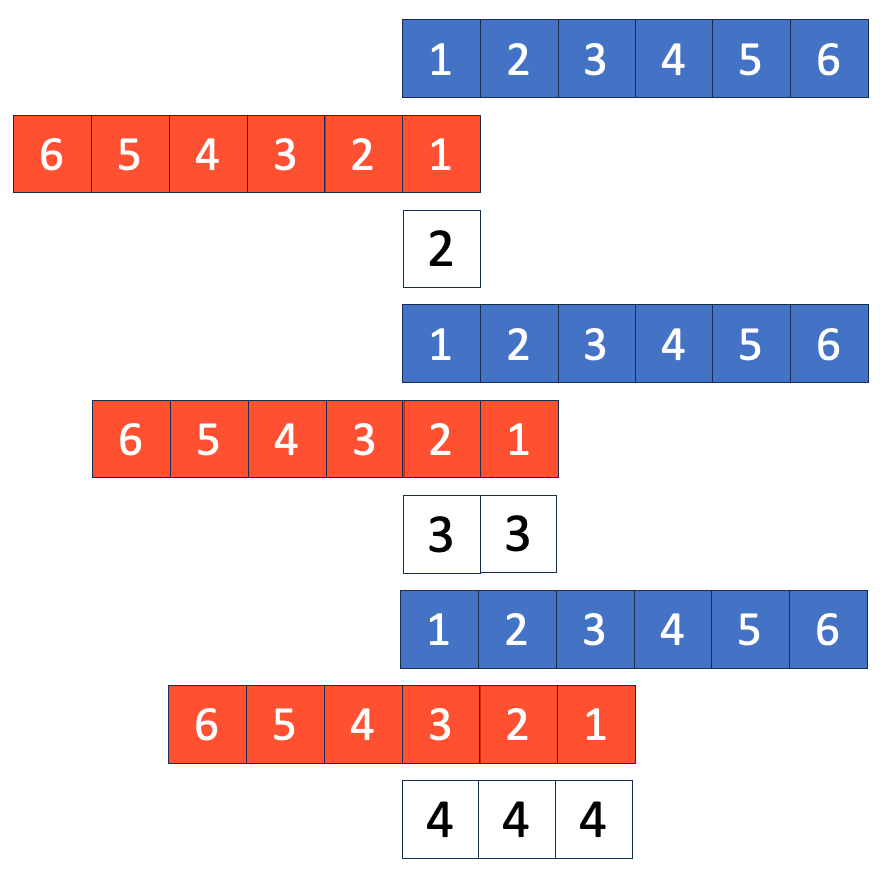}
    \caption{}
    \label{fig:ch1_dist_uniforme_conv}
    \end{subfigure}
    \begin{subfigure}{.5\textwidth}
    \centering
    \includegraphics[width=\linewidth]{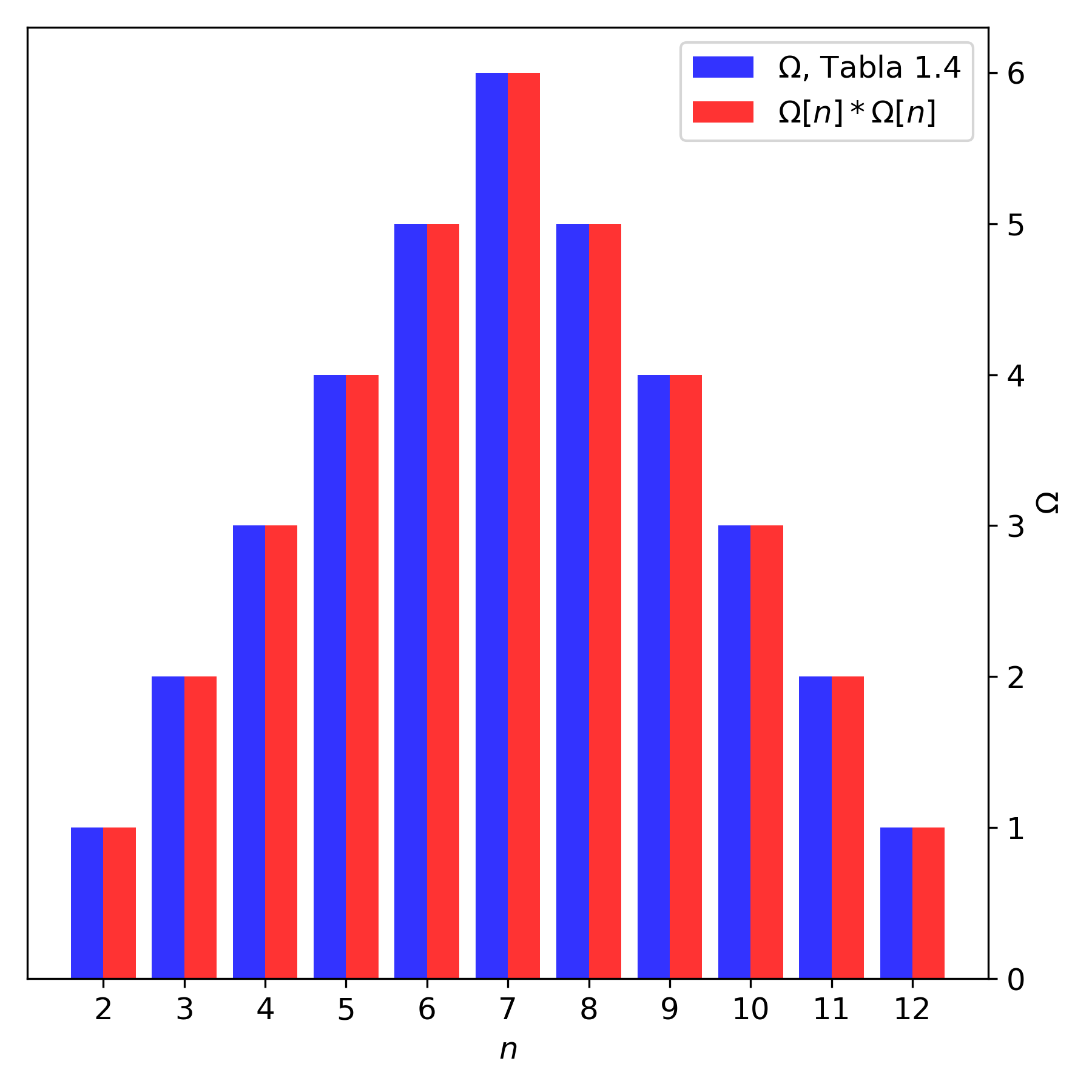}
    \caption{}
    \label{fig:ch1_dist_uniforme_Omega}
    \end{subfigure}
    \caption{(a) Representación de posibilidades al sumar caras de dos dados (uno azul y otro rojo) y (b) $\Omega$ obtenido en la Tabla~\ref{tab:ch1_dist_uniforme_Omega} (representado por las barras azules), que corresponde con la convolución del número de estados $\Omega[n]$ para un dado (uno para cada posible cara) con el número de estados $\Omega[n]$ del otro dado (representados por las barras rojas).}
\end{figure}

Con los valores de $\Omega$ de la Tabla \ref{tab:ch1_dist_uniforme_Omega} se puede calcular la probabilidad de ocurrencia de cada uno de los posibles resultados de la columna suma, para ello conviene enumerar cada valor de suma con un número entero $n$, entonces, el número de estados accesibles para el valor $n$ estará dado por $\Omega_n$. De esta manera, la probabilidad $p_n$ de obtener $n$ al sumar los valores de las caras superiores de dos dados lanzados de tal manera que el valor en cada dado es estadísticamente independiente del obtenido en el otro dado, se puede obtener como se muestra en la ecuación \eqref{eq:ch1_dist_uniforme_Omega}. En la Figura~\ref{fig:ch1_dist_uniforme_Omega} se pueden observar las probabilidades calculadas con la ecuación \eqref{eq:ch1_dist_uniforme_Omega}, donde la probabilidad mayor corresponde para el valor $n=7$ y las menores para $n=2$ y $n=12$.

El proceso descrito arriba se puede plantear de otra manera si se representan los valores en las caras del dado en arreglos lineales que se van solapando de un valor a la vez, como se muestra esquemáticamente en la Figura \ref{fig:ch1_dist_uniforme_conv}. Cada cuadro relleno de color (azul y rojo) en la Figura \ref{fig:ch1_dist_uniforme_conv} corresponde con el número en una de las caras del dado, por lo tanto, el proceso que se muestra en la figura evidencia los casos en los cuales se encuentra cada posible valor de la suma de las caras (ver cuadro negro). Es de notar que cada uno de los valores posibles tiene la misma probabilidad, por lo tanto, como las variables aleatorias son estadísticamente independientes, la probabilidad conjunta será el producto de las dos probabilidades, en este caso $\frac{1}{6}\frac{1}{6}=\frac{1}{36}$; como se tienen el número de posibilidades señaladas en el cuadro sin color, entonces, al sumar dichas probabilidades conjuntas para cada caso se obtiene la probabilidad de un determinado valor, por ejemplo, la probabilidad así calculada para el número tres será $2\frac{1}{6^2}=\frac{1}{18}$, que concuerda con el valor obtenido en el cálculo anterior. 

A manera de síntesis de lo analizado arriba, la probabilidad de obtener un cierto valor $n$ al sumar los números en las caras de los dados está dada por la Ecuación \ref{eq:ch1_dist_uniforme_Omega}, si $\Omega[n]$ representa el número de estados para las caras del dado, es decir, uno para valores entre uno y seis, cero para los demás valores, entonces, como el Teorema~\ref{lema:probabilidad_suma_variables_aleatorias} establece que la probabilidad de sumar el valor de las dos caras está dada por la ecuación \eqref{eq:ch1_convolucion_discreta}, el número de estados accesibles del sistema combinado será $\Omega[n] * \Omega[n]$, que está en concordancia con los valores obtenidos en la Tabla~\ref{tab:ch1_dist_uniforme_Omega}, como se muestra en la Figura~\ref{fig:ch1_dist_uniforme_Omega}. 

\begin{definition}
    Sean $x_1[n]$ y $x_2[n]$ dos variables discretas, la convolución de las dos variables se define como,
    \begin{equation}
        \label{eq:ch1_convolucion_discreta}
        x_1[n] * x_2[n] = \sum_{k} x_1[k] x_2[n-k] = \sum_k x_1[n-k] x_2[k]
    \end{equation}
    \index{Convolución de variables discretas}
\end{definition}

\begin{definition}
    Sean $f(x)$ y $g(x)$ dos variables continuas, la \textbf{convolución} de las dos variables aleatorias se define como,
\begin{equation}
        \label{eq:ch1_convolucion_continua}
        f(x) * g(x) = \int_{-\infty}^\infty f(y) g(x-y) dy = \int_{-\infty}^\infty f(x-y) g(y) dy
    \end{equation}
    \index{Convolución de variables continuas}
\end{definition}

\begin{lema}
    \label{lema:probabilidad_suma_variables_aleatorias}
    Sean dos variables aleatorias $X$, $Y$, estadísticamente independientes. La probabilidad de la variable aleatoria $Z = X+Y$ está dada por la convolución de las probabilidades de las dos variables aleatorias.
\end{lema}

\section{Distribución de Poisson}
\index{Distribución de Poisson}

Una manera de obtener la distribución de Poisson es considerar el caso límite de la distribución binomial para $n\to\infty$, haciendo $p=\lambda/n$, donde $\lambda$ es el parámetro de Poisson que corresponde a la cantidad media de ocurrencia de eventos. La probabilidad asociada a la distribución de Poisson está dada por,

\begin{equation}
\label{eq:ch1_lim_dist_binomial_poisson}
    \prob{X=k} = \lim_{n\to\infty} W_n[k] = \lim_{n\to\infty}\frac{n!}{k! (n-k)!} \left( \frac{\lambda}{n} \right)^k \left( 1 - \frac{\lambda}{n} \right)^{n-k}
\end{equation}

Es importante tener en cuenta que para $k$ finito se cumple, 

\begin{equation}
\label{eq:ch1_exponencial_lim}
    \lim_{n\to\infty} \left( 1 - \frac{\lambda}{n} \right)^{n-k} = \lim_{n\to\infty} \left( 1 - \frac{\lambda}{n} \right)^n = e^{-\lambda}    
\end{equation}

Por otra parte, se puede analizar el comportamiento asintótico de la parte izquierda, para lo cual es conveniente utilizar logaritmo natural del límite,

$$\ln\left[ \lim_{n\to\infty}\frac{n!}{k! (n-k)!} \right]  = \lim_{n\to\infty} \ln\left[ \frac{n!}{k! (n-k)!} \right] $$

Para $n\to\infty$ se puede utilizar la \textit{aproximación de Stirling}\index{Aproximación de Stirling}\footnote{La \textit{aproximación de Stirling}, $\ln n! \approx n\ln n - n$, para valores muy grandes de $n$ permite encontrar el valor asintótico deseado, en tal caso, la aproximación (descrita con el símbolo $\approx$) tiende a ser una igualdad.}, $\ln n! \approx n\ln n - n$, con lo cual,

$$\ln\left[ \lim_{n\to\infty}\frac{n!}{k! (n-k)!} \right]  \approx \lim_{n\to\infty} \left[ n \ln{n} - \ln (k!) - (n-k) \ln (n-k) - n + k \right]$$

Al simplificar términos se obtiene,

\begin{equation}
\label{eq:ch1_combinatoria_lim}
    \ln\left[ \lim_{n\to\infty}\frac{n!}{k! (n-k)!} \right]  \approx \ln\left[ \lim_{n\to\infty}\frac{n^k}{k!} \right]
\end{equation}

Al tener en cuenta los comportamientos asintóticos (para $n\to \infty$) dados por \eqref{eq:ch1_exponencial_lim} y \eqref{eq:ch1_combinatoria_lim}, de la ecuación \eqref{eq:ch1_lim_dist_binomial_poisson} se obtiene la probabilidad para la distribución de Poisson,

$$\prob{X=k} = \lim_{n\to\infty} W_n[k] = \lim_{n\to\infty}\frac{n^k}{k!} \frac{\lambda^k}{n^k} e^{-\lambda} = \frac{\lambda^k}{k!} e^{-\lambda} $$

Esta distribución fue propuesta por primera vez en 1838 por Siméon-Denis Poisson quien la utilizó en una investigación sobre la probabilidad de los juicios en materias criminales y civiles. \index{Variable aleatoria!Poisson}

\begin{definition}
La probabilidad $W_\lambda[k]$ de observar $k$ realizaciones de un evento cuyo valor medio de ocurrencia es $\lambda$, está dada por,
\begin{equation}
\label{eq:ch1_dist_Poisson}
    W_\lambda[k] = \frac{\lambda^k}{k!} e^{-\lambda}
\end{equation}

\end{definition}

\subsection*{Ejemplo ilustrativo: emisión de partículas $\alpha$ por isótopos radiactivos}

\begin{figure}[th]
    \centering
    \includegraphics[width=\textwidth]{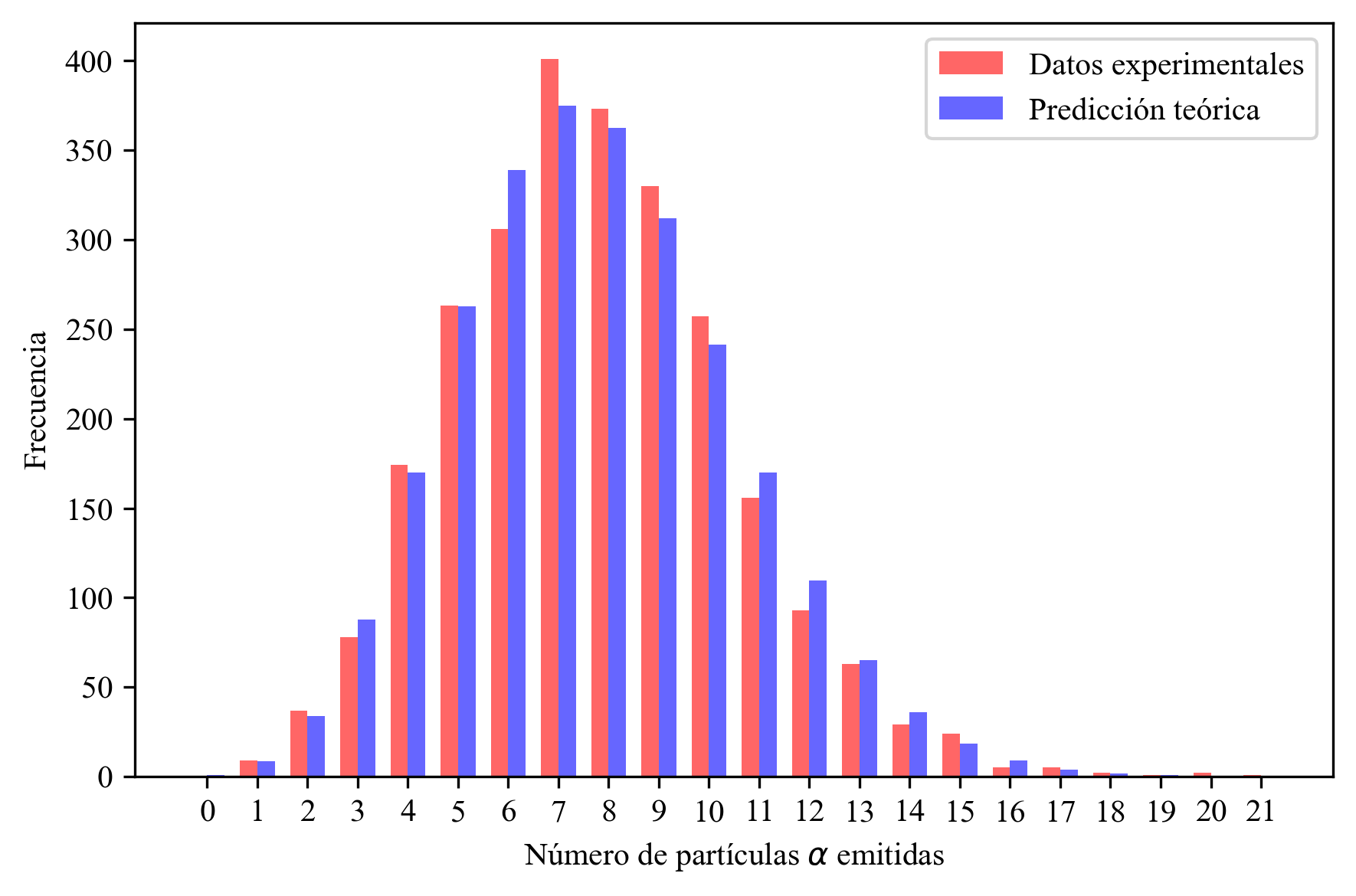}
    \caption{Comparación entre la frecuencia experimental y teórica para la emisión de partículas $\alpha$, según datos reportados por Rutherford, Geiger y Bateman \cite{Rutherford1910}.}
    \label{fig:ch1_ch1_rutherford-geiger}
\end{figure}

En 1910 Rutherford, Geiger y Bateman mostraron que la emisión de partículas $\alpha$ por isótopos radioactivos, con baja emisión, tienen un valor medio aproximadamente constante y dicha emisión puede ser modelada por la distribución de Poisson \cite{Rutherford1910}. Es importante notar que no toda la emisión radioactiva, sobre todo si es de alta intensidad, sigue la distribución de Poisson \cite{Frigerio1974}. Supóngase que, siguiendo el trabajo de Rutherford y Geiger, el valor medio de partículas $\alpha$ está dado por $\lambda=7.4$ partículas por intervalos de $1/8$ de minuto. En la Tabla \ref{tab:ch1_rutherford-geiger} se muestran los datos reportados en \cite{Rutherford1910} y los valores calculados a partir de reemplazar $\lambda=7.4$ en la ecuación \eqref{eq:ch1_dist_Poisson}, multiplicados por el número total de partículas contadas, es decir, la frecuencia de partículas $\alpha$. Esta información se ve mejor en la Figura \ref{fig:ch1_ch1_rutherford-geiger}, donde se puede apreciar que la predicción de la frecuencia de emisión con una distribución de Poisson concuerda con los datos experimentales.

\begin{table}[ht]
\centering
\caption{Comparación entre la frecuencias experimental y teórica para la emisión de partículas $\alpha$, según datos reportados por Rutherford, Geiger y Bateman \cite{Rutherford1910}.}
\label{tab:ch1_rutherford-geiger}
\begin{tabular}{|c|c|c|}
\hline 
Número de partículas & Frecuencia experimental & Frecuencia teórica \\
\hline \hline
0 & 0 & 1 \\
1 & 9 & 9 \\
2 & 37 & 34 \\
3 & 78 & 88 \\
4 & 174 & 170 \\
5 & 263 & 263 \\
6 & 306 & 339 \\
7 & 401 & 375 \\
8 & 373 & 363 \\
9 & 330 & 312 \\
10 & 257 & 241 \\
11 & 156 & 170 \\
12 & 93 & 110 \\
13 & 63 & 65 \\
14 & 29 & 36 \\
15 & 24 & 19 \\
16 & 5 & 9 \\
17 & 5 & 4 \\
18 & 2 & 2 \\
19 & 1 & 1 \\
20 & 2 & 0 \\
21 & 1 & 0 \\
\hline
\hline
Total & 2609 & 2609 \\
\hline

\end{tabular}
\end{table}

\section{Distribución normal}
\index{Distribución normal}

La distribución normal, también conocida como gaussiana, se observa en una gran variedad de fenómenos, lo cual hace de esta una distribución de gran importancia. Así como en el estudio de la distribución de Poisson se partió de la distribución binomial, donde se realizó la aproximación $p\to0$ cuando $n\to\infty$, en un contexto similar, De Moivre y Laplace demostraron que la distribución binomial para valores de $n$ grandes cuando $k$ es del orden de $np$, se puede aproximar por una distribución normal; a este resultado se le conoce como teorema de Moivre-Laplace, ver ecuación \eqref{eq:ch1_Moivre-Laplace_theorem}. 

\begin{lema}
    \textbf{Teorema de Moivre-Laplace.}
    Sea $X$ una variable aleatoria binomial caracterizada por la probabilidad $W_n[k]$. Cuando $n$ crece, para $k$ en la vecindad de $np$ y $pq>0$, \index{Teorema de Moivre-Laplace}
    \begin{equation}
    \label{eq:ch1_Moivre-Laplace_theorem}
        W_n[k] = \frac{n!}{k!(n-k)!}p^kq^{n-k} \approx \frac{1}{\sqrt{2\pi npq}} e^{-\frac{(k-np)^2}{2npq}}    
    \end{equation}
\end{lema}

\begin{definition}
    La probabilidad $f_X(x)dx$ de observar $x \leq X < x+dx$ para una variable aleatoria normal $X$ con valor esperado $\mu$ y desviación estándar $\sigma$ está dada por el producto entre el intervalo $dx$ y la densidad de probabilidad $f_X(x)$, ecuación \eqref{eq:ch1_dist_normal}.\index{Distribución normal}\index{Variable aleatoria!normal o gaussiana}
    
    \begin{equation}
    \label{eq:ch1_dist_normal}
        f_X(x)dx = \frac{1}{\sigma\sqrt{2\pi}} e^{\frac{-(x-\mu)^2}{2\sigma^2}}dx
    \end{equation}
\end{definition}

Esto no quiere decir que la distribución normal sea un caso particular de la binomial, lo que indica es que la distribución binomial, para valores de $n$ grande y $k$ en la vecindad\footnote{En este contexto vecindad significa que $k$ toma valores alrededor de $np$.} de $np$, se puede aproximar con una distribución normal. De manera formal, una distribución normal está caraceterizada por la probabilidad definida por la ecuación \eqref{eq:ch1_dist_normal} y es independiente de la variable aleatoria binomial.

El teorema del límite central, del cual el teorema de Moivre-Laplace se considera un caso particular, nos muestra cómo a partir de una variable aleatoria $X$, con cualquier distribución de probabilidad, es posible construir una variable $Z_n$ cuya distribución tiende a la normal normal, ver ecuación \eqref{eq:ch_teorema_limite_central_Z}.

Para complementar el estudio del teorema del límite central, se sugiere responder las preguntas de autoexplicación del ejemplo trabajado titulado: \href{https://colab.research.google.com/github/davidalejandromiranda/StatisticalPhysics/blob/main/notebooks/es_SumaVariablesAleatorias.ipynb}{variables Aleatorias y el Teorema del Límite Central.}

\begin{lema}
    \textbf{Teorema del límite central.}
    Sean $X_1, X_2, X_3, \cdots, X_n$ variables aleatorias estadísticamente independientes e idénticamente distribuidas con $\E{X_i}=\mu$ y $\E{(X_i-\mu)^2} = \sigma^2 < \infty$. La variable aleatoria $Z_n$, dada por la ecuación \eqref{eq:ch_teorema_limite_central_Z}, tiende a la distribución normal cuando $n$ tiende a infinito. \index{Teorema del límite central}
    \begin{equation}
        \label{eq:ch_teorema_limite_central_Z}
        Z_n = \frac{1}{\sigma \sqrt{n}} \sum_{i=1}^n {X_i - n\mu}
    \end{equation}
    \begin{equation}
        \label{eq:ch_teorema_limite_central}
        \lim_{n\to\infty} \prob{Z_n < z} = \int_{-\infty}^z \frac{1}{\sqrt{2\pi}} e^{-x^2/2} dx
    \end{equation}
\end{lema}

\begin{lema}
    Sean $X_1, X_2, X_3, \cdots, X_n$ variables aleatorias estadísticamente independientes e idénticamente distribuidas con $\E{X_i}=\mu$ y $\E{(X_i-\mu)^2} = \sigma^2 < \infty$. La variable aleatoria $S_n = X_1 + X_2 + \cdots + X_n$ tiene media $\E{S_n} = n\mu$ y varianza $\E{(S_n - \E{S_n})^2} = n\sigma$.
\end{lema}

\subsection*{Ejemplo ilustrativo: movimiento aleatorio en dos dimensiones}

\begin{figure}[th]
    \centering
    
    \begin{subfigure}{.5\textwidth}
        \centering
        \includegraphics[width=.8\linewidth]{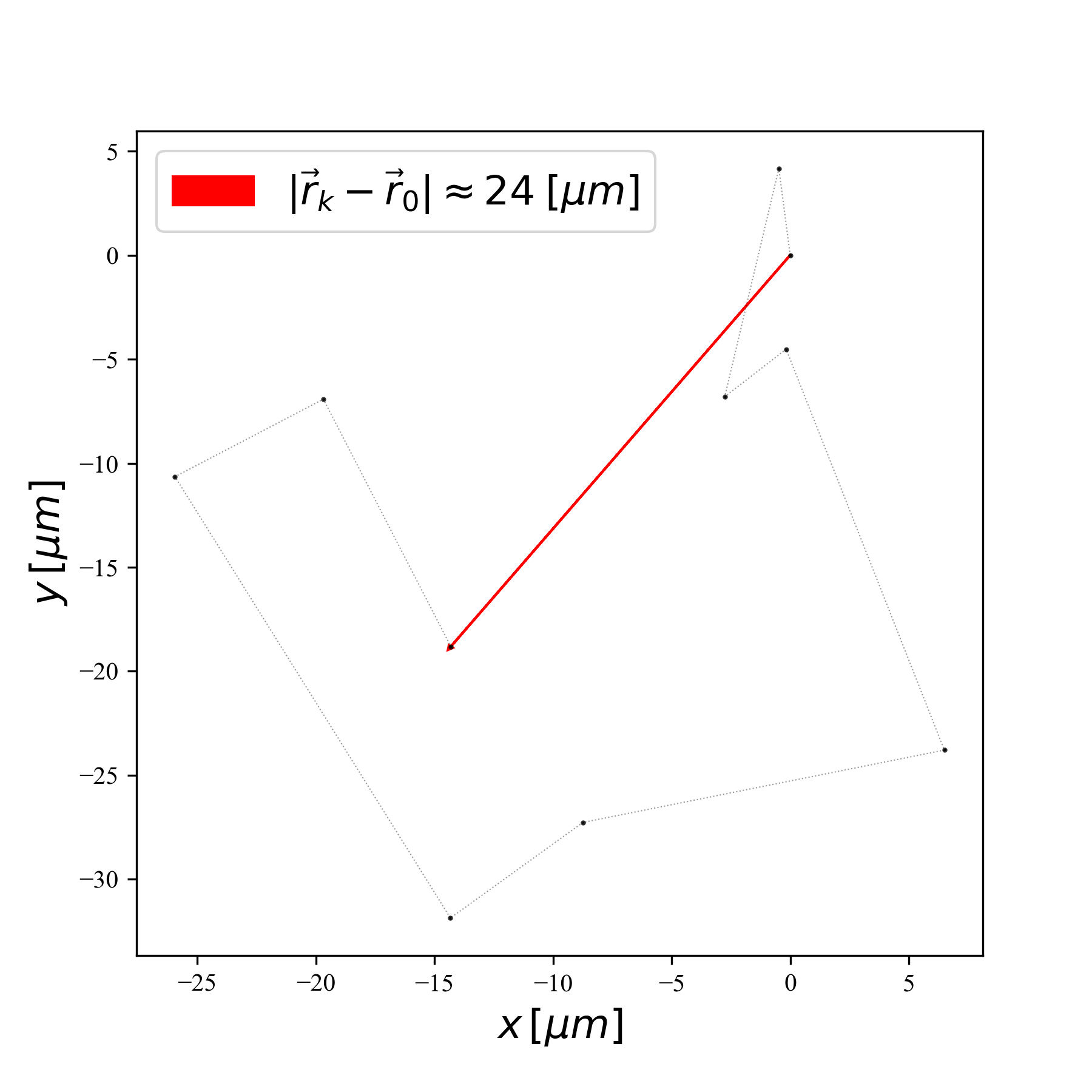}
        \caption{}
        \label{fig:ch1_ejemplo_dist_normal_10pasos}
    \end{subfigure}%
    \begin{subfigure}{.5\textwidth}
        \centering
        \includegraphics[width=.8\linewidth]{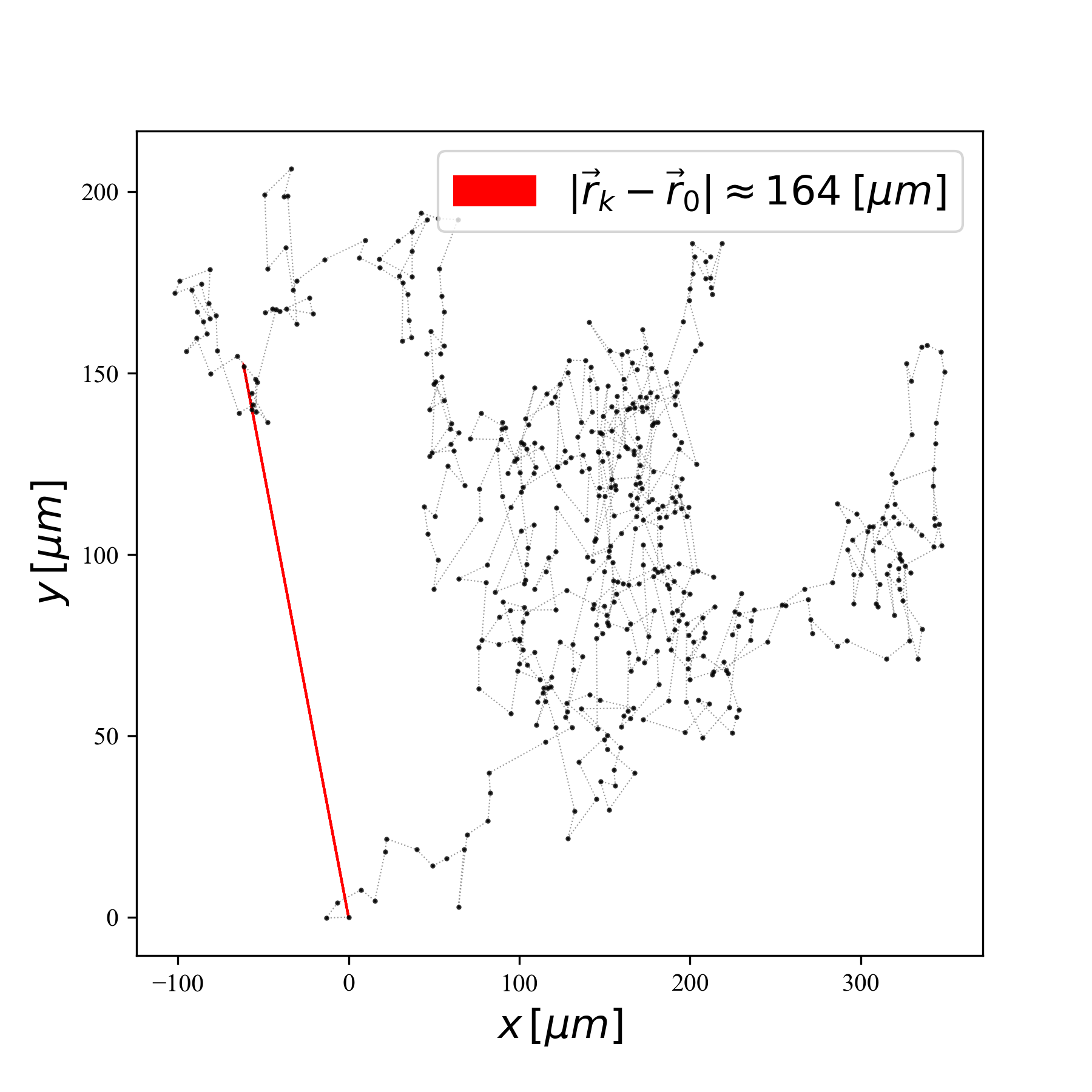}
        \caption{}
        \label{fig:ch1_ejemplo_dist_normal_500pasos}
    \end{subfigure}
    \caption{Trayectoria $\vec{r}=X\hat{i} + Y\hat{j}$ de una partícula que describe un movimiento aleatorio, donde cada una de sus componentes ($X$, $Y$) son variables aleatorias normales. Se puede observar cuando una partícula da (a) diez pasos, (b) 500 pasos.}
\end{figure}

\begin{figure}[th]
    \centering
    \begin{subfigure}{.5\textwidth}
        \centering
        \includegraphics[width=.8\linewidth]{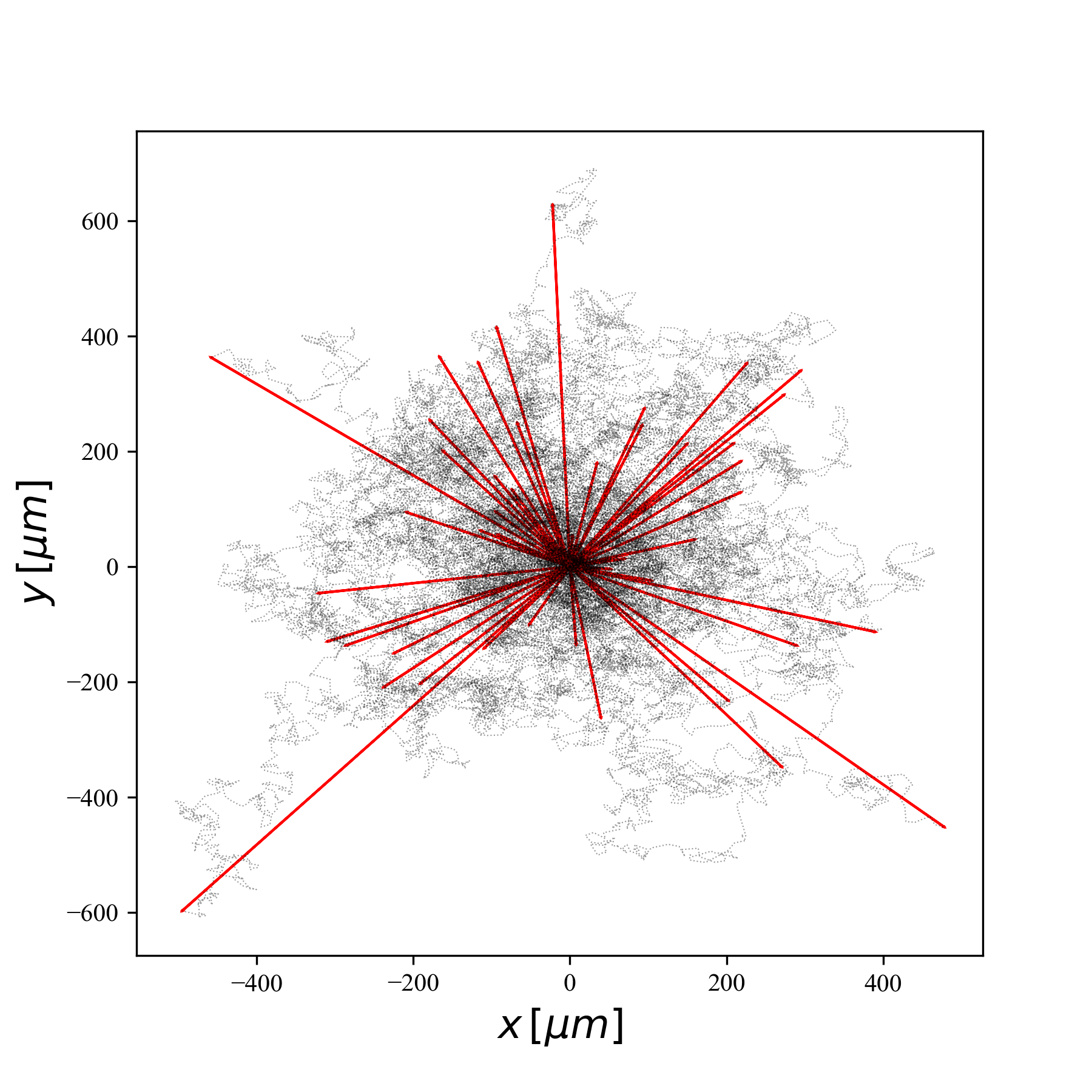}
        \caption{}
        \label{fig:ch1_ejemplo_dist_normal_ensamble}
    \end{subfigure}%
    \begin{subfigure}{.5\textwidth}
        \centering
        \includegraphics[width=.8\linewidth]{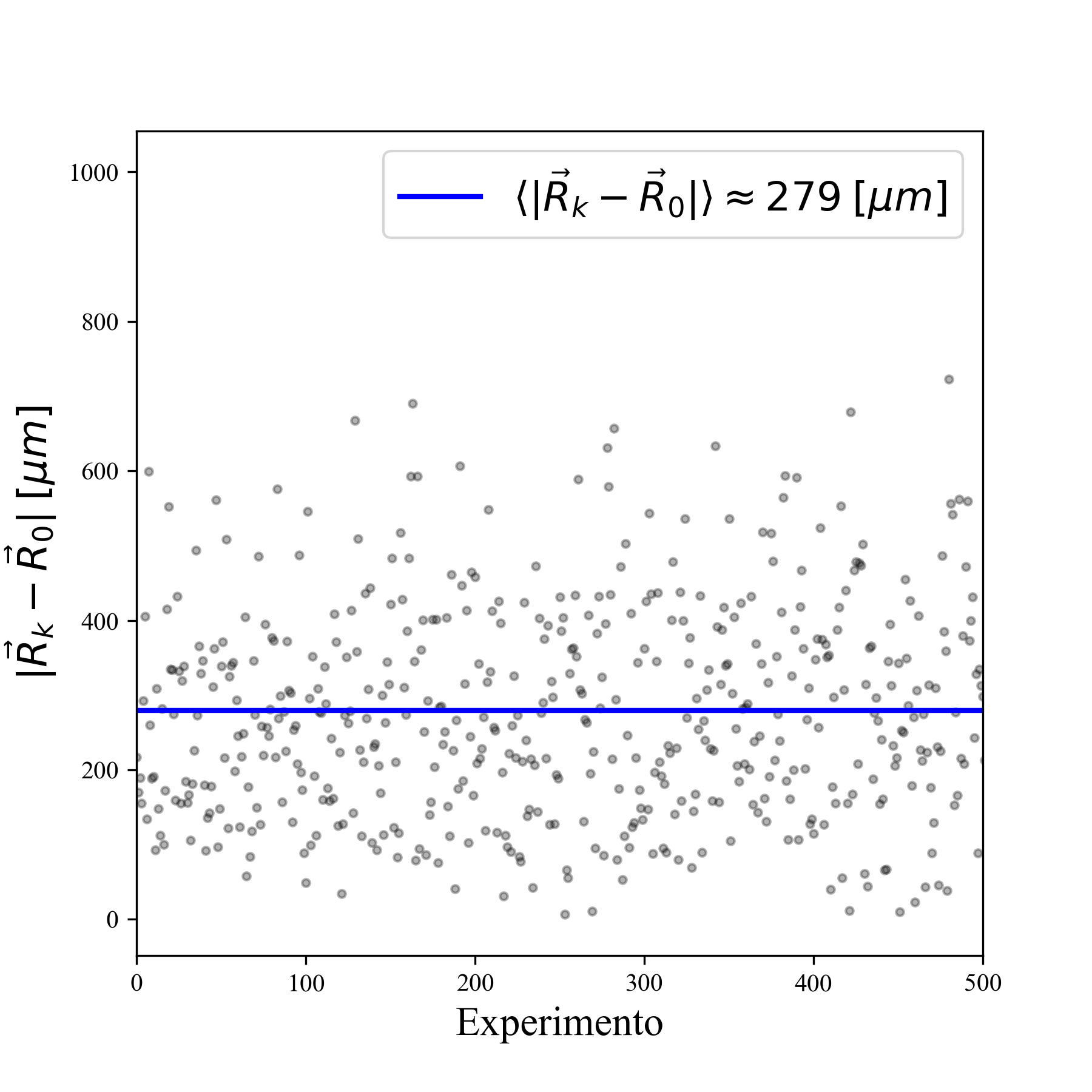}
        \caption{}
        \label{fig:ch1_ejemplo_dist_normal_valor_medio}
    \end{subfigure}
    
    \caption{Trayectoria $\vec{r}=X\hat{i} + Y\hat{j}$ de una partícula que describe un movimiento aleatorio, donde cada una de sus componentes ($X$, $Y$) son variables aleatorias normales. Se pueden observar (a) las trayectorias de 50 experimentos, donde por cada experimento se tiene la trayectoria para 500 pasos de una partícula. También se  observa (b) el valor medio de la distancia recorrida obtenido de $10^4$ experimentos.}
\end{figure}

\begin{figure}
    \centering
    \includegraphics[width=\linewidth]{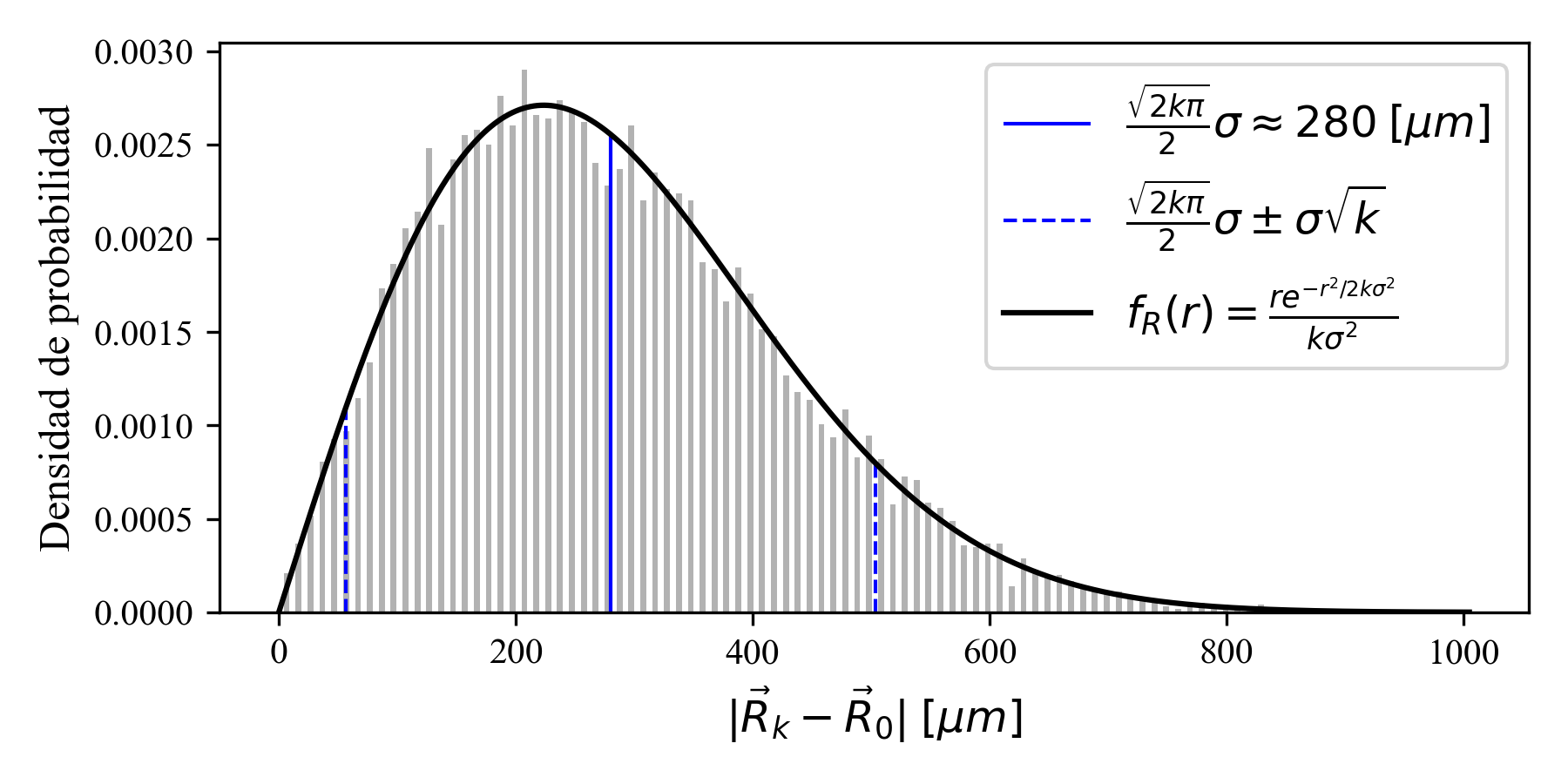}
    \caption{Densidad de probabilidad obtenida a partir de $10^4$ experimentos de camino aleatorio en dos dimensiones (representada por las barras grises). La línea sólida corresponde con la predicción teórica.}
    \label{fig:ch1_ejemplo_dist_normal_valor_densidad}
\end{figure}

Suponga que se tiene una partícula clásica que se mueve en dos dimensiones de tal manera que cada paso está determinado por dos variables aleatorias normales $\Delta X$, $\Delta Y$, ambas con media $\mu = 0$ y desviación estándar $\sigma = 10 \,\mu m$. ¿Cuál es la distancia media recorrida por la partícula en 500 pasos?

Al simular el experimento para diez pasos se obtiene un resultado como el que se muestra en la Figura \ref{fig:ch1_ejemplo_dist_normal_10pasos}. Al aumentar el número de pasos a 500 se observa una trayectoria como muestra Figura \ref{fig:ch1_ejemplo_dist_normal_500pasos}. Si se repite el experimento varias veces y se superponen las trayectorias, se obtiene la Figura \ref{fig:ch1_ejemplo_dist_normal_ensamble}, donde la distancia media recorrida, calculada a partir de los diferentes experimentos, se puede observar en la Figura \ref{fig:ch1_ejemplo_dist_normal_valor_medio} y la densidad de probabilidad en la Figura \ref{fig:ch1_ejemplo_dist_normal_valor_densidad}.

A partir del análisis inicial de las simulaciones, es posible formular el problema de manera general con el objetivo de encontrar una solución analítica a la pregunta planteada. Sean $\Delta X_k$, $\Delta X_k$ para $k=1, 2, \cdots, 500$, las componentes en $x$ e $y$ de la contribución del $k$-ésimo paso al desplazamiento de la partícula. Como la partícula parte del origen, $x_0 = y_0 = 0$, entonces, el primer paso estará dado por las componentes $X_1 = x_0 + \Delta X_1$, $Y_1 = y_0 + \Delta Y_1$; el segundo paso por, $X_2 = X_1 + \Delta X_2$, $Y_2 = Y_1 + \Delta Y_2$. Note que $X_k = X_{k+1} + \Delta X_k$, donde $X_0 = \Delta X_0 = 0$, por lo tanto,

$$X_k = \sum_{i=1}^k \Delta X_i, \quad Y_k = \sum_{i=1}^k \Delta Y_i$$

De esta manera, la posición para el $k$-ésimo paso está dada por,

\begin{equation}
    \vec{R}_k = \hat{i} \sum_{i=1}^k \Delta X_i + \hat{j} \sum_{i=1}^k \Delta Y_i
\end{equation}

El valor esperado del recorrido de la partícula, para el $k$-ésimo paso, está dado por,

$$\E{|\vec{R}_k - \vec{R}_0|} = \E{|\vec{R}_k|} = \E{\sqrt{\left(\sum_{i=1}^k \Delta X_i\right)^2 + \left(\sum_{i=1}^k \Delta Y_i\right)^2}}$$

Como cada paso está determinado por dos variables aleatorias normales $\Delta X_i$, $\Delta Y_i$, con igual media ($\mu$) y varianza ($\sigma^2$), el teorema del límite central implica que la suma de estas variables aleatorias normales, para cada componente, corresponderá con una variable aleatoria normal, $\Delta X$ e $\Delta Y$, cuya media es $\mu'=k\mu$ y su varianza, $\sigma'^2 = k\sigma^2$, por lo tanto,

$$\E{|\vec{R}_k - \vec{R}_0|} = \E{\sqrt{\Delta X^2 + \Delta Y^2}}$$

Como cada paso es estadísticamente independiente de los otros y cada coordenada es estadísticamente independiente de la otra, la densidad de probabilidad para calcular el valor esperado está dada por el producto de dos densidades gaussianas, una para cada coordenada,

$$\E{|\vec{R}_k - \vec{R}_0|} = \frac{1}{2\pi k\sigma^2}  \int_{-\infty}^\infty \int_{-\infty}^\infty \sqrt{x^2 + y^2} e^{-\frac{(x-\mu)^2}{2k\sigma^2}} e^{-\frac{(y-\mu)^2}{2k\sigma^2}} dx dy $$

Como $\mu=0$, se puede definir una nueva variable $r = \sqrt{x^2 + y^2} \geq 0$, en tal caso $dxdy$ se puede reemplazar por $r d\theta dr$, donde $0\leq\theta \leq 2\pi$. Con este cambio de variables se obtiene,

$$\E{|\vec{R}_k - \vec{R}_0|} = \frac{1}{2\pi k\sigma^2} \int_0^{2\pi} \int_0^\infty r^2 e^{-\frac{r^2}{2k\sigma^2}} d\theta dr = \frac{1}{k\sigma^2} \int_0^\infty r^2 e^{-\frac{r^2}{2k\sigma^2}} dr$$

La varianza $\sigma'^2 = k\sigma^2$ de una variable aleatoria $X$ normal, con media $\mu=0$, está dada por,

\begin{equation}
    \sigma'^2 = \E{X^2} = \frac{1}{\sigma' \sqrt{2\pi}} \int_{-\infty}^\infty x^2 e^{-\frac{x^2}{2\sigma'^2}} dx = \frac{2}{\sigma' \sqrt{2\pi}} \int_{0}^\infty x^2 e^{-\frac{x^2}{2\sigma'^2}} dx     
\end{equation}

Por lo tanto, el valor esperado de la distancia recorrida por la partícula, calculado con la ecuación \eqref{eq:ch1_dist_normal_desplazamiento_mean}, es $\E{|\vec{R}_k - \vec{R}_0|} = \frac{\sqrt{(2\pi)(500)}}{2} 10 \, \mu m \approx 280 \, \mu m$.

\begin{equation}
\label{eq:ch1_dist_normal_desplazamiento_mean}
    \E{|\vec{R}_k - \vec{R}_0|} = \frac{\sqrt{2 k \pi}}{2} \sigma
\end{equation}

Como $\E{|\vec{R}_k - \vec{R}_0|} = \E{r} = \int_0^\infty r f_R(r) dr$, se puede deducir que la densidad de probabilidad para el problema estudiado está dada por la ecuación \eqref{eq:ch1_dist_normal_densidad}, la cual concuerda con las simulaciones mostradas en la Figura \ref{fig:ch1_ejemplo_dist_normal_valor_densidad}.

\begin{equation}
    \label{eq:ch1_dist_normal_densidad}
    f_R(r) = \frac{r e^{-\frac{r^2}{2k\sigma^2}}}{k\sigma^2}
\end{equation}





\section{Material complementario para profundizar lo aprendido}

Se sugiere responder las preguntas de autoexplicación de los ejemplos trabajados que se presentan a continuación. Estos ejemplos pueden ser ejecutados en \textit{Google Colab} (haga clic en el enlace). En cada ejemplos se aborda un caso de estudio específico, diseñado con el propósito de orientar al estudiante y facilitar una comprensión más profunda de los conceptos abordados en este capítulo.

\begin{itemize}
    \item \href{https://colab.research.google.com/github/davidalejandromiranda/StatisticalPhysics/blob/main/notebooks/es_VariableAleatoria.ipynb}{Variables aleatorias y momentos centrales.}
    \item \href{https://colab.research.google.com/github/davidalejandromiranda/StatisticalPhysics/blob/main/notebooks/es_AproximacionStirling.ipynb}{Estadística de números grandes, aproximación de Stirling}
    \item \href{https://colab.research.google.com/github/davidalejandromiranda/StatisticalPhysics/blob/main/notebooks/es_CaminoAleatorio.ipynb}{El problema del camino aleatorio.}
    \item \href{https://colab.research.google.com/github/davidalejandromiranda/StatisticalPhysics/blob/main/notebooks/es_DistribucionesYNormalidad.ipynb}{Distribuciones y normalidad.}

\end{itemize} 

\section{Problemas propuestos}

\begin{exercise}
    En un cierto sistema se tienen ocho lugares disponibles donde se puede ubicar una partícula por cada uno. Si la probabilidad que una partícula se ubique en un cierto lugar es $1/2$ y que en dicho lugar no haya partícula es también $1/2$, ¿cuál es la probabilidad que $r$ partículas estén ocupando lugares disponibles? 
\end{exercise}
    
\begin{exercise}
    Obtenga la probabilidad que al lanzar $n$ monedas con igual probabilidad para cara y sello, se obtengan $k$ caras.
\end{exercise}

\begin{exercise}
    Un sistema físico puede contener hasta tres partículas, donde cada una aporta $E_0$ a su energía. Si la presencia o no de una partícula en el sistema se modela como un ensayo de Bernoulli con probabilidad $p=0.5$, obtenga la probabilidad que la energía del sistema sea $kE_0$, donde $k$ es un entero entre cero y tres.
\end{exercise}

\begin{exercise}
    Sean dos $(2)$ dados de $10$ caras, tales que un número natural único se utiliza para identificar cada cara. Demuestre que la probabilidad de obtener la suma de las caras igual a $E$ está dada por:
       
        \[
        p(E) = 
        \begin{cases} 
        \frac{E-1}{100} & \text{si } 2 \leq E \leq 10 \\
        \frac{20-E+1}{100} & \text{si } 10 < E \leq 20 \\
        0 & \text{en otro caso}
        \end{cases}
        \]
\end{exercise}

\begin{exercise}
    Sean dos $(2)$ dados de $60$ caras, tales que un número natural único se utiliza para identificar cada cara. Demuestre que la probabilidad de obtener la suma de las caras igual a $E$ está dada por:
        \[
        p(E) = 
        \begin{cases} 
        \frac{E-1}{3600} & \text{si } 2 \leq E \leq 60 \\
        \frac{120-E+1}{3600} & \text{si } 60 < E \leq 120 \\
        0 & \text{en otro caso}
        \end{cases}
        \]
\end{exercise}

\begin{exercise}
    Explique en qué condiciones la variable aleatoria $Y = \sum\limits_{i=1}^{N} X_i$, donde $X_i$ es una variable aleatoria no necesariamente normal, se puede aproximar como una variable aleatoria normal.
\end{exercise}

\begin{exercise}
    Sean $\{x_n^{(i)}\}$, para $i=1,2,\ldots,100$ y $n=1,2,\ldots,500$, cien conjuntos, cada uno con 500 observaciones de una cierta variable aleatoria $X$ con distribución estadística diferente a la normal. ¿Cuál esperaría que fuera la distribución de la suma de las observaciones, es decir, del conjunto formado al hacer $\sum\limits_i \{x_n^{(i)}\} = \{y_n\}$, donde $y_n = \sum_i x_n^{(i)}$? Argumente su respuesta.
\end{exercise}

\begin{exercise}
    Sean $x_1$ y $x_2$ dos series de observaciones de la variable aleatoria $X$. Demuestre si cada una de las siguientes expresiones es falsa o verdadera. Argumente cada respuesta.
    
    \begin{itemize}
        \item[a)] $x_1 + x_2 \neq 0$.
        \item[b)] $x_1 - x_2 = 0$.
        \item[c)] $\langle x_1 + x_2 \rangle \neq 0$.
        \item[d)] $\langle x_1 - x_2 \rangle = 0$.
    \end{itemize}

\end{exercise}

\begin{exercise}
    Sea $ax^2$ la densidad de probabilidad para una variable aleatoria continua $X$ con posibles observaciones entre $-1$ y $1$. Calcule,
    \begin{itemize}
        \item[a)] El valor esperado de $x$.
        \item[b)] La desviación estándar de $x$.
    \end{itemize}

\end{exercise}

\begin{exercise}
    Sean dos variables aleatorias discretas $X, Y$, tales que $X$ tiene probabilidad $p[n]=an+bn^2$ para $0 \leq n \leq 10$ y cero para otro $n$; por su parte, $Y$ tiene probabilidad $q[n]=c$, para $3 \leq n \leq 5$ y cero para otro $n$; donde $a,b,c$ son constantes. Explique (incluya las ecuaciones requeridas) cómo se calcula la probabilidad para una tercera variable aleatoria $Z=X+Y$ si las variables aleatorias $X$ e $Y$ son estadísticamente independientes.
\end{exercise}

\chapter{Capítulo II. Sistemas aislados y su descripción estadística: ensamble microcanónico}
En la naturaleza se presentan diversos fenómenos donde un gran número de partes (por lo general asumidas como partículas) están presentes. Al conjunto formado por las partes que describen un fenómeno en la naturaleza se suele llamar sistema\footnote{Por ejemplo, el sistema formado por las partículas clásicas que ocupan un cierto volumen, donde las partículas son partes del conjunto formado por todas aquellas en el volumen, relacionadas entre sí por la condición de no interactuar entre si y cuyo movimiento está gobernado por las leyes de Newton (ordenamiento entre las partes) conforman un gas ideal (fenómeno).} y una aproximación razonable es considerar a este aislado del resto del universo. Incluso en el caso de tener un sistema que interactúa con otro, se puede considerar a cada uno como una parte de un gran sistema formado por los dos, donde el gran sistema se puede asumir aislado del resto del universo; de esta manera, considerar sistemas aislados es una aproximación razonable para describir fenómenos de la naturaleza.

Un sistema aislado, en un determinado tiempo, se encuentra en un cierto estado, conocido como microestado. La descripción del comportamiento de las partes constituyentes del sistema se suele realizar en términos de alguna teoría de la Física (por ejemplo, la mecánica cuántica), sin embargo, la descripción del sistema completo implica complicaciones tales que se vuelve restrictiva la generalización de dichas teorías. En este contexto, la combinación de la Física con la Estadística, bajo las siguientes suposiciones, permite describir el comportamiento del sistema aislado como un todo: \textit{primero}, existe una distribución estadística con la cual se pueden describir los microestados del sistema; \textit{segundo}, cada parte constituyente se describe por medio de alguna teoría Física; \textit{tercero}, se conoce cómo enumerar los diferentes microestados en que se puede encontrar el sistema y \textit{cuarto}, se conoce cómo calcular las probabilidades de ocurrencia para hacer predicciones sobre el sistema.

En este capítulo se abordará el estudio de sistemas aislados con energía en un rango comprendido entre $E$ y $E+dE$, donde cada microestado del sistema tiene la misma probabilidad de ocurrencia. Se analizarán dos situaciones generales, una donde la energía del sistema es analizada en términos de su evolución temporal (proceso aleatorio) y la otra, donde se considera un gran número de sistemas preparados de la misma manera (ensamble), en cuyo caso la energía del sistema se trata en términos estadísticos a partir de lo observado en cada sistema; estas dos situaciones generales se asumirán equivalentes (principio ergódico), por lo tanto, se puede centrar el estudio de los sistemas aislados con solo una de ellas, el estudio de ensambles, que para el caso de considerar solo sistemas aislados con microestados cuya energía está comprendida entre $E$ y $E+dE$, corresponde con un ensamble microcanónico.

\begin{definition}
    Al conjunto de partes (por lo general, partículas), que relacionadas entre sí ordenadamente\footnote{En este contexto, \textit{ordenadamente} hace referencia a la existencia de algún patrón, simetría, ley, principio o modelo con el cual se pueda determinar la relación entre las partes constituyentes.} contribuyen a determinado fenómeno, se le denomina \textbf{sistema}.\index{Sistema}
\end{definition}

\section{Proceso aleatorio, ensamble y el principio ergódico}

En el capítulo anterior se estudió el movimiento aleatorio, en dos dimensiones, de una partícula clásica, para ello se consideró como variables aleatorias normales, con media $\mu$ y varianza $\sigma^2$, a las proyecciones del desplazamiento (paso), sobre los ejes coordenados $x$, $y$, dadas por $\Delta X$ y $\Delta y$. De esta manera, cada paso está determinado por un desplazamiento aleatorio que es independiente de los demás pasos. Cuando la partícula recorre $k$ pasos, su posición es,

$$\vec{R}_k = \hat{i} \sum_{i=1}^k \Delta X_i + \hat{j} \sum_{i=1}^k \Delta Y_i$$

Supóngase que se quiere realizar una observación del movimiento de la partícula en la cual se cuenten tres pasos, entonces, la posición final de la partícula está dada por,

$$\vec{r}_{3}^{\,(1)} = \hat{i} \left[ \Delta x_1^{(1)} + \Delta x_2^{(1)} + \Delta x_3^{(1)}\right] + \hat{j} \left[ \Delta y_1^{(1)} + \Delta y_2^{(1)} + \Delta y_3^{(1)}\right] $$

Donde el superíndice $^{(1)}$ indica el número de la observación. Al realizar una segunda observación, bajo las mismas condiciones, la posición final de la partícula está dada por,

$$\vec{r}_{3}^{\,(2)} = \hat{i} \left[ \Delta x_1^{(2)} + \Delta x_2^{(2)} + \Delta x_3^{(2)}\right] + \hat{j} \left[ \Delta y_1^{(2)} + \Delta y_2^{(2)} + \Delta y_3^{(2)}\right] $$

Como cada desplazamiento $\Delta x$ es una observación de la variable aleatoria normal $\Delta X$, es de esperarse que el primer paso de la primera observación, $\vec{r}_1^{\,(1)} = \hat{i} \Delta x_1^{(1)} + \hat{j} \Delta y_1^{(1)}$, sea diferente del primer paso de la segunda observación, $\vec{r}_1^{\,(2)} = \hat{i} \Delta x_1^{(2)} + \hat{j} \Delta y_1^{(2)}$; es decir, $\Delta x_1^{(1)} \neq \Delta x_1^{(2)}$ y $\Delta y_1^{(1)} \neq \Delta y_1^{(2)}$. De igual manera $\vec{r}_2^{\,(1)} \neq \vec{r}_2^{\,(2)}$ y $\vec{r}_3^{\,(1)} \neq \vec{r}_3^{\,(2)}$. 

Es importante notar que al suponer pasos discretos para describir el movimiento de una partícula en el espacio-tiempo, implícitamente se está suponiendo un cierto intervalo de tiempo $\tau$ entre los pasos. Esto implica que al tener en cuenta el tiempo, como variable independiente, se puede simplificar la notación de tal manera que se utilice un solo índice para señalar el número de la observación y en vez de especificar el número de pasos se utiliza el tiempo transcurrido entre ellos; por ejemplo, $x_1^{\,(2)}$ se escribiría como $x_2(\tau)$, donde ahora el tiempo $t=\tau$ se utiliza para identificar el primer paso, mientras que el subíndice en $x_2(\tau)$ indica el número de la observación. En esta notación, $\vec{r}_{1000}^{\,(n)}$ se escribiría como $\vec{r}_n(1000\tau)$ y para cualquier tiempo\footnote{De manera general, el tiempo $t$ transcurrido desde el inicio del proceso aleatorio ($t=0$) hasta el $k$-ésimo paso se puede expresar como $t=k\tau_k$, donde cada paso tiene asociado un intervalo de tiempo $\tau_k$. Por simplicidad, se asume que $\tau_k=\tau$; en otras palabras, se asume que el intervalo de tiempo en todos los intervalos es constante.}, $\vec{r}_n(t)$, donde $t=k\tau$.

\begin{figure}[th]
    \centering
    \begin{subfigure}{0.48\textwidth}
        \includegraphics[width=\linewidth]{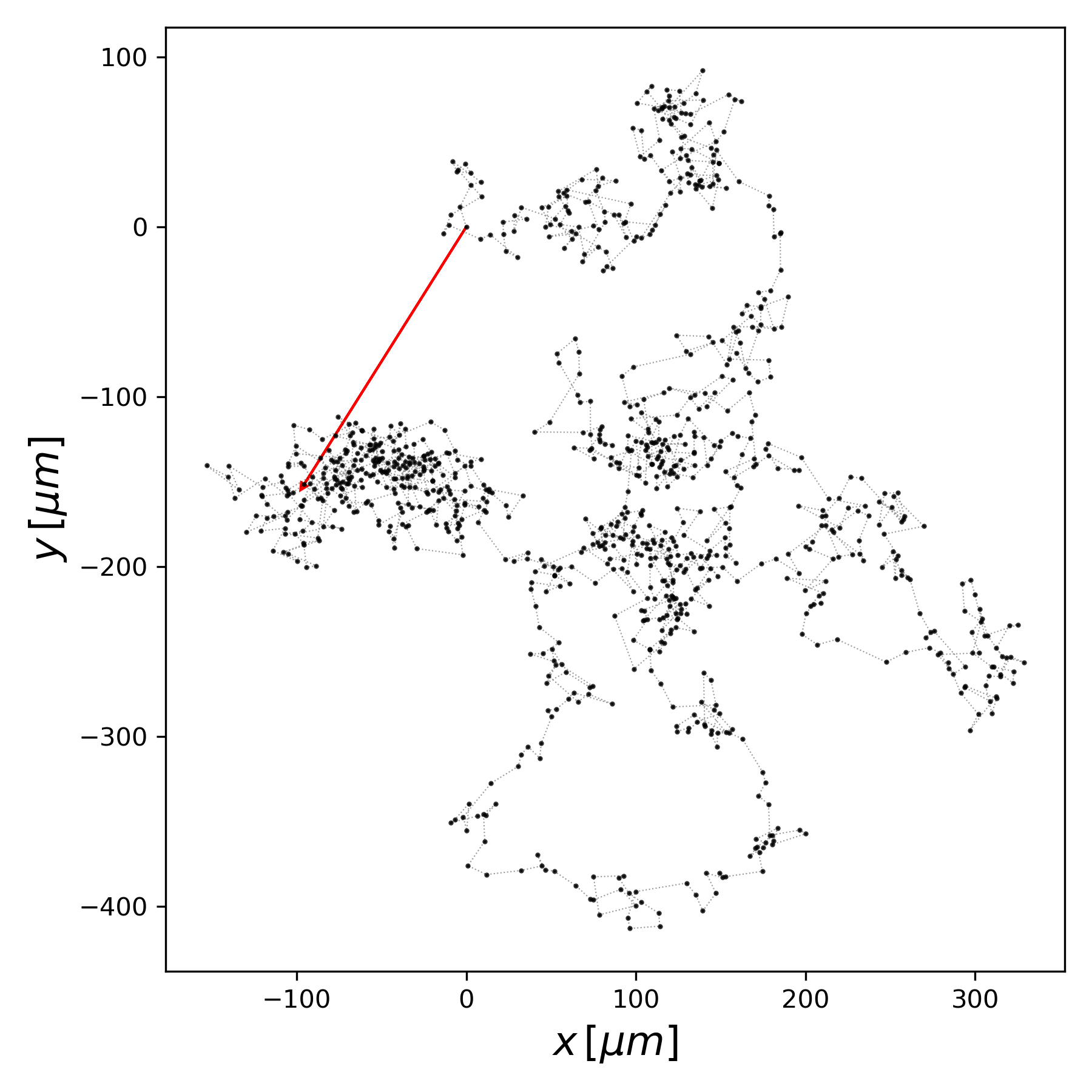}
        \caption{\label{fig:ch2_proceso_aleatorio_1000pasos_1observacion}}
    \end{subfigure}
    \hfill
    \begin{subfigure}{0.48\textwidth}
        \includegraphics[width=\linewidth]{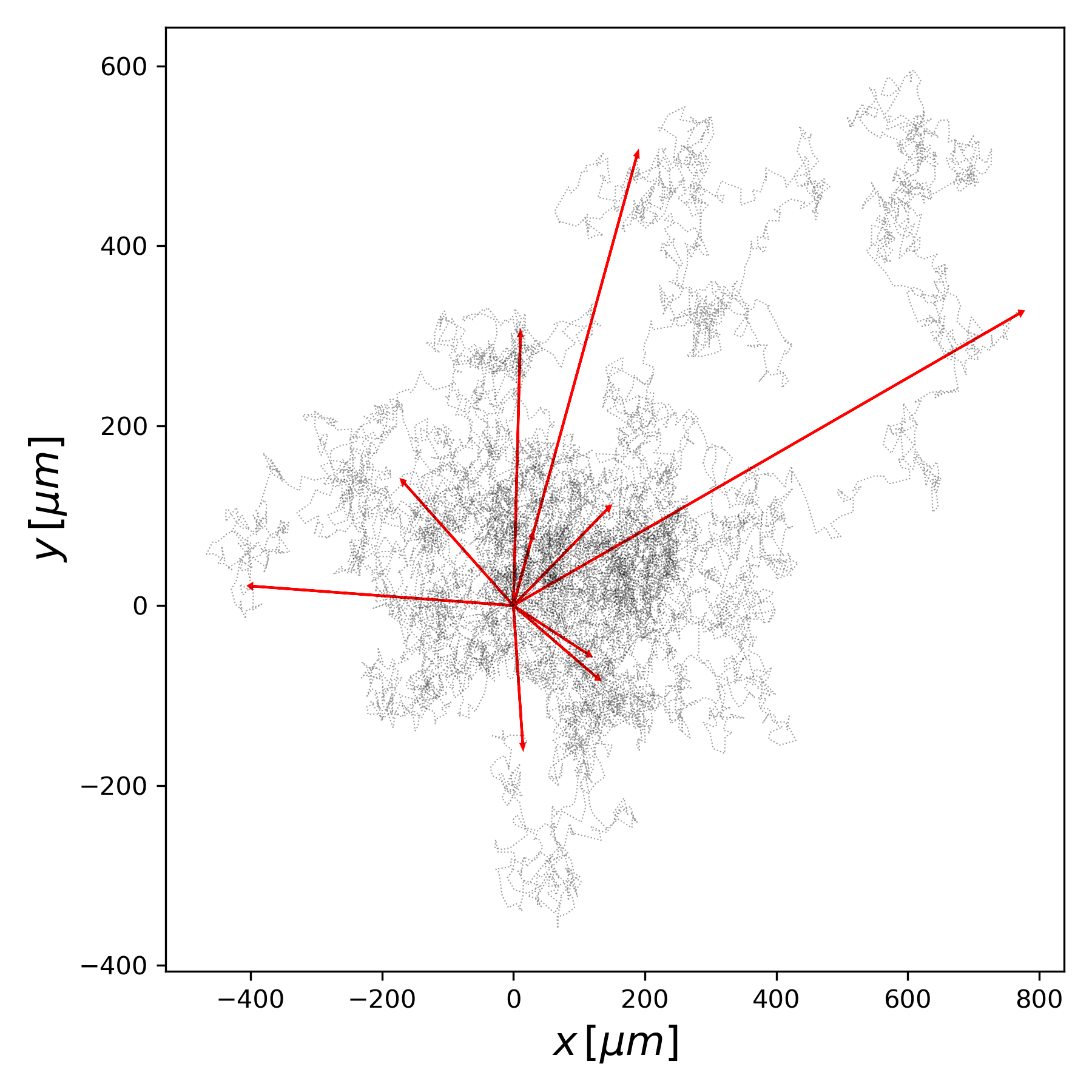}
        \caption{\label{fig:ch2_proceso_aleatorio_1000pasos_10observacion}}
    \end{subfigure}
    \caption{(a) Trayectoria formada por $10^3$ desplazamientos aleatorios de una partícula para una observación y (b) superposición de las trayectorias para diez observaciones diferentes. Las líneas punteadas representan las trayectorias y las flechas rojas, el vector desplazamiento.}
\end{figure}

Si se aumenta el número de pasos, entonces, se observa una trayectoria como la que se muestra en la Figura \ref{fig:ch2_proceso_aleatorio_1000pasos_1observacion}, donde una partícula recorre $10^3$ pasos, cada uno descrito por $\Delta X$, $\Delta Y$ con $\mu = 0$ y $\sigma = 10\,\mu m$; al superponer las trayectorias de diez observaciones, Figura \ref{fig:ch2_proceso_aleatorio_1000pasos_10observacion}, se nota que el vector desplazamiento $\vec{r}_n(1000\tau)$, para $n=1, 2, \cdots,\, 10$ es diferente para cada observación. El desplazamiento (aleatorio) en un tiempo arbitrario $t=k\tau$, para la $n$-ésima observación, se puede escribir como $\vec{r}_n(t)$, lo cual implica un proceso seguido por la partícula, consistente en $k$ pasos recorridos durante un tiempo $t=k\tau$. A este tipo de procesos, donde cada evento (o estado) varia en el tiempo de manera aleatoria se conoce como proceso aleatorio.

\begin{definition}
    Sea $t$ una variable independiente, por lo general el tiempo, entonces, se dice que $Y(t)$ es un \textbf{proceso aleatorio} si $Y(t)$ no depende de manera completamente definida de la variable independiente. Por lo tanto, al realizar varias observaciones, diferentes funciones $y(t)$ son observadas, una por cada observación.\index{Proceso aleatorio}
\end{definition}

\begin{figure}[th]
  \centering
  \begin{subfigure}{0.48\textwidth}
    \centering
    \includegraphics[width=\linewidth]{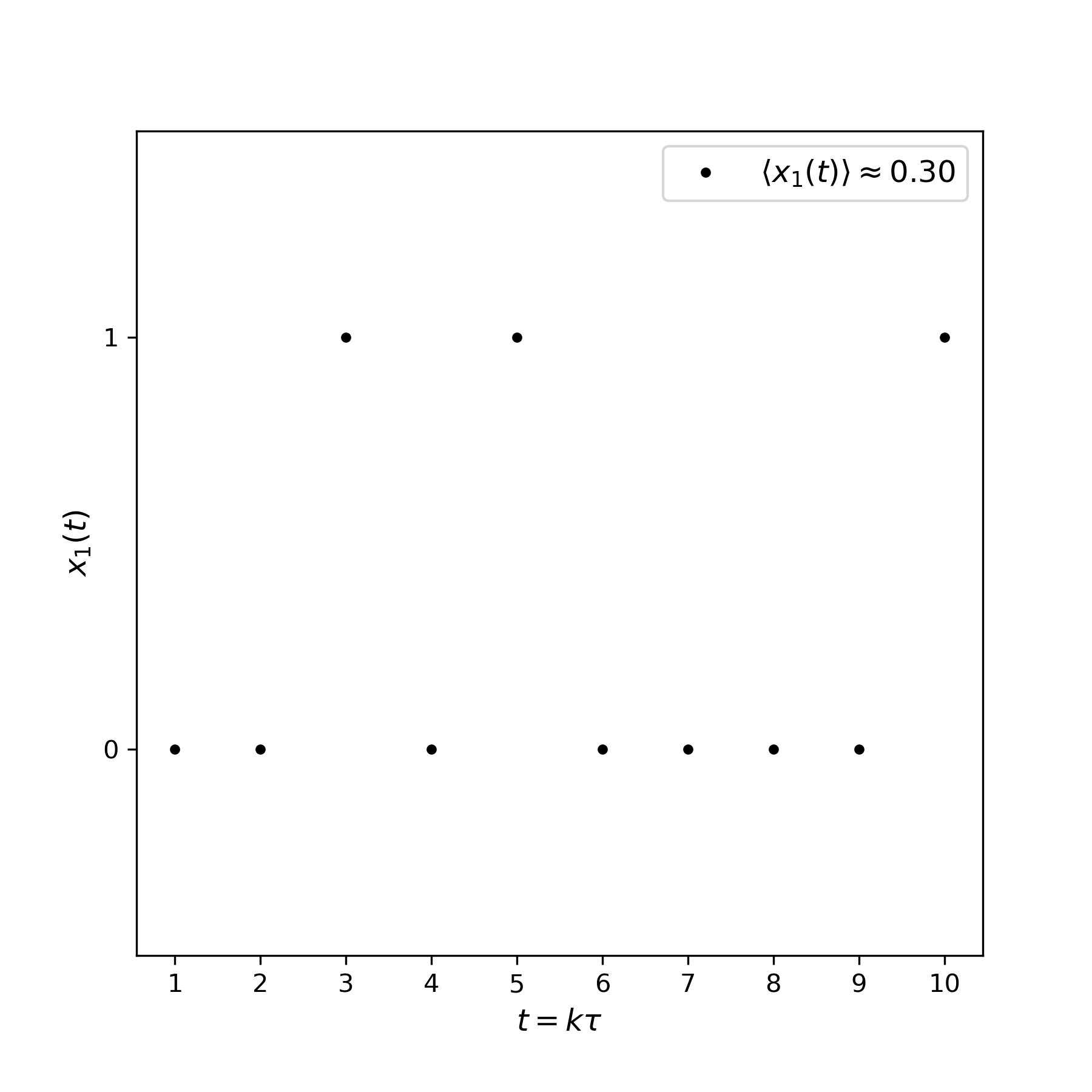}
    \caption{}
    \label{fig:ch2_proceso_aleatorio_binomial_10observaciones}
  \end{subfigure}
  \begin{subfigure}{0.48\textwidth}
    \centering
    \includegraphics[width=\linewidth]{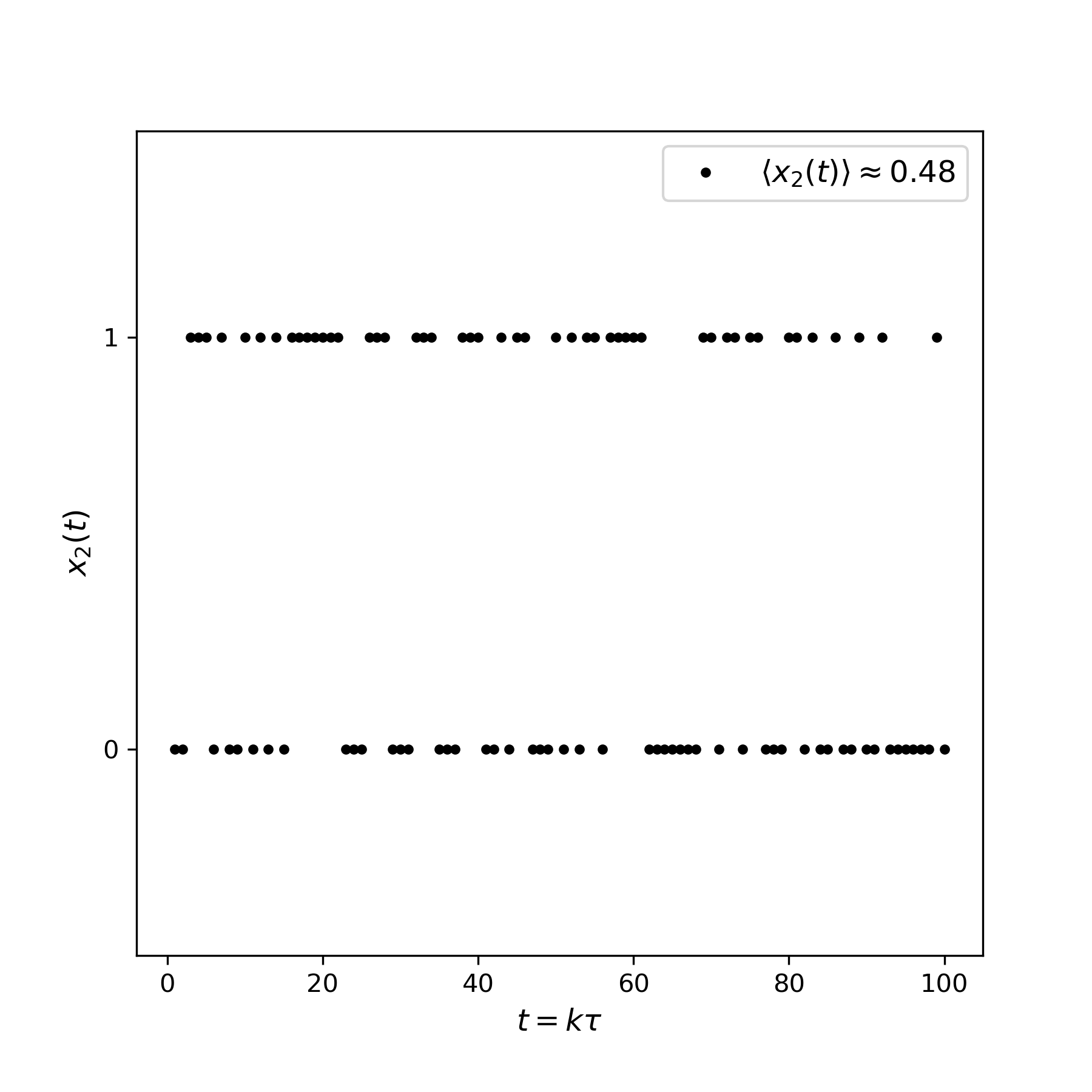}
    \caption{}
    \label{fig:ch2_proceso_aleatorio_binomial_1000observaciones}
  \end{subfigure}
  \caption{Proceso aleatorio binomial que representa el lanzamiento de una moneda (a) diez veces y (b) $10^3$ veces.}
\end{figure}

Otro ejemplo de proceso aleatorio consiste en observar el resultado del lanzamiento de una moneda no cargada cuyas caras están marcadas con los números $0$, $1$ y la probabilidad de ocurrencia de ambas caras es la misma ($p=1/2$), como se estudió en el capítulo anterior, ver Figura \ref{fig:ch1_lanzamiento_una_moneda}. Supongamos que se lanza $10$ veces la moneda dejando un intervalo de tiempo $\Delta t = \tau = 10s$ entre lanzamiento, como se muestra en la Figura \ref{fig:ch2_proceso_aleatorio_binomial_10observaciones}; si se nombra esta primera observación del proceso aleatorio $x_1(t)$, donde $t=k\tau$, entonces, el valor esperado está dado por,

$$\E{x_1(t)}=\sum_{i=1}^k x_1(i\tau)$$

El valor esperado para la primera observación del proceso aleatorio es $\E{x_1(t)}\approx 0.3$. Supóngase ahora una segunda observación, esta vez para un tiempo más largo, como se muestra en la Figura \ref{fig:ch2_proceso_aleatorio_binomial_1000observaciones}; en este caso, $\E{x_2(t)}\approx 0.48$, que concuerda con el valor esperado teórico ($0.5$). Es importante restar la diferencia entre el valor esperado obtenido en la primera observación que solo tuvo en cuenta diez intervalos de tiempo, $\E{x_1(t)}$ y la segunda observación del proceso aleatorio, con mil intervalos de tiempo; esto sugiere que cuando se aumenta el intervalo de tiempo, por ejemplo de $10\tau$ a $1000\tau$, el valor esperado $\E{x(t)}$ tiende al predicho teóricamente.

Otra forma de abordar el problema es preparar $k$ monedas de la misma manera, es decir, que sus caras estén marcadas con los números $0$, $1$ y la probabilidad de ocurrencia de las caras sea igual ($p=0.5$).  En tal caso, el experimento consistirá en observar el resultado obtenido al lanzar las monedas, sin que ninguna altere el resultado de la otra, es decir, garantizando independencia estadística. El conjunto de monedas preparadas bajo las mismas condiciones (se podrían pensar que son copias idénticas) se conoce como ensamble estadístico o simplemente ensamble. Es de esperar que al construir un ensamble con $10$ monedas se obtenga un valor esperado similar al obtenido a partir del proceso aleatorio $x_1(t)$ y un ensamble con $1000$ monedas, a $x_2(t)$. A esta idea intuitiva que el valor esperado calculado sobre ensamble y el calculado sobre el proceso aleatorio sean similares se conoce como la hipótesis ergódica. Cuando se supone válida la hipótesis ergódica, se pueden calcular las cantidades estadísticas o bien sobre el ensamble o sobre el proceso aleatorio, en ambos casos se esperaría obtener el mismo resultado.

\begin{definition}
    Un \textbf{ensamble} es un conjunto de sistemas idénticos preparados bajo las mismas condiciones.\index{Ensamble}
\end{definition}

\begin{definition}
    Los estados en que se puede encontrar un sistema aislado se conocen como \textbf{microestados}.\index{Microestado}
\end{definition}

\begin{definition}
    \textbf{Hipótesis ergódica.} En un periodo de tiempo suficientemente largo, todos los posibles microestados accesibles de un sistema tienen la misma probabilidad.\index{Hipótesis ergódica}
\end{definition} 

\section{Ensamble microcanónico}

Después de analizar nuevamente el problema del camino aleatorio y el lanzamiento de monedas, se puede pensar en generalizar dicho análisis para cualquier \textit{sistema aislado}. Supongamos que la hipótesis ergódica es válida y que es conveniente analizar el problema desde el punto de vista de ensambles. En tal caso, es necesario definir de manera precisa cómo organizar el ensamble para con ello poder calcular el número de estados accesibles $\Omega$ y las probabilidades. 

Sea un sistema en el microestado $\ket{r, k}$ con energía $E_r$, donde $k$ es un índice que identifica a cada uno de los posibles microestados con energía $E_r$; esto quiere decir que para un cierto valor de energía $E_r$ pueden existir muchos valores $k$ y con ello, muchos posibles microestados con la misma energía. Si se preparan muchos sistemas bajo las mismas condiciones, existen diversas maneras de organizar el ensamble, una de ellas consiste en incluir en el ensamble solo aquellos sistemas que tienen energía $E_r$ en un cierto intervalo comprendido entre $E$ y $E+dE$; si se supone que la probabilidad de cada microestado en el intervalo de energía es la misma\footnote{Esto quiere decir que la probabilidad para cada microestado obedece a una distribución uniforme.}, a este ensamble de \textit{sistemas aislados} se le conoce como microcanónico. Esto implica que en un \textit{ensamble microcanónico} el número total de estados accesibles $\Omega_T(E)$ determina la probabilidad de cada uno de los $\Omega(E_r)$ microestado con energía $E_r$, la cual corresponde con la probabilidad de una distribución uniforme dada por la ecuación \eqref{eq:ch2_prob_microestados}, donde $\Omega_T(E)$ es la suma de todos los posibles microestados, ecuación \eqref{eq:ch2_num_estados_accesibles_microcanonico}. 

\begin{equation}
    \label{eq:ch2_prob_microestados}
    \prob{\ket{\psi}=\ket{r, k}} = 
    \begin{cases}
        \frac{1}{\Omega_T(E)}, & \text{si } E \leq E_r < E + dE \\
        0, & \text{otro } E_r    
    \end{cases}
\end{equation}

\begin{equation}
    \label{eq:ch2_num_estados_accesibles_microcanonico}
    \Omega_T(E) = \sum_r \Omega(E_r)
\end{equation}

\begin{definition}
    Un \textbf{ensamble microcanónico} es aquel formado por sistemas aislados tal que sus microestados accesibles, para un cierto intervalo de energía entre $E$ y $E+dE$, tienen la misma probabilidad de ocurrencia.\index{Ensamble microcanónico}
\end{definition}

Con el anterior resultado se puede encontrar una manera general para calcular valores esperados e incertidumbres\footnote{La incertidumbre de una variable aleatoria $X$ corresponde con la raíz cuadrada de la varianza, es decir, con su desviación estándar.} de una cantidad física. 

Sea una cierta cantidad física $X$, modelada como una variable aleatoria y sea un ensamble microcanónico donde cada microestado tiene probabilidad $p_r$;  la probabilidad asociada a los microestados con energía $E_r$, dentro del rango comprendido entre $E$ y $E+dE$, se obtiene a partir del cociente entre el número de microestados con energía $E_r$ y la suma, para todos los valores de $E_r$, de los números de estados accesibles, ecuación \eqref{eq:ch2_prob_ensamble_microcanónico}. \index{Ensamble microcanónico!probabilidad}

\begin{equation}
    \label{eq:ch2_prob_ensamble_microcanónico}
	p(E_r) = p_r = \prob{E=E_r} = \frac{\Omega(E_r)}{\sum\limits_r \Omega(E_r)} = \frac{\Omega(E_r)}{\Omega_T}
\end{equation}

Por lo tanto, el valor esperado $\E{X}$ está dado por la ecuación \eqref{eq:ch2_valor_esperado_microcanonico}; la incertidumbres $\sqrt{\E{X^2} - \E{X}^2}$ , por la raíz cuadrada de la la varianza descrita por la ecuación \eqref{eq:ch2_varianza_microcanonico} y, en general, el $n$-ésimo momento central, por la ecuación \eqref{eq:ch2_momento_centrale_microcanonico}.\index{Ensamble microcanónico!valores esperados}\index{Ensamble microcanónico!incertidumbres}

\begin{equation}
    \label{eq:ch2_valor_esperado_microcanonico}
    \E{X} = \sum_{r} X_r \prob{E=E_r} = \frac{1}{\Omega_T} \sum_{r} X_r \Omega(E_r)
\end{equation}

\begin{equation}
    \label{eq:ch2_varianza_microcanonico}
    \E{(X-\E{X})^2} = \left[ \frac{1}{\Omega_T} \sum_{r} X_r^2 \Omega(E_r) \right] - \E{X}^2
\end{equation}

\begin{equation}
    \label{eq:ch2_momento_centrale_microcanonico}
    \E{(X-\E{X})^n} = \frac{1}{\Omega_T} \sum_{r} (X_r - \E{X})^n \Omega(E_r)
\end{equation}

\newpage

\section{Sistema aislado formado por $r$ partículas idénticas con energía entre $E_1$ y $9E_1$}
\label{sec:ch2_ejemplo_ilustrativo_sistema_aislado_9E_1}

\begin{figure}[th]
    \centering
    \begin{subfigure}{0.48\textwidth}
        \centering
        \includegraphics[width=0.9\linewidth]{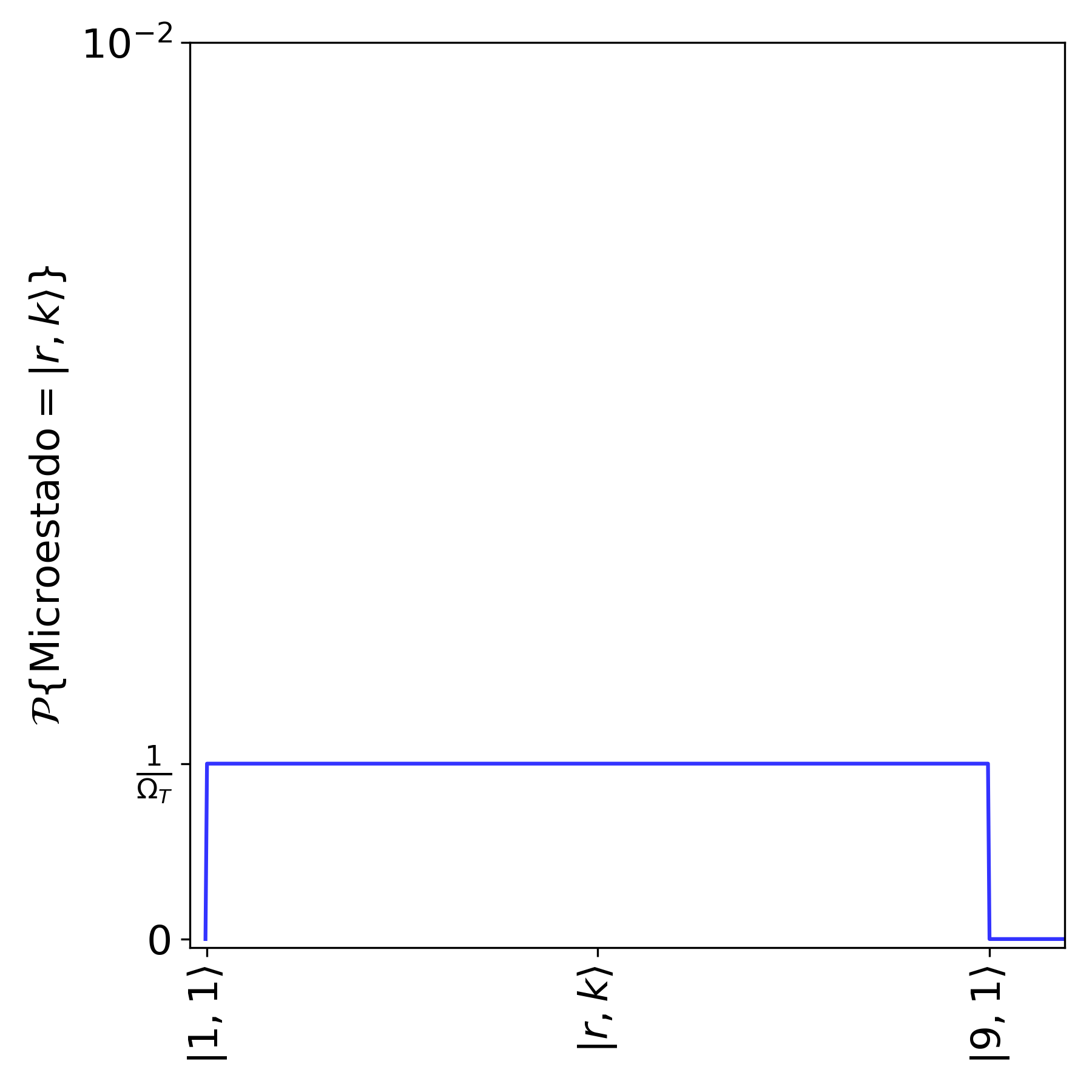}
        \caption{}
        \label{fig:ch2_Omega_microestados_probabilidad}
    \end{subfigure}
    \begin{subfigure}{0.48\textwidth}
        \centering
        \includegraphics[width=0.9\linewidth]{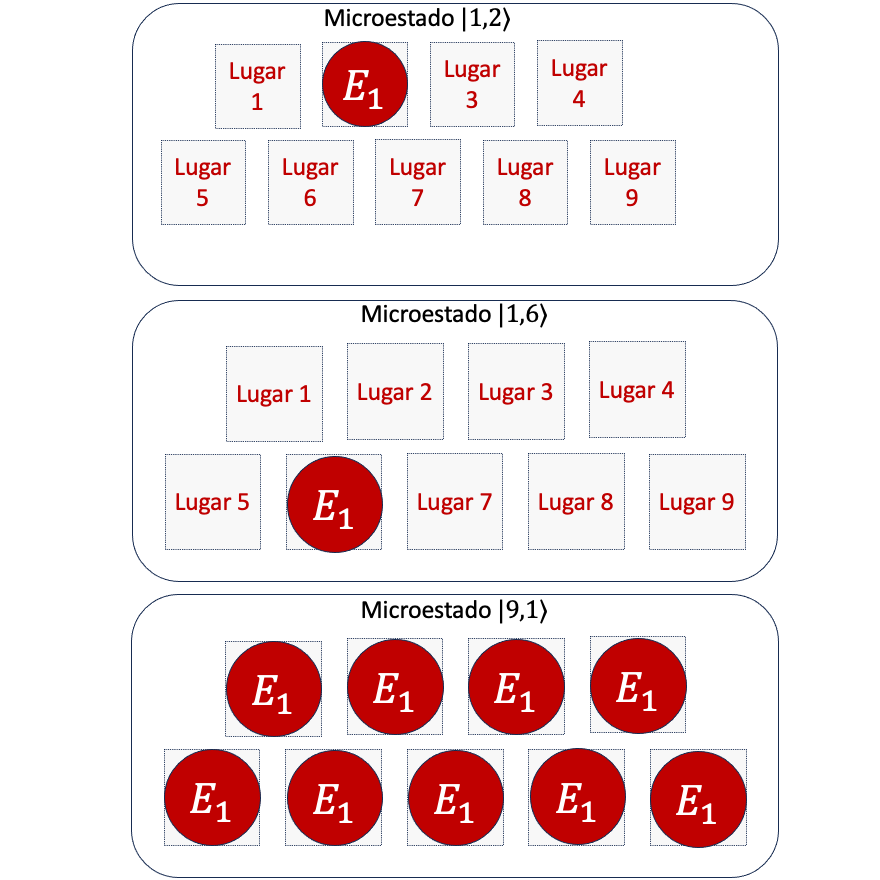}
        \caption{}
        \label{fig:ch2_dos_sistemas_binomial}
    \end{subfigure}
    \caption{(a) Probabilidad para cada uno de los posibles microestados $\ket{r,k}$ de un ensamble microcanónico formado por sistemas de nueve partículas idénticas, cada una de las cuales aporta $E_1$ de energía al sistema y se pueden (b) representar esqumáticamente como círculos que pueden (o no) ocupar uno de nueve lugares.}
\end{figure}

\begin{figure}[ht]
    \begin{subfigure}{0.48\textwidth}
        \centering
        \includegraphics[width=0.9\linewidth]{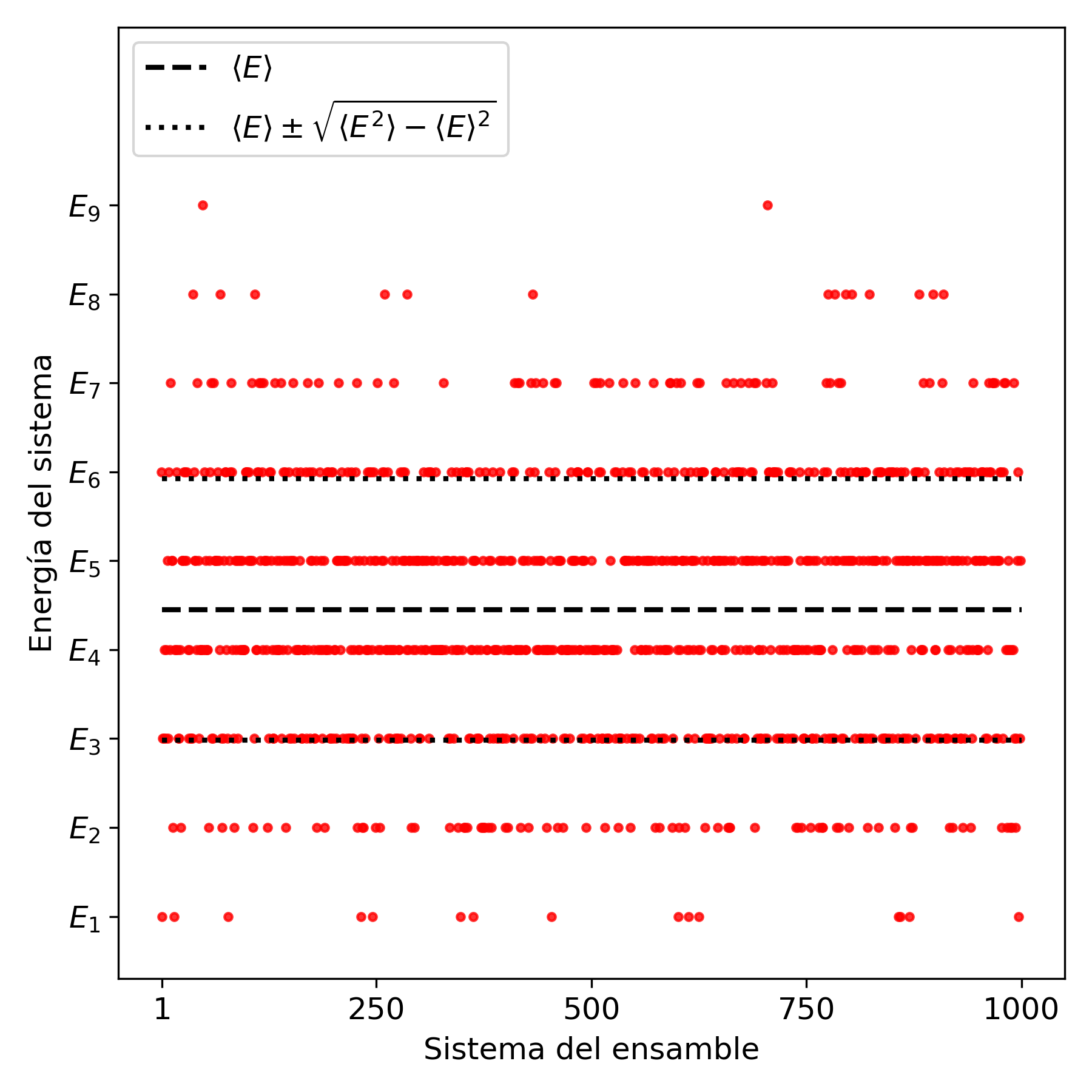}
        \caption{}
        \label{fig:ch2_Omega_microestados_energia_binomial_obs}
    \end{subfigure}
    \begin{subfigure}{0.48\textwidth}
        \centering
        \includegraphics[width=0.9\linewidth]{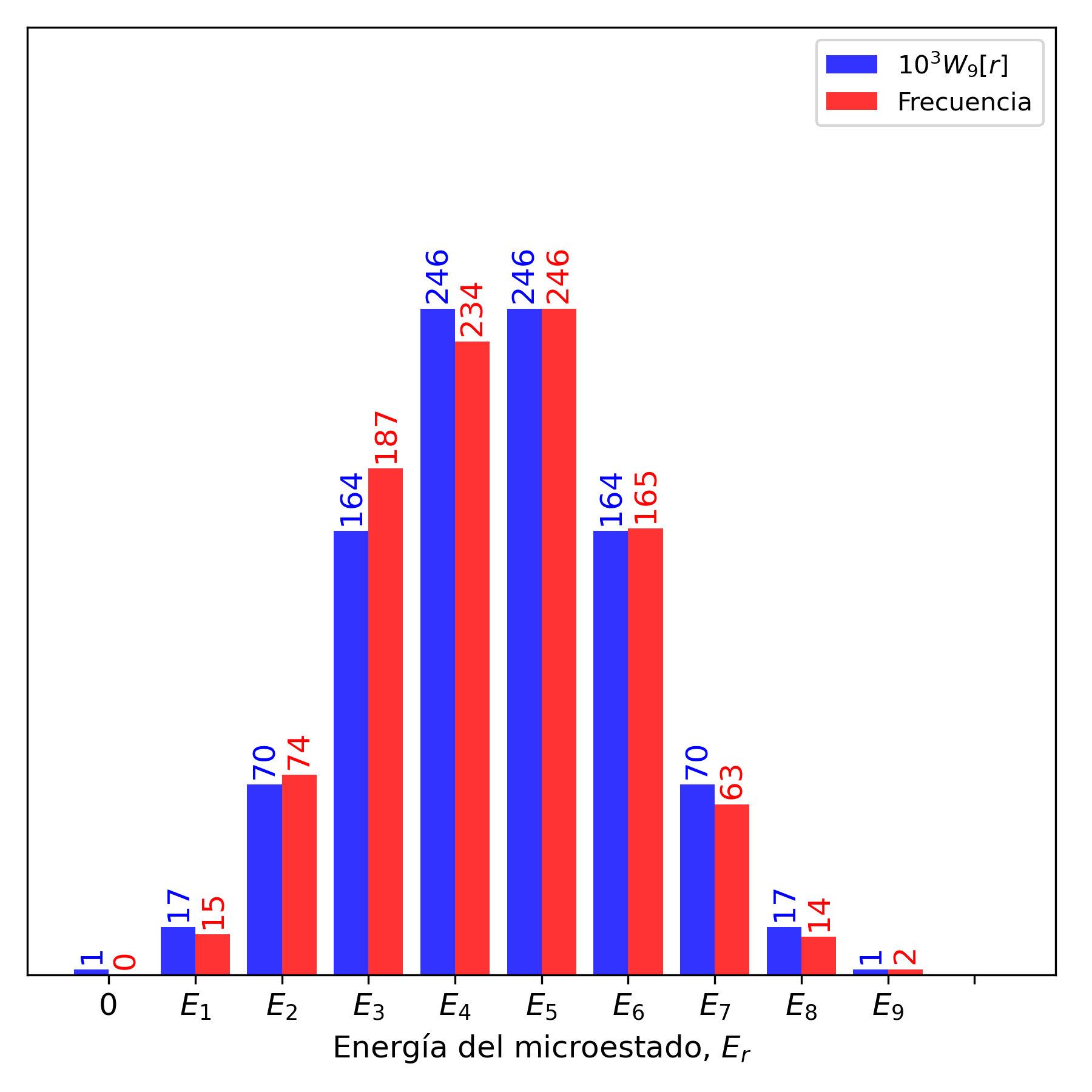}
        \caption{}
        \label{fig:ch2_Omega_microestados_energia_binomial}
    \end{subfigure}
    \caption{(a) Energía de $10^3$ sistemas en un ensamble microcanónico y (b) número de estados accesibles $\Omega(E_r)$ obtenido al contar los sistemas con energía en cada rango (barras rojas), lo cual es consistente con $10^3 W_9[r]$ (barras azules).}
\end{figure}

Sea un sistema aislado formado por nueve partículas idénticas, cada una de las cuales aporta $E_1$ de energía al sistema, con microestados $\ket{r,k}$, donde el número cuántico $r$ determina su energía y $k$, identifica los estados degenerados para $E_r$. Como para cada energía $E_r$ existen varios posibles valores $k$, es de notar que los microestados $\ket{r, k}$ tienen igual probabilidad de ocurrencia, como se muestra en la Figura \ref{fig:ch2_Omega_microestados_probabilidad}. Para el rango de energía $E_r$ entre $E_1$ y $9E_1$, los microestados pueden tomar nueve posibles valores de energía, donde $r=1, 2, \cdots, 9$.  ¿Cuánto es el valor esperado de la energía del sistema aislado y cuánto es su incertidumbre?

Antes de poder calcular el valor esperado y la incertidumbre, es necesario obtener la probabilidad $p_r$ y el número de estados accesibles $\Omega(E_r)$ para los microestados con energía $E_r$; con estas cantidades y la ecuación \eqref{eq:ch2_valor_esperado_microcanonico} se obtiene el valor esperado de la energía y con la raíz cuadrada de la ecuación \eqref{eq:ch2_varianza_microcanonico}, la incertidumbre. Con este fin, y para entender mejor el comportamiento del sistema, consideremos un ensamble formado por $10^3$ sistemas aislados, preparados bajo las mismas condiciones\footnote{En este caso, la condición es que la energía del microestado esté en el intervalo comprendido entre $E=E_1$ y $E+dE = 9E_1$.}. 

Es de notar que cada posible microestado tiene igual probabilidad de ocurrencia, sin embargo, el número de microestados con energía $E_r$ es diferente para cada valor de $r$. Lo anterior se puede comprender mejor si se utiliza un modelo visual; supongamos que las partículas se organizan en un arreglo de nueve elementos, entonces, solo existe una posibilidad de obtener la energía $E_9$, como se observa en la Figura \ref{fig:ch2_dos_sistemas_binomial}, mientras que la energía $E_1$ tiene nueve posibilidades diferentes que corresponde con la combinatoria $\binom{9}{1}$. Con esto en mente y considerando que en el sistema se tienen nueve posibles lugares para albergar a las partículas, es posible construir el ensamble suponiendo que cada sistema en el ensamble tiene energía igual a $E_r = rE_1$, donde $r$ es el número de partículas presentes en los lugares disponibles; es razonable suponer que cada lugar tiene 50\% de probabilidad que la partícula se encuentre presente y 50\%, que no se encuentre en dicho lugar, por lo tanto, ubicar las partículas en dichos lugares corresponde con una variable aleatoria binomial con $p=0.5$ y nueve ensayos de Bernoulli. 

En consecuencia, el numero de estados accesibles en el ensamble microcanónico para el sistema en estudio se modela por medio de una distribución binomial con probabilidad de éxito $p=0.5$ y nueve ensayos, con ello, la probabilidad de observar el $r$-ésimo estado es $W_9[r]$, donde $W_n[r]$ está dada por la ecuación \eqref{eq:ch1_probabilidad_binomial}. 

Con el anterior análisis, se ha mostrado que el número de estados accesibles $\Omega(E_r)$, para el microestado con $r$ partículas, del sistema aislado, es consistente con una distribución binomial con probabilidad de éxito $p=0.5$, dado por la ecuación \eqref{eq:ch2_ejemplo_microcanonico_omega1} y el número total de microestados es 512, ver ecuación \eqref{eq:ch2_ejemplo_microcanonico_omegaT1}.

\begin{equation}
    \label{eq:ch2_ejemplo_microcanonico_omega1}
    \Omega(E_r) = \binom{9}{r} = \frac{9!}{r!(9-r)!}
\end{equation}

\begin{equation}
    \label{eq:ch2_ejemplo_microcanonico_omegaT1}
    \Omega_T(E) = \sum_{r=1}^9 \binom{9}{r} = \sum_{r=1}^9 \frac{9!}{r!(9-r)!} = (1+1)^9 = 512    
\end{equation}

Antes de calcular el valor esperado de la energía y la incertidumbre, analicemos el caso del ensamble microcanónico formado por $10^3$ sistemas; con el análisis anterior se puede afirmar que la energía $E$ se modela por una variable aleatoria binomial con nueve posibles resultados, cada uno con probabilidad $p=0.5$ de ocurrencia, como se muestra en la Figura \ref{fig:ch2_Omega_microestados_energia_binomial_obs}. El número de estados se obtiene al contar los sistemas del ensamble en cada rango de energía ($E=E_r$), barras rojas en la Figura \ref{fig:ch2_Omega_microestados_energia_binomial}, lo cual es consistente con $10^3 W_9[r]$, barras azules en la Figura \ref{fig:ch2_Omega_microestados_energia_binomial}.

Teniendo en cuenta que $\frac{1}{512}=\left( \frac{1}{2} \right)^9$ se puede escribir como $\left( \frac{1}{2} \right)^r \left( \frac{1}{2} \right)^{9-r}$, el método asociado al uso de la ecuación \eqref{eq:ch1_identidad_dist_binomial} y los valores esperados $\E{r}=np$, $\E{r^2}=npq+n^2p^2=npq+\E{r}^2$ para una distribución binomial, entonces, el valor esperado de la energía $\E{E}$ y la incertidumbre $\sigma = \sqrt{\E{E^2} - \E{E}^2}$ se obtienen como se muestra a continuación.

$$\E{E} = \frac{1}{512}\sum_{r=1}^9 \binom{9}{r} r E_1 = \frac{E_1}{512}\sum_{r=1}^9 \frac{9!}{r!(9-r)!} r = E_1\sum_{r=1}^9 \frac{9!}{r!(9-r)!} r \left( \frac{1}{2} \right)^r \left( \frac{1}{2} \right)^{9-r}$$

Al anterior resultado se aplica el método asociado al uso de la ecuación \eqref{eq:ch1_identidad_dist_binomial} y se obtiene el valor esperado.

$$\E{E} = E_1 \E{r} = \frac{9}{2} E_1$$

En el cálculo de la incertidumbre se parte de $\E{E^2}$.

$$\E{E^2}  = \frac{1}{512}\sum_{r=1}^9 \binom{9}{r} r^2 E_1^2 = \frac{E_1^2}{512}\sum_{r=1}^9 \frac{9!}{r!(9-r)!} r^2 = E_1^2\sum_{r=1}^9 \frac{9!}{r!(9-r)!} r^2 \left( \frac{1}{2} \right)^r \left( \frac{1}{2} \right)^{9-r}$$

Al anterior resultado se aplica el método asociado al uso de la ecuación \eqref{eq:ch1_identidad_dist_binomial} y se obtiene la incertidumbre.

$$\E{E^2} = E_1^2 \E{r^2} = \frac{9}{4}E_1^2 + \E{E}^2$$

$$\sigma = \sqrt{\E{E^2} - \E{E}^2} = \frac{3}{2}E_1$$

\newpage

\section{Número de estados accesibles $\Omega(E_i; \{n_i\})$ para un sistema con $N$ partículas e $I$ niveles de energía}
\label{sec:omega_sistema_N_particulas_I_niveles}

Hemos avanzado en la comprensión de sistemas compuestos por partículas y su análisis en el marco de la física estadística a partir del uso de ensambles. Ahora vamos a generalizar lo estudiado a un caso más general. El objetivo de esta generalización es aplicar la teoría que hemos desarrollado hasta el momento a la mayor variedad posible de sistemas. Para lograr este objetivo se procederá de la siguiente manera: primero se plantea el problema y se formula una pregunta a resolver; segundo, se sintetizan los resultados obtenidos hasta ahora; tercero, se establecen unas definiciones orientadas a enriquecer nuestro vocabulario y facilitar del proceso de aprendizaje y por último, se resuelve la pregunta planteada.

Sea un sistema formado por $N$ partículas distribuidas en $I$ estados con energías $E_i$, donde $i=1, 2, \cdots, I$. Si en el estado $i$-ésimo hay $n_i$ partículas, ¿cuánto es el número de estados accesibles $\Omega(E_i)$ con energía $E_i$?

Del ejemplo estudiado en el apartado anterior se obtuvo que cuando se tienen $N$ partículas, cada una con energía $E_1$, el número de estados accesibles está dado por,

$$\Omega(E_r) = \binom{N}{r} = \frac{N!}{r!(N-r)!}$$

Si se suponen dos posibles estados de energía, $E_0 = 0$ y $E_1 \neq 0$, cuando se tienen $r$ partículas en el sistema, es equivalente a decir que se tienen $n_1 = r$ partículas en el estado $E_1$ y $n_0 = N - n_1$, en el estado $E_0$; en esta descripción $n_1$ es el número de ocupación para el estado con energía $E_1$ y $n_0$, el número de ocupación para las partículas con energía $E_0$. De esta manera, el número de estados accesibles para la energía $E_r$ se puede re-escribir de la siguiente manera,

$$\Omega(E_r) = \frac{N!}{n_1!n_0!}$$

La anterior forma del número de estados accesibles se puede interpretar como el cociente entre el número total de posibles permutaciones de las partículas, dado por $N!$ y el producto entre el número total de posibles permutaciones de partículas con energía $E_1$, dado por $n_1!$ y las partículas con energía $E_0$, dado por $n_0$.  

Si llamamos configuración al conjunto ordenado formado por los números de ocupación, $\{ n_1, n_2, \cdots, n_I \}$, donde la energía del sistema está dada por la ecuación \eqref{eq:ch2_energia_configuracion}, en el caso de dos posibles estados de energía, donde la energía de uno de estos es cero, solo existe una configuración para cada valor $E_r$: $\{n_0, n_1\} = \{r, N-r\}$. Sin embargo, al agregar más posibles valores de energía en los que pueda encontrarse la partícula, habrá más posibles configuraciones asociadas con un mismo valor de energía.

\begin{definition}
    Sea un sistema formado por $N$ partículas, tal que $n_i$ partículas tienen energía $E_i$, donde $i=1, 2, \cdots, I$. Al número de partículas $n_i$ con energía $E_i$ se le conoce como \textbf{número de ocupación} del estado $i$ y cumple con la ecuación \eqref{eq:ch2_numero_ocupacion}.
    \begin{equation}
        \label{eq:ch2_numero_ocupacion}
        N = \sum_{i=1}^I n_i
    \end{equation}
    \index{Número de ocupación}
\end{definition}

\begin{definition}
    Al conjunto ordenado $\{ n_1, n_2, \cdots, n_I \}$ formado por los números de partículas con energías $E_i$ se le conoce como \textbf{configuración} y se denota como \gls{configuration}.
    \index{Configuración}
\end{definition}

\begin{definition}
    Se conoce como \textbf{energía de la configuración} \gls{configuration} a la suma de la energía de las partículas que conforman el sistema y está dada por la ecuación \eqref{eq:ch2_energia_configuracion}.
    \begin{equation}
        \label{eq:ch2_energia_configuracion}
        E = \sum_{i=1}^I n_i E_i
    \end{equation}
    \index{Energía de la configuración}
\end{definition}

Esto quiere decir que para $N!$ partículas en la configuración $\{ n_1, n_2, \cdots, n_I \}$, el número de estados accesibles para el sistema con energía $E$ está dado por la ecuación \eqref{eq:ch2_numero_estados_accesibles_ocupacion}.\index{Ensamble microcanónico!número de estados accesibles para una configuración}

\begin{equation}
    \label{eq:ch2_numero_estados_accesibles_ocupacion}
    \Omega(E; \{n_i\}) =\frac{N!}{n_1!n_2!\cdots n_I!} 
\end{equation}

Como hay más de una posible configuración con energía $E$, el número de estados accesibles para una energía $E$ está dado por la suma de todas las configuraciones, expresada por la ecuación \eqref{eq:ch2_numero_estados_accesibles_N_particulas}, donde el término $\{n_i\}$ en la suma indica que esta se realiza para todas las posibles configuraciones con energía $E$.

\begin{equation}
    \label{eq:ch2_numero_estados_accesibles_N_particulas}
    \Omega(E) = \sum_{\{n_i\}} \Omega(E; \{n_i\}) = \sum_{\{n_i\}} \frac{N!}{n_1!n_2!\cdots n_I!}
\end{equation}

\newpage
\section{Postulados de la física estadística}

Después de analizar sistemas aislados y aplicar de manera conjunta conceptos de la Física y la Estadística, se pueden enunciar tres postulados en los cuales basar nuestro estudio. Estos postulados se formulan suponiendo que se tiene un sistema aislado con energía entre $E$ y $E+dE$, cuyos microestados son descritos por la Física y la distribución de los mismos, por la Estadística.

\begin{itemize}
    \item \textbf{Primer postulado}. Se asume válida la hipótesis ergódica.\index{Hipótesis ergódica}
    \item \textbf{Segundo postulado}. Principio de equiprobabilidad de microestados: en equilibrio termodinámico, todo microestado tiene la misma probabilidad de ocurrencia.\index{Principio de equiprobabilidad}
    \item \textbf{Tercer postulado}. El equilibrio termodinámico corresponde con el macroestado con máxima probabilidad.\index{Equilibrio termodinámico}
\end{itemize}
\index{Postulados de la física estadística}

De manera general, la aplicación de estos postulados implica cuatro pasos a seguir en la formulación de un problema en Física Estadística, \textit{primero}, se determina la regla para enumerar los estados del sistema, a esto se le conoce como especificación del estado del sistema; \textit{segundo}, se construye un ensamble de sistemas: \textit{tercero} se calcula la probabilidad y \textit{cuarto}, se calculan las cantidades físicas de interés.

\begin{definition}
	El \textbf{macroestado} de un sistema es el estado determinado por todos sus parámetros macroscópicos (o externos).\index{Macroestado}.
\end{definition}

\begin{definition}
	Al conjunto de parámetros que determinan el macroestado de un sistema se les conoce como \textbf{parámetros macroscópicos} o parámetros externos. Un ejemplo de ellos corresponde con la presión, volumen y temperatura que definen el macroestado de un gas ideal.\index{Parámetros macroscópicos}\index{Parámetros externos}
\end{definition}

\section{Sistema formado por D dados con C caras}
\label{sec:D_dados_C_caras}

Sean $D$ dados, cada uno de ellos con $C$ caras, tales que un número natural único se utiliza para identificar cada cara, ¿cuál es el valor más probable obtenido al sumar los números en cada cara después de lanzar los $D$ dados?

\subsection{Primer postulado: hipótesis ergódica} 

Se asume que al conformar un ensamble de sistemas aislados, las cantidades estadísticas calculadas en el ensamble son iguales que las obtenidas en un proceso aleatorio para tiempos largos.

\subsection{Segundo postulado: principio de equiprobabilidad}

Antes de aplicar el principio de equiprobabilidad es necesario definir cómo se especifica el estado del sistema. El sistema en estudio corresponde a $D$ dados con $C$ caras y se asume aislado del resto del universo. La regla para especificar el estado del sistema consiste en identificar el número en la cada de cada dado; por ejemplo, si se tienen dos dados ($D=2$) de cuatro caras ($C=4$), el estado del sistema se determina por dos números $\{c_1, c_2\}$, correspondientes al resultado de lanzar cada dado. Esto quiere decir que para identificar un microestado se requieren $D$ números, correspondientes a los valores obtenidos para los dados, $\{c_1, c_2, \cdots, c_D\}$, donde $c_d$ es el valor obtenido al lanzar el dado. Además, se puede asociar a la energía la suma de las caras de los dados, es decir, 

\begin{equation}
    \label{eq:ch2_ejemplo_dados_energia}
    E = \sum_{d=1}^D c_d
\end{equation}

Cada microestado, identificado por el conjunto de números naturales $\{c_1, c_2, \cdots, c_D\}$, tiene igual probabilidad de ocurrencia. Además, todos los valores $c_d$ se modelan como variables aleatorias uniformes discretas, con probabilidad $\prob{c_d = k} = 1/C$, siendo $k = 1, 2, \cdots, C$ y se asume que el lanzamiento de cada dado es estadísticamente independiente.

\subsection{Tercer postulado: equilibrio termodinámico}

El tercer postulado establece que el valor de energía $E$ con mayor número de microestados será el más probable. Esto es equivalente a decir que el valor esperado de la energía $\E{E}$ representa el estado en equilibrio termodinámico.

\subsection{Construcción del ensamble y obtención del valor esperado: solución numérica}

La solución numérica se obtiene al modelar el lanzamiento de cada dado con una variable aleatoria uniforme discreta, con probabilidad $p_d[k]=\prob{c_d = k}$, dada por,

\[
p_d[k] =
\begin{cases}
  \prob{c_d = k} = \frac{1}{C}, & \text{si } 1 \leq k \leq C \\
  0, & \text{en otro caso}
\end{cases}
\]

Al tener las variables aleatorias que modelan a cada dado, se hacen $N+1$ observaciones y para cada observación se obtiene la energía $E[n]$ como la suma de los valores de las caras, ecuación \eqref{eq:ch2_ejemplo_dados_energia}. 

Con el resultado de las observaciones se construye un ensamble, formado por el conjunto de valores $E[n]$. Para obtener la probabilidad asociada a cada valor de energía se realiza la siguiente aproximación: se define valor intervalo de energía $dE$ y se determina la frecuencia $f[n]$ de la energía en los rangos dados por $[ndE, ndE + dE)$, donde $n=0, 1, 2, \cdots, N$, de esta manera, la probabilidad $p(E)$ que la energía del microestado se encuentre entre $E$ y $E+dE$, donde $E=ndE$, se aproxima por,

$$p(E)=\frac{\Omega(E)}{\sum\limits_{E} \Omega(E)} \approx \frac{f[n]}{\sum\limits_{n=0}^N f[n]}$$

Entonces, el valor esperado de la energía se obtiene con la siguiente expresión,

$$\E{E} \approx \frac{\sum\limits_{n=0}^N  E[n] f[n]}{\sum\limits_{n=0}^N f[n]}$$

\begin{figure}[t]
    \begin{subfigure}{0.48\textwidth}
        \centering
        \includegraphics[width=0.95\linewidth]{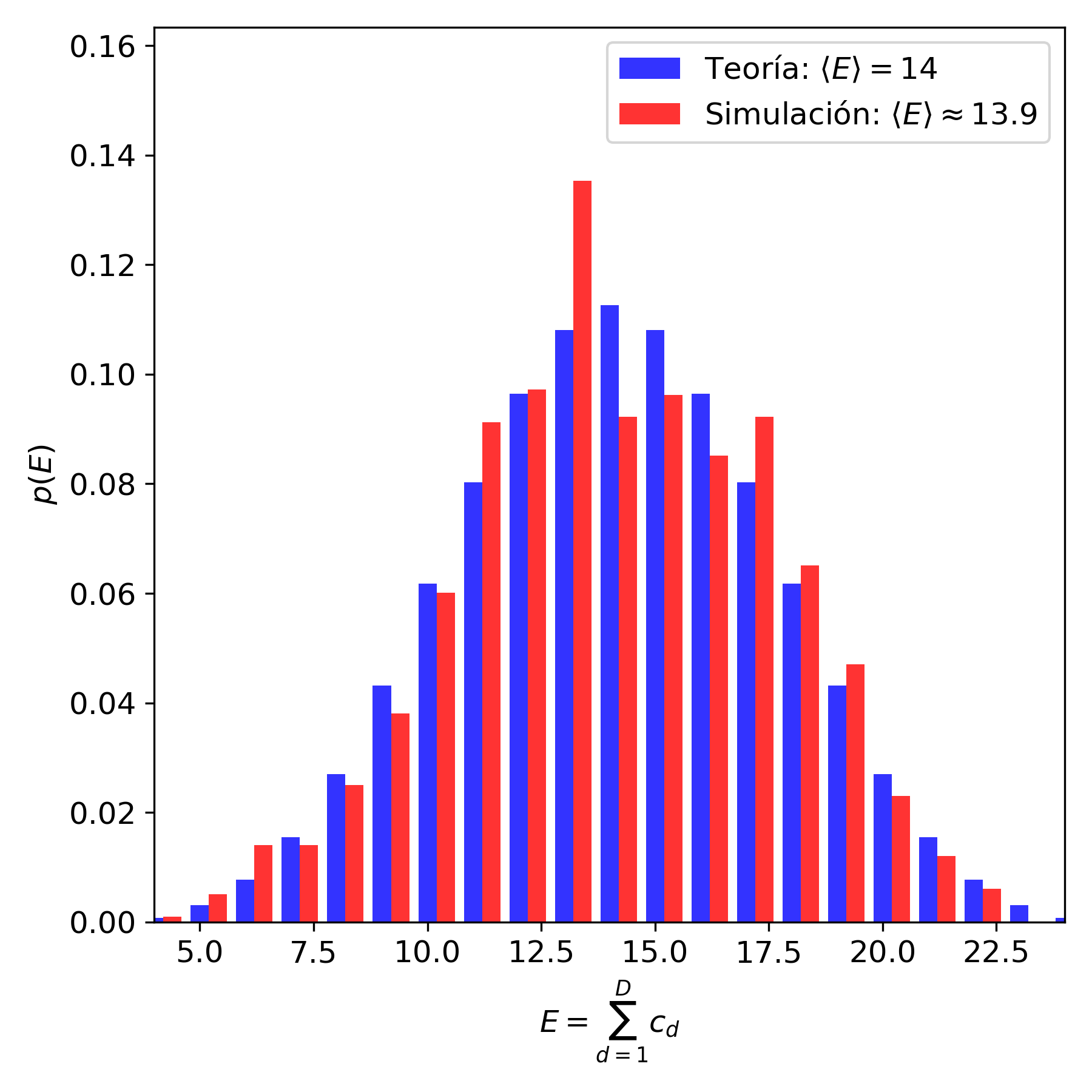}
        \caption{}
        \label{fig:ch2_ejemplo1_dados_distribucion_energia}
    \end{subfigure}
    \begin{subfigure}{0.48\textwidth}
        \centering
        \includegraphics[width=0.95\linewidth]{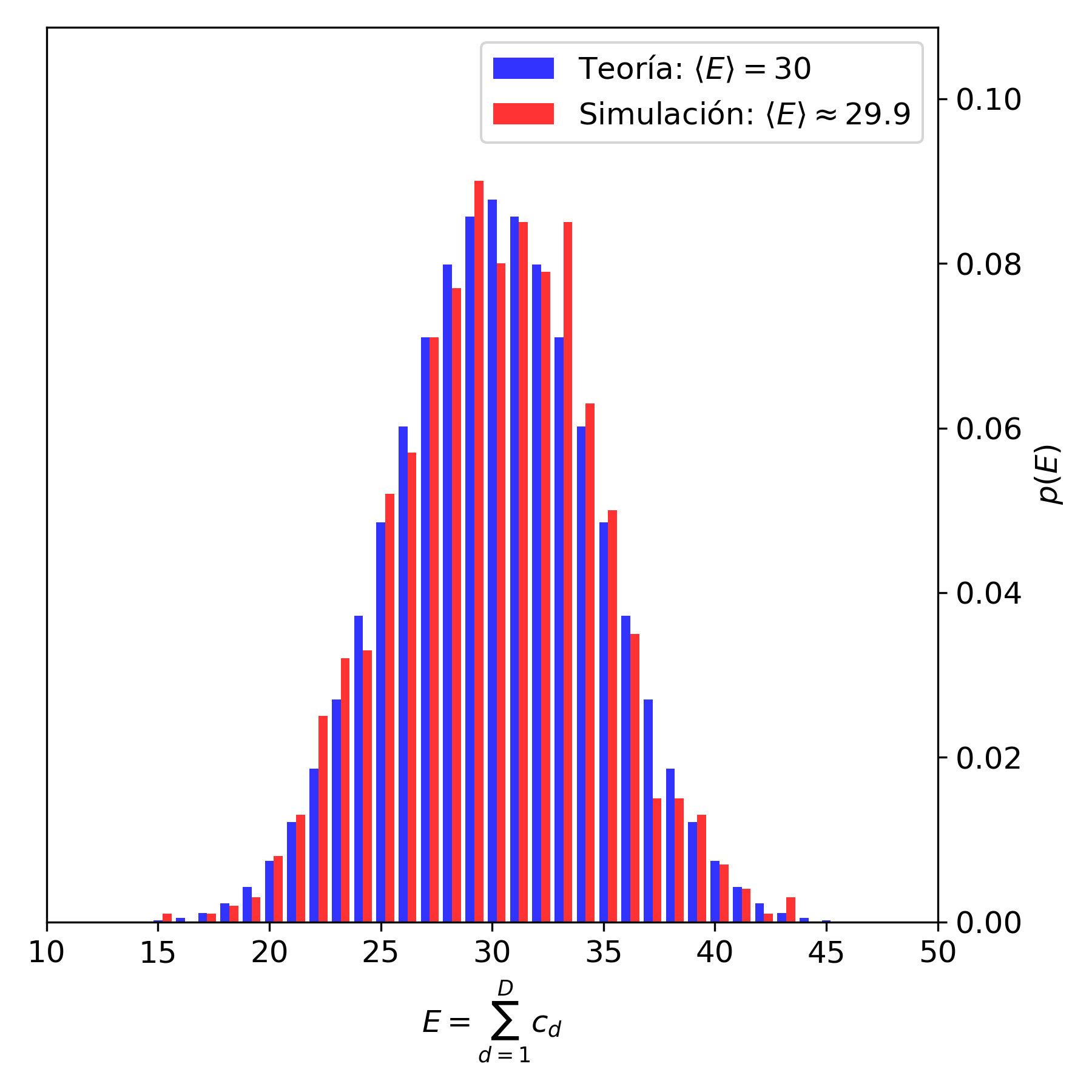}
        \caption{}
        \label{fig:ch2_ejemplo2_dados_distribución_energía}
    \end{subfigure}
    \caption{Probabilidad que al sumar las $C$ caras de $D$ dados se obtenga el valor $E$ para (a) $D=4$, $C=6$ y (b) $D=10$, $C=5$. Las barras en color azul muestran la predicción teórica obtenida analíticamente y las barras en color rojo, el valor numérico obtenido al realizar $10^3$ observaciones del experimento aleatorio.}
    \label{fig:ch2_ejemplo_dados_distribucion_energia}
\end{figure}

En la Figura~\ref{fig:ch2_ejemplo_dados_distribucion_energia} se observan los resultados obtenidos al realizar dos simulaciones en las que se siguió el proceso descrito en esta sección. Con el objetivo que el lector adquiera un mayor nivel de comprensión, se le sugiere que experimente con el ejemplo trabajado titulado \href{https://colab.research.google.com/github/davidalejandromiranda/StatisticalPhysics/blob/main/notebooks/es_DadosExperimento.ipynb}{descripción de un sistema formado por dados}, el cual está disponible en el repositorio \href{https://colab.research.google.com/github/davidalejandromiranda/StatisticalPhysics}{GitHub}.

\subsection{Solución analítica}

El Teorema \ref{lema:probabilidad_suma_variables_aleatorias} implica que, como cada dado se modela por una variable aleatoria uniforme $c_d$, la probabilidad $p(E=c_1+c_2+\cdots +c_D)$ está dada por la convolución de las probabilidades $p_d[k] = \prob{c_d=k}$,

$$p(E=c_1+c_2+ \cdots +c_D) = p_1[n] * p_2[n] * \cdots * p_D[n]$$

La propiedad asociativa de la convolución implica que esta se puede realizar por pares, donde el primer par está dado por la ecuación \eqref{eq:ch1_convolucion_discreta} y se resuelve de la siguiente manera,

\[
p_1[n] * p_2[n] = \sum_{k=-\infty}^\infty p_1[k] p_2[n-k] = \frac{1}{C} \sum_{k=1}^C p_2[n-k]
\]

Donde,

\[
p_2[n-k] =
\begin{cases}
  \frac{1}{C}, & \text{si } n-C \leq k \leq n - 1  \\
  0, & \text{en otro caso}
\end{cases}
\]

Con lo cual,

\[
p_1[n] * p_2[n] = \frac{1}{C^2} \sum_{k=\max\{1, n - C\}}^ {\min\{n-1, C\}} p_2[n-k] =  \begin{cases}
  \frac{n-1}{C^2}, & \text{si } 2 \leq n \leq C  \\
  \frac{2C - n +1}{C^2}& \text{si } C < n \leq 2C  \\
  0, & \text{en otro caso}
\end{cases}
\]

El siguiente paso es obtener $\left(p_1[n]*p_2[n]\right)*p_3[n]$ y al repetir este proceso hasta completar las $D-1$ convoluciones, se obtiene la probabilidad analítica, de la cual se muestran, como barras azules, dos ejemplos en la Figura \ref{fig:ch2_ejemplo_dados_distribucion_energia}. Con la probabilidad, el valore esperado para la energía, que corresponde con el valor más probable, se obtiene de la manera usual, $\E{E} = \sum\limits_{E} E p(E)$. 
\section{Teorema $\HBoltzmann$}
\index{Teorema H}

Se suele atribuir a Boltzmann identificar la relación entre la descripción estadística de los sistemas y la entropía. En 1872 Boltzmann postuló el teorema $\HBoltzmann$, que plantea que la cantidad $\HBoltzmann$, definida como el valor esperado del logaritmo natural de una probabilidad, puede permanecer constante o disminuir en el tiempo \cite{Boltzmann1872}. Dos décadas después que Boltzmann postuló el teorema $\HBoltzmann$, Zermelo (1896) presentó argumentos para demostrar la invalidez general de dicho teorema \cite{Zermelo1896}; el principal argumento de Zermelo se basó en el teorema de recurrencia de Poincaré (ver Teorema \ref{lema:teorema_recurrencia_Poincare}), el cual establece que un sistema encerrado en un volumen finito alcanzará nuevamente un estado, en el espacio de fase, en el que estuvo previamente, por lo tanto, si la entropía del sistema se asocia al estado del sistema, entonces, esta no necesariamente siempre aumentará (disminución de la cantidad $\HBoltzmann$). Zermelo fue el primero en presentar contra-argumentos al teorema $\HBoltzmann$, sin embargo, a lo largo de la historia, el teorema $\HBoltzmann$ ha seguido siendo objeto de diversas críticas, como se analiza detalladamente en el trabajo de Brown y colaboradores \cite{Brown2009}. Las críticas más destacadas incluyen señalamientos sobre la falta de universalidad y la inconsistencia con el principio de reversibilidad en el marco de la mecánica clásica. No obstante, a pesar de los múltiples argumentos sobre la invalidez, en un contexto general, del teorema $\HBoltzmann$, es conveniente abordarlo en el contexto de la mecánica cuántica para partículas no interactuantes, siguiendo el enfoque descrito por Reif \cite[pp. 624-626]{Reif1965}; con este enfoque, se pueden encontrar argumentos (con validez limitada, pero académicamente valiosos) que evidencian la relación de la entropía con el número de estados accesibles $\Omega(E)$. 

Antes de continuar, es importante anotar que el siguiente análisis tiene un rango de validez limitado a las suposiciones realizadas, sin embargo, hasta donde se conoce, la segunda ley de la termodinámica es válida para todo sistema aislado; esto quiere decir que el análisis presentado a continuación es un buen ejercicio académico para dilucidar relaciones entre la concepción mecánico cuántica de la naturaleza y la concepción mecánica estadística de la misma. Algunos autores, como Barbara Drossel, señalan que la concepción errónea radica en tratar de deducir la segunda ley de la termodinámica a partir de concepciones determinísticas de la realidad\footnote{Drossel incluye tanto la descripción clásica como la cuántica en las concepciones determinísticas de la realidad, en el sentido que la dinámica del estado de un sistema en ambas teorías obedecen a unas ecuaciones determinísticas: ecuaciones canónicas de Hamilton para la descripción hamiltoniana de sistemas clásicos y la ecuación de Scrhödinger, para la mecánica cuántica.}, lo cual hace que se parta de la suposición errónea que es posible, de alguna manera, identificar con precisión el estado inicial (o cualquier estado) de un sistema en el espacio de fase \cite{Drossel2015}. Ante estos argumentos, se podría pensar que la especificación del estado de un sistema no necesariamente debe ser realizado en el espacio de fase, dado que el estado de un sistema se puede especificar, en mecánica cuántica, por los números cuánticos de un conjunto completo de observables compatibles\footnote{En mecánica cuántica, un conjunto completo de observables compatibles a un conjunto de operadores hermíticos que conmutan entre si y definen una base para especificar cualquier estado del sistema.} asociados al sistema, sin necesidad de definir dicho estado en el espacio de fase; esta es una discusión actual que está abierta y seguramente suscitará debate académico en los próximos años.

\begin{definition}
    Sea $p_r$ la probabilidad que un sistema se encuentre en un microestado con energía $E_r$. La \textbf{cantidad H de Boltzmann} se define como el valor esperado del logaritmo natural de la probabilidad $p$ y está dada por la ecuación \eqref{eq:ch3_cantidad_H_Boltzmann}.
    \begin{equation}
        \HBoltzmann = \E{\ln p} = \sum_r p_r\ln p_r
        \label{eq:ch3_cantidad_H_Boltzmann}
    \end{equation} \index{Cantidad H de Boltzmann}
\end{definition}

\begin{lema}
    \label{lema:teorema_H_Boltzmann}
    \textbf{Teorema $\HBoltzmann$ de Boltzmann}. En el tiempo la cantidad $\HBoltzmann$, dada por la ecuación \eqref{eq:ch3_cantidad_H_Boltzmann}, permanece constante (la probabilidad no cambia) o disminuye en el tiempo.
    \begin{equation}
        \frac{d\HBoltzmann}{dt} \leq 0
    \end{equation}
\end{lema}

Supongamos que se tiene un sistema descrito por hamiltoniano mecánico cuántico $\hat{H}$, invariante en el tiempo\footnote{La condición de invarianza en el tiempo permite expresar la evolución temporal de un microestado como $\ket{\psi_r(t)} = e^{-iE_rt/\hbar}\ket{\psi_r(0)}$.}, tal que $p_r(t)$ es la probabilidad que, en un cierto instante de tiempo $t$, el sistema se encuentre en un microestado $\ket{\psi_r}$ y $W_{rs}$, la probabilidad que el sistema pase, con el tiempo, del microestado $\ket{\psi_r}$ a $\ket{\psi_s}$. La variación respecto al tiempo de la cantidad $\HBoltzmann$ está dada por,

\[
\frac{d\HBoltzmann}{dt} = \frac{d}{dt} \sum_r p_r \ln p_r = \sum_r \frac{dp_r}{dt} ( 1 + \ln p_r)  
\]

\begin{figure}[th]
    \centering
    \includegraphics[width=0.6\linewidth]{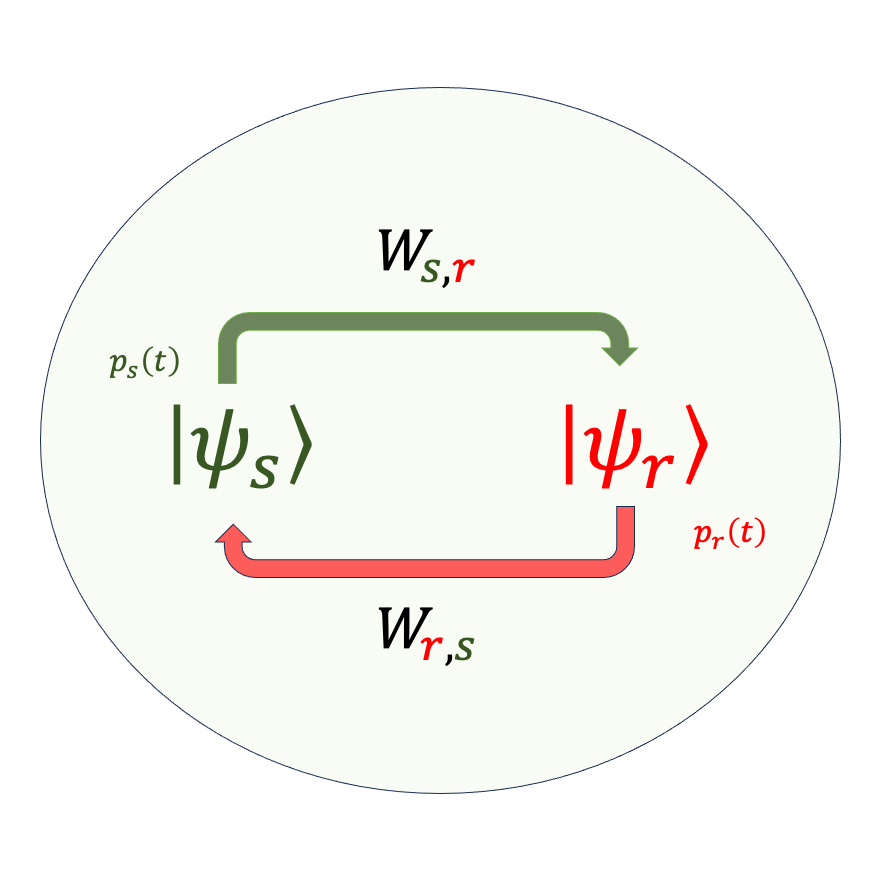}
    \caption{Representación esquemática de un sistema aislado donde se muestran dos estados, $\ket{\psi_r}$, con probabilidad $p_r(t)$ y $\ket{\psi_s}$, con probabilidad $p_s(t)$. Si el sistema tiene probabilidad $W_{s,r}$ que el microestado cambie de $\ket{\psi_s}$ a $\ket{\psi_r}$ y probabilidad $W_{r,s}$ que el microestado cambie de $\ket{\psi_r}$ a $\ket{\psi_s}$, entonces, la variación de la probabilidad $p_r(t)$ respecto al tiempo corresponde con la diferencia entre el aumento de probabilidad debido a $W_{s,r}$ menos la disminución dada por $W_{r,s}.$}
    \label{fig:ch2_variacion_probabilidad_teorema_H_Boltzmann}
\end{figure}

Donde $\frac{dp_r}{dt}$ corresponde con la diferencia entre el aumento de la probabilidad de encontrar el sistema en el microestado $\ket{\psi_r}$ debida a cambios de estados desde $\ket{\psi_s}$ a $\ket{\psi_r}$, menos la probabilidad que el microestado del sistema pase de $\ket{\psi_r}$ a $\ket{\psi_s}$, como se muestra esquemáticamente en la Figura \ref{fig:ch2_variacion_probabilidad_teorema_H_Boltzmann},

\[
\frac{dp_r}{dt} = \sum_s ( p_s W_{s,r} - p_r W_{r, s})
\]

Como se está asumiendo que el sistema se encuentra aislado, si se supone que cada microestado es aproximadamente independiente de los otros (interacción despreciable entre microestados) la energía $E_r$, según la mecánica cuántica, corresponde con uno de los valores propios del hamiltoniano $\hat{H}$, por lo tanto, $\hat{H}\ket{\psi_r} = E_r \ket{\psi_r}$ y la probabilidad $W_{r,s}$ es aproximadamente igual a $W_{r,s}$, como se muestra a continuación.\footnote{Al asumir que el sistema está aislado, se puede considerar que su energía está especificada, $E = E_r$, por lo tanto, el índice $s$ denotaría un microestado dentro de los posibles microestados con energía $E_r$. Otra posibilidad, más general, es definir un rango de energías entre $E$ y $E+dE$, en tal caso, en la expresión para $W_{r,s}$ se debe utilizar el proyector de estados para determinar la proyección entre estados con diferentes energía; sin embargo, en dicho caso más general, la aproximación $W_{r,s}\approx W_{s,r}$ sigue siendo válida.}

\[ 
W_{r,s} \approx \abs{\braket{\psi_s}{\psi_r}}^2 = \abs{\braket{\psi_r}{\psi_s}}^2 \approx W_{s,r}
\]

Con el anterior resultado, la variación en el tiempo de la cantidad $\HBoltzmann$ se puede escribir de dos maneras, como muestran las ecuaciones \eqref{eq:ch3_variacion_cantidad_H_1} y \eqref{eq:ch3_variacion_cantidad_H_2}. La justificación de estas dos formas equivalentes radica en que no debe existir diferencias si se cambian los índices de las ecuaciones; al sumar ambas ecuaciones, teniendo en cuenta que $W_{r,s} \approx W_{s, r}$ y despejando la variación en el tiempo de la cantidad $\HBoltzmann$, se obtiene la ecuación \eqref{eq:ch3_variacion_cantidad_H_3}

\begin{equation}
    \label{eq:ch3_variacion_cantidad_H_1}
    \frac{d\HBoltzmann}{dt} = \sum_{r,s} W_{r,s} (p_s - p_r) ( 1 + \ln p_r)  
\end{equation}

\begin{equation}
    \label{eq:ch3_variacion_cantidad_H_2}
    \frac{d\HBoltzmann}{dt} = \sum_{r,s} W_{s,r} (p_r - p_s) ( 1 + \ln p_s)  
\end{equation}

\begin{equation}
    \label{eq:ch3_variacion_cantidad_H_3}
    \frac{d\HBoltzmann}{dt} = -\frac{1}{2}\sum_{r,s} W_{r,s} (p_r - p_s) ( \ln p_r - \ln p_s)  
\end{equation}

La ecuación \eqref{eq:ch3_variacion_cantidad_H_3} tiene tres posibles soluciones, primero, si $p_r = p_s$, entonces, $d\HBoltzmann/dt = 0$. Segundo, si $p_r > p_s$, entonces, $\ln p_r > \ln p_s$, por lo tanto, el signo del producto $(p_r - p_s) ( \ln p_r - \ln p_s)$ es positivo y $d\HBoltzmann/dt < 0$. Tercero, si $p_r < p_s$, entonces, $\ln p_r < \ln p_s$, por lo tanto, el signo del producto $(p_r - p_s) ( \ln p_r - \ln p_s)$ es positivo y $d\HBoltzmann/dt < 0$. En conclusión,

\[
\frac{d\HBoltzmann}{dt} \leq 0
\]

\section{Entropía de Boltzmann}
\index{Ensamble microcanónico!entropía de Boltzmann}

\begin{definition}
    La \textbf{entropía de Boltzmann}, $S$, para un sistema aislado con energía entre $E$ y $E+dE$, cuyo número de estados accesibles es $\Omega(E)$, está dada por la ecuación \eqref{eq:ch2_entropia_Boltzmann}, donde $k_B$ es la constante de Boltzmann.
    \begin{equation}
        \label{eq:ch2_entropia_Boltzmann}
        S = k_B \ln \Omega(E)
    \end{equation} \index{Entropía de Botlzmann}
\end{definition}

Boltzmann postuló que la entropía de un sistema aislado con energía entre $E$ y $E+dE$ es proporcional al logaritmo natural del número de estados accesibles. Los pasos lógicos que condujeron a Boltzmann a realizar esta definición se basan en el teorema $\HBoltzmann$. 

En esta sección se hará un análisis basado en el enfoque de Boltzmann, pero considerando un sistema con energía cuantizada, como se estudió en la sección \ref{sec:omega_sistema_N_particulas_I_niveles}, donde se encontró que el número de estados accesibles para una cierta configuración $\{n_i\}$ está dado por la ecuación $\eqref{eq:ch2_numero_estados_accesibles_ocupacion}$ y el número total de estados accesibles con energía entre $E$ y $E+dE$, por la ecuación \eqref{eq:ch2_numero_estados_accesibles_N_particulas}; por lo tanto, la probabilidad de encontrar al sistema en una configuración $\{n_i\}$ está dada por la ecuación \eqref{eq:ch2_probabilidad_estados_accesibles_ocupacion}, donde $r$ ahora representa la configuración $\{n_i\}$. 

\begin{equation}
    \label{eq:ch2_probabilidad_estados_accesibles_ocupacion}
    \prob{n_1 = \eta_1, n_2 = \eta_2, \cdots, n_I = \eta_I} = \frac{\Omega(E; \{n_i\})}{\Omega(E)}= \frac{\Omega(E; r)}{\Omega(E)} = p(\{n_i\})
\end{equation}

Para calcular la cantidad $\HBoltzmann$ se necesita calcular $\ln p_r$ para $N$ grandes. Al utilizar la aproximación de Stirling se obtiene,  

\[
\ln p(\{n_i\}) = \ln \left[ \frac{N!}{n_1!n_2!\cdots n_I!} \right] - \ln \Omega(E) \approx N\ln N - N - \sum_{i=1}^{I} ( n_i \ln n_i - n_i ) - \ln \Omega(E)
\]

Como $\sum\limits_{i=1}^{I} n_i = N$,

\[
\ln p(\{n_i\}) \approx N \ln N - \ln \Omega(E) - \sum_{i=1}^{I} n_i \ln n_i 
\]

De esta manera, la cantidad $\HBoltzmann$ se obtiene como,

\[
\HBoltzmann = \E{\ln p_r} \approx N \ln N - \ln \Omega(E) - \E{ \sum_{i=1}^{I} n_i \ln n_i }
\]

Donde $\E{ \sum\limits_{i=1}^{I} n_i \ln n_i }$ corresponde con el valor esperado de la suma $\sum\limits_{i=1}^{I} n_i \ln n_i$, el cual se obtiene para la configuración $\{n_i^{*}\}$ con mayor número de estados accesibles $\Omega(E, \{n_i^{*}\})$. Como el número de partículas $N = \sum\limits_{i=1}^I n_i$ es una constante en el tiempo, y $d\HBoltzmann / dt \leq 0$, se obtiene que,

\[
   \frac{d \ln \Omega(E)}{dt} + \sum_{i=1}^{I} \frac{d }{dt} (n_i^{*} \ln n_i^{*}) \geq 0
\]

Esta expresión implica dos casos, \textit{primero}, que el número de estados $\Omega(E)$ permanezca constante en el tiempo, entonces, el valor esperado $\E{ \sum\limits_{i=1}^{I} n_i \ln n_i }$ se mantiene constante y se obtiene el caso extremo $d \ln \Omega(E)/dt = 0$. \textit{Segundo}, si el número de estados $\Omega(E)$ aumenta en el tiempo, se obtiene una nueva condición de equilibrio y el valor esperado $\E{ \sum\limits_{i=1}^{I} n_i \ln n_i }$ cambia de $\sum\limits_{i=1}^{I} n_i^{*} \ln n_i^{*}$ a un nuevo valor esperado $\sum\limits_{i=1}^{I} n_i^{**} \ln n_i^{**}$, el cual permanecerá constante a menos que vuelva a cambiar la energía del sistema. Por esta razón y teniendo en cuenta la definición de la entropía de Boltzmann, ecuación \eqref{eq:ch2_entropia_Boltzmann}, se tiene que el teorema $\HBoltzmann$ implica que la entropía de Boltzmann permanece igual o aumenta en el tiempo, consistente con la segunda ley de la termodinámica,

\begin{equation}
    \frac{dS}{dt} \geq 0
\end{equation}

\section{Entropía de Shannon}
\index{Entropía de Shannon}

A partir de la aplicación de los conceptos de la física estadística, en 1948 Shannon propuso la teoría matemática de la información \cite{Shannon1948}. En su teoría, Shannon definió la entropía $H(X)$ como menos el valor esperado del logaritmo en base $b$ de la probabilidad $p(x) = \prob{X=x}$ que al observar una variable aleatoria $X$ se observe un valor $x$. Dependiendo la base $b$ del logaritmo, la entropía de Shannon se expresa en \textit{bits}, para $b=2$; \textit{nepit} (o \textit{nat}) si $b=e$ y \textit{dits} (también conocida como \textit{bans} o \textit{hartleys}), para $d=10$. La entropía de Shannon para una señal $x(t)$ es una medida de la cantidad de información que esta contiene.

\begin{definition}
    Sea una variable aleatoria $X$ con probabilidad $p(x) = \prob{X=x}$ que al observar una variable aleatoria $X$ se observe un valor $x$. La \textbf{entropía de Shannon} $H(X)$ se define como,
    \begin{equation}
        H(X) = -\E{\log_b p(x)} = - \sum\limits_{x}p(x) \log_b p(x)
    \end{equation} \index{Entropía de Shannon}
\end{definition}

\section{Interacción entre sistemas}

En la naturaleza la energía se manifiesta de maneras diferentes, entre ellas se encuentra el calor, que es una forma de energía que fluye entre sistemas a diferente temperatura. La interacción entre sistemas donde solo se intercambia calor se conoce como interacción térmica. Cuando el intercambio de energía se produce sin transferencia de calor, pero uno de los sistemas realiza trabajo sobre el otro, se dice que se tiene una interacción mecánica; la combinación de ambos tipos de interacción corresponde con las interacciones termodinámicas. 

\begin{figure}[t]
    \centering
    \includegraphics[width=0.55\linewidth]{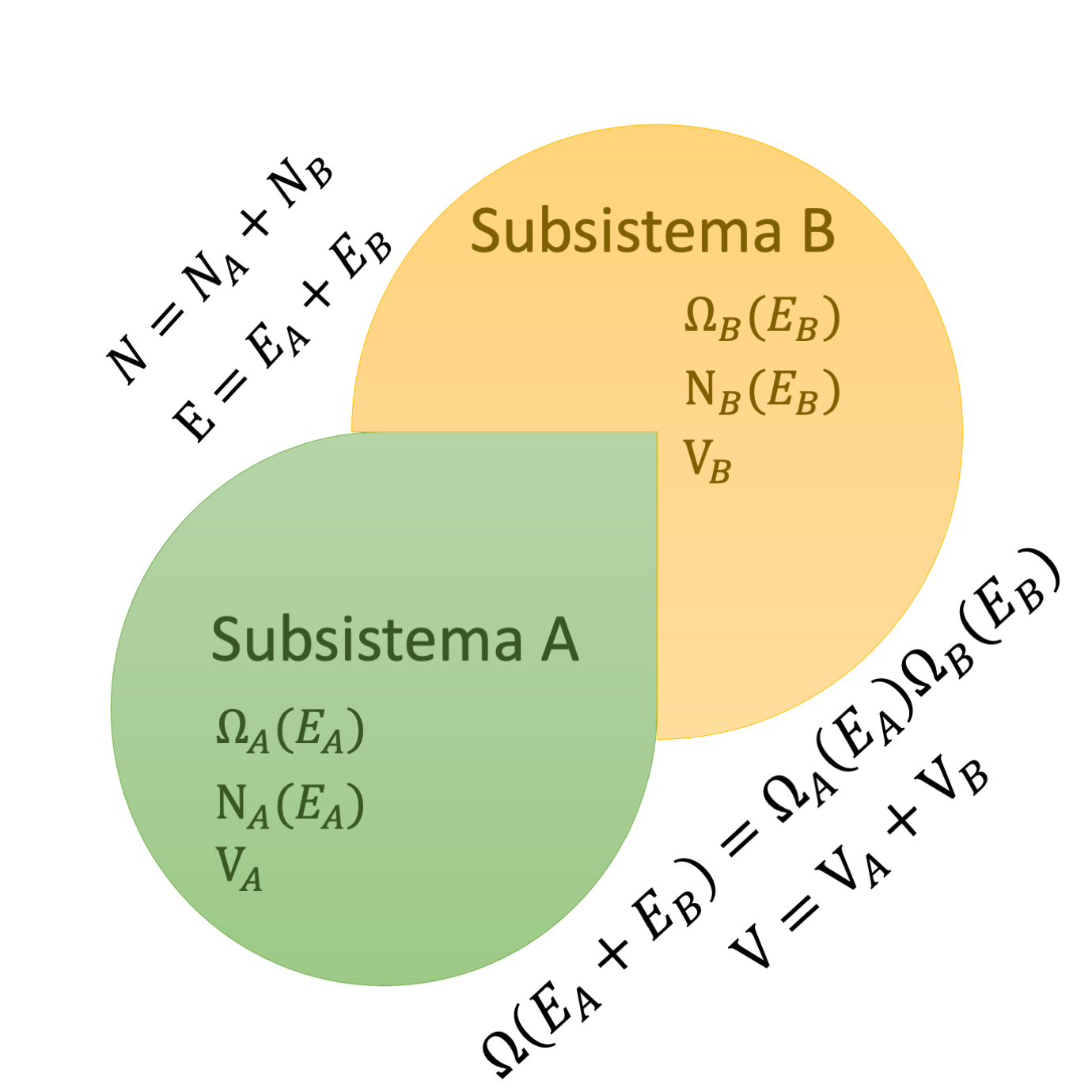}
    \caption{Sistema con dos subsistemas, $A$ y $B$. El subsistema $A$ tiene energía $E_A$, número de partículas $N_A$ y número de estados accesibles $\Omega_A(E_A)$. El sistema $B$ tiene energía $E_B$, número de partículas $N_B$ y número de estados accesibles $\Omega_B(E_B)$. La energía total del sistema $E$, es la suma de la energía de sus subsistemas, $E=E_A+E_B$; de la misma manera, el número total de partículas $N$ en el sistema, es la suma del número de partículas en los subsistemas, $N=N_A+N_B$. El número de estados accesibles $\Omega(E)$ del sistema corresponde al producto del número de estados accesibles de los subsistemas, $\Omega(E=E_A+E_B) = \Omega_A(E_A) \Omega_B(E_B)$.}
    \label{fig:ch2_sistema_dos_subsistemas}
\end{figure}

En esta sección vamos a estudiar la interacción entre sistemas, para ello asumiremos que tenemos un sistema aislado con energía $E$, número de partículas $N$, número de estados accesibles $\Omega(E)$ y volumen $V$, el cual está formado por dos subsistemas\footnote{En este planteamiento, la interacción entre sistemas independientes la modelamos como la interacción entre dos subsistemas de un sistema aislado.}, $A$ y $B$, como se muestra esquemáticamente en la Figura~\ref{fig:ch2_sistema_dos_subsistemas}, donde el subsistema $A$ tiene energía $E_A$, número de partículas $N_A$, número de estados accesibles $\Omega_A(E_A)$ y volumen $V_A$; el subsistema $B$ tiene energía $E_B$, número de partículas $N_B$, número de estados accesibles $\Omega_A(E_B)$ y volumen $V_B$. Los parámetros de estos subsistemas se relacionan con los del sistema aislado por las siguientes ecuaciones,

\begin{subequations}
\label{eq:ch2_relaciones_interaccion_sistemas}
\begin{align}
    E & = E_A + E_B \label{eq:energia} \\
    N & = N_A + N_B \label{eq:numero} \\
    V & = V_A + V_B \label{eq:volumen} \\
    \Omega(E) & = \Omega_A(E_A) \Omega_B(E_B) \label{eq:omega}
\end{align}
\end{subequations}

\begin{definition}
    Un \textbf{subsistema} es una parte de un sistema que por si misma forma un sistema; en tal sentido, un subsistema es un sistema que hace parte de otro sistema. \index{Subsistema}
\end{definition}

\begin{definition}
    El \textbf{calor} $Q$ es una forma de energía presente en la interacción entre sistemas a diferentes temperaturas o cuando se transfiere energía entre sistemas a una misma temperatura con variación de su entropía.
\end{definition}

\begin{definition}
    La \textbf{interacción térmica} ocurre cuando dos sistemas intercambian energía en forma de calor.
\end{definition}

\begin{definition}
    La \textbf{interacción mecánica} ocurre cuando dos sistemas intercambian energía en forma de trabajo; en tal caso, existe un parámetro externo $x$ que es modificado en el intercambio de energía.
\end{definition}

\begin{definition}
    La \textbf{interacción termodinámica} implica la presencia tanto de interacción térmica como mecánica.
\end{definition}

Antes de continuar, analicemos el número de estados accesibles $\Omega(E)$. Supongamos que el subsistema $A$ se encuentra en el microestado $r$ con energía $E_A$ y que los microestados del sistema $B$ son estadísticamente independientes de los del subsistema $A$; esto implica que para el microestado $r$ en el subsistema $A$ se tienen $\Omega_B(E_B)$ microestados en el subsistema $B$, entonces, el número de estados accesibles $\Omega(E)$ del sistema se obtendrá al sumar todos los $r$ posibles microestados del sistema $A$, teniendo en cuenta los microestados del sistema $B$, es decir,

\[
\Omega(E=E_A+E_B) = \sum\limits_{r=1}^{\Omega(E_A)} \Omega_B(E_B) = \Omega(E_A)\Omega_B(E_B)
\]

A continuación, se aplicará los elementos de la Física Estadística estudiados hasta el momento para describir sistemas en equilibrio termodinámico, además de interacciones térmicas, mecánicas y termodinámicas.

\subsection{Ejemplo ilustrativo: sistema con cuatro partículas y seis posibles niveles de energía}

\begin{table}[th]
\centering
\caption{Microestados (17 de los 80) de un sistema aislado con energía $10E_0$, formado por cuatro partículas que pueden tomar energía entre $E_0$ y $6E_0$ tal que se tienen dos subsistemas, $A$ y $B$, cada uno con dos partículas, cuyos microestados se identifican por los números naturales $n_{1}^{(A)}$, $n_{2}^{(A)}$, para el subsistema $A$ y $n_{1}^{(B)}$, $n_{2}^{(B)}$, para el subsistema $B$. En este caso (a) se considera que el microestado $(n_1, n_2)$ se puede distinguir de $(n_2, n_1)$, para un subsistema dado.}
\label{tab:ch2_microestados_10Eo_estados_distinguibles}
\begin{tabular}{|ccccccc|}
\hline
$n_{1}^{(A)}$ & $n_{2}^{(A)}$ & $n_{1}^{(B)}$ & $n_{2}^{(B)}$ & $E_A$ & $E_B$ & $E_T = E_A + E_B$ \\
\hline \hline
1 & 1 & 2 & 6 & 2$E_0$ & 8$E_0$ & 10$E_0$ \\
1 & 1 & 3 & 5 & 2$E_0$ & 8$E_0$ & 10$E_0$ \\
1 & 1 & 4 & 4 & 2$E_0$ & 8$E_0$ & 10$E_0$ \\
1 & 1 & 5 & 3 & 2$E_0$ & 8$E_0$ & 10$E_0$ \\
1 & 1 & 6 & 2 & 2$E_0$ & 8$E_0$ & 10$E_0$ \\ \hline
1 & 2 & 1 & 6 & 3$E_0$ & 7$E_0$ & 10$E_0$ \\
1 & 2 & 2 & 5 & 3$E_0$ & 7$E_0$ & 10$E_0$ \\
1 & 2 & 3 & 4 & 3$E_0$ & 7$E_0$ & 10$E_0$ \\
1 & 2 & 4 & 3 & 3$E_0$ & 7$E_0$ & 10$E_0$ \\
1 & 2 & 5 & 2 & 3$E_0$ & 7$E_0$ & 10$E_0$ \\
1 & 2 & 6 & 1 & 3$E_0$ & 7$E_0$ & 10$E_0$ \\
2 & 1 & 1 & 6 & 3$E_0$ & 7$E_0$ & 10$E_0$ \\
2 & 1 & 2 & 5 & 3$E_0$ & 7$E_0$ & 10$E_0$ \\
2 & 1 & 3 & 4 & 3$E_0$ & 7$E_0$ & 10$E_0$ \\
2 & 1 & 4 & 3 & 3$E_0$ & 7$E_0$ & 10$E_0$ \\
2 & 1 & 5 & 2 & 3$E_0$ & 7$E_0$ & 10$E_0$ \\
2 & 1 & 6 & 1 & 3$E_0$ & 7$E_0$ & 10$E_0$ \\
\hline
\end{tabular}
\end{table}

\begin{table}[th]
\centering
\caption{Número de estados accesibles para el sistema con microestados descritos en la Tabla \ref{tab:ch2_microestados_10Eo_estados_distinguibles}.}
\label{tab:ch2_estados_accesibles_10Eo_estados_distinguibles}
\begin{tabular}{|cccccc|}
\hline
$E_A / E_0$ & $E_B / E_0$ & $(E_A + E_B)/E_0$ & $\Omega_A(E_A)$ & $\Omega_B(E_B)$ & $\Omega_T=\Omega_A \Omega_B$ \\
\hline  \hline
2 & 8 & 10 & 1 & 5 & 5 \\
3 & 7 & 10 & 2 & 6 & 12 \\
4 & 6 & 10 & 3 & 5 & 15 \\
5 & 5 & 10 & 4 & 4 & 16 \\
6 & 4 & 10 & 5 & 3 & 15 \\
7 & 3 & 10 & 6 & 2 & 12 \\
8 & 2 & 10 & 5 & 1 & 5 \\
\hline
\end{tabular}
\end{table}

\begin{table}[ht]
\centering
\caption{Microestados (16 de los 28) de un sistema aislado con energía $10E_0$, formado por cuatro partículas que pueden tomar energía entre $E_0$ y $6E_0$ tal que se tienen dos subsistemas, $A$ y $B$, cada uno con dos partículas, cuyos microestados se identifican por los números naturales $n_{1}^{(A)}$, $n_{2}^{(A)}$, para el subsistema $A$ y $n_{1}^{(B)}$, $n_{2}^{(B)}$, para el subsistema $B$. En este caso (b) se considera que el microestado $(n_1, n_2)$ se puede distinguir de $(n_2, n_1)$, para un subsistema dado.}
\label{tab:ch2_microestados_10Eo_estados_indistinguibles}
\begin{tabular}{|ccccccc|}
\hline
$n_{1}^{(A)}$ & $n_{2}^{(A)}$ & $n_{1}^{(B)}$ & $n_{2}^{(B)}$ & $E_A$ & $E_B$ & $E_T = E_A + E_B$ \\
\hline \hline
1 & 1 & 2 & 6 & 2$E_0$ & 8$E_0$ & 10$E_0$ \\
1 & 1 & 3 & 5 & 2$E_0$ & 8$E_0$ & 10$E_0$ \\
1 & 1 & 4 & 4 & 2$E_0$ & 8$E_0$ & 10$E_0$ \\ \hline
1 & 2 & 1 & 6 & 3$E_0$ & 7$E_0$ & 10$E_0$ \\
1 & 2 & 2 & 5 & 3$E_0$ & 7$E_0$ & 10$E_0$ \\
1 & 2 & 3 & 4 & 3$E_0$ & 7$E_0$ & 10$E_0$ \\ \hline
1 & 3 & 1 & 5 & 4$E_0$ & 6$E_0$ & 10$E_0$ \\
1 & 3 & 2 & 4 & 4$E_0$ & 6$E_0$ & 10$E_0$ \\
1 & 3 & 3 & 3 & 4$E_0$ & 6$E_0$ & 10$E_0$ \\
2 & 2 & 1 & 5 & 4$E_0$ & 6$E_0$ & 10$E_0$ \\
2 & 2 & 2 & 4 & 4$E_0$ & 6$E_0$ & 10$E_0$ \\
2 & 2 & 3 & 3 & 4$E_0$ & 6$E_0$ & 10$E_0$ \\ \hline
1 & 4 & 1 & 4 & 5$E_0$ & 5$E_0$ & 10$E_0$ \\
1 & 4 & 2 & 3 & 5$E_0$ & 5$E_0$ & 10$E_0$ \\
2 & 3 & 1 & 4 & 5$E_0$ & 5$E_0$ & 10$E_0$ \\
2 & 3 & 2 & 3 & 5$E_0$ & 5$E_0$ & 10$E_0$ \\
\hline
\end{tabular}
\end{table}

\begin{table}[ht]
\centering
\caption{Número de estados accesibles para el sistema con microestados descritos en la Tabla \ref{tab:ch2_microestados_10Eo_estados_indistinguibles}.}
\label{tab:ch2_estados_accesibles_10Eo_estados_indistinguibles}
\begin{tabular}{|cccccc|}
\hline
$E_A / E_0$ & $E_B / E_0$ & $(E_A + E_B)/E_0$ & $\Omega_A(E_A)$ & $\Omega_B(E_B)$ & $\Omega_T(E_A + E_B)$ \\
\hline \hline
2 & 8 & 10 & 1 & 3 & 3 \\
3 & 7 & 10 & 1 & 3 & 3 \\
4 & 6 & 10 & 2 & 3 & 6 \\
5 & 5 & 10 & 2 & 2 & 4 \\
6 & 4 & 10 & 3 & 2 & 6 \\
7 & 3 & 10 & 3 & 1 & 3 \\
8 & 2 & 10 & 3 & 1 & 3 \\
\hline
\end{tabular}
\end{table}

Sea un sistema aislado del resto del universo, con energía $10E_0$, formado por dos subsistemas, $A$ y $B$, cada uno con dos partículas, tal que cada partícula aporta al subsistema una energía $nE_0$ para $n=1,2,3,4,5,6$. Si los subsistemas pueden intercambiar energía, sin variar ni su número de partículas ni su volumen, y el microestado de cada subsistema es identificado por el par ordenado $(n_1, n_2)$ (a) si es posible diferenciar el microestado $(n_1, n_2)$ de $(n_2, n_1)$, ¿cuál es la energía de los subsistemas cuando estos se encuentran en equilibrio termodinámico? (b) Si es imposible distinguir entre los microestados $(n_1, n_2)$ y $(n_2, n_1)$, ¿cuál es la energía de los subsistemas cuando estos están en equilibrio termodinámico?

Como el microestado de cada subsistema se puede identificar por el par ordenado de números naturales $(n_1, n_2)$, con posibles valores $1, 2, 3, 4, 5, 6$, donde la suma de los dos números determina la energía del subsistema, $E = (n_1+n_2)E_0$, el caso es análogo al estudiado en la sección \ref{sec:D_dados_C_caras}, con la restricción de los posibles valores de energía para cada partícula (de $E_0$ a $6E_0$) y la energía total del sistema aislado ($10E_0$). Todos los microestados del sistema como un todo deberán cumplir con la restricción que la suma de la energía de los dos subsistemas debe ser $10E_0$; esto debido a que el sistema está aislado del resto del universo, lo cual le impide recibir o perder energía. En la Tabla~\ref{tab:ch2_microestados_10Eo_estados_distinguibles} se muestran 17 de los 80 microestados para el sistema en el caso (a), donde es posible identificar entre los microestados $(n_1, n_2)$ y $(n_2, n_1)$ en un subsistema dado, con lo cual, el número de estados accesibles está dado por los valores mostrados en la Tabla~\ref{tab:ch2_estados_accesibles_10Eo_estados_distinguibles}, donde se puede notar que el mayor número de microestados se obtiene cuando ambos subsistemas tienen energía $5E_0$, es decir, cuando los subsistemas están en equilibrio termodinámico. 

En el caso (a) se puede observar claramente un único máximo para el número de estados accesibles del sistema como un todo, respecto a la energía de uno de los subsistemas. Sin embargo, cuando se asume que es imposible distinguir entre los microestados $(n_1, n_2)$ y $(n_2, n_1)$, caso (b), como se muestra en la Tabla~\ref{tab:ch2_microestados_10Eo_estados_indistinguibles}, se observan dos máximos en el número de estados del sistema como un todo, ver Tabla~\ref{tab:ch2_estados_accesibles_10Eo_estados_indistinguibles}; este resultado no es tan extraño como parecería a primera vista, porque al ser imposible distinguir entre microestados de un mismo subsistema, ¿no debería haber también alguna imposibilidad de diferenciar el estado de equilibrio entre estos? Nótese que la condición de equilibrio termodinámico en el caso (b) corresponde con que uno de los subsistemas (cualquiera de los dos) tenga energía $4E_0$ y el otro $6E_0$.

\subsection{¿El número de estados accesibles tiene máximos?}

En el ejemplo ilustrativo anterior se observa que el número de estados accesibles tiene por lo menos un máximo. ¿Se espera que este sea un comportamiento generalizado para el número de estados accesibles de un sistema físico? A continuación, se estudiarán algunos argumentos a favor de una respuesta afirmativa a esta pregunta. Para iniciar, es necesario resaltar la relación entre el número de estados accesibles $\Omega(E)$ y la probabilidad $\prob{E \leq \text{Energía} < E + dE}$ que el sistema se encuentre con una energía entre $E$ y $E+dE$,

\[
\prob{E \leq \text{Energía} < E + dE} = \frac{\Omega(E)}{\sum\limits_{E'} \Omega(E')}
\]

Nótese que $\sum\limits_{E} \Omega(E)$ es una constante, por lo tanto, se puede afirmar que la probabilidad es proporcional a $\Omega(E)$, con lo cual la forma de la función de probabilidad y la función número de estados accesibles será la misma. Por lo tanto, toda conclusión sobre la forma de la función $P(E \leq \text{Energía} < E + dE)$ aplica a $\Omega(E)$ y viceversa. Dado un cierto sistema físico, aislado del resto del universo, como es el caso de un sistema que hace parte de un ensamble microcanónico, es razonable pensar que hay valores de energía suficientemente bajos para los cuales la probabilidad (y por ende, el número de estados accesibles) es despreciable o cero. El mismo argumento se puede dar para valores de energía muy altos, donde se esperaría que fuera poco probable (o completamente improbable) encontrar al sistema.

Al tener valores de probabilidad (y, por ello, número de estados accesibles) cercanos a cero para energías muy bajas y muy altas, y dado que la probabilidad solo puede tomar valores positivos y la suma de todas las probabilidades posibles es uno, entonces es necesario que exista al menos un valor de energía, con una probabilidad mayor que cero. De igual manera, a energías muy altas se esperan valores de probabilidad que tienden a cero. Por consiguiente, es razonable suponer que al analizar la función de probabilidad desde energías suficientemente bajas, donde los valores son cercanos a cero, hasta energías suficientemente altas, donde también se esperan valores próximos a cero, la probabilidad debe alcanzar al menos un máximo entre ambos extremos para que al sumar todos los posibles valores de probabilidad se obtenga uno. Para finalizar, es conveniente mencionar que para una función positivamente valuada, como la probabilidad, al imponer las condiciones que tienda a cero en los límites de energía muy pequeña y muy alta, implica la existencia de un máximo que permita cumplir con la condición que la suma de probabilidades es uno; por consiguiente, se espera que el número de estados accesibles tenga al menos un máximo y, con ello, una condición de equilibrio termodinámico. 

Es importante notar que si el número de estados accesibles de un sistema aislado carece de un máximo, entonces, sería imposible encontrar un estado de equilibrio termodinámico.

\subsection{Equilibrio termodinámico}
\index{Ensamble microcanónico!equilibrio termodinámico}

El tercer postulado de la Física Estadística, para sistemas aislados, establece la relación entre condición de equilibrio termodinámico y la máxima probabilidad. En tal caso, para una energía $E=E_A + E_B$ constante, el sistema compuesto por dos subsistemas $A$ y $B$ alcanzará el equilibrio termodinámico cuando $\Omega(E)$ sea máxima ante el intercambio de energía entre los subsistemas. Supongamos que se toma como referencia el subsistema $A$, entonces, el máximo de $\Omega(E)$ respecto al intercambio de energía (en forma de calor) entre los dos subsistemas se obtiene cuando,

\[
\frac{d\Omega(E)}{dE_A} = 0
\]

Como $\Omega(E) = \Omega(E_A)\Omega_B(E_B)$,

\[
\frac{d}{dE_A} [\Omega(E_A)\Omega_B(E_B)] = \Omega_B(E_B) \frac{\partial \Omega_A(E_A)}{\partial E_A} + \Omega_A(E_A) \frac{\partial \Omega_B(E_B)}{\partial E_B} \frac{dE_B}{dE_A} = 0
\]

Al asumir fija la energía total $E=E_A + E_B$, se tiene la relación $dE_B/dE_A = -1$, con lo cual,

\[
\Omega_B(E_B) \frac{\partial\Omega_A(E_A)}{\partial E_A} - \Omega(E_A) \frac{\partial \Omega_B(E_B)}{\partial E_B} = 0
\]

Reordenando términos y teniendo en cuenta que $\frac{d\ln f(x)}{dx} = \frac{1}{f(x)}\frac{df(x)}{dx}$, se obtiene la condición de equilibrio termodinámico,

\begin{subequations}
\label{eq:ch2_equilibrio_termodinamico}
\begin{align}
     \frac{\partial \ln \Omega_A(E_A)}{\partial E_A} & = \frac{\partial \ln \Omega_B(E_B)}{\partial E_B} \\
     \beta & = \frac{1}{k_B T} \\
     T_A & = T_B
\end{align}
\end{subequations}

Notese que la relación entre el parámetro $\beta$ y la temperatura absoluta, tal como se define en termodinámica, se puede comprobar fácilmente al analizar la estructura matemática del parámetro $\beta$ y compararla con la definición de la entropía de Boltzmann, teniendo en cuenta la relación termodinámica $\partial S / \partial E = T^{-1}$,

\[
\beta = \frac{\partial \ln \Omega(E)}{\partial E} = \frac{1}{k_B}\frac{\partial S}{\partial E} = \frac{1}{k_B T}
\]

En conclusión, al aplicar la Física Estadística al estudio de sistemas en contacto cuando estos alcanzan el equilibrio termodinámico se obtiene una condición que corresponde con la ley cero de la termodinámica. 

\begin{definition}
    Se define el \textbf{parámetro $\beta$} por la ecuación \eqref{eq:ch2_parametro_beta}, donde $k_B$ es la constante de Boltzmann y $T$ es un parámetro que se denomina temperatura absoluta.
    \begin{equation}
    \label{eq:ch2_parametro_beta}
        \beta = \frac{\partial \ln \Omega(E)}{\partial E} = \frac{1}{k_B T}
    \end{equation} \index{Parámetro $\beta$}
\end{definition}

\begin{lema}
    Sea un sistema $A$ con energía $E_A$ y número de estados accesibles $\Omega_A$, en contacto con un sistema $B$, con energía $E_B$ y número de estados accesibles $\Omega_B$, tales que $E=E_A+E_B$, donde $E$ es una constante. Los dos sistemas alcanzan el \textbf{equilibrio termodinámico} cuando se cumple la ecuación \eqref{eq:ch2_equilibrio_termodinamico}, es decir, cuando su temperatura absoluta es igual.
\end{lema} 
\subsection{Interacción térmica}
\index{Ensamble microcanónico!interacción térmica}

Supongamos que entre los subsistemas $A$ y $B$ solo ocurre interacción térmica, esto implica que el volumen de los sistemas, así como cualquier otro parámetro externo asociado a la realización de trabajo sobre uno de los sistemas, permanece constante. En tal caso, la variación de la entropía del subsistema $A$ cuando ocurre intercambio de calor con el subsistema $B$ está dada por, 

\[
dS_A(E_A, V_A) = \frac{\partial k_B \ln \Omega_A(E_A)}{\partial E_A} dE_A = k_B \beta_A dE_A = \frac{dE_A}{T}
\]

Como la energía intercambiada entre los dos sistemas es calor, $dE_A = \dbar Q$, entonces, se obtiene que la variación del calor se expresa en términos de la variación de la entropía, donde el símbolo $\dbar$ denota el carácter de diferencial inexacta del calor, es decir, que la variación del calor depende de la trayectoria seguida para alcanzar el estado final, partiendo del estado inicial en que se encontraba el sistema. Un detalle muy importante al definir el calor intercambiado entre los subsistemas $A$ y $B$ consiste en definir una referencia, para ello se asumirá que el calor absorbido por el sistema es positivo, mientras que el calor cedido, es negativo.

\begin{equation}
    \label{eq:ch2_calor_entropia}
    \dbar Q = T dS
\end{equation}

\begin{definition}
    Sea $f(x_1, x_2, \cdots, x_n)$ una función continua y diferenciable; además, sea $\vec{r} = x_1 \hat{u}_1 + x_2 \hat{u}_2 + \cdots + x_n \hat{u}_n$, donde $\hat{u}_i$ son vectores unitarios ortonormales entre si. Se dice que $df$ es una \textbf{diferencial exacta} si se puede expresar como,
    \begin{equation}
    \label{eq:ch2_diferencial_exacta} 
    df = \frac{\partial f}{\partial x_1} dx_1 + \frac{\partial f}{\partial x_2} dx_2 + \cdots + \frac{\partial f}{\partial x_n} dx_n = \vec{\nabla}f \cdot d\vec{r}
    \end{equation}
\end{definition}

\begin{definition}
    Cuando el diferencial de una función $g(x_1, x_2, \cdots, x_n)$ no se puede expresar como la ecuación \eqref{eq:ch2_diferencial_exacta} se dice que dicha \textbf{diferencial es inexacta} y se expresa como $\dbar g$.
\end{definition}

\begin{definition}
    El \textbf{calor absorbido} $\dbar Q$ por un sistema es positivo, mientras que el calor cedido, es negativo. \index{Calor absorbido}
\end{definition}

\subsection{Interacción mecánica}
\index{Ensamble microcanónico!interacción mecánica}

\begin{figure}[ht]
    \centering
    \includegraphics[width=0.6\linewidth]{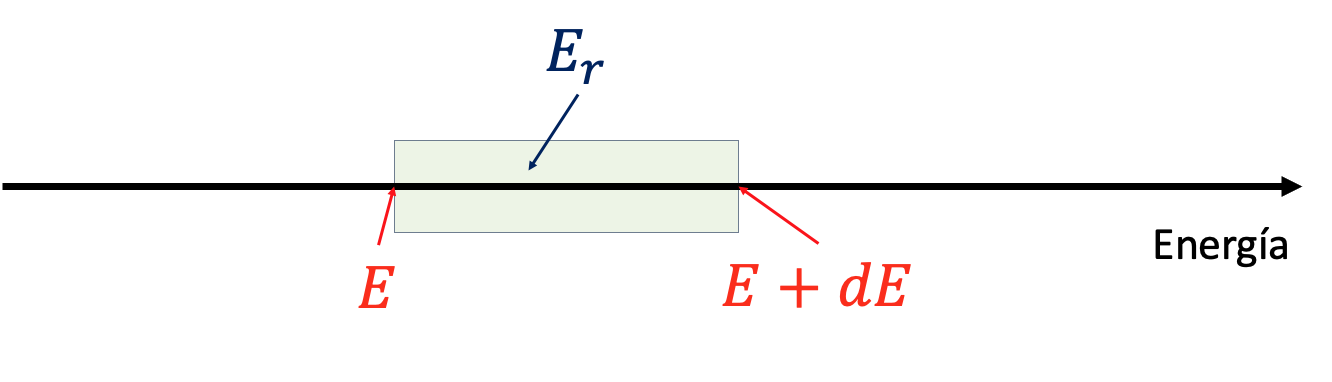}
    \caption{Representación esquemática de la energía del sistema.}
    \label{fig:ch2_energia_Er_microestado}
\end{figure}

Dos sistemas en contacto, ambos aislados del resto del universo, pueden interactuar sin necesidad del intercambio de calor, en tal caso, la energía de los sistemas cambia por la variación de un cierto parámetro externo $x$, como, por ejemplo, el volumen, haciendo que uno de los sistemas realice un trabajo sobre el otro sistema; a esta interacción se le conoce como mecánica. Si el parámetro externo varía una cantidad infinitesimal $dx$, el trabajo realizado por un sistema en el microestado con energía $E_r$, sobre el otro sistema, estará dado por $\frac{\partial E_r}{\partial x} dx$, donde $X=-\frac{\partial E_r}{\partial x}$ se conoce como fuerza generalizada conjugada a $x$. Para encontrar el valor esperado del trabajo realizado por la variación infinitesimal del parámetro externo $x$ necesitamos estudiar cómo depende el número de estados accesibles de dicho parámetro. En el estudio del trabajo $\dbar W$ debido a una variación infinitesimal del parámetro externo $x$ se asumirá que este es positivo cuando dicho trabajo es realizado por el sistema, mientras que se considerará negativo, cuando es realizado sobre el sistema; si bien es cierto que esta es una convención y se podría asumir lo contrario (positivo cuando el trabajo es realizado sobre el sistema), es necesario definir una convención, de manera análoga a como se define que la carga de un electrón es negativa o que la corriente es positiva cuando se refiere a cargas positivas.

\begin{figure}[t]
    \centering
    \begin{subfigure}{0.49\textwidth}
        \includegraphics[width=\linewidth]{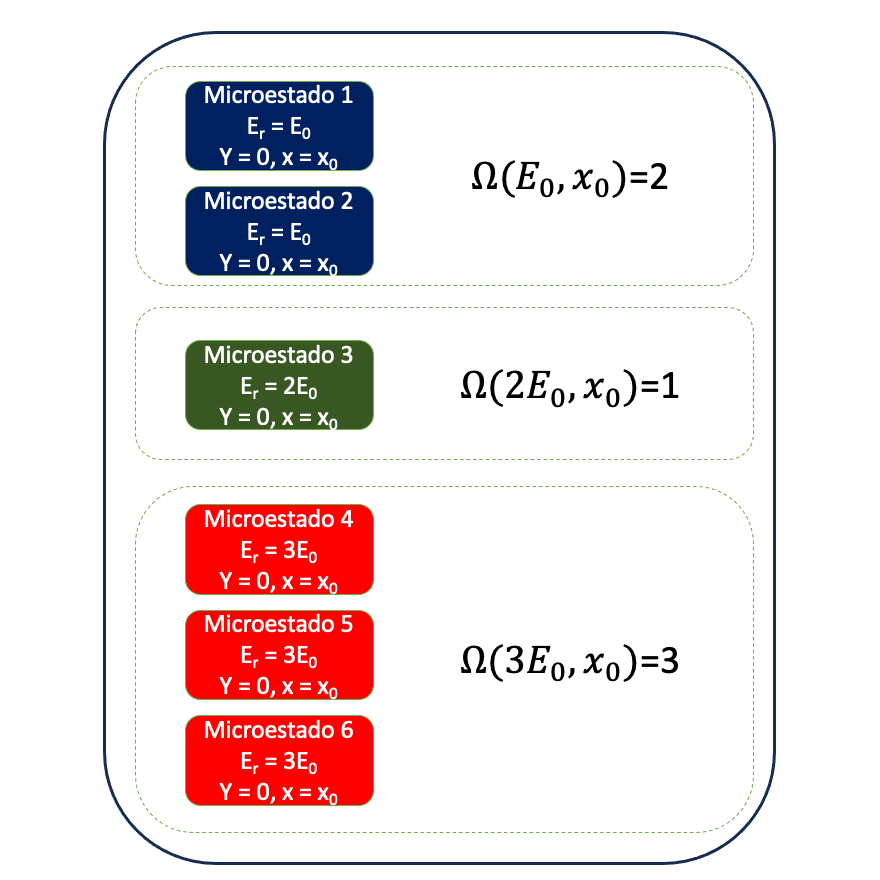}
        \caption{\label{fig:ch2_Omega_E_x_Omega_Y_a}}
    \end{subfigure}
    \hfill
    \begin{subfigure}{0.49\textwidth}
        \includegraphics[width=\linewidth]{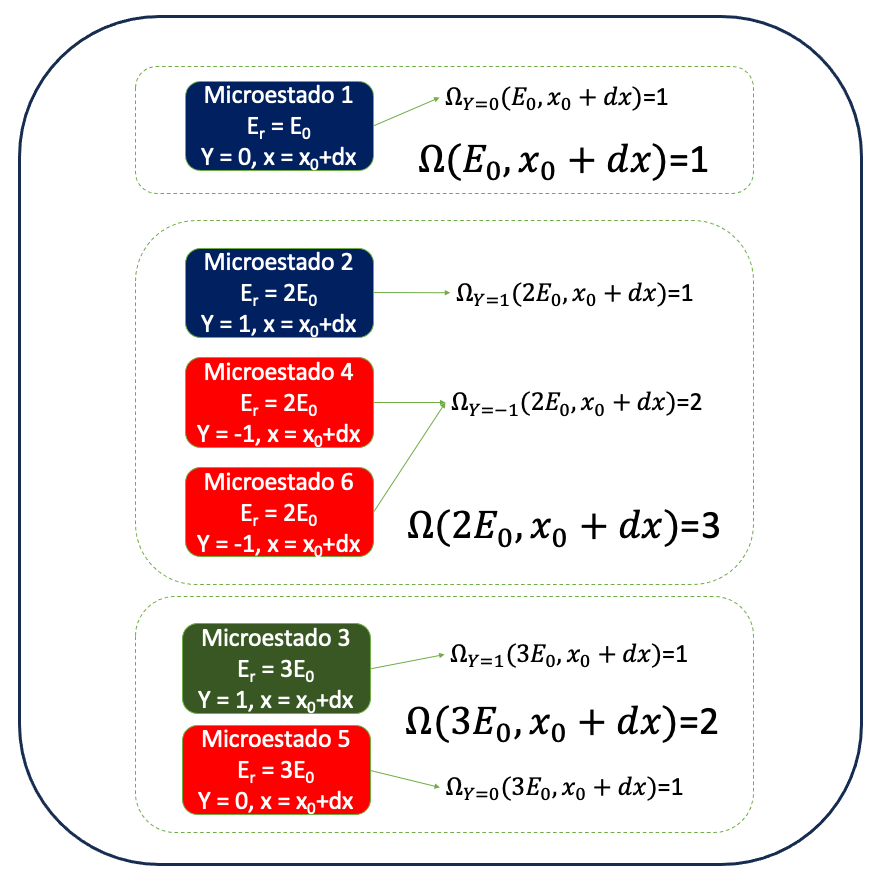}
        \caption{\label{fig:ch2_Omega_E_x_Omega_Y_b}}
    \end{subfigure}
    \caption{Representación esquemática de un sistema con seis microestados cuando al parámetro externo $x$ es (a) $x_0$ y (b) $x_0 + dx$. El número de estados accesibles $\Omega(E_r; x)$, cuando $x$ varía en $dx$, cambia debido a que la energía de los microestados varía una cantidad $\Delta_xE_r = Ydx$; esto hace que los microestados se reorganicen y (b) dentro del conjunto formado por los que tienen energía $E_r$ estos se pueden organizar de acuerdo al valor de $Y$, dando lugar al número de estados $\Omega_Y(E_r; x)$.}
    \label{fig:ch2_Omega_E_x_Omega_Y}
\end{figure}

Supongamos que el sistema se encuentra en un cierto microestado con energía $E_r$ entre $E$ y $E+dE$, como se ilustra en la Figura \ref{fig:ch2_energia_Er_microestado}, donde el parámetro externo $x$ condiciona el número de estados accesibles, $\Omega(E_r; x)$; además, $Y=\frac{\partial E_r}{\partial x}$. Una variación infinitesimal $dx$ del parámetro externo $x$ produce una variación $\Delta_x E_r$ en la energía de los microestados con energía $E_r$,

\begin{equation}
    \Delta_x E_r = \frac{\partial E_r}{\partial x} dx = Y dx = -X dx
\end{equation}

El número de estados accesibles con energía entre $E$ y $E+dE$, cuando el parámetro $x$ varía una cantidad infinitesimal $dx$, que introduce variaciones $\Delta_x E_r$ en el energía del microestado para $\frac{\partial E_r}{\partial x}$ entre $Y$ e $Y+dY$, está dada por $\Omega_Y(E_r; x)$, como se muestra esquemáticamente en la Figura \ref{fig:ch2_Omega_E_x_Omega_Y}, donde la suma de todos los estados asociados a la variación corresponderá con el número de estados accesibles $\Omega(E_r; x)$ para el microestado con energía $E_r$, ecuación \eqref{eq:ch2_sum_Omega_Y}.

\begin{equation}
    \label{eq:ch2_sum_Omega_Y}
    \Omega(E_r; x) = \sum\limits_{Y} \Omega_Y(E_r; x)
\end{equation}

La probabilidad $p_Y(E_r; x)$ que el sistema se encuentre en un microestado con energía $E_r$ entre $E$ y $E+dE$ para un valor $x$ del parámetro externo y variaciones $\frac{\partial E_r}{\partial x}$ entre $Y$ e $Y+dY$ está dada por la ecuación \eqref{eq:ch2_probabilidad_Y} y el valor esperado $\E{Y}$, por la ecuación \eqref{eq:ch2_valor_esperado_Y}.

\begin{equation}
    \label{eq:ch2_probabilidad_Y}
    p_Y = \frac{\Omega_Y(E_r; x)}{\Omega(E_r; x)}
\end{equation}

\begin{equation}
    \label{eq:ch2_valor_esperado_Y}
    \E{Y} = \sum\limits_Y Y p_Y = \frac{1}{\Omega(E_r; x)} \sum\limits_Y Y \Omega_Y(E_r; x) = -\E{X}
\end{equation}

\begin{figure}[t]
    \centering
    \includegraphics[width=\linewidth]{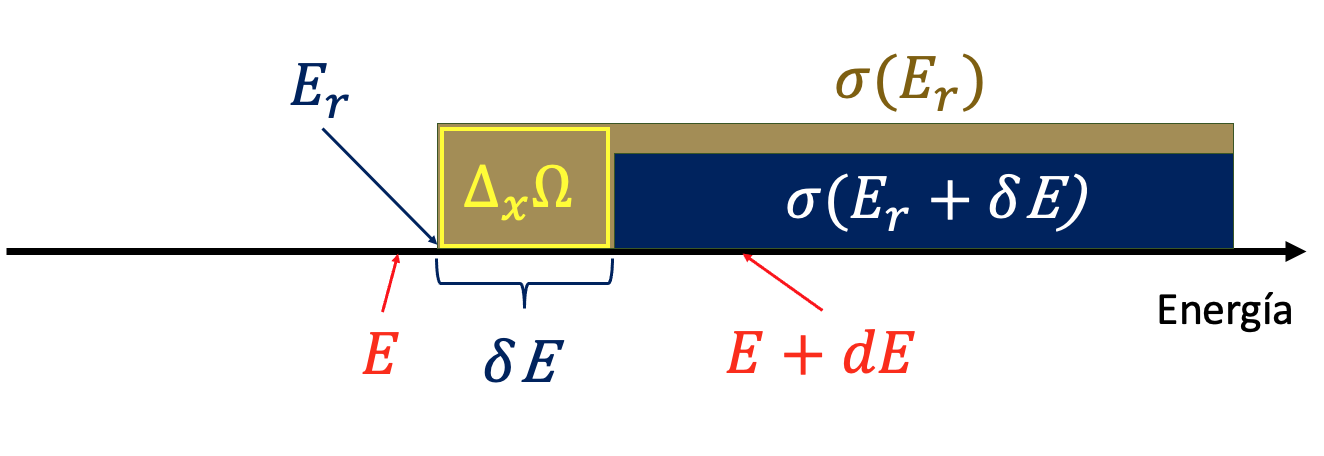}
    \caption{Representación esquemática de la relación entre la variación $\Delta_x \Omega(E_r; x)$ del número de estados accesibles con energía $E_r$ debido a variación infinitesimal $dx$ del parámetro externo, el número de microestados $\sigma$ que cambian su energía desde un valor menor que $E_r$ a uno mayor que $E_r$ cuando el parámetro externo $x$ cambia de $x$ a $x+dx$ y $\sigma(E_r + \delta E)$.}
    \label{fig:ch2_numero_estados_Omega_x}
\end{figure}

La variación $\Delta_x \Omega(E_r; x)$ del número de estados accesibles con energía $E_r$ debido a variación infinitesimal $dx$ del parámetro externo $x$ y $\frac{\partial E_r}{\partial x}$ entre $Y$ e $Y+dY$ se expresa, por la ecuación \eqref{eq:ch2_variacion_x_Omega}, en términos del número de estados $\sigma(E_r)$ cuya energía cambia de un valor menor a $E_r$ a un valor mayor a este cuando el parámetro externo $x$ cambia de $x$ a $x+dx$ y el número de estados $\sigma(E_r+\delta E)$ cuya energía cambia de un valor menor a $E_r + \delta E$ a un valor mayor a este debido a la variación del parámetro externo $x$, como se muestra esquemáticamente en la Figura \ref{fig:ch2_numero_estados_Omega_x}.

\begin{equation}
    \label{eq:ch2_variacion_x_Omega}
    \Delta_x \Omega(E_r; x) = \frac{\partial\Omega(E_r; x)}{\partial x} dx = \sigma(E_r) - \sigma(E_r + \delta E)
\end{equation}

De la misma manera que en la ecuación \eqref{eq:ch2_sum_Omega_Y} se definió la relación entre $\Omega(E_r; x)$ y $\Omega_Y(E_r; x)$, en la ecuación \eqref{eq:ch2_sum_sigma_Y} se define la relación entre $\sigma(E_r)$ y $\sigma_Y(E_r)$.

\begin{equation}
    \label{eq:ch2_sum_sigma_Y}
    \sigma(E_r; x) = \sum\limits_{Y} \sigma_Y(E_r; x)
\end{equation}

Donde,

\[
\sigma_Y(E_r) = \Omega_Y(E_r; x) \frac{\Delta_x E_r}{\delta E} = \frac{\Omega_Y(E_r; x)}{\delta E} Y dx
\]

Por lo tanto, $\sigma(E_r; x)$ toma la siguiente forma,

\[
\sigma(E_r; x) = \left( \frac{dx}{\delta E} \right) \sum\limits_{Y} Y \Omega_Y(E_r; x)
\]

Al multiplicar por uno de la forma $\Omega(E_r; x) / \Omega(E_r; x)$, 

\[
\sigma(E_r; x) = \left( \frac{dx}{\delta E} \right) \Omega(E_r; x) \left( \frac{1}{\Omega(E_r; x)} \sum\limits_{Y} Y \Omega_Y(E_r; x) \right) = \left( \frac{dx}{\delta E} \right) \Omega(E_r; x) \E{Y}
\]

Al reemplazar este resultado en la expansión en series de potencia de la ecuación \eqref{eq:ch2_variacion_x_Omega}, dada por,

\[
\Delta_x \Omega(E_r; x) = \frac{\partial\Omega(E_r; x)}{\partial x} dx = \sigma(E_r) - \sigma(E_r + \delta E)= -\frac{\partial \sigma}{\partial E_r} \delta E + \mathcal{O}(\delta E^2)
\]

Se obtiene,

\[
\frac{\partial\Omega(E_r; x)}{\partial x} \approx - \frac{\partial}{\partial E_r} \left( \Omega(E_r; x) \E{Y} \right) = \Omega(E_r; x) \frac{\partial \E{Y}}{\partial E_r} - \E{Y} \frac{\partial \Omega(E_r; x)}{\partial E_r}
\]

Dividiendo por $\Omega(E_r; x)$, teniendo en cuenta que $\frac{1}{\Omega(E_r; x)} \frac{\partial \Omega(E_r; x)}{\partial E_r} = \frac{\partial \ln \Omega(E_r; x)}{\partial E_r}$, la definición del parámetro $\beta$ y dado que se espera que la variación de $\E{Y}$ respecto a la energía sea despreciable comparada con la variación de $\ln \Omega(E_r; x)$, se obtiene,

\[
\E{Y} = - k_B T \frac{\partial \ln \Omega(E_r; x)}{\partial x} 
\]

Este resultado nos indica que el valor esperado $\E{X}$ de la fuerza generalizada $X$ conjugada al parámetro externo $x$ está determinado por la variación del número de estados accesibles respecto al parámetro externo $x$; además, al variar el parámetro externo $x$ una cantidad infinitesimal $dx$, uno de los subsistemas realiza un trabajo $\dbar W = \E{X} dx$ sobre el otro subsistema. Se debe resaltar el comportamiento como diferencial inexacta del trabajo realizado, debido a que este depende de la manera en que se realiza la variación del parámetro externo $x$.

\begin{definition}
    La \textbf{fuerza generalizada $X$ conjugada} a un cierto parámetro externo $x$, para un sistema con energía $E$, está dada por la ecuación \eqref{eq:ch2_fuerza_generalizada}.
    \begin{equation}
        \label{eq:ch2_fuerza_generalizada}
        X = - \frac{\partial E}{\partial x}
    \end{equation}\index{Fuerza generalizada conjugada a un parámetro $x$}
\end{definition}

\begin{definition}
    El trabajo $\dbar W$ realizado por una fuerza generalizada $X$ conjugada a $x$, cuando el parámetro externo varía un valor infinitesimal $dx$ está dado por la ecuación \eqref{eq:ch2_trabajo_fuerza_generalizada}.
    \begin{equation}
        \label{eq:ch2_trabajo_fuerza_generalizada}
        \dbar W = X dx = - \frac{\partial E}{\partial x} dx
    \end{equation}\index{Trabajo de una fuerza generalizada}
\end{definition}

\begin{definition}
    Sea $\Omega(E_r; x)$ el número de estados accesibles para el microestado con energía $E_r$ con parámetro externo entre $x$ y $x+dx$. La probabilidad $p_r(x)$ que el microestado, con energía $E_r$ tenga parámetro externo $x$ está dada por la ecuación \eqref{eq:ch2_probabilidad_r_x}.
    \begin{equation}
        \label{eq:ch2_probabilidad_r_x}
        p_r(x) = \frac{\Omega(E_r; x)}{\sum\limits_{x} \Omega(E_r; x) }
    \end{equation}
\end{definition}

\begin{definition}
    Sea $\Omega(E_r; x)$ el número de estados accesibles para el microestado con energía $E_r$ con parámetro externo entre $x$ y $x+dx$. El valor esperado $\E{x}$ del parámetro externo $x$ está dado por la ecuación \eqref{eq:ch2_valor_esperado_x}.
    \begin{equation}
        \label{eq:ch2_valor_esperado_x}
        \E{x} = \sum\limits_{x} x p_r(x) = \frac{\sum\limits_{x} x \Omega(E_r; x)}{\sum\limits_{x} \Omega(E_r; x) }
    \end{equation}
\end{definition}

\begin{lema}
    Sea $\Omega(E_r; x)$ el número de estados accesibles con energía $E_r$ entre $E$ y $E+dE$, donde $x$ es un parámetro externo que puede variar y realizar un trabajo dado por la ecuación \eqref{eq:ch2_trabajo_fuerza_generalizada}. El valor esperado de la fuerza generalizada $X$ conjugada a $x$ está dado por la ecuación \eqref{eq:ch2_valor_esperado_fuerza_generalizada}.
    \begin{equation}
        \label{eq:ch2_valor_esperado_fuerza_generalizada}
        \E{X} = k_B T \frac{\partial \ln \Omega(E_r; x)}{\partial x} = T \frac{\partial S}{\partial x} 
    \end{equation}\index{Ensamble microcanónico!valor esperado de la fuerza generalizada}
\end{lema}

\begin{definition}
    El trabajo $\dbar W$ realizado por un sistema, cuando hay una variación infinitesimal $dx$ del parámetro externo $x$, es positivo. \index{Trabajo realizado}
\end{definition}

\subsection{Interacción termodinámica}
\index{Ensamble microcanónico!interacción termodinámica}

Supongamos que en el sistema en estudio se presentan simultáneamente interacciones térmicas y mecánicas, en tal caso, se dice que el sistema experimenta interacciones termodinámicas, las cuales están descritas por la variación del número de estados accesibles debido al intercambio de energía en forma de calor y al trabajo realizado por uno de los subsistemas cuando un parámetro externo (por ejemplo, la presión) varia.

Sea $\Omega_A(E_A, x)$ el número de estados accesibles del subsistema $A$ para el estado con energía $E_A$ en el rango entre $E$ y $E+dE$, con parámetro externo $x$. Supongamos que el sistema $A$ se encuentra en un microestado con energía alrededor de la energía $\E{E_A}$ del estado de equilibrio termodinámico y que el parámetro $x$ varía una cantidad infinitesimal $dx$. La variación del número de estados se puede expresar en términos del $\ln \Omega_A(E_A, x)$, con el fin de aplicar los resultados obtenidos al analizar las interacciones térmicas y mecánicas, ecuaciones \eqref{eq:ch2_calor_entropia} y \eqref{eq:ch2_valor_esperado_fuerza_generalizada},

\[
d\ln \Omega_A(E_A, x) = \frac{\partial \ln \Omega_A(E_A, x)}{\partial E_A} dU + \frac{\partial \ln \Omega_A(E_A, x)}{\partial x} dx
\]

Como estamos estudiando la variación de la energía alrededor de la energía $\E{E_A}$ del estado de equilibrio termodinámico, entonces, el diferencia de energía del sistema $A$ está dado por $dU = d\E{E_A}$. Reemplazando $\E{X}$ y $\dbar Q$ dados por las ecuaciones \eqref{eq:ch2_calor_entropia} y \eqref{eq:ch2_valor_esperado_fuerza_generalizada} se obtiene,

\[
d\ln \Omega_A(E_A, x) = \beta dU + \beta \E{X} dx
\]

Como $\dbar W = \E{X} dx$ y $\beta = 1/k_BT$,

\[
T d k_B \ln \Omega_A(E_A, x) = dU + \dbar W
\]

Teniendo en cuenta la entropía de Boltzmann, dada por la ecuación \eqref{eq:ch2_entropia_Boltzmann} y reordenando términos se obtiene la relación entre las energías involucradas en el proceso de interacción termodinámica, consistente con la primera ley de la termodinámica,

\[
    dU = TdS_A - \dbar W
\]

Este resultado nos indica que la variación del valor esperado $\E{E_A}$ de la energía del subsistema $A$ es igual a la diferencia entre el calor absorbido  $\dbar Q = T d S_A$ de $B$ y el trabajo $\dbar W$ realizado por $A$ sobre $B$, lo cual concuerda con la formulación de la primera ley de la termodinámica, donde $dU$ corresponde con la variación infinitesimal de la energía interna del subsistema $A$.

\begin{definition}
    El valor esperado de la energía de un sistema se denomina \textbf{energía interna} y se expresa como $U=\E{E}$. \index{Energía interna}
\end{definition}

\begin{lema}
    La variación de la energía interna $dU$ de un sistema que interactúa termodinámicamente con otro, del cual absorbe una cantidad de calor $\dbar Q = T dS$ y sobre el que realiza un trabajo $\dbar W = \E{X} dx$ al variar el parámetro externo $x$ una cantidad $dx$, está dada por la ecuación 
    \begin{equation}
    \label{eq:ch2_primera_ley_termdinamica}
        dU = \dbar Q - \dbar W
    \end{equation}
\end{lema}

\section{Física estadística y termodinámica}
\label{sec:ch2_fisica_estadistica_temperatura_cero}

En este capítulo se estudió el comportamiento de los sistemas aislados con energía en un rango comprendido entre $E$ y $E+dE$. Al conjunto de sistemas aislados en dicho rango de energía, preparados de la misma manera, lo denominamos ensamble microcanónico y se plantearon tres postulados, con los cuales se obtuvo la condición de equilibrio termodinámico, ecuación \eqref{eq:ch2_equilibrio_termodinamico}, consistente con la ley cero de la termodinámica. Al analizar las interacciones termodinámicas entre sistemas en contacto, ambos aislados del resto del universo\footnote{Esta interacción fue modelada como dos subsistemas de un sistema aislado, con energía en un rango entre $E$ y $E+dE$.}, se encontró que el valor esperado de la energía del sistema, obtenida a partir de la física estadística, coincide con la energía interna de la termodinámica y se relaciona con el calor absorbido y el trabajo realizado por la ecuación \eqref{eq:ch2_primera_ley_termdinamica}, consistente con la primera ley de la termodinámica.

El anterior resultado se obtuvo al asumir que la entropía de Boltzman, ecuación \eqref{eq:ch2_entropia_Boltzmann}, corresponde con la entropía del sistema y que el teorema $\HBoltzmann$ de Boltzman, interpretado en el marco de la mecánica cuántica, es válido. Con ello, se encontró que la entropía o permanece constante o aumenta en el tiempo, consistente con la segunda ley de la termodinámica. En este resultado se obtuvo el parámetro $\beta$, ecuación \eqref{eq:ch2_parametro_beta}, el cual se relacionó con la temperatura absoluta por medio de la siguiente ecuación,

\index{Ensamble microcanónico!temperatura absoluta}

\[
    T = \left( k_B \frac{\partial \ln \Omega}{\partial E} \right)^{-1}
\]

\begin{figure}[t]
    \centering
    \includegraphics[width=\linewidth]{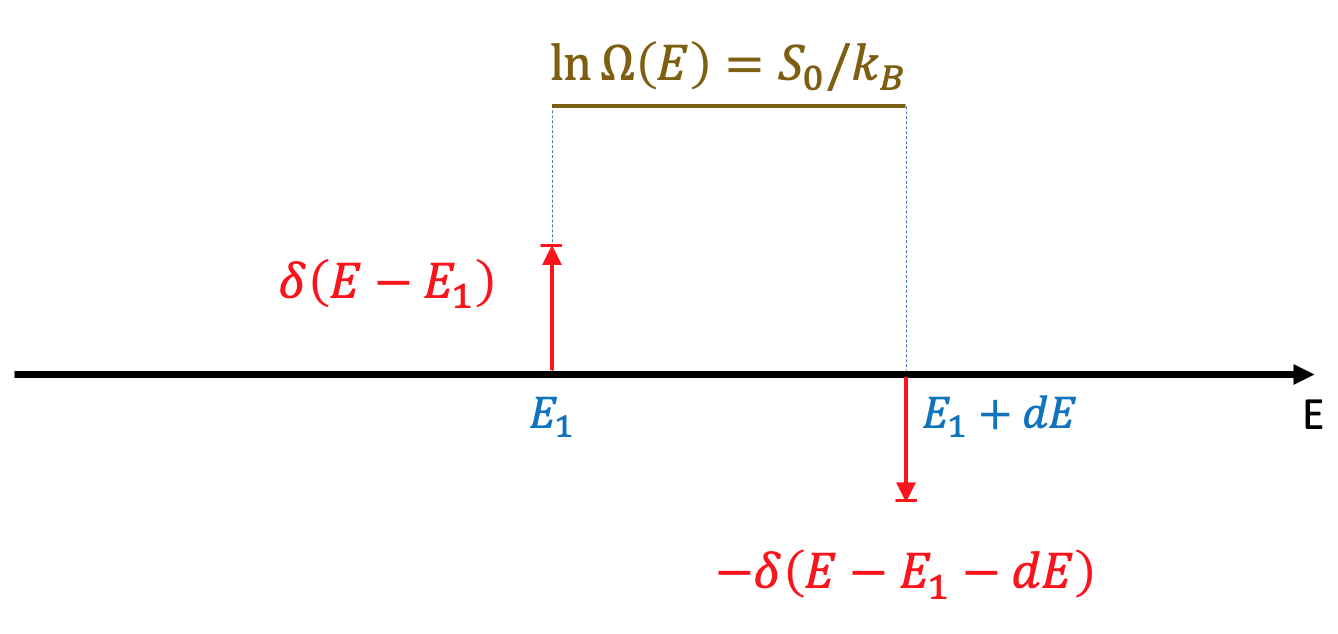}
    \caption{Representación esquemática de un número de estados accesibles constante en un intervalo de energías entre $E_1$ y $E_1+dE$. Las flechas rojas representan los deltas de Dirac obtenidos al realizar la derivada $\frac{\partial \ln \Omega}{\partial E}$.}
    \label{fig:ch2_ln_Omega_constante}
\end{figure}

Para estudiar el caso extremo, cuando la temperatura del sistema tiende a cero absoluto, es conveniente analizar los niveles de energía predichos para un cierto sistema mecánico cuántico aislado. En un sistema mecánico cuántico aislado, donde existen niveles de energía bien definidos, el número de estados accesibles, para un determinado nivel de energía $E_1$ está determinado por la degenerancia del estado. Para analizar este tipo de sistemas, supongamos que $ \ln \Omega(E)=S_0/k_B$ es constante en un rango de energías comprendido entre $E_1$ y $E_1 + dE$ mientras que su valor es cero en $|E - E_1| > dE$, como se muestra esquemáticamente en la Figura \ref{fig:ch2_ln_Omega_constante}, entonces,

\[
\frac{\partial \ln \Omega}{\partial E} = \gamma_0 \delta(E-E_1) - \gamma_0 \delta(E-E_1-dE)
\]

Donde $\gamma_0$ es una constante de proporcionalidad. Al evaluar $\partial \ln \Omega / \partial E$ cuando $E$ tiende a $E_1$ (al igual que cuando $E$ tiende a $E_1+dE$), debido al delta de Dirac, la temperatura absoluta tiende a cero, $T=0$; como se están asumiendo niveles de energía cuantizados, entonces $dE \to 0$. Esto quiere decir que cuando el sistema mecánico cuántico aislado\footnote{Como el sistema está aislado, se asume que no interactúa con ningún otro sistema.} se encuentra en un estado cuantizado,\footnote{Las energías están separadas unas de otras en valores definidos (cuantización).} la temperatura del sistema coincide con el cero absoluto y tendrá una entropía $S=k_B \ln \Omega(E_1) = S_0$ constante, resultado consistente con la tercera ley de la termodinámica.

\section{Magnetización de sistema formado por partículas paramagnéticas con espín 1/2}

Sea un sistema aislado formado por $N$ partículas no interactuantes con espín $s=1/2$, donde cada partícula tiene un momento dipolar magnético $\vec{\gls{muB}}$ cuya magnitud está dada por, la ecuación \eqref{eq:ch2_momento_magnetico_paramag}, donde $g_s = 2.00231930436$ es el factor giromagnético o de Landé\index{Factor giromagnético}\index{Factor de Landé}; $q_e$, la carga del electrón; $m_e$, la masa del electrón y $S=\hbar/2$, la magnitud del momento espín.

\begin{equation}
    \label{eq:ch2_momento_magnetico_paramag}
    \gls{muB} = g_s \frac{q_e}{2 m_e}S
\end{equation}

\begin{figure}[t]
    \centering
    \includegraphics[width=0.5\linewidth]{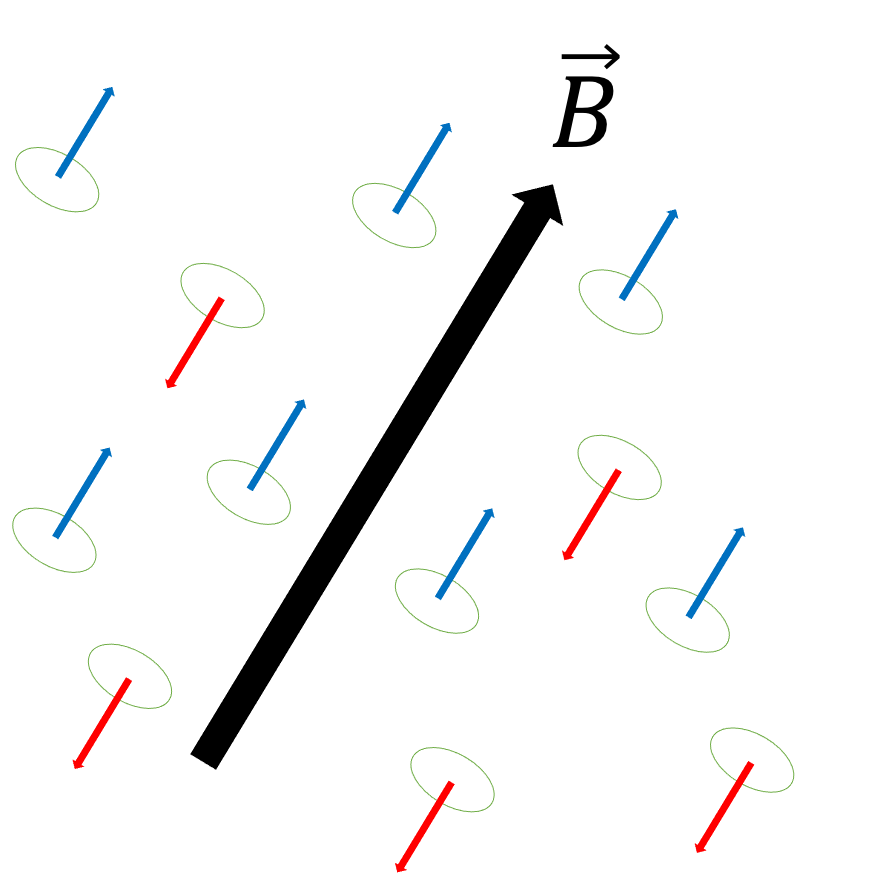}
    \caption{Partículas paramagnéticas con magnitud del momento dipolar magnético $\gls{muB}$ alineadas de manera paralela ({\color{blue}azul}) y antiparalela ({\color{red}rojo}) con el campo magnético externo $\vec{B}$ (negro).}
    \label{fig:ch2_particulas_alienacion_cammpo_paramag}
\end{figure}

Al aplicar un campo magnético externo $\vec{B}$, debido a que el espín es 1/2, las partículas se pueden alinear con el campo magnético de manera paralela o antiparalela a este, como se muestra esquemáticamente en la Figura \ref{fig:ch2_particulas_alienacion_cammpo_paramag}.  Si $n_1$ es el número de partículas alineadas paralelas al campo y $n_2 = N - n_1$, las antiparalelas, la energía del sistema estará dada por la ecuación \eqref{eq:ch2_energia_paramag}. ¿Cuánto es el número de estados accesibles $\Omega(E)$ con energía entre $E$ y $E + \delta E$? ¿Cuál es la relación entre el valor esperado de la energía $\E{E}$ y la temperatura absoluta $T$? ¿Bajo cuáles consideraciones la temperatura absoluta $T$ es negativa? ¿Cómo depende la magnetización $M$ de la temperatura absoluta?

\begin{equation}
    \label{eq:ch2_energia_paramag}
    E = - \gls{muB} B (n_1 - n_2) = - \gls{muB} B (2n_1 - N)
\end{equation}

\subsection{Número de estados accesibles}

Como la suma de partículas alineadas con el campo magnético externo de manera paralela $n_1$ y antiparalela $n_2$ es igual al número total de partículas, $n_1+n_2=N$, y dada la ecuación \eqref{eq:ch2_energia_paramag}, los números $n_1$ y $n_2$ se pueden expresar como,

\[
n_1 = \frac{N}{2} - \frac{E}{2\gls{muB} B}
\]

\[
n_2 = \frac{N}{2} + \frac{E}{2\gls{muB} B}
\]

De las expresiones anteriores se tiene que, para un valor fijo de $n_1$ existe un único valor $n_2 = N - n_1$, con $\frac{N!}{n_1! n_2!}$ microestados. Para contar todos los microestados con energía entre $E$ y $E+\delta E$, se deben incluir aquellos debidos a las posibles variaciones de $n_1$ en el intervalo de energía $\delta E$, 

\[
|\Delta n_1| = \left| \frac{\partial n_1}{\partial E}  \delta E \right| = \frac{\delta E}{2\gls{muB}B} = |\Delta n_2|
\]

\begin{figure}[t]
    \centering
    \begin{subfigure}{0.49\textwidth}
        \includegraphics[width=\linewidth]{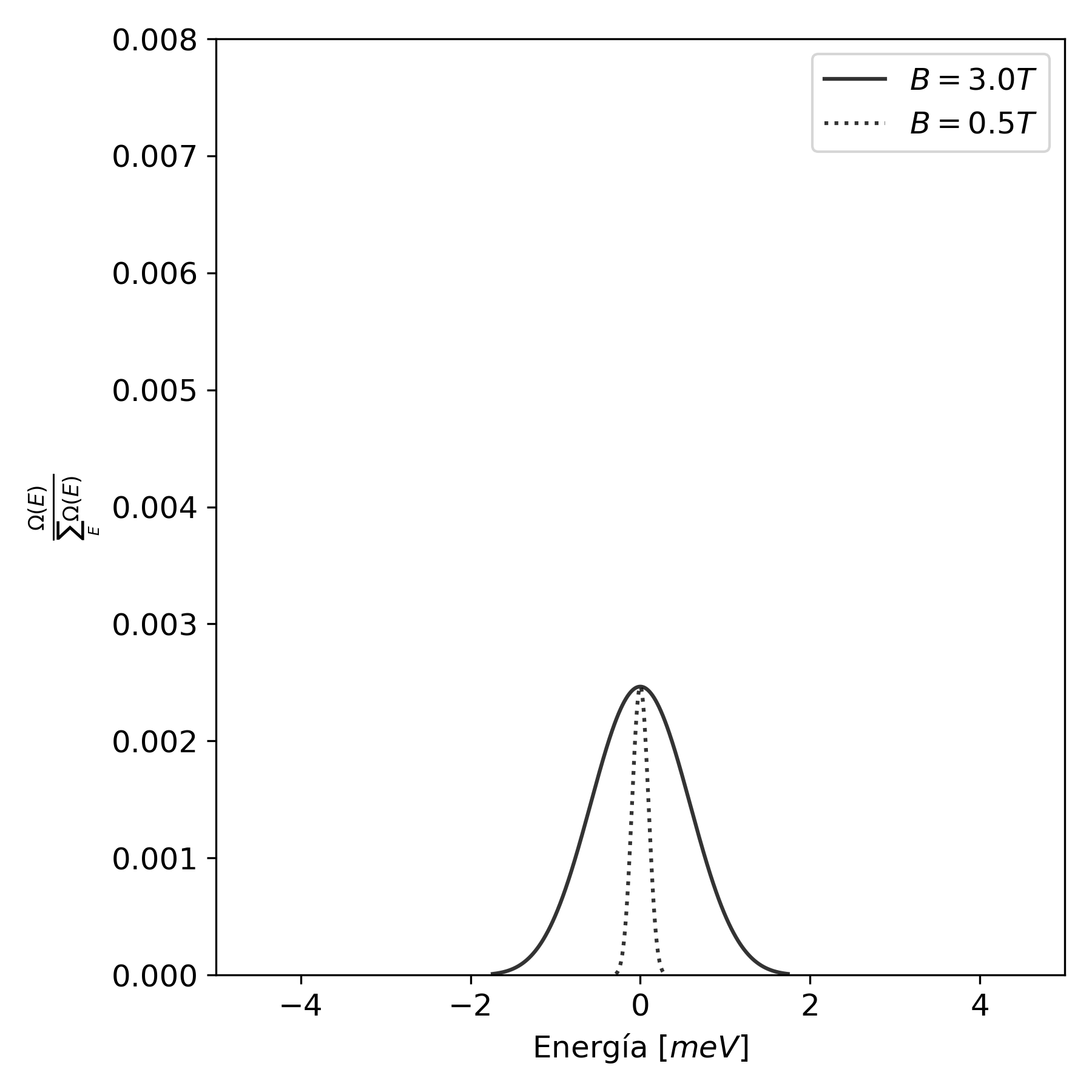}
        \caption{\label{fig:ch2_Omega_param1}}
    \end{subfigure}
    \hfill
    \begin{subfigure}{0.49\textwidth}
        \includegraphics[width=\linewidth]{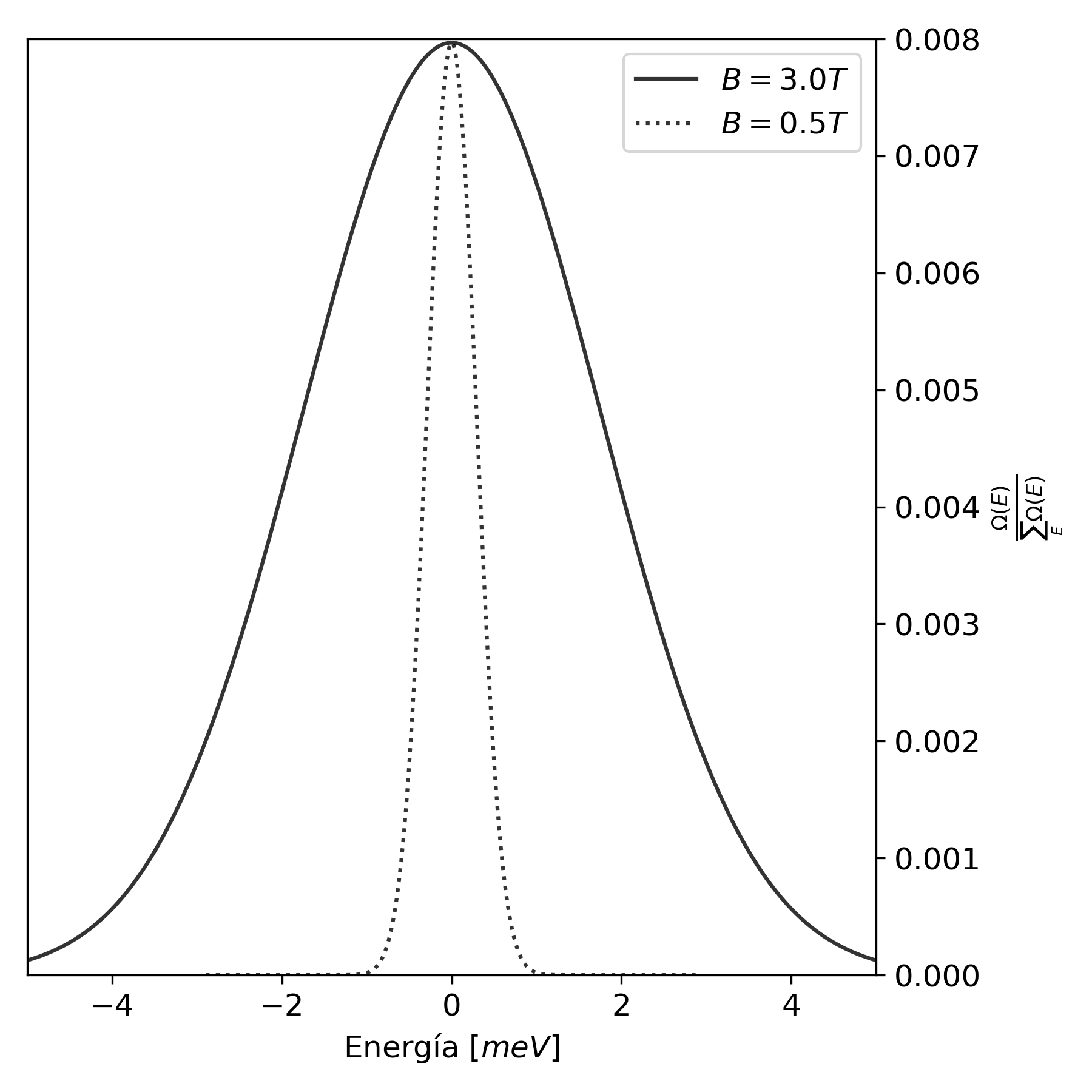}
        \caption{\label{fig:ch2_Omega_param2}}
    \end{subfigure}
    \caption{Número de estados accesibles $\Omega(E)$ para (a) 10 y (b) 100 partículas paramagnéticas con espín 1/2, carga igual a la del electrón, con energía entre $E$ y $E+\delta E$ y $\delta E = 0.2 meV$, en presencia de un campo magnético $B$.}
    \label{fig:ch2_Omega_param}
\end{figure}

Entonces, el número de estados accesibles con energía entre $E$ y $E + \delta E$ está dado por la ecuación \eqref{eq:ch2_Omega_paramag}. En la Figura \ref{fig:ch2_Omega_param} se observa el número de estados accesibles para $\delta E = 10^{-6} eV$, para diferentes valores de campo magnético y número total de partículas.

\begin{equation}
    \label{eq:ch2_Omega_paramag}
    \Omega(E) = \frac{N!}{\left(\frac{N}{2}-\frac{E}{2\gls{muB} B} \right)! \left( \frac{N}{2} + \frac{E}{2\gls{muB} B} \right)!} \frac{\delta E}{2\gls{muB} B}
\end{equation}

\subsubsection{Número total de estados accesibles}

El número total de estados accesibles se obtiene al sumar todos los posibles valores de energía dados por la ecuación \eqref{eq:ch2_energia_paramag}, que corresponde a energías entre $-N\gls{muB}$ y $N\gls{muB}$,

\[
\Omega_T = \sum\limits_{E=-\gls{muB} B N}^{\gls{muB} B N} \Omega(E) = \sum\limits_{E=-\gls{muB} B N}^{\gls{muB} B N} \frac{N!}{\left(\frac{N}{2}-\frac{E}{2\gls{muB} B} \right)! \left( \frac{N}{2} + \frac{E}{2\gls{muB} B} \right)!} \frac{\delta E}{2\gls{muB} B}
\]
Al tener en cuenta la ecuación \eqref{eq:ch2_energia_paramag} y dado que $n_1+n_2 = N$, la sumatoria para calcular $\Omega_T$ se puede realizar en términos del número de partículas $n_1$ alineadas paralelas al campo, 

\[
\Omega_T = \sum\limits_{n_1=1}^{N} \Omega(n_1)
\]

Donde $\Omega(n_1)$ está dado por,

\[
\Omega(n_1) = \frac{N!}{n_1! (N-n_1)!} \frac{\delta E}{2\gls{muB} B}
\]

Por lo tanto, 

\[
\Omega_T = \frac{\delta E}{2\gls{muB} B} \sum\limits_{n_1=1}^{N} \frac{N!}{n_1! (N-n_1)!} = \frac{\delta E}{2\gls{muB} B} \sum\limits_{n_1=1}^{N} \frac{N!}{n_1! (N-n_1)!} (1)^{n_1} (1)^{N-n_1}
\]

Por el teorema del binomio,

\[
\Omega_T = \frac{\delta E}{2\gls{muB}} (1+1)^N = \frac{\delta E}{2\gls{muB}} 2^N
\]
\subsubsection{Valor esperado del número de partículas alineadas paralelas al campo externo}

Cuando se tiene un sistema aislado con energía definida en un cierto rango entre $E$ y $E+\delta E$, el valor esperado  $\E{n_1}$ del número de partículas alineadas paralelas al campo mangnético externo se puede calcular directamente del número de estados accesibles; para ello, conviene utilizar el método fisicomatemático empleado en la sección \ref{sec:ch2_ejemplo_ilustrativo_sistema_aislado_9E_1}, el cual se basa en la identidad dada por la ecuación \eqref{eq:ch1_identidad_dist_binomial}. En tal sentido, al hacer $p=q=1$ se obtiene,

\[
\E{n_1} = \frac{\sum\limits_{n_1=1}^{N} n_1 \Omega(n_1)}{\Omega_T} = 2^{-N} \frac{\delta E}{2\gls{muB} B} \sum\limits_{n_1=1}^{N} \frac{N!}{n_1! (N-n_1)!} n_1 p^{n_1}q^{N-n_1} 
\]
Al aplicar la identidad dada por la ecuación \eqref{eq:ch1_identidad_dist_binomial}, además del teorema del binomio, se obtiene,
\[
\E{n_1} = 2^{-N} \frac{\delta E}{2\gls{muB} B} p \frac{\partial}{\partial p}\sum\limits_{n_1=1}^{N} \frac{N!}{n_1! (N-n_1)!} p^{n_1}q^{N-n_1} = 2^{-N} \frac{\delta E}{2\gls{muB} B} p \frac{\partial}{\partial p} (p + q)^N
\]

\[
\E{n_1} = 2^{-N} \frac{\delta E}{2\gls{muB} B} N (p + q)^{N-1}
\]

Con lo cual, el valor esperado $\E{n_1}$ está dado por,

\begin{equation}
    \label{eq:ch2_valor_esperado_n1_param_1}
    \E{n_1} = \frac{\delta E}{4\gls{muB} B} N
\end{equation}

\subsection{Energía y temperatura absoluta}

\subsubsection{Valor esperado de la energía}

Cuando se tiene al sistema aislado, con energía en un cierto rango entre $E$ y $E+\delta E$, el valor esperado de la energía está dado por,

\[
\E{E} = \frac{1}{\Omega_T} \sum\limits_{E=-\gls{muB} B N}^{\gls{muB} B N} E \Omega(E)
\]

Como la energía está relacionada con $n_1$ por la ecuación \eqref{eq:ch2_energia_paramag},

\[
\E{E} = -  \frac{\gls{muB} B}{\Omega_T} \sum\limits_{n_1=1}^{N} \Omega(n_1) [n_1 -(N-n_1)] = - \frac{\gls{muB} B}{\Omega_T} \sum\limits_{n_1=1}^{N} \Omega(n_1) (2n_1 - N)
\]

Al utilizar el resultado anterior para $\E{n_1}$ se obtiene,

\[
\E{E} = - \gls{muB} B (2\E{n_1} - N)
\]

\subsubsection{Temperatura absoluta}

Supongamos que el sistema se encuentra a una cierta temperatura absoluta $T$ que satisface la ecuación \eqref{eq:ch2_parametro_beta}. Como en equilibrio el sistema se encuentra en el estado con mayor probabilidad, que para este caso corresponde con $U = \E{E}$ y $\eta_1 = \E{n_1}$, 

\[
\beta = \frac{1}{k_B T} = \frac{\partial \ln \Omega(U)}{\partial U} = \frac{\partial \ln \Omega(\eta_1)}{\partial \eta_1} \frac{\partial \eta_1}{\partial U}
\]

Al aplicar la aproximación de Stirling, teniendo en cuenta que $N = \eta_1 + \eta_2$,

\[
\ln \Omega(\eta_1) \approx \ln\left( \frac{\delta E}{2\gls{muB} B} \right) + N \ln N - \eta_1 \ln \eta_1 - (N-\eta_1) \ln (N - \eta_1)
\]

Además, como $U = - \gls{muB} B ( 2\eta_1 - N)$,
\[
\frac{\partial \eta_1}{\partial U} = -\frac{1}{2\gls{muB} B} 
\]
Por lo tanto,

\[
\frac{\partial \ln \Omega(\eta_1)}{\partial \eta_1} \approx - \ln \eta_1 + \ln (N-\eta_1) = \ln \left( \frac{N-\eta_1}{\eta_1} \right)
\]
Con este resultado se puede obtener una expresión para la temperatura,

\[
\frac{1}{k_B T} \approx -\frac{1}{2\gls{muB} B} \ln \left( \frac{N-\eta_1}{\eta_1} \right)
\]

Al despejar $\eta_1$ se obtiene,

\[
\eta_1 \approx \frac{N}{1+e^{-\frac{2\gls{muB} B}{k_B T}}} = \frac{Ne^{\frac{\gls{muB} B}{k_B T}}}{e^{\frac{\gls{muB} B}{k_B T}}+e^{-\frac{\gls{muB} B}{k_B T}}}
\]

\begin{figure}[t]
    \centering
    \begin{subfigure}{0.49\textwidth}
        \includegraphics[width=\linewidth]{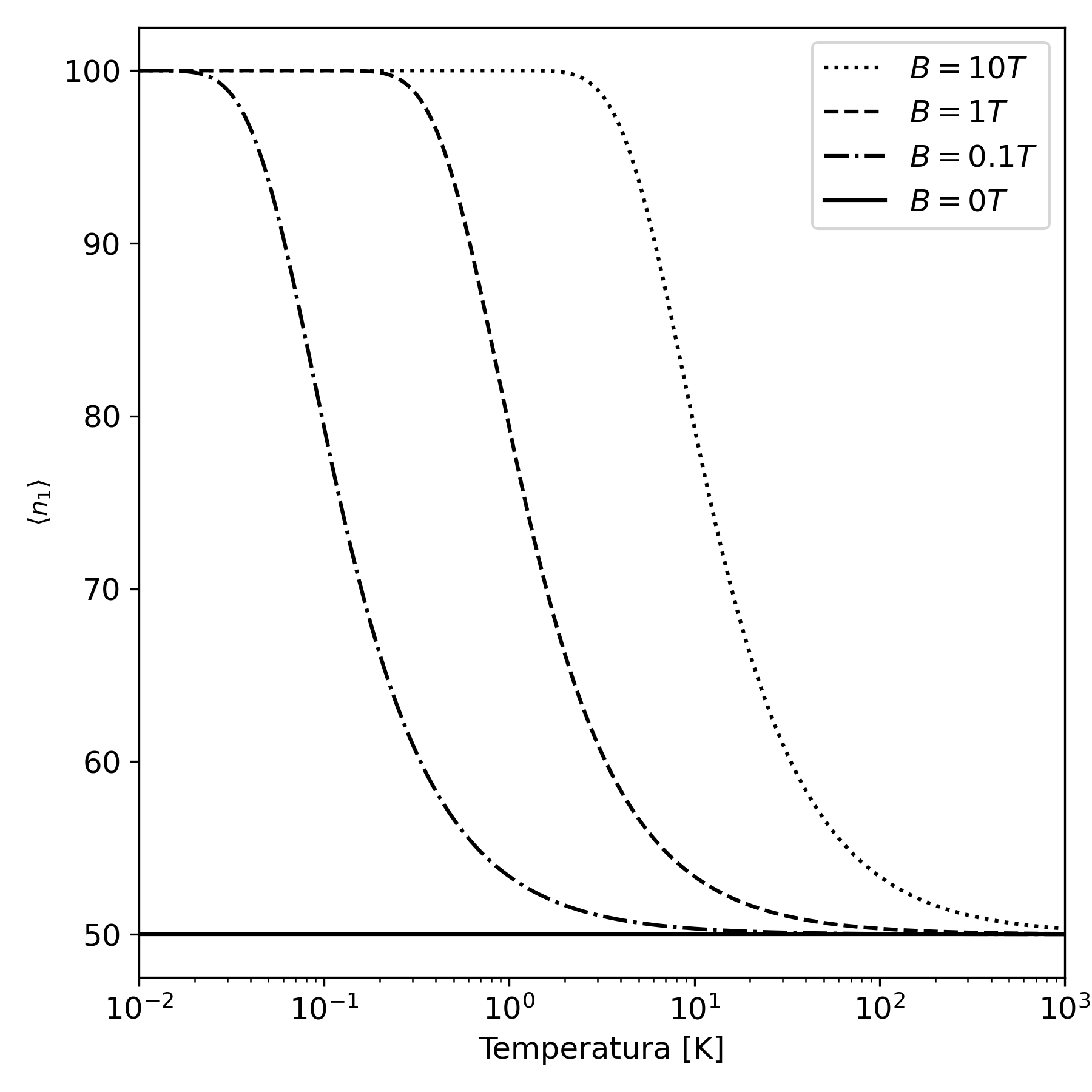}
        \caption{\label{fig:ch2_valor_esperado_n1_param1}}
    \end{subfigure}
    \hfill
    \begin{subfigure}{0.49\textwidth}
        \includegraphics[width=\linewidth]{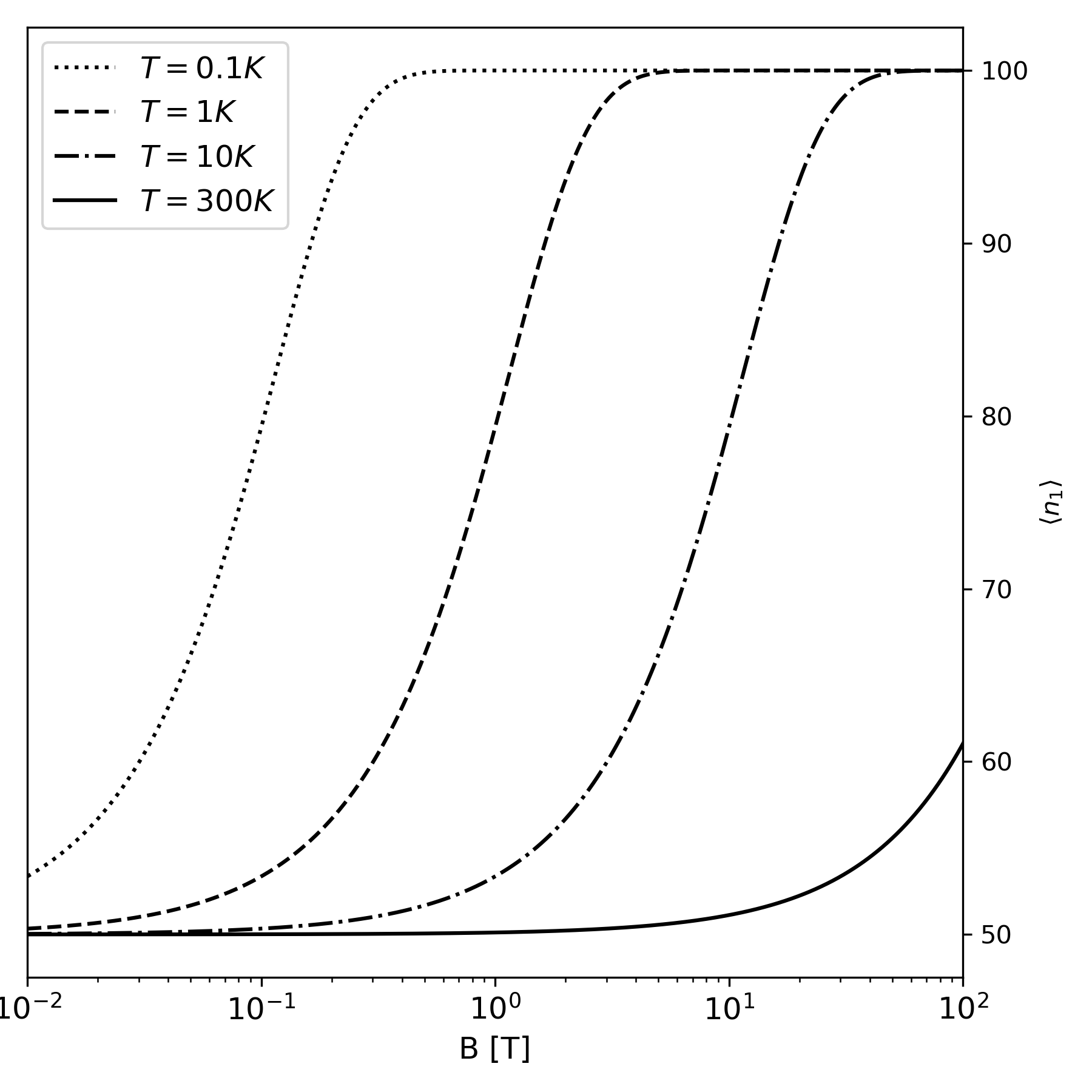}
        \caption{\label{fig:ch2_valor_esperado_n1_param2}}
    \end{subfigure}
    \caption{Valor esperado $\E{n_1}$ del número de partículas alineadas con un campo magnético externo $\vec{B}$, para un sistema formado por $N=100$ partículas paramagnéticas con espín 1/2, carga igual a la del electrón, con energía entre $E$ y $E+\delta E$, como función de (a) la temperatura para diferentes valores de campo magnético y (b) del campo magnético para diferentes valores de temperatura.}
    \label{fig:ch2_valor_esperado_n1_param}
\end{figure}

Como $\eta_1 = \E{n_1}$ y usando la definición de coseno hiperbólico, $\cosh$, se obtiene el valor esperado del número de partículas alienadas paralelas al campo magnético externo dado por la ecuación \eqref{eq:ch2_valor_esperado_n1_param}, el cual se muestra en la Figura \ref{fig:ch2_valor_esperado_n1_param} para diferentes valores del campo magnético y la temperatura.

\begin{equation}
    \label{eq:ch2_valor_esperado_n1_param}
    \E{n_1} \approx \frac{ Ne^{\frac{\gls{muB} B}{k_B T}}}{2 \cosh \left( \frac{\gls{muB} B}{k_B T} \right)}
\end{equation}

\begin{figure}[t]
    \centering
    \begin{subfigure}{0.49\textwidth}
        \includegraphics[width=\linewidth]{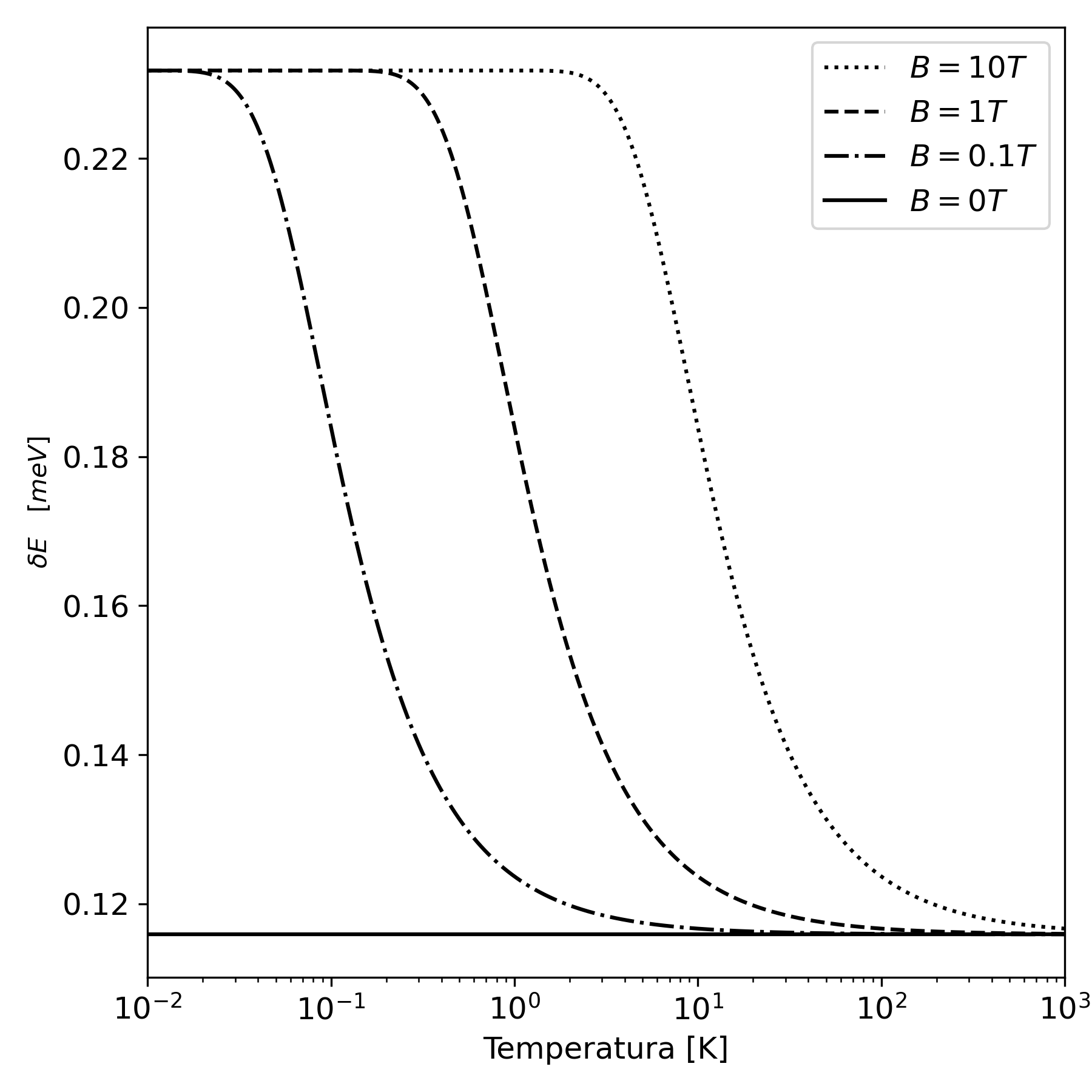}
        \caption{\label{fig:ch2_deltaE_param1}}
    \end{subfigure}
    \hfill
    \begin{subfigure}{0.49\textwidth}
        \includegraphics[width=\linewidth]{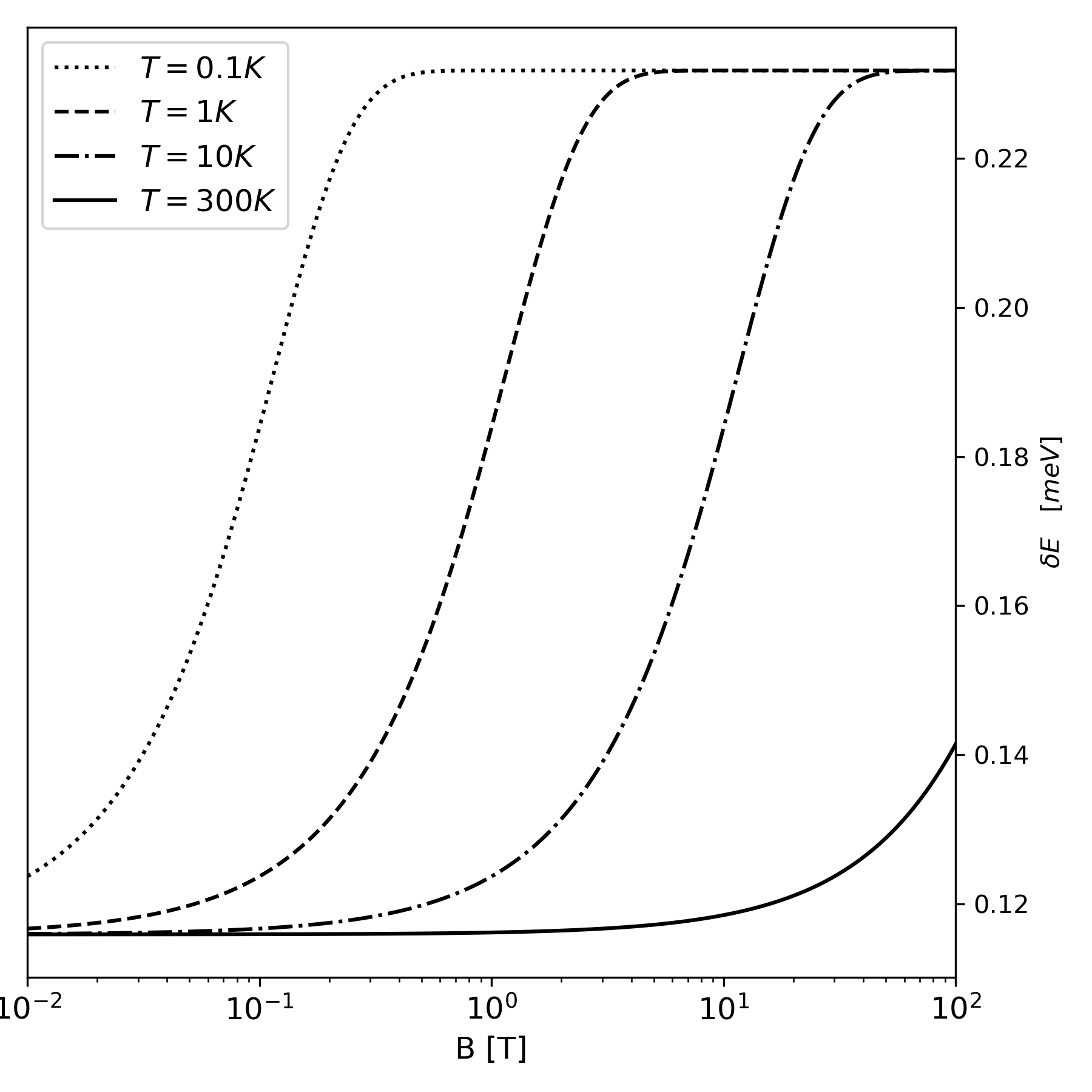}
        \caption{\label{fig:ch2_deltaE_param2}}
    \end{subfigure}
    \caption{Intervalo de energía $\delta E$ para un sistema formado por $N=100$ partículas paramagnéticas con espín 1/2, carga igual a la del electrón, con energía entre $E$ y $E+\delta E$, en presencia de un campo magnético externo, como función de (a) la temperatura para diferentes valores de campo magnético y (b) del campo magnético para diferentes valores de temperatura.}
    \label{fig:ch2_deltaE_param}
\end{figure}

Al comparar el resultado anterior con la ecuación \eqref{eq:ch2_valor_esperado_n1_param_1} se obtiene que $\delta E$ depende de la temperatura y el campo magnético, ecuación \eqref{eq:ch2_deltaE_param}, como se muestra en la Figura \ref{fig:ch2_deltaE_param}.

\begin{equation}
    \label{eq:ch2_deltaE_param}
    \delta E \approx \frac{ 2 \gls{muB} B e^{\frac{\gls{muB} B}{k_B T}}}{\cosh \left( \frac{\gls{muB} B}{k_B T} \right)}
\end{equation}

Y el valor esperado para la energía $U = \E{E} = - \gls{muB} B ( 2\eta_1 - N)$, cuando el sistema tiene temperatura $T$, está dado por,

\[
U \approx - \gls{muB} B \left( 2 \frac{Ne^{\frac{\gls{muB} B}{k_B T}}}{e^{\frac{\gls{muB} B}{k_B T}}+e^{-\frac{\gls{muB} B}{k_B T}}} - N \right) = - \gls{muB} B N \left( \frac{e^{\frac{\gls{muB} B}{k_B T}}- e^{-\frac{\gls{muB} B}{k_B T}}}{e^{\frac{\gls{muB} B}{k_B T}}+e^{-\frac{\gls{muB} B}{k_B T}}} \right)
\]

\begin{figure}[t]
    \centering
    \begin{subfigure}{0.49\textwidth}
        \includegraphics[width=\linewidth]{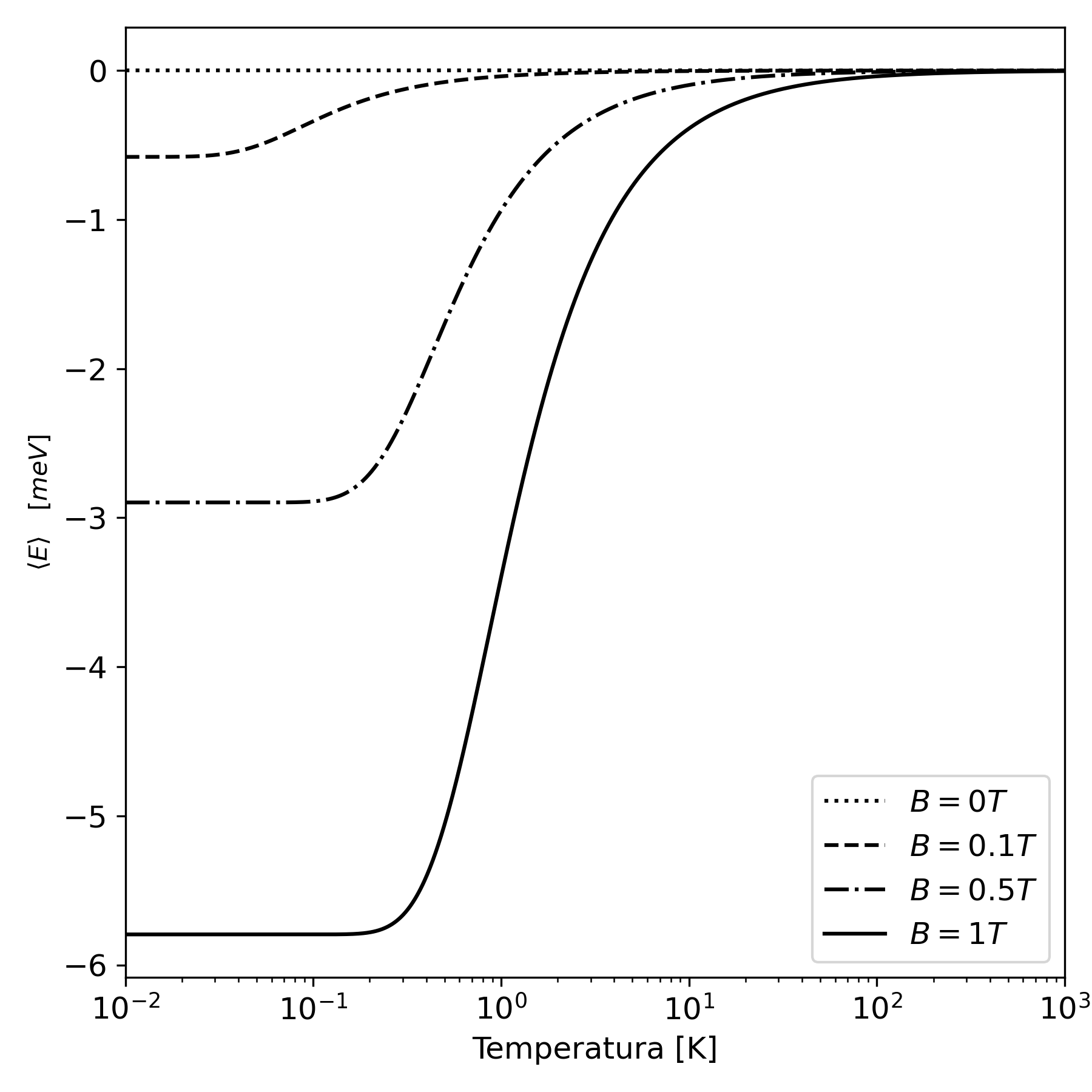}
        \caption{\label{fig:h2_valor_esperado_energía_param2}}
    \end{subfigure}
    \hfill
    \begin{subfigure}{0.49\textwidth}
        \includegraphics[width=\linewidth]{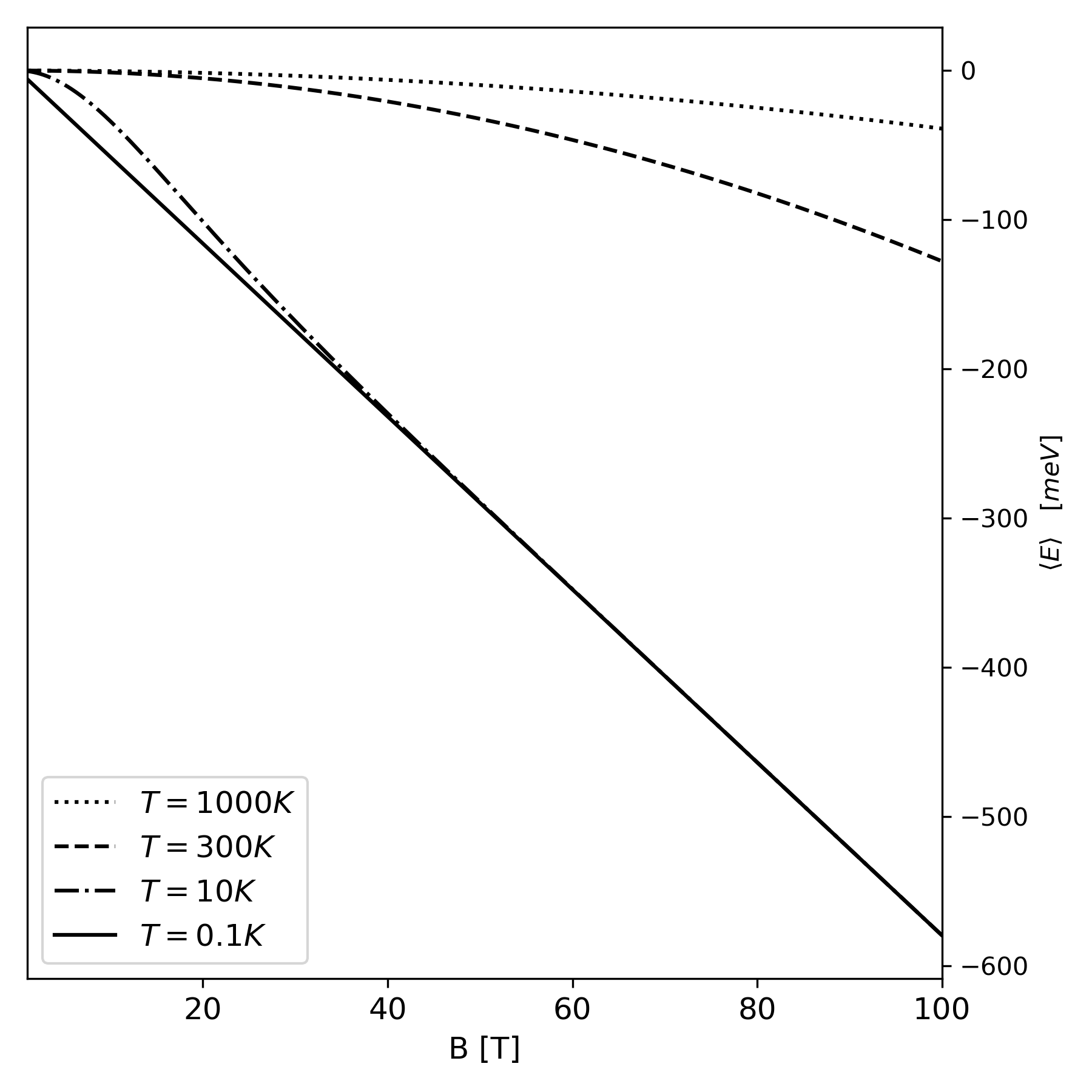}
        \caption{\label{fig:ch2_valor_esperado_energia_param2}}
    \end{subfigure}
    \caption{Valor esperado $\E{E}$ de la energía para un sistema formado por $N=100$ partículas paramagnéticas con espín 1/2, carga igual a la del electrón, con energía entre $E$ y $E+\delta E$, en presencia de un campo magnético externo, como función de (a) la temperatura para diferentes valores de campo magnético y (b) del campo magnético para diferentes valores de temperatura.}
    \label{fig:ch2_valor_esperado_energia_param}
\end{figure}

Al usar la definición de la tangente hiperbólica, $\tanh$, y teniendo en cuenta que $U = \E{E}$, se obtiene el valor esperado de la energía, ecuación \eqref{eq:ch2_valor_esperado_energia_param}, como se observa en la Figura \ref{fig:ch2_valor_esperado_energia_param}.

\begin{equation}
    \label{eq:ch2_valor_esperado_energia_param}
    \E{E} \approx - \gls{muB} B N \tanh \left( \frac{\gls{muB} B}{k_B T} \right)
\end{equation}

\subsection{Entropía y temperatura absoluta negativa}

\begin{figure}[t]
    \centering
    \includegraphics[width=\linewidth]{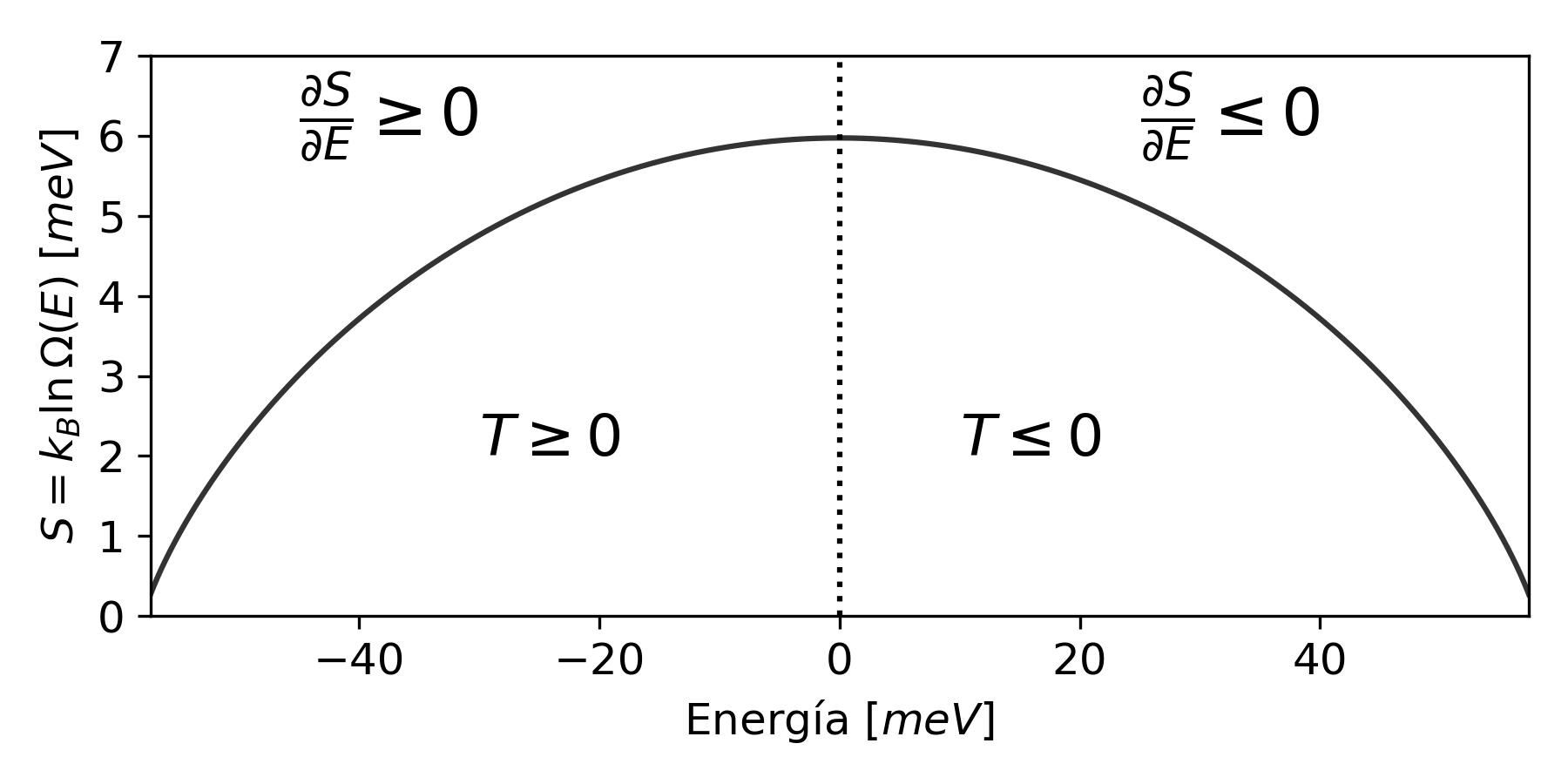}
    \caption{Entropía para un sistema formado por $N=100$ partículas paramagnéticas con espín 1/2 y carga igual a la del electrón, en presencia de un campo magnético externo de $10 T$.}
    \label{fig:ch2_Entropy_param}
\end{figure}

Uno de los resultados más interesantes del estudio de partículas con propiedades magnéticas, es la aparente existencia de temperaturas absolutas negativas, originada por el aumento en la entropía cuando se disminuye la energía del sistema, como se muestra en la Figura \ref{fig:ch2_Entropy_param}. La entropía del sistema está dada por la ecuación \eqref{eq:ch2_entropia_Boltzmann}, donde el número de estados accesibles está dado por la ecuación \eqref{eq:ch2_Omega_paramag}. Al utilizar la aproximación de Stirling, teniendo en cuenta que $\frac{\delta E}{2\gls{muB} B}$  es mucho menor que $N\ln N$ y $n_1\ln n_1$, además de simplificar términos, se obtiene la ecuación \eqref{eq:ch2_Entropia_param}, donde $n_1 = \frac{N}{2} - \frac{E}{2\gls{muB} B}$.

\begin{equation}
    \label{eq:ch2_Entropia_param}
    S \approx k_B \left[ NlnN - n_1\ln n_1 - (N-n_1) \ln (N-n_1) \right]
\end{equation}

Dado que $k_B \beta = 1/T = \partial S / \partial E$, en la gráfica de la entropía como función de la energía, que se muestra en la Figura \ref{fig:ch2_Entropy_param}, se pueden distinguir dos zonas para la temperatura absoluta: una zona para temperaturas absolutas positivas (a la izquierda) y otra para temperaturas absolutas negativas (a la derecha). Si consideramos la entropía como una función continua de la energía, una temperatura nula implica una discontinuidad en la curva de entropía versus energía, como se analizó en la sección \ref{sec:ch2_fisica_estadistica_temperatura_cero}. Sin embargo, la Figura \ref{fig:ch2_Entropy_param} sugiere temperaturas negativas, es decir, a medida que la energía aumenta, la entropía (y por consiguiente, el número de estados) disminuye. Esta observación la realizaron por primera vez en 1950 Pursell y Pound, quienes realizaron experimentos del espín nuclear en LiF \cite{Purcei1950} y observaron magnetizaciones negativas atípicas asociadas, según sus análisis, a un aumento en la entropía cuando disminuye la energía interna del sistema; este resultado fue interpretado por Pursell y Pound como consecuencia de una temperatura absoluta negativa. En su estudio, Pursell y Pound reportaron que un sistema en un estado de temperatura absoluta negativa no está frió, por el contrario, se encuentra muy caliente, con capacidad de dar energía a cualquier sistema a temperatura positiva con que se coloque en contacto. 

Desde el descubrimiento de Pursell y Pound, ha iniciado un amplio debate en la comunidad científica, sobre la existencia o no de las temperaturas absolutas negativas \cite{Onsageir1949, Purcei1950, Klein1956, Ramsey1956, Ramsey1956a, Coleman1959, Hecht1960, Tremblay1976, Hakonen1992, Mandt2013, Romero-Rochin2013, Dunkel2013, Swendsen2015, Swendsen2015a, Frenkel2015, Buonsante2017, Abraham2017, Hama2018, Hou2019, Miceli2019}. La controversia sobre la existencia de temperaturas negativas alcanzó su punto máximo en 2013 con el trabajo de Dunkel y Hilbert, donde argumentaron su imposibilidad \cite{Dunkel2013}. Sin embargo, Abraham y Penrose en 2017 revisaron la física de las temperaturas absolutas negativas y concluyeron que \lq\lq a pesar de los argumentos en contra, las temperaturas absolutas negativas tenían sentido teórico y sí se producían en experimentos diseñados para crearlas\rq\rq \cite{Abraham2017}. Este debate es un ejemplo del principio de falibilidad de la ciencia y, a pesar de los grandes esfuerzos dedicados a su estudio, es un tema sobre el cual sigue habiendo controversias.

\subsection{Magnetización y temperatura absoluta}

La magnetización $M$ es igual al número de momentos magnéticos por unidad de volumen. Si $V$ es el volumen ocupado por las partículas con momento magnético $\gls{muB}$, la magnetización, para partículas paramagnéticas, se puede expresar como una fuerza generalizada conjugada a la magnitud del campo magnético,

\[
X = V M = - \frac{\partial E}{\partial B} 
\]

Como $E=-\gls{muB} B (2 n_1 - N)$,

\[
\frac{\partial E}{\partial B} = \gls{muB} (2 n_1 - N) = - \frac{E}{B} = V M
\]

\begin{figure}[t]
    \centering
    \begin{subfigure}{0.49\textwidth}
        \includegraphics[width=\linewidth]{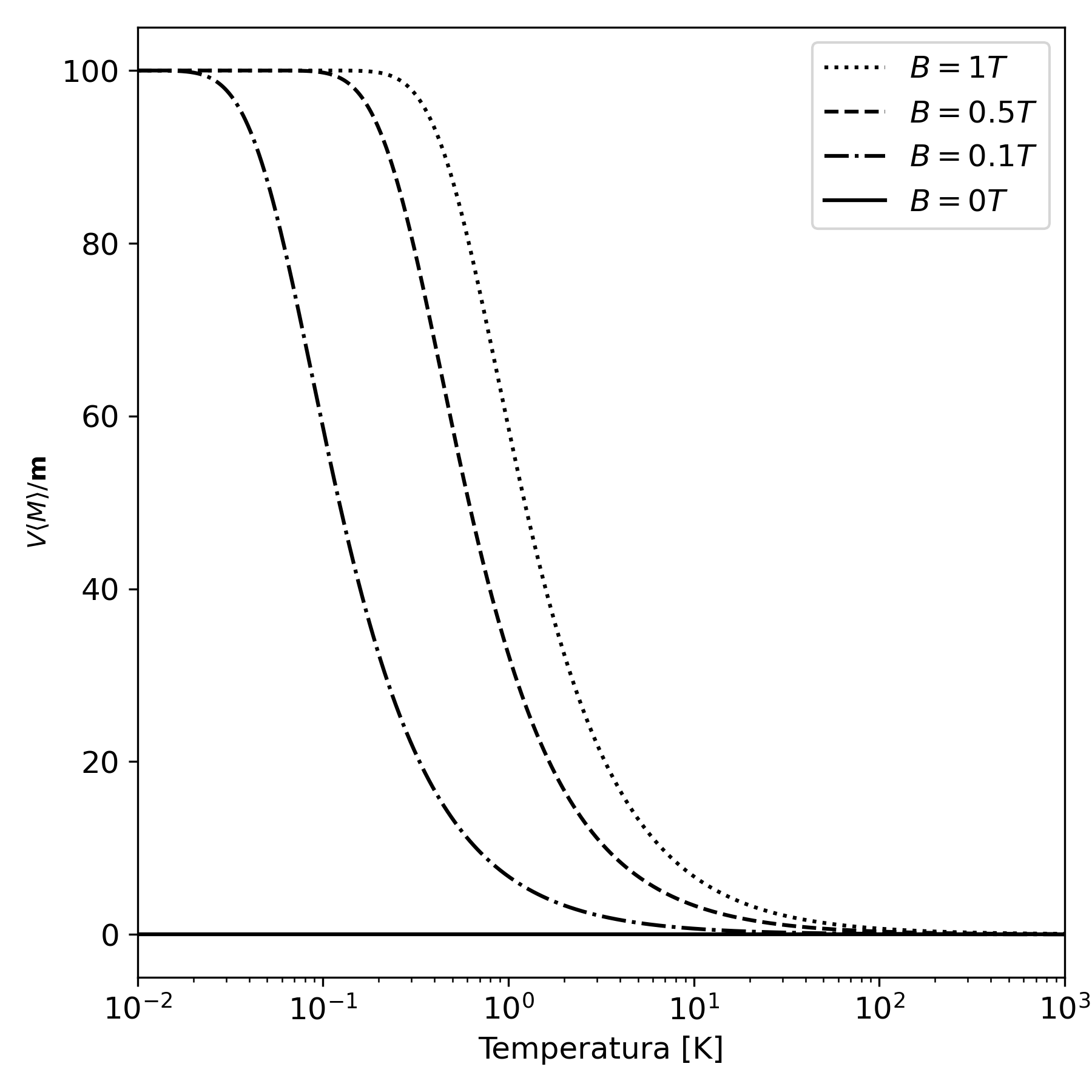}
        \caption{\label{fig:ch2_magnetizacion_param1}}
    \end{subfigure}
    \hfill
    \begin{subfigure}{0.49\textwidth}
        \includegraphics[width=\linewidth]{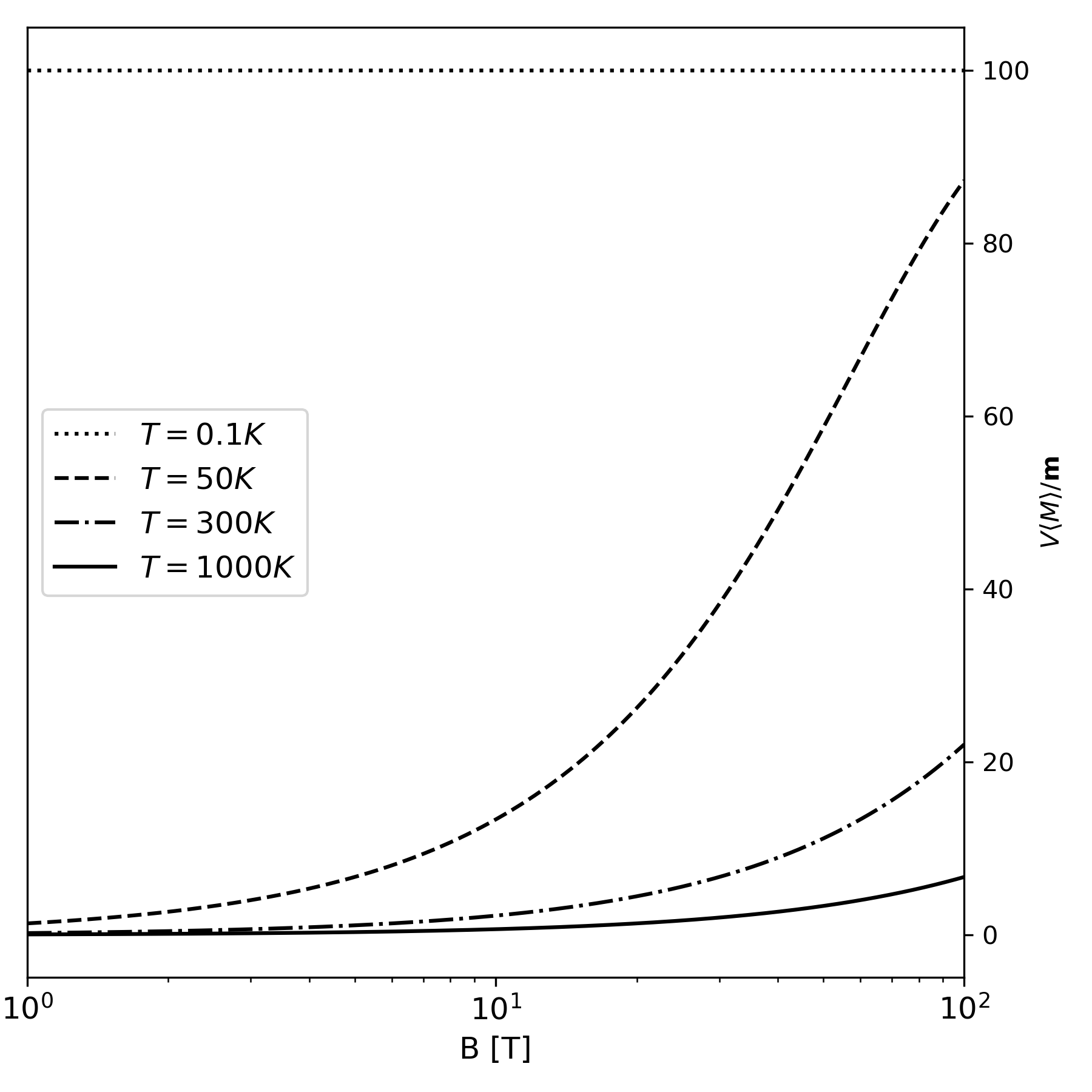}
        \caption{\label{fig:ch2_magnetizacion_param2}}
    \end{subfigure}
    \caption{Valor esperado $\E{M}$ de la magnetización para un sistema formado por $N=100$ partículas paramagnéticas contenidas en un volumen $V$, con espín 1/2, carga igual a la del electrón, con energía entre $E$ y $E+\delta E$, en presencia de un campo magnético externo, como función de (a) la temperatura para diferentes valores de campo magnético y (b) del campo magnético para diferentes valores de temperatura.}
    \label{fig:ch2_magnetizacion_param}
\end{figure}

Por lo tanto, el valor esperado $\E{M}$ de la magnetización estará dado por el cociente entre el valor esperado $\E{E}$ de la energía y la magnitud del campo magnético, ecuación \eqref{eq:ch2_magnetizacion_param}, como se muestra en la Figura \ref{fig:ch2_magnetizacion_param}.

\begin{equation}
    \label{eq:ch2_magnetizacion_param}
    \E{M} = \frac{\gls{muB} N}{V} \tanh \left( \frac{\gls{muB} B}{k_B T} \right)
\end{equation}

Con el objetivo que el lector adquiera un mayor nivel de comprensión, se le sugiere resolver las preguntas de autoexplicación del ejemplo trabajado titulado \href{https://colab.research.google.com/github/davidalejandromiranda/StatisticalPhysics/blob/main/notebooks/es_ParticulasConEspinEnCapoH_Temperatura.ipynb}{número de estados accesibles para de partículas paramagnéticas con espín 1/2}.

\section{Interacción entre subsistemas formados por dados}

Sea un sistema formado por $D$ dados, cada uno con $C$ caras marcadas con $C$ números naturales consecutivos iniciando en uno, que se lanzan sobre una superficie de área $2A$ dividida en dos partes iguales. Al lanzar los dados, algunos podrán quedar en una de las dos partes de área $A$, de tal manera que a los dados en una de las áreas se consideran parte del subsistema $A_1$ y los de la otra, al subsistema $A_2$. Si cada dado aporta al sistema una energía $E_0$, independiente del valor de cada cara, donde los números en las caras de los dados identifican el microestado para un cierto valor de energía $DE_0$,  ¿cuánto es el número de estados accesibles, la temperatura y la entropía si se considera que cada dado es distinguible de los demás? ¿Cuánto es el número de estados accesibles, la temperatura y la entropía si se considera que los dados son indistinguibles (dados idénticos)?

\subsection{Dados distinguibles}

\begin{figure}[t]
    \centering
    \includegraphics[width=\linewidth]{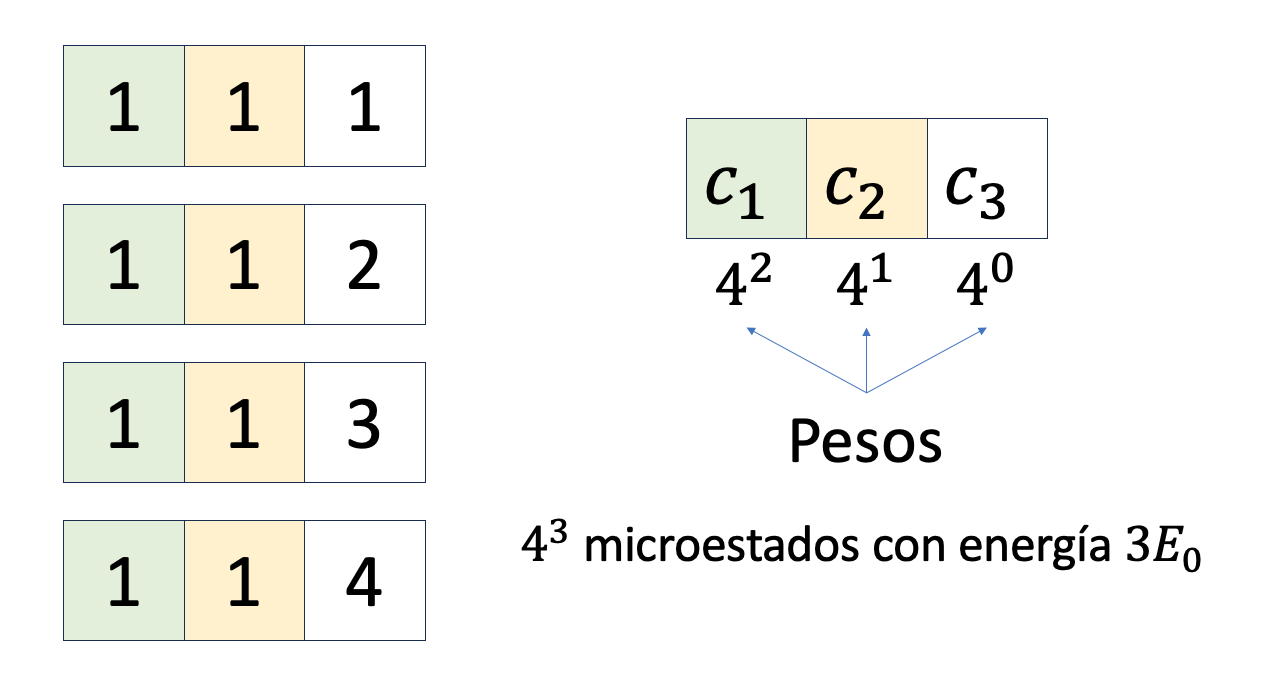}
    \caption{Representación esquemática de microestados para un sistema formado por tres dados ($D=3$) distinguibles, cada uno con cuatro caras ($C=4$), cuya energía es $E=3E_0$.}
    \label{fig:ch2_Omega_distinguibles_dados}
\end{figure}

Cuando los dados son distinguibles, un dado se puede identificar del otro, por ejemplo, por medio de un código único (puede ser un color, un número, barras, o cualquier forma de identificar a cada dado). El número de estados accesibles $\Omega(E; C)$ para un sistema con $D=3$ dados con $C=4$ caras, cuya energía $E=2E_0$, sería el número total de microestados con números de caras diferentes, como se muestra en la Figura \ref{fig:ch2_Omega_distinguibles_dados}.

En general, al tener un sistema con $D$ dados, cada uno con $C$ caras, se podrá tener una configuración con $n_1$ dados en $A_1$ y $n_2$ dados en $A_2$, donde $n_1 + n_2 = D$. El número de estados accesibles $\Omega_1(E; C)=\Omega_1(n_1 E_0; C)$ para el subsistema $A_1$ con $n_1$ dados con $C$ caras, cuya energía energía es $n_1 E_0$ está dado por $C^{n_1}$ y $\Omega(n_2 E_0; C)=C^{n_2}$. Por lo tanto, el número de estados accesibles $\Omega(n_1+n_2; C)$ para el sistema, será el producto de los números de estados accesibles de los subsistemas, 

\[
\Omega(n_1+n_2; C)=\Omega_1(n_1 E_0; C)\Omega_2(n_2 E_0; C)\]

Donde, $\Omega(n_1+n_2; C)$ es independiente de los valores $n_1$ y $n_2$, como muestra le ecuación \eqref{eq:ch2_Omega_distinguibles_dados}.

\begin{equation}
    \label{eq:ch2_Omega_distinguibles_dados}
    \Omega(D E_0; C) = C^{D}
\end{equation}

De manera general, la entropía del sistema $S$, en términos de las entropías de los subsistemas, $S_1$ y $S_2$, está dada por, 

\begin{align}
S &= k_B \ln \Omega(E; C) \nonumber \\
&= k_B \ln \Big[ \Omega_1(n_1 E_0; C) \Omega_2(n_1 E_0; C) \Big] \nonumber \\
&= k_B \ln \Omega_1(n_1 E_0; C) + k_B \ln \Omega_2(n_2 E_0; C)
\end{align}

\begin{equation}
    S = S_1 + S_2
\end{equation}

\begin{figure}[t]
    \centering
    \begin{subfigure}{0.49\textwidth}
        \includegraphics[width=\linewidth]{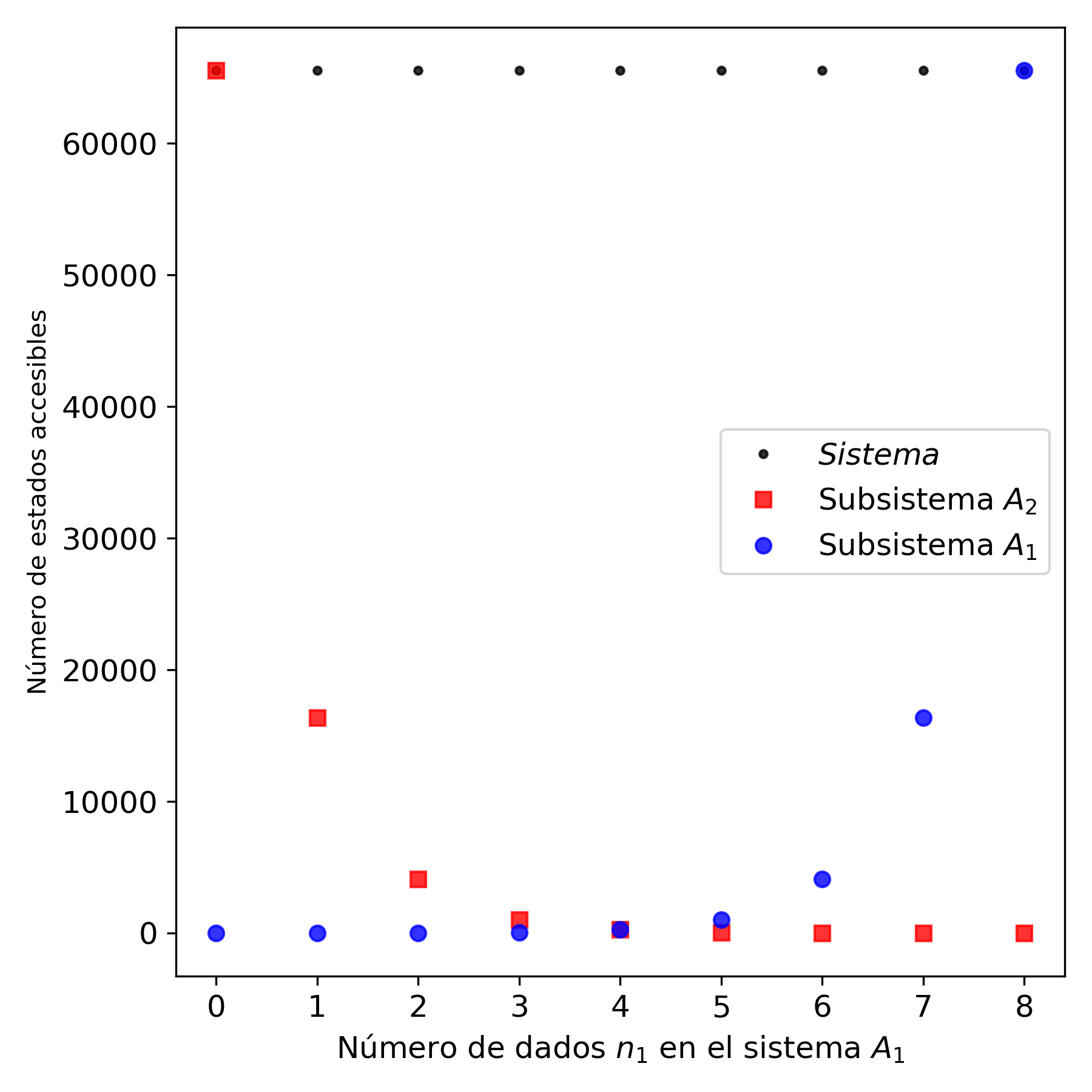}
        \caption{}
    \end{subfigure}
    \begin{subfigure}{0.49\textwidth}
        \includegraphics[width=\linewidth]{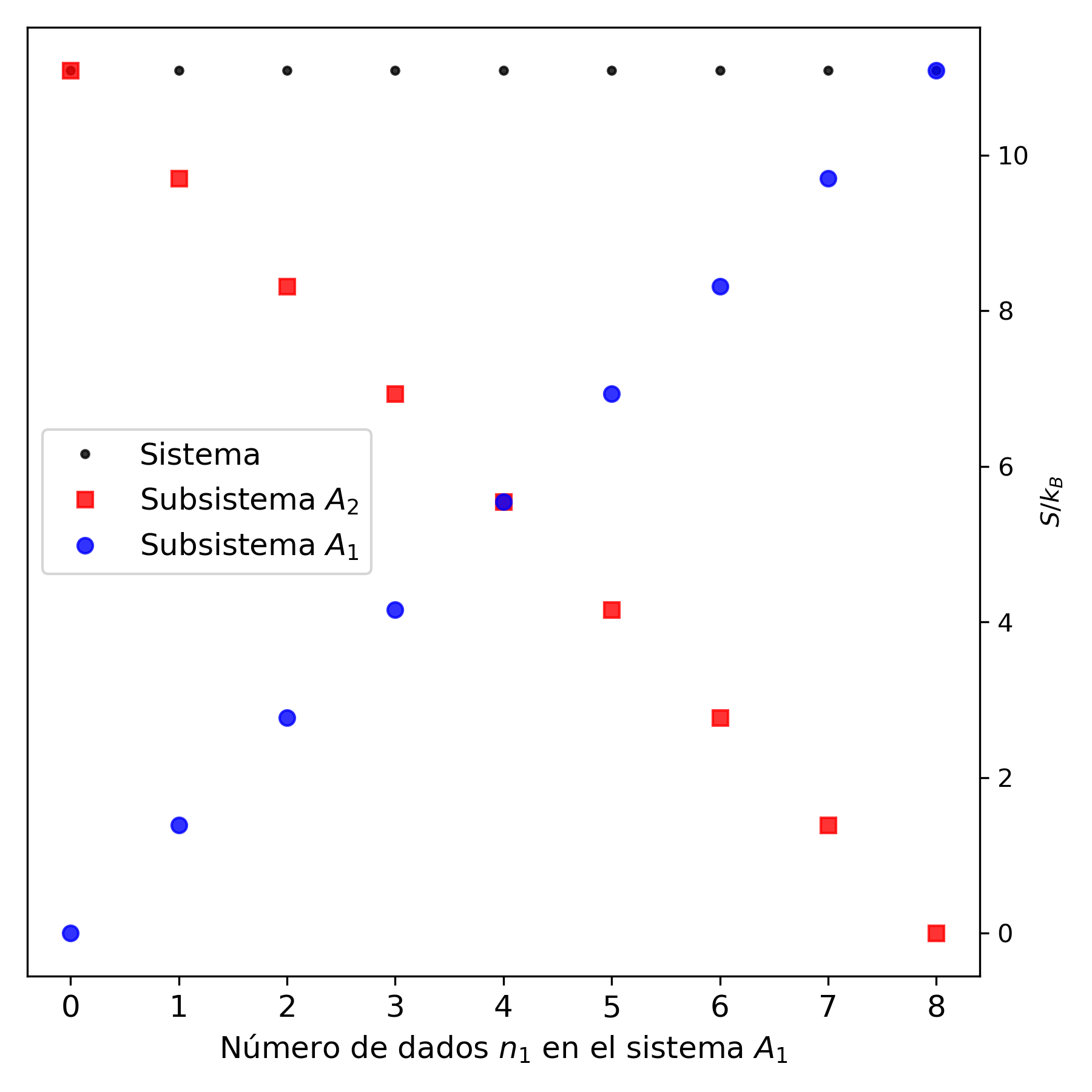}
        \caption{}
    \end{subfigure}
    \caption{(a) Número de estados accesibles y (b) entropía para dos subsistemas de un sistema formado por $D=8$ dados de $C=4$ caras cada uno de ellos.}
    \label{fig:ch2_Omega_Entropia_distinguibles_dados}
\end{figure}

En el caso de dados distinguibles, la entropía de cada subsistema dependerá de manera lineal respecto al número de dados, $S_i = k_B n_i \ln C$, como se muestra en la Figura \ref{fig:ch2_Omega_Entropia_distinguibles_dados} donde se presenta un sistema con $D=8$ dados, cada un con $C=4$ caras.

Por otra parte, el estado de equilibro termodinámico entre los sistemas se observa cuando $\beta_1=\beta_2$, donde $\beta$ está dado por la ecuación \eqref{eq:ch2_parametro_beta}, que para el i-ésimo sistema está dada por,

\[
\beta_i = \frac{1}{E_0} \frac{\partial}{\partial n_i} [n_i \ln C] = \frac{\ln C}{E_0}
\]
Por lo tanto, la temperatura para cada subsistema es $T=E_0 / \left( k_B \ln C \right)$, que es independiente del número de dados. Este resultado implica que todos los posibles microestados corresponden a condiciones de equilibrio termodinámico y son equiprobables.

\begin{definition}
    Un conjunto de partículas se conoce como \textbf{distinguibles} si es posible identificar a una de la otra.
\end{definition}

\subsection{Número de estados accesibles para dados indistinguibles}

\begin{figure}[t]
    \centering
    \begin{subfigure}{0.49\textwidth}
        \includegraphics[width=\linewidth]{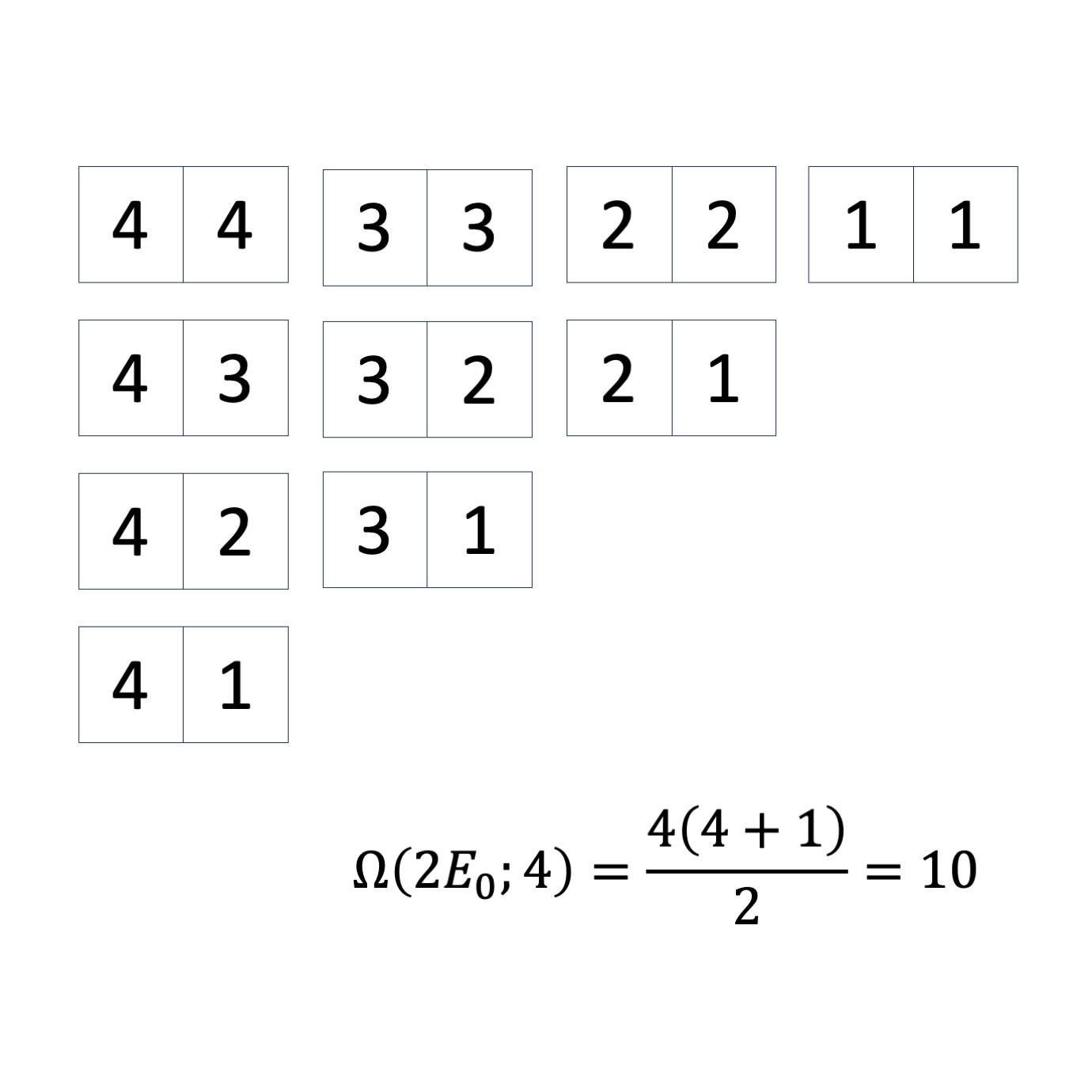}
        \caption{}
        \label{fig:ch2_Omega_indistinguibles_2dados_4caras}
    \end{subfigure}
    \begin{subfigure}{0.49\textwidth}
        \includegraphics[width=\linewidth]{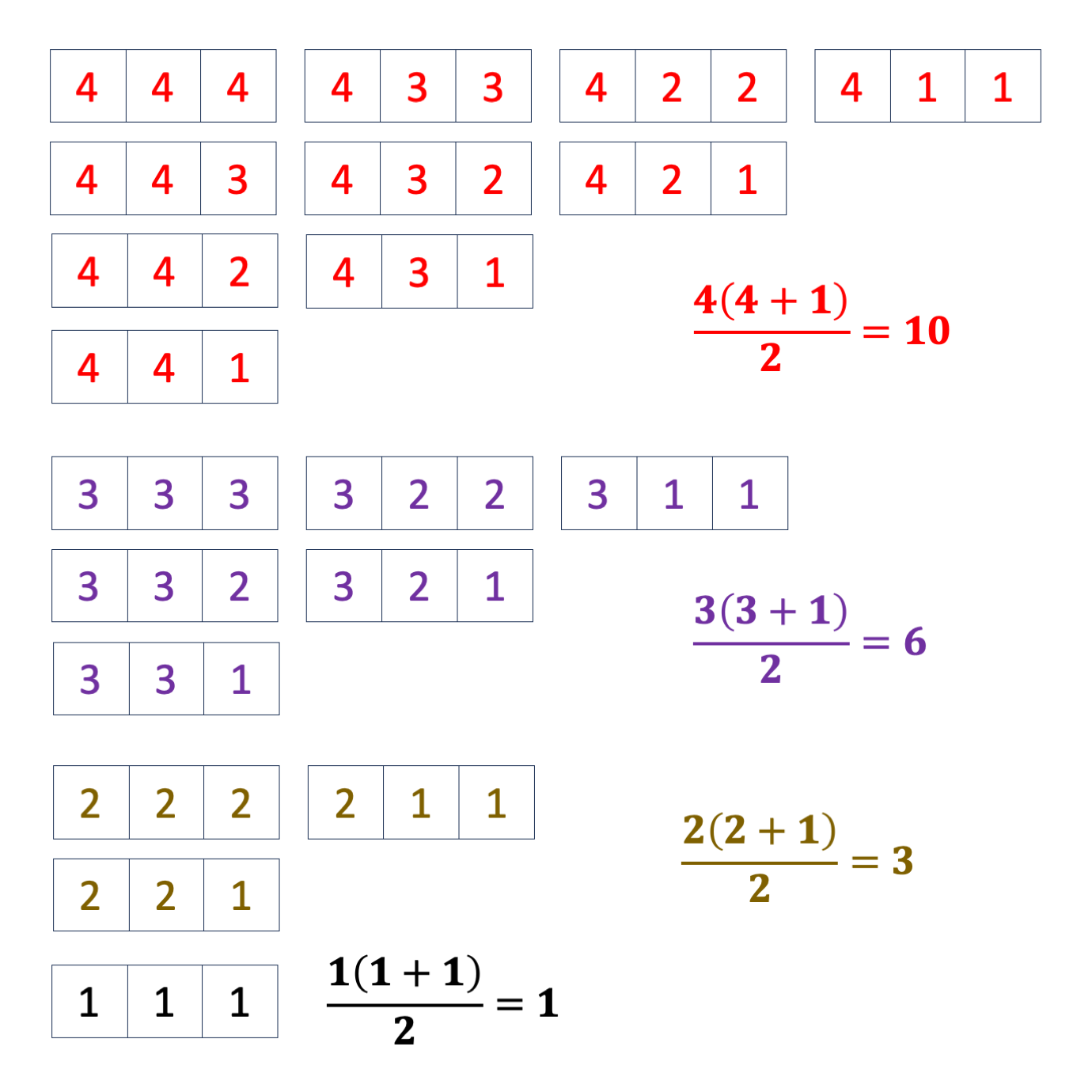}
        \caption{}
        \label{fig:ch2_Omega_indistinguibles_3dados_4caras}
    \end{subfigure}
    \caption{Microestados para un sistema con (a) $D=2$ dados con $C=4$ caras y (b) $D=3$ dados con $C=4$ caras.}
\end{figure}

Cuando los dados son indistinguibles, es imposible diferenciar un dado del otro, sin embargo, si se puede distinguir el número en cada una de las caras. La combinación de los diferentes posibles números de las caras de los dados da lugar a diferentes microestados para un mismo número $D$ de dados con $C$ caras. En la Figura \ref{fig:ch2_Omega_indistinguibles_2dados_4caras} se observan los posibles microestados para dos dados con cuatro caras y en la Figura \ref{fig:ch2_Omega_indistinguibles_3dados_4caras}, tres dados de cuatro caras. 

De la Figura \ref{fig:ch2_Omega_indistinguibles_2dados_4caras} se puede inferir que el número de estados $\Omega(2E_0; C)$ para un sistema con dos dados de $C$ caras está dado por,

\[
\Omega(2E_0; C) = \frac{C(C+1)}{2}
\]

De la misma manera, a partir de la Figura \ref{fig:ch2_Omega_indistinguibles_3dados_4caras} se puede inferir el número de estados $\Omega(3E_0; C)$ para un sistema con tres dados, cada uno con $C$ caras está dado por,

\[
\Omega(3E_0; C) = \sum\limits_{i=1}^C \frac{i(i+1)}{2} = \frac{C(C+1)(C+2)}{3!}
\]

Al seguir la secuencia, el número de estados accesibles $\Omega(4E_0; C)$ para un sistema con cuatro dados, cada uno con $C$ caras está dado por,

\[
\Omega(4E_0; C) = \sum\limits_{j=1}^C \sum\limits_{i=1}^j \frac{i(i+1)}{2} = \frac{C(C+1)(C+2)(C+3)}{4!}
\]

De esta manera, para $D$ dados, cada uno con $C$ caras se tienen,

\[
\Omega(DE_0; C) = \frac{\prod\limits_{i=0}^{D-1} (C+i)}{D!} = \frac{C(C+1)(C+2)\cdots(C+D-1)}{D!}
\]

La anterior ecuación se puede escribir en términos de factoriales como se muestra a continuación,

\begin{equation}
    \label{eq:ch2_Omega_indistinguibles_dados}
    \Omega(DE_0; C) = \frac{(C+D-1)!}{D!(C-1)!}
\end{equation}

La entropía, para números grandes de dados $D$ y caras $C$, se puede obtener al aplicar la aproximación de Stirling,

\begin{align}
S(nE_0; C) &= k_B \ln \Omega(nE_0; C) \nonumber \\
&\approx k_B \Big[ (C+D-1)\ln(C+D-1) - (C+D-1) + \nonumber \\
&\quad\quad - D\ln D + D - (C-1)\ln(C-1) + (C-1) \Big]
\end{align}

En equilibrio termodinámico el sistema se encuentra en el estado con mayor probabilidad, que corresponde también con el estado para el cual $\beta_1=\beta_2$, donde $\beta$ está dado por la ecuación \eqref{eq:ch2_parametro_beta}, que para el i-ésimo sistema está dada por,

\[
\beta_i = \frac{1}{k_BE_0} \frac{\partial S(n_iE_0; C)}{\partial n_i} \approx \frac{1}{E_0} \ln \left[ \frac{C+n_i-1}{n_i} \right]
\]

\begin{figure}[t]
    \centering
    \begin{subfigure}{0.49\textwidth}
        \includegraphics[width=\linewidth]{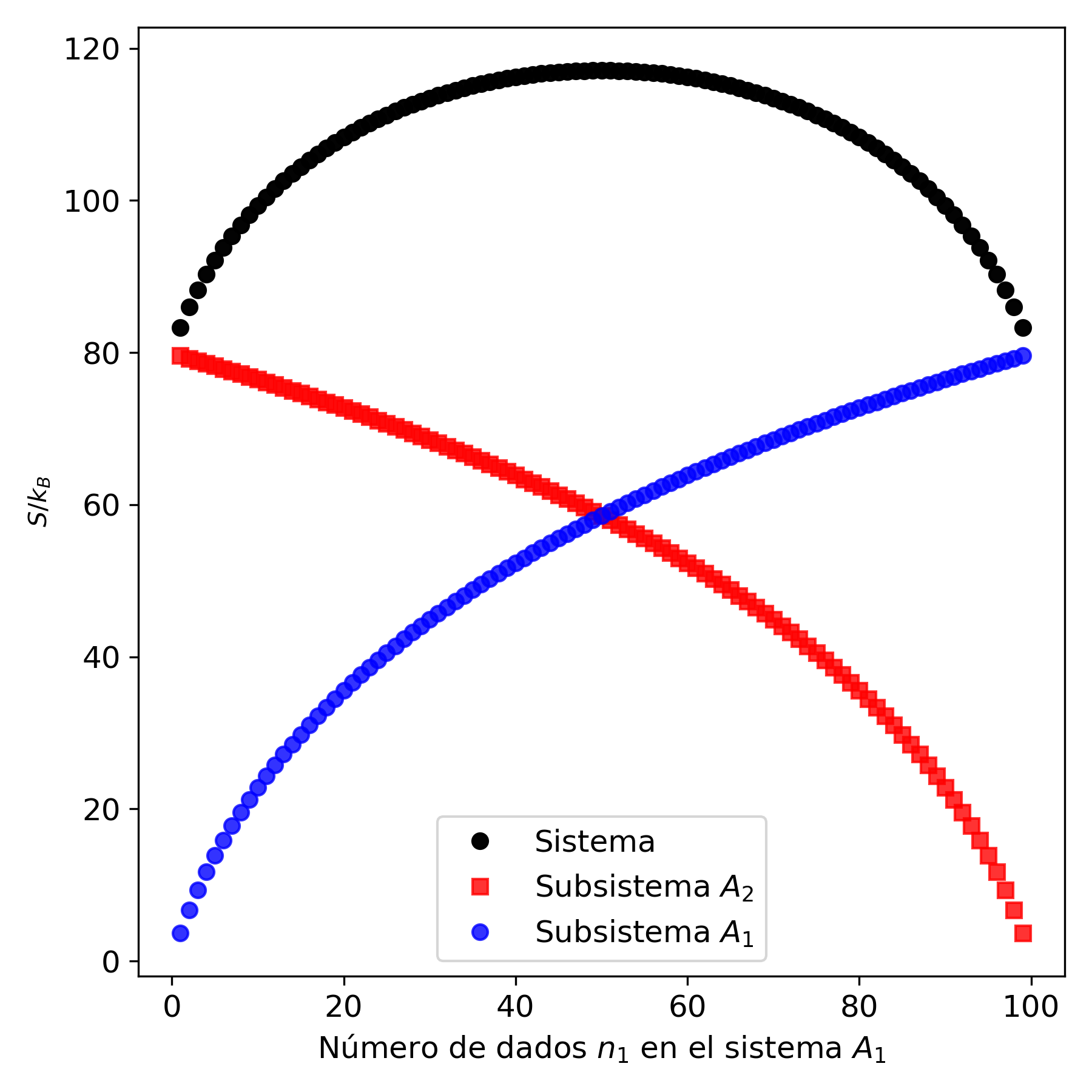}
        \caption{}
        \label{fig:ch2_Omega_Entropia_indistinguibles_dados}
    \end{subfigure}
    \begin{subfigure}{0.49\textwidth}
        \includegraphics[width=\linewidth]{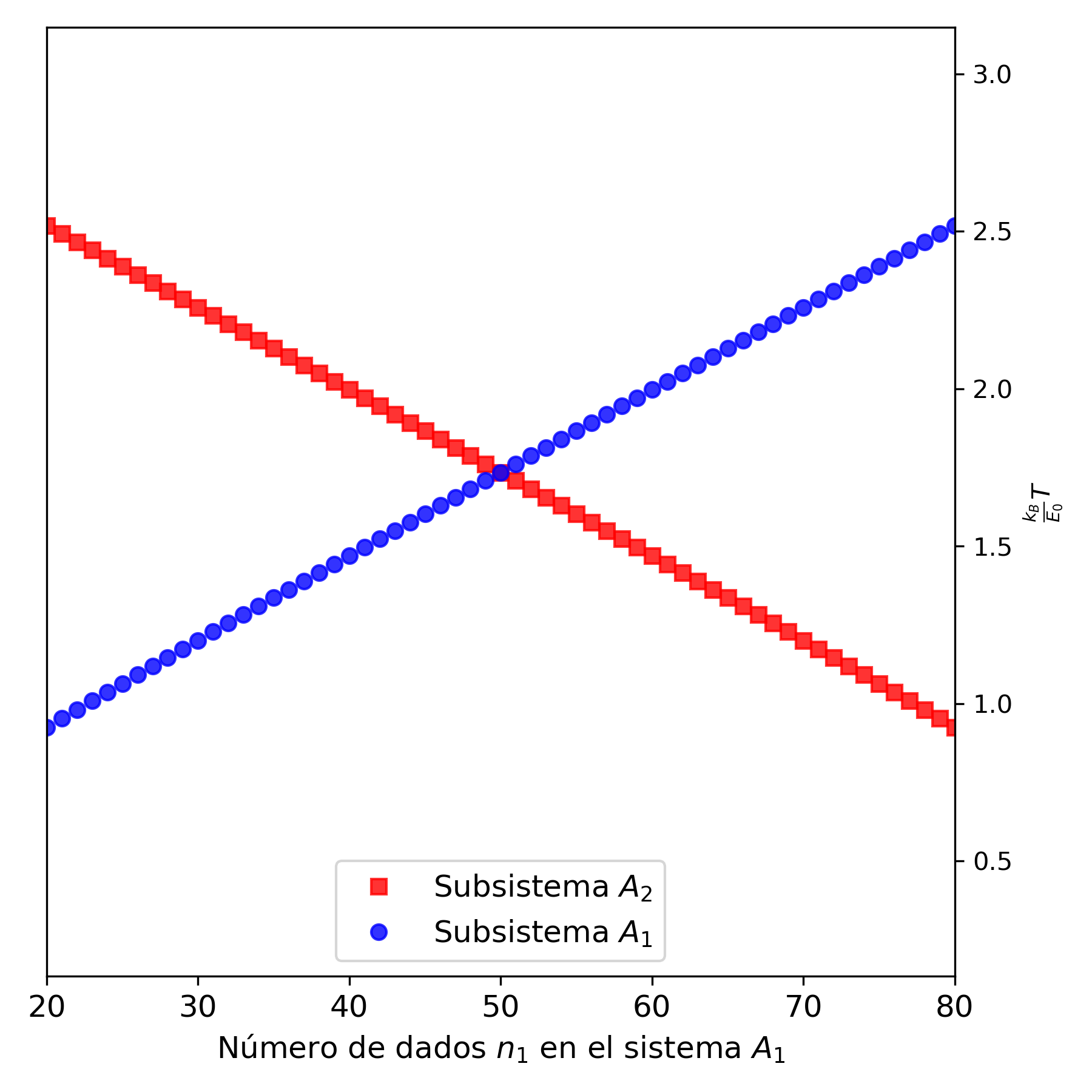}
        \caption{}
        \label{fig:ch2_Omega_Temperatura_indistinguibles_dados}
    \end{subfigure}
    \caption{(a) Entropía y (b) temperatura absoluta para un sistema con $D=100$ dados con $C=40$ caras cada uno.}
    \label{fig:ch2_Omega_Temperatura_Entropia_indistinguibles_dados}
\end{figure}

Por lo tanto, la temperatura para cada subsistema es $T_i \approx E_0 / \left[ k_B \ln  \left(\frac{C+n_i-1}{n_i}\right) \right]$, donde la condición de equilibrio termodinámico se alcanza cuando el número de partículas es igual en ambos subsistemas, como se muestra en la Figura \ref{fig:ch2_Omega_Temperatura_Entropia_indistinguibles_dados}.

Con el objetivo que el lector adquiera un mayor nivel de comprensión, se le sugiere resolver las preguntas de autoexplicación del ejemplo trabajado titulado \href{https://colab.research.google.com/github/davidalejandromiranda/StatisticalPhysics/blob/main/notebooks/es_InteraccionSistemasDeDados.ipynb}{interacción entre dos sistemas formados por dados}

\begin{definition}
    Un conjunto de partículas se conoce como \textbf{indistinguibles} si es imposible identificar a una de la otra.
\end{definition}

\section{Sistemas aislados en el límite clásico}

El microestado de un sistema clásico aislado, conformado por $N$ partículas, está determinado por $6N$ cantidades en el espacio de fase, $3N$ de posición y $3N$ de momentum. De esta manera, una posición en el espacio de fase permite identificar un microestado del sistema y el conjunto formado por todas las posiciones en las cuales se puede encontrar al sistema, corresponde con todos los posibles microestados de este. 

Con el fin de introducir los conceptos a estudiar se partirá de un caso simple, dado por un sistema formado por una única partícula confinada en todas excepto una de las dimensiones espaciales, en tal caso, el sistema se describe por solo dos cantidades, $x$ y $p_x$, las cuales definen una trayectoria en el espacio de fase cuyo perímetro representa el hipervolumen ocupado por el sistema; por ejemplo, un oscilador armónico clásico describe una trayectoria elíptica en el espacio de fase y el perímetro de esta elipse corresponde con el hipervolumen ocupado por el sistema en dicho espacio. Un microestado del sistema confinado a una dimensión espacial ocupa un hipervolumen en el espacio de fase $h=\delta x \delta p_x$ donde tanto $\delta x$ como $\delta p_x$ son cantidades que tienden a cero. 

Es de resaltar que el principio de incertidumbre de Heisenberg establece un límite para la precisión con la que se puede determinar, simultáneamente, $\delta q$ y $\delta p_x$, con lo cual se puede concluir que es imposible determinar con precisión un punto en el espacio de fase y, con ello, se evidencia la imposibilidad de predecir con precisión el estado inicial (o cualquier otro estado) de un sistema. El desconocer las implicaciones de esta indeterminación conduce a la idea equivocada de precisión infinita, descrita por Barbara Drossel \cite{Drossel2015}. En tal sentido, durante el estudio de sistemas representados en el espacio de fase debemos mantener presente la imposibilidad de determinar con precisión el microestado del sistema en dicho espacio.\footnote{Si está interesado en ampliar sus conocimientos sobre la idea equivocada de precisión infinita, se sugiere estudiar el capítulo de Barbara Drossel en el libro titulado \textit{on the relation between the second law of thermodynamics and classical and quantum mechanics} \cite{Drossel2015}.}

Para sistemas con mayores grados de libertad, por ejemplo, un sistema formado por una partícula confinada a dos dimensiones, con grados de libertad $x_1$, $x_2$, $p_{x_1}$ y $p_{x_2}$, el espacio de fase tendrá una mayor dimensión y el hipervolumen ocupado por el sistema en dicho espacio ya no será un perímetro, por ello se le ha nombrado como hipervolumen a la porción del espacio de fase ocupada por las configuraciones de un sistema. En este sistema se dice que se tiene $f=2$ grados de libertad espacial y el hipervolumen de un microestado está dado por $h^f = (\delta x \delta p_x)^2$. 

A partir del análisis anterior se puede concluir que el número de estados accesibles $\Omega(E)$ con energía entre $E$ y $E+dE$ se obtiene al realizar el cociente entre el hipervolumen en el espacio de fase ocupado por el sistema cuando su energía se encuentra en el intervalo comprendido entre $E$ y $E+dE$ y el hipervolumen $h^f$ que ocupa un microestado, donde $h=\delta q \delta p$ y $f$ es el número de grados de libertad espaciales. Además, al conocer el número de estados accesibles, se puede obtener la probabilidad que el sistema se encuentre en un estado con energía entre $E$ y $E+dE$; con ello, es posible calcular los valores esperados e incertidumbres.

A continuación, se analizará el ensamble microcanónico de sistemas clásicos. Se plantearán y considerarán las implicaciones de los teoremas de Poincaré y Liouville en la determinación de la probabilidad de encontrar al sistema en un cierto microestado; también se analizará un sistema que describe un movimiento armónico simple clásico, además del gas ideal clásico y se derivará tanto su ecuación de estado como el valor de la entropía para el gas ideal clásico.

\begin{definition}
	A cada una de las variables independientes que especifican el microestado de un sistema se le conoce como \textbf{grado de libertad}.\index{Grado de libertad}
\end{definition}

\begin{definition}
	El número de \textbf{grados de libertad espacial} de un sistema se denota por la letra $f$ (del vocablo inglés, \textit{freedom}). \index{Número de grados de libertad espacial}
\end{definition}

\begin{definition}
	Sean $f$ grados de libertad espacial, $q_i$ y $f$ de momentum, $p_i$. Al espacio formado por todos los posibles puntos $(q_1, q_2, \cdots, q_f, p_1, p_2, \cdots, p_f) = (\{q_i\}, \{p_i\})$ se le conoce como \textbf{espacio de fase}.\index{Espacio de fase}
\end{definition}

\begin{definition}
	Sea un sistema formado por $N$ partículas, tal que existen $3N$ coordenadas espaciales $q_i$ y $3N$ momentum $p_i$, donde $i=1, 2, \cdots, 3N$. Al conjunto formado por los $6N$ valores $(q_1, q_2, \cdots, q_{3N}, p_1, p_2, \cdots, p_{3N})=(\{q_i\}, \{p_i\})$ se le conoce como \textbf{configuración en el espacio de fase}. \index{Espacio de fase!configuración}
\end{definition}

\begin{definition}
	Un microestado corresponde con una configuración en el espacio de fase.\index{Microestado en el espacio de fase}
\end{definition}

\begin{definition}
	El espacio ocupado por todas las configuraciones de un sistema en el espacio de fase define el \textbf{hipervolumen} del sistema en dicho espacio.\index{Espacio de fase!hipervolumen}
\end{definition}

\begin{definition}
	El hipervolumen ocupado por un microestado de un sistema con $f$ grados de libertad se denota por $h^f$, donde $h$ es una cantidad infinitesimal que se puede escribir como $h=\delta q \delta p$. \index{Espacio de fase!hipervolumen de un microestado}
\end{definition}

\section{Espacio de fase y número de estados accesibles}

\begin{lema}
	\textbf{Teorema de recurrencia de Poincaré}. Cualquier configuración del espacio de fase $(\{q_i\}, \{p_i\})$ de un sistema en un volumen finito se repite después de un intervalo de tiempo finito (que puede ser corto o largo).
	\index{Teorema de recurrencia de Poincaré}\label{lema:teorema_recurrencia_Poincare}
\end{lema}

\begin{definition}
	La \textbf{densidad de microestados en el espacio de fase} $\rho$ es el número de configuraciones en el espacio de fase por unidad de hipervolumen para un microestado.\index{Espacio de fase!densidad de microesatdos}
\end{definition}

\begin{definition}
	Sean dos funciones \( f(\{q_i\}, \{p_i\}) \) y \( g(\{q_i\}, \{p_i\}) \) de las coordenadas generalizadas \( \{q_i\} \) y los momentos conjugados \( \{p_i\} \) de un sistema, donde $i=1, 2, \cdots, 3N$, el \textbf{corchete de Poisson} entre \( f \) y \( g \) se representa como $\{f, g\}$ y se definen como,
	
	\begin{equation}
		\{f, g\} = \sum_{i=1}^{3N} \left( \frac{\partial f}{\partial q_i} \frac{\partial g}{\partial p_i} - \frac{\partial f}{\partial p_i} \frac{\partial g}{\partial q_i} \right)
	\end{equation}
	\index{Corchete de Poisson}
\end{definition}

\begin{lema}
	\textbf{Teorema de Liouville}. La densidad de microestados en el espacio de fase para un sistema con hamiltoniano $H$ permanece invariante en el tiempo y se cumple la ecuación \eqref{eq:ch3_teorema_Liouville}.\index{Teorema de Liouville}
	\begin{equation}
		\label{eq:ch3_teorema_Liouville}
		\frac{d\rho}{dt} = \frac{\partial \rho}{\partial t} + \{\rho, H\} = 0
	\end{equation}
\end{lema}

Sea un sistema aislado formado por $N$ partículas, donde cada punto $(\{q_i\}, \{p_i\})$ en el espacio de fase corresponde con una configuración, a la cual se le puede asociar un microestado. Si en una unidad de hipervolumen $h^{3N}$ se tienen $C$ configuraciones, entonces, la densidad de microestados en el espacio de fase, alrededor de dicho hipervolumen, será $\rho \approx C/h^{3N}$; de manera general, la densidad $\rho = \rho(E; \{q_i\}, \{p_i\})$, es decir, puede variar de punto en punto en el espacio de fase. 

Si el sistema tiene una energía entre $E$ y $E+dE$, la cual define un diferencial de hipervolumen $dq_1 \cdots dq_{3N} dp_1 \cdots dp_{3N}$ en el espacio de fase, entonces, el número de estados accesibles para un sistema aislado, está dado por,

\begin{equation}
	\label{eq:ch3_numero_estados_accesibles_sistemas_clasicos}
	\Omega(E; \{q_i\}, \{p_i\}) = \rho(E; \{q_i\}, \{p_i\}) dq_1 \cdots dq_{3N} dp_1 \cdots dp_{3N}
\end{equation}

\section{Oscilador armónico clásico unidimensional}
\index{Oscilador armónico clásico}

Sea un sistema formado por una partícula de masa $m$ sujeta a un resorte con constante elástica $k$ que describe un movimiento armónico simple clásico. ¿Cuánto es el número de estados accesibles para el sistema con energía $E$ cuando la masa se encuentra en una posición entre $x$ y $x+dx$? ¿Cuánto es la probabilidad de encontrar el sistema en una cierta posición entre $x$ y $x+dx$ cuando su energía es $E$? ¿Cuánto es la entropía del sistema? ¿Cuánto es la temperatura absoluta del sistema?

\subsection{Espacio de fase}

El hamiltoniano del sistema está dado por la ecuación \eqref{eq:ch3_oscilador_armonico_clasico_hamiltoniano}, donde la energía total sistema $E=H$.  

\begin{equation}
	\label{eq:ch3_oscilador_armonico_clasico_hamiltoniano}
	H = \frac{p^2}{2m} + \frac{1}{2}kx^2
\end{equation}

Al aplicar las ecuaciones canónicas de Hamilton se obtiene para el momentum,

\[
\frac{dp}{dt} = -\frac{\partial H}{\partial x} = - kx \to \frac{dp}{dt} + kx = 0
\]

Y para la posición,

\[
\frac{dx}{dt} = \frac{\partial H}{\partial p} = \frac{p}{m} \to p = m\frac{dx}{dt}
\]

Estas ecuaciones diferenciales acopladas se pueden escribir como,

\[
\frac{d^2x}{dt^2} + \frac{k}{m}x = 0
\]

\begin{figure}[t]
    \centering
    \begin{subfigure}{0.49\textwidth}
        \includegraphics[width=\linewidth]{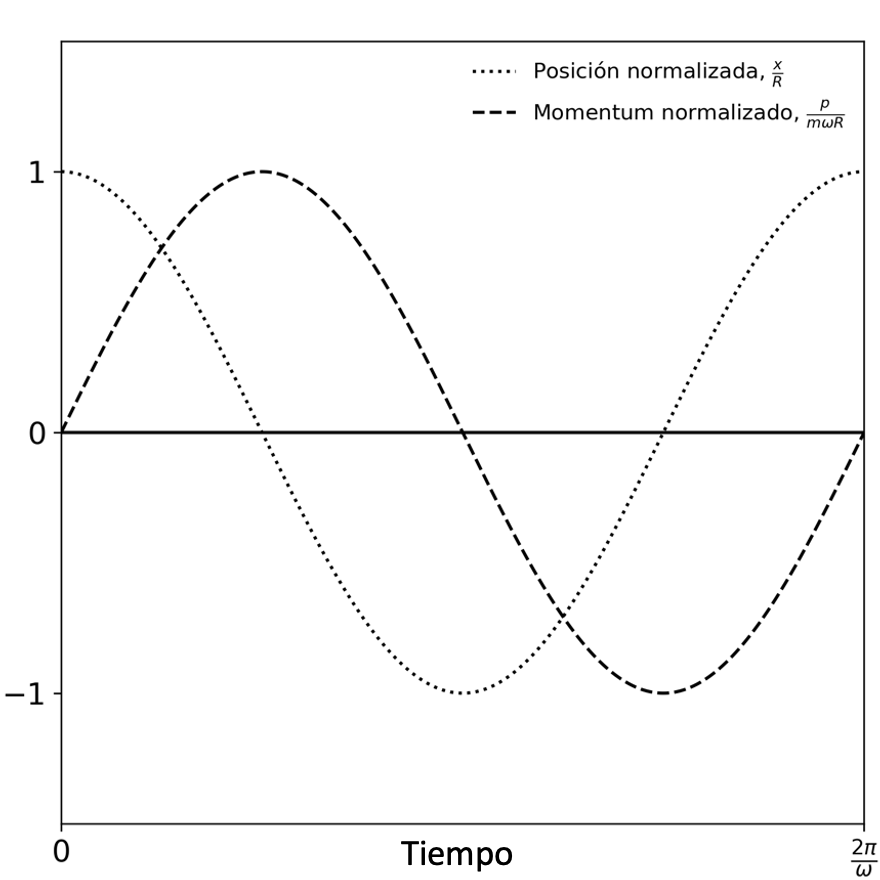}
        \caption{}
        \label{fig:ch3_oscilador_armonico_clasico_x_p}
    \end{subfigure}
    \begin{subfigure}{0.49\textwidth}
        \includegraphics[width=\linewidth]{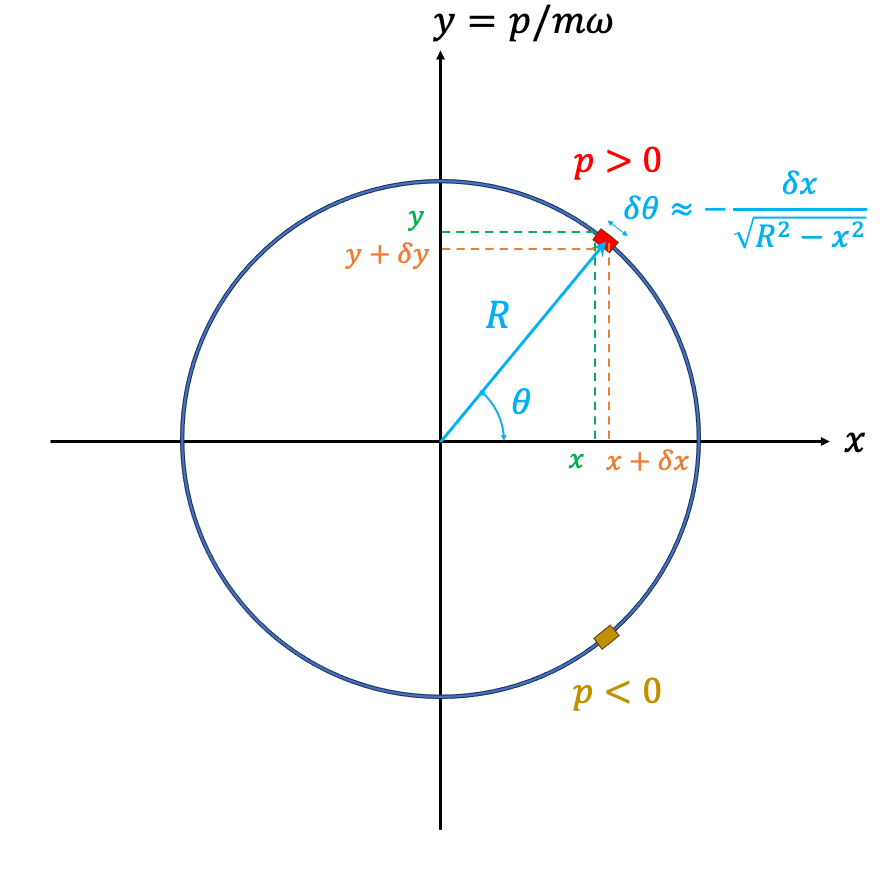}
        \caption{}
        \label{fig:ch3_oscilador_armonico_clasico_espacio_fase}
    \end{subfigure}
    \caption{(a) Posición $x$ y momentum $p$ de una partícula que describe un movimiento armónico simple y (b) trayectoria en el espacio de fase.}
    \label{fig:ch3_oscilador_armonico_clasico_espacio_fase_x_p}
\end{figure}

Esta es la ecuación característica de un movimiento armónico simple, como se muestra en la Figura \ref{fig:ch3_oscilador_armonico_clasico_x_p}, cuya solución analítica está dada por,

\begin{equation}
	\label{eq:ch3_oscilador_armonico_clasico_frecuencia}
	\omega^2 = \frac{k}{m}
\end{equation}

\begin{equation}
	\label{eq:ch3_oscilador_armonico_clasico_posición}
	x = R \cos (\omega t)
\end{equation}

\begin{equation}
	\label{eq:ch3_oscilador_armonico_clasico_momentum}
	p = m \omega R \sen (\omega t) = m\omega \sqrt{R^2 - x^2}
\end{equation}

Como $k=m\omega^2$, la energía del sistema, en términos de la posición y el momentum se puede expresar como se muestra en la ecuación \eqref{eq:ch3_oscilador_armonico_clasico_energia_px}.

\begin{equation}
	\label{eq:ch3_oscilador_armonico_clasico_energia_px}
	E = \frac{p^2}{2m} + \frac{1}{2}m\omega^2 x^2
\end{equation}

Al reemplazar las ecuaciones \eqref{eq:ch3_oscilador_armonico_clasico_posición} y \eqref{eq:ch3_oscilador_armonico_clasico_momentum} en \eqref{eq:ch3_oscilador_armonico_clasico_energia_px}, se obtiene,

\begin{equation}
	\label{eq:ch3_oscilador_armonico_clasico_energía}
	E = \frac{1}{2} m\omega^2 R^2
\end{equation}

Con el anterior resultado se puede hacer un cambio de variables en la ecuación \eqref{eq:ch3_oscilador_armonico_clasico_energia_px} multiplicándola por $\frac{2}{m\omega^2}$ para obtener la representación con unidades espaciales del sistema en el espacio de fase,

\begin{equation}
	y = \frac{p}{m\omega}
\end{equation}

\begin{equation}
	R^2 = x^2 + y^2
\end{equation}

Esto nos dice que la representación en el espacio de fase (en dimensiones espaciales) corresponde con una circunferencia de radio $R=\sqrt{2E/m\omega^2}$, como se muestra en la Figura \ref{fig:ch3_oscilador_armonico_clasico_espacio_fase}.

\subsection{Número de estados accesibles y probabilidad}

El número de estados accesibles $\Omega(E; x)$ para el sistema con energía $E$ cuya posición se encuentra entre $x$ y $x+dx$ está dada por,

\[
\Omega(E; x) = \rho(E; x) \delta s = \rho(E; x) R |\delta \theta| = \frac{2}{h} R |\delta \theta|
\]

Donde $\delta s$ es la longitud de arco sobre la circunferencia para el intervalo $\delta x$; $h=ds=Rd\theta$, es el diferencial de arco asociado a un microestado y la densidad de microestados $\rho$ está determinada por las dos posibles configuraciones asociadas a cada valor de $x$, una con valor positivo de $p$ y la otra, con valor negativo, como se muestra en la Figura \ref{fig:ch3_oscilador_armonico_clasico_espacio_fase}. Para obtener $|\delta \theta|$ debemos recordar que $x(\theta)=R \cos \theta$, por lo tanto,

\[
\delta x = x(\theta + \delta \theta) - x(\theta) = \frac{\partial x}{\partial \theta} \delta \theta + \mathcal{O}(\delta \theta^2) = -R \delta \theta \sen \theta + \mathcal{O}(\delta \theta^2)
\]

Dado que $y=R\sen\theta = \sqrt{R^2 - x^2}$,

\[
|\delta \theta| \approx \frac{\delta x}{\sqrt{R^2 - x^2}}
\]

Si $\delta \theta \to 0$, entonces, $\delta\theta\to d\theta$ y $\delta x \to dx$, con lo cual, 

\begin{equation}
	\label{eq:ch3_oscilador_armonico_clasico_estados_accesibles}
	\Omega(E; x) = \frac{2R dx}{h\sqrt{R^2 - x^2}}
\end{equation} 

Además, el número total de estados accesibles del sistema $\Omega_T(E)$, está dado por el perímetro de la circunferencia,

\begin{equation}
	\label{eq:ch3_oscilador_armonico_clasico_estados_accesibles_total}
	\Omega_T(E) = \frac{2\pi R}{h}
\end{equation} 

De esta manera, la probabilidad $\varrho(E; x)dx$ que la posición del sistema con energía $E$ se encuentre entre $x$ y $x+dx$ está dada por el cociente entre $\Omega(E; x)$ y $\Omega_T(E)$,

\begin{equation}
	\varrho(E; x)dx = \frac{\Omega(E; x)}{\Omega_T(E)} = \frac{1}{\pi \sqrt{R^2 - x^2}} dx
\end{equation}

\subsection{Entropía y temperatura absoluta}

La entropía del sistema con energía entre $E$ y $E+dE$ está dada por el logaritmo natural de la ecuación \eqref{eq:ch3_oscilador_armonico_clasico_estados_accesibles_total},

\[
	S = k_B \ln R + k_B \ln \left( \frac{2\pi}{h} \right)
\]

Como $R^2 = 2E / m\omega^2$,

\[
S = \frac{1}{2}k_B \ln \left( \frac{2E}{m\omega^2} \right) + k_B \ln \left( \frac{2\pi}{h} \right)
\]

Al separar los términos constantes de la energía,

\begin{equation}
	S = \frac{1}{2}k_B \ln E + k_B \ln \left( \frac{2\pi}{h} \sqrt{\frac{2}{m\omega^2}} \right)
\end{equation}

Por lo tanto, la temperatura absoluta está dada por,

\begin{equation}
	T = \left( \frac{\partial S}{\partial E} \right)^{-1} = \frac{2E}{k_B}
\end{equation}

Con el objetivo que el lector adquiera un mayor nivel de comprensión, se le sugiere resolver las preguntas de autoexplicación del ejemplo trabajado titulado \href{https://colab.research.google.com/github/davidalejandromiranda/StatisticalPhysics/blob/main/notebooks/es_PenduloSimpleEspacioFase.ipynb}{oscilador armónico clásico - análisis en el espacio de fase}.

\section{El gas ideal clásico}
\index{Gas ideal clásico!ensamble microcanónico}

Sea un gas ideal clásico formado por $N$ partículas no interactuantes con masa $m$, cuyo microestado está determinado por $6N$ coordenadas, $3N$ de posición ,$q_i$ y $3N$ de momentum, $p_i$. ¿Cuánto es el número de estados accesibles para el sistema con energía entre $E$ y $E+dE$? ¿Cuánto es la entropía del sistema? ¿Cuánto es la temperatura absoluta del sistema? ¿Cuál es la ecuación de estado del sistema?

\begin{definition}
	Un sistema formado por partículas no interactuantes se conoce como \textbf{gas ideal}.\index{Gas ideal}
\end{definition}

\begin{definition}
    Un sistema formado por $N$ partículas se considera un \textbf{gas ideal clásico} si su hamiltoniano se puede aproximar con la ecuación \eqref{eq:ch4_hamiltoniano_gas_ideal_clásico}.\index{Gas ideal clásico}
    \begin{equation}
        \label{eq:ch4_hamiltoniano_gas_ideal_clásico}
        H = \sum_{i=1}^{3N} \frac{p_i^2}{2m}
    \end{equation}
\end{definition}

\subsection{Espacio de fase}
\index{Gas ideal clásico!espacio de fase}
El hamiltoniano del gas ideal clásico está dado por la ecuación \eqref{eq:ch4_hamiltoniano_gas_ideal_clásico}, donde $H$ corresponde con la energía $E$ del sistema. Es importante notar que en un gas ideal el potencial de interacción entre sus partes constituyentes es despreciable.

Al hacer el cambio de variable dado por $R^2=2mE$, la ecuación \eqref{eq:ch4_hamiltoniano_gas_ideal_clásico} toma la forma de una hiperesfera en un espacio con dimensión $3N$,

\[
R^2 = p_1^2 + p_2^2 + \cdots + p_{3N}^2
\]

Cuyo hipervolumen está dado por la ecuación \eqref{eq:ch3_gas_ideal_clasico_hipervolumen_esfera}, donde $\Gamma$ es la función gamma.

\begin{equation}
	\label{eq:ch3_gas_ideal_clasico_hipervolumen_esfera}
	\mho_{3N} = \frac{\pi^{3N/2}R^{3N}}{\Gamma\left( \frac{3N}{2} + 1 \right)} = \frac{\pi^{3N/2}(2mE)^{3N/2}}{\Gamma\left( \frac{3N}{2} + 1 \right)}
\end{equation}

\subsection{Número de estados accesibles y probabilidad}
\index{Gas ideal clásico!número de estados accesibles}
Cada microestado del gas ideal clásico tiene una configuración en el espacio de fase única, por lo tanto, la densidad de microestados en el espacio de fase $\rho = 1/h^{3N}$, donde $h=\delta q \delta p$, por lo tanto, el número de estados accesibles $\Omega(E)$, con energía entre $E$ y $E+dE$ se puede expresar en términos del número de estados accesibles $\Phi(E)$ con energía menor que $E$,

\begin{equation}
	\Omega(E) = \Phi(E+dE) - \Phi(E) = \frac{\partial \Phi}{\partial E} dE + \mathcal{O}(dE^2)
\end{equation}

Donde, 

\begin{equation}
	\Phi(E) = \frac{1}{h^{3N}} \int\limits_{\text{Energía} < E} {dq_1 \cdots dq_{3N} dp_1 \cdots dp_{3N}}
\end{equation}

Como cada partícula que conforma el gas está asociada a una posición $\vec{r}_i = q_i \hat{u}_i + q_{i+1} \hat{u}_{i+1} + q_{i+2} \hat{u}_{i+2}$, donde $\hat{u}_j$ son vectores unitarios asociados a las coordenadas $q_j$, el volumen ocupado por cada partícula está dado por $V= \int d^3\vec{r} = \int dq_idq_{i+1}dq_{i+2}$, por lo tanto, el volumen ocupado por las $N$ partículas será $V^N$,

\[
\Phi(E) = \frac{V^N}{h^{3N}} \int\limits_{\text{energía} < E} {dp_1 \cdots dp_{3N}}
\]

Además, el hipervolumen que ocupan los momentum en el espacio de momentum corresponde con el hipervolumen de una esfera, dado por la ecuación \eqref{eq:ch3_gas_ideal_clasico_hipervolumen_esfera},

\begin{equation}
	\Phi(E) = \frac{V^N}{h^{3N}} \frac{\pi^{3N/2}(2mE)^{3N/2}}{\Gamma\left( \frac{3N}{2} + 1 \right)}
\end{equation}

En consecuencia, el número de estados accesibles del gas ideal está dado por,

\begin{equation}
	\Omega(E) \approx \frac{\partial \Phi}{\partial E} dE = \frac{3mNV^N}{h^{3N}}\frac{\pi^{3N/2}(2mE)^{\frac{3N}{2}-1}}{\Gamma\left( \frac{3N}{2} + 1 \right)} dE
\end{equation}

\subsection{Entropía y temperatura}
\index{Gas ideal clásico!entropía (ensamble microcanónico)}
La entropía para el gas ideal está dada por,

\[
	S = k_B \ln \Omega(E) \approx k_B\left( \frac{3N}{2}-1 \right) \ln E + k_B N\ln V + k_B \ln \left[ \frac{3N(2m)^{3N/2}\pi^{3N/2}}{2h^{3N}\Gamma\left( \frac{3N}{2} + 1 \right)} dE \right]
\]

Como $N$ es un número muy grande, del orden de $10^{23}$, $\frac{3N}{2}-1 \approx \frac{3N}{2}$,

\begin{equation}
S \approx k_B\left( \frac{3N}{2} \right) \ln E + k_B N\ln V + k_B \ln \left[ \frac{3N(2m)^{3N/2}\pi^{3N/2}}{2h^{3N}\Gamma\left( \frac{3N}{2} + 1 \right)} dE \right]
\end{equation}

Y la temperatura,

\begin{equation}
	T = \left( \frac{\partial S}{\partial E} \right)^{-1} \approx \left[ k_B \left( \frac{3N}{2} \right) \frac{1}{E} \right]^{-1} = \frac{2}{3N}\frac{E}{k_B}
\end{equation}

\subsection{Ecuación de estado}
\index{Gas ideal clásico!ecuación de estado}

La presión $P$ es la fuerza generalizada conjugada al volumen $V$, por lo tanto, su valor esperado se puede obtener a partir de la ecuación \eqref{eq:ch2_valor_esperado_fuerza_generalizada},

\begin{equation}
    \label{eq:ch3_ecuación_estado_gas_ideal_clásico}
    \E{P} = T\frac{\partial S}{\partial V} \approx T \frac{k_B N}{V}
\end{equation}

Este resultado corresponde con la ecuación de estado para el gas ideal clásico.






Con el objetivo que el lector adquiera un mayor nivel de comprensión, se le sugiere resolver las preguntas de autoexplicación del ejemplo trabajado titulado \href{https://colab.research.google.com/github/davidalejandromiranda/StatisticalPhysics/blob/main/notebooks/es_GasIdealLeyes.ipynb}{leyes del gas ideal clásico}.

\section{Problemas propuestos}

\begin{exercise}
    Sea un sistema con niveles de energía $E_i$ para $i=1,2,\ldots,I$, donde para el i-ésimo nivel de energía se tienen $n_i$ partículas, tal que $\sum_{i=1}^{I} n_i = N$.

    \begin{itemize}
        \item[a)] Muestre que el número de estados accesibles para una configuración $\{n_i\}$ está dado por 
    \[
    \Omega(E;\{n_i\}) = \frac{N!}{n_1! n_2! \cdots n_I!}
    \]

        \item[b)] Obtenga la entropía de Boltzmann para la configuración.
    \end{itemize}

\end{exercise}

\begin{exercise}
    Sea un sistema aislado cuyos microestados están dados por $\vert n,l \rangle$, tal que el operador hamiltoniano $\hat{H}$ y momento angular $\hat{L}^2$ cumplen con $\hat{H}\vert n,l \rangle = E_n \vert n,l \rangle$ y $\hat{L}^2 \vert n,l \rangle = \hbar^2 l(l+1)\vert n,l \rangle$. Obtenga la expresión para el número de estados accesibles del sistema.    
\end{exercise}

\begin{exercise}
    Sea un sistema aislado formado por dos osciladores armónicos clásicos unidimensionales desacoplados y con la misma frecuencia angular, cada uno de ellos descrito por su propia coordenada interna.

    \begin{itemize}
        \item[a)] Describa gráficamente el espacio de fase del sistema. Explique sus gráficos.
        
        \item[b)] Obtenga la expresión para el número de estados accesibles del sistema.
    \end{itemize}

\end{exercise}

\begin{exercise}
    Demuestre que el número de estados accesibles $\Omega(E)$ para un gas ideal con energía entre $E$ y $E+dE$ está dado por,
\[
\Omega(E) = \frac{3NV^N}{2h^{3N}} \cdot \frac{\pi^{N/2} (2mE)^{3N/2}}{\Gamma(3N/2 + 1)} \cdot \frac{dE}{E}
\]

\end{exercise}

\begin{exercise}
    El número de estados de un gas ideal clásico con energía entre $E$ y $E+\delta E$ está dado por $\Omega(E) \approx \left(\frac{3N}{2 \gamma \delta E}\right) V^N E^{(3N/2)}$, donde $\gamma$ es una constante; $N$, el número de partículas del gas ideal y $V$, el volumen que ocupan dichas partículas. Calcule el valor esperado de la presión.
\end{exercise}

\begin{exercise}
    A partir del número de estados accesibles $\Omega(E)$ para energías entre $E$ y $E+dE$, interprete las leyes de la termodinámica.
\end{exercise}

\begin{exercise}
    Sea $\Omega(E)$ el número de estados accesibles de un cierto sistema aislado con energía entre $E$ y $E+dE$. Asuma que la entropía del sistema aislado está dada por $S = k_B \ln \Omega(E)$.

    \begin{itemize}
        \item[a)] Encuentre la condición de equilibrio termodinámico para la interacción entre sistemas aislados.

        \item[b)] Explique cuál es el sentido físico de la temperatura absoluta y bajo qué circunstancia se podría considerar que un sistema tiene temperatura absoluta negativa.

        \item[c)] Interprete la tercera ley de la termodinámica en términos de $\Omega(E)$.
    \end{itemize}
\end{exercise}

\begin{exercise}
    Calcule el valor esperado de la magnetización para átomos paramagnéticos con espín $1/2$ en presencia de un campo magnético externo.
\end{exercise}

\begin{exercise}
    Sea una partícula clásica que describe un movimiento armónico simple dado por la ecuación $x = x_0 \sin(\omega t + \varphi)$. Demuestre dónde es más probable encontrar la partícula:
    \begin{itemize}
        \item[a)] Es mucho más probable encontrar la partícula en el centro.
        \item[b)] Es mucho más probable encontrar la partícula en los bordes.
        \item[c)] Se conoce con precisión, por lo tanto, es 100\% para cualquier posición.
        \item[d)] Es imposible de determinar dónde es más probable encontrar la partícula.
    \end{itemize}

\end{exercise}

\begin{exercise}
    Se dice que al resolver la ecuación de Schrödinger para un cierto sistema mecánico cuántico la energía obtenida y las funciones de onda corresponden con la solución a temperatura absoluta cero. Utilice argumentos de la Física Estadística que permitan verificar la validez o falsedad de esta afirmación.
\end{exercise}

\begin{exercise}
    ¿Cuántos estados accesibles tiene un sistema si su entropía es $86.17\,eV$? Argumente su respuesta.
\end{exercise}

\begin{exercise}
    Sea un gas ideal clásico formado por partículas con carga eléctrica $q_e$ sometidas a un campo eléctrico armónico que oscila a una frecuencia angular $\omega$.
    \begin{itemize}
        \item[a)] Estime el valor de la energía interna.
        \item[b)] Describa, cualitativamente, cómo cambia el resultado anterior en el caso de tener, en vez de un gas ideal clásico, uno cuántico formado por electrones, a temperatura ambiente.
    \end{itemize}
\end{exercise}

\begin{exercise}
    Calcule la magnetización de un sistema formado por $10^3$ partículas paramagnéticas con número cuántico espín $1/2$ y magnitud del momento angular $2\hbar$.
\end{exercise}

\chapter{Capítulo III. Sistemas en contacto con un reservorio de calor a temperatura T: ensamble canónico}
Un ensamble se construye con sistemas preparados bajo las mismas condiciones, por ejemplo, que los sistemas estén aislados y cuya energía se encuentre en un rango comprendido entre $E$ y $E+dE$, lo cual da lugar a un ensamble microcanónico. Otro tipo de condiciones que se pueden seleccionar para construir un ensamble es preparar los sistemas tal que su temperatura esté fija en un cierto valor $T$ y que puedan intercambiar energía con un sistema más grande, conocido como reservorio de calor, estando ambos (el sistema y el reservorio) aislados del resto del universo; a tal ensamble se le conoce como canónico y es el objeto de estudio de este capítulo.

\begin{definition}
    La \textbf{temperatura absoluta} se define como un parámetro externo que controla el intercambio de energía entre sistemas. Sea $S$ la entropía del sistema y $E$, su energía, entonces, la temperatura absoluta está dada por la ecuación \eqref{eq:ch4_temperatura_absoluta_def}.\index{Temperatura absoluta}
    \begin{equation}
        \label{eq:ch4_temperatura_absoluta_def}
        T = \left( \frac{\partial S}{\partial E} \right)^{-1}
    \end{equation}
\end{definition}

\begin{definition}
    Un \textbf{reservorio de calor} es un sistema lo suficientemente grande en comparación con otros sistemas con los que pueda interactuar, a una temperatura fija $T$ y con suficiente capacidad calorífica tal que al colocarse en contacto con otro sistema el reservorio de calor puede suministrar o absorber tanta energía para que al alcanzar el equilibrio termodinámico la temperatura de ambos sistemas sea $T$.\index{Reservorio de calor}
\end{definition}

\begin{definition}
    Un \textbf{ensamble canónico} consiste en un conjunto de sistemas preparados bajo las mismas condiciones tal que su temperatura está definida y el sistema puede intercambiar energía con un reservorio con el que se encuentra en contacto térmico tal que el gran sistema, formado por el sistema y el reservorio, están aislados del resto del universo.\index{Ensamble canónico}
\end{definition}

\section{Descripción estadística del sistema y cantidades termodinámicas en un ensamble canónico}
\index{Ensamble canónico}

\begin{figure}[t]
    \centering
    \includegraphics[width=0.55\linewidth]{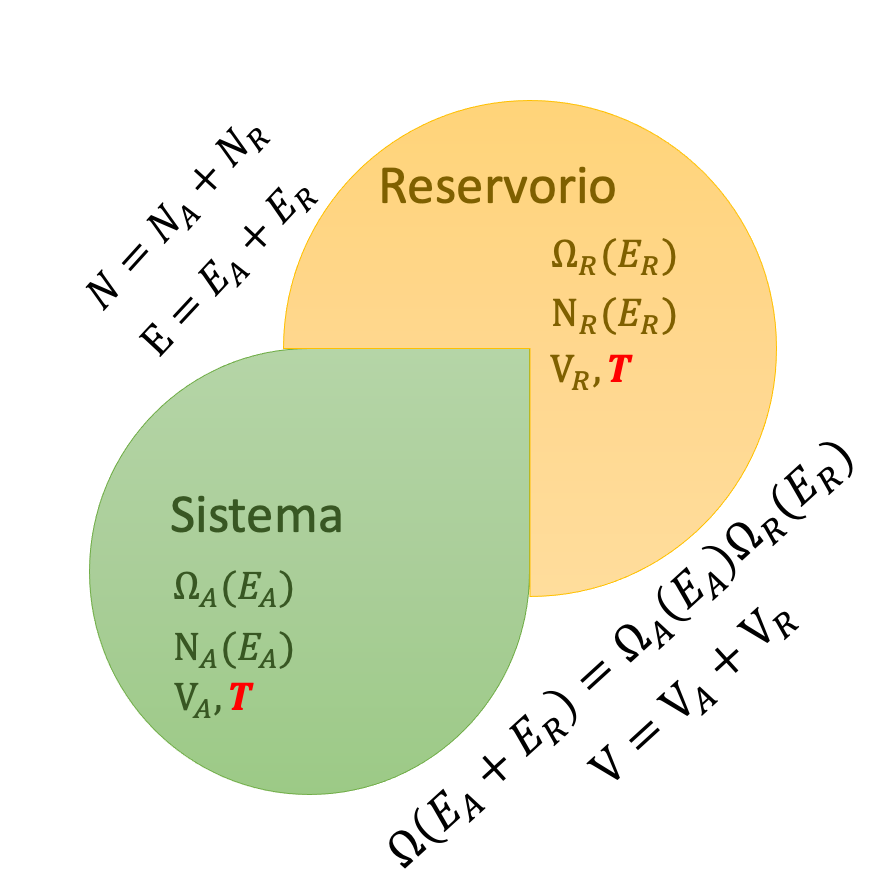}
    \caption{Sistema $A$ en contacto con un reservorio de calor $R$. El sistema $A$ tiene energía $E_A$, número de partículas $N_A$, número de estados accesibles $\Omega_A(E_A)$ y está a una temperatura $T$, en equilibrio termodinámico con el reservorio de calor $R$. El reservorio $R$ tiene energía $E_R \gg E_A$, número de partículas $N_R\gg N_A$ y número de estados accesibles $\Omega_R(E_R) \gg \Omega_A(E_A)$. La energía total $E$ del gran sistema formado por el reservorio y el sistema $A$, es la suma de las energías, $E=E_A+E_R$; de la misma manera, el número total de partículas $N$ está dado por $N=N_A+N_R$. El número de estados accesibles $\Omega(E)$, corresponde al producto del número de estados accesibles, $\Omega(E=E_A+E_R) = \Omega_A(E_A) \Omega_R(E_R)$.}
    \label{fig:ch3_sistema_reservorio}
\end{figure}

Sea un sistema $A$ en contacto con un reservorio de calor $R$ con el cual se encuentra en equilibrio termodinámico a una temperatura $T$, como se muestra en la Figura \ref{fig:ch3_sistema_reservorio}. El reservorio de calor es un sistema con la suficiente capacidad calorífica para absorber tanto calor del sistema $A$ sin cambios significativos en su temperatura; de la misma manera, tiene tanta energía para suministrar al sistema $A$ sin modificar significativamente su temperatura, haciendo que esta permanezca aproximadamente constante en un valor $T$ a pesar de la interacción con el sistema $A$. 

\subsection{Condición de equilibrio termodinámico}

Como el gran sistema formado por el reservorio y el sistema $A$ están aislados, el número total de estados accesibles $\Omega(E=E_A+E_R) = \Omega_A(E_A) \Omega_R(E_R)$ es constante y se cumple que,

\[
	\frac{\partial \Omega(E=E_A+E_R)}{\partial E_A} = 0
\]

Debido a que el número total de estados del gran sistema permanece constante cuando varía la energía del sistema $A$, se tiene la condición de equilibrio termodinámico entre el sistema $A$ y el reservorio de calor,

\[
    \Omega_R(E_A) \frac{\partial \Omega_A(E_A)}{\partial E_A} = - \Omega_A(E_A) \frac{\partial \Omega_R(E_R)}{\partial E_R} \frac{\partial E_R}{\partial E_A}
\]

Como $E = E_A + E_R$ y $E$ se mantiene constante, entonces, $\partial E_R / \partial E_A = -1$. Al reordenar términos,

\[
    \frac{1}{\Omega_R(E_A)} \frac{\partial \Omega_A(E_A)}{\partial E_A} = \frac{1}{\Omega_R(E_R)} \frac{\partial \Omega_R(E_R)}{\partial E_R}
\]

Al expresar la anterior ecuación en términos de logaritmo,

\[
    \frac{\partial \ln \Omega_A(E_A)}{\partial E_A} = \frac{\partial \ln \Omega_R(E_R)}{\partial E_R}
\]

Como $\beta_i = \partial \ln \Omega_i / \partial E_i$, se encuentra que la expresión anterior corresponde con la condición de equilibrio termodinámico,

\[
\beta_A = \beta_R \to T_A = T_R
\]

Donde, 

\begin{equation}
	\label{eq:ch3_beta_equilibrio}
	\beta = - \frac{\partial \ln \Omega_R(E-E_A)}{\partial E_A}	
\end{equation}

\subsection{Número de estados $\Omega_A(E_A)$ y probabilidad $p(E_A)$}

Para encontrar la probabilidad que el sistema $A$ se encuentre en un estado con energía $E_A$ se debe calcular $\Omega(E_A)$. Partiendo del logaritmo natural del número total de estados del gran sistema, se tiene,

\begin{equation}
	\label{eq:ch3_prob_eq1}
	\ln \Omega(E) = \ln \Omega_A(E_A) + \ln \Omega_R(E - E_A)
\end{equation}

Como $E_R \gg E_A$, se puede expandir $\ln \Omega_R(E - E_A)$ alrededor de $E_A$ y teniendo en cuenta la ecuación \eqref{eq:ch3_beta_equilibrio},

\[
	\ln \Omega_R(E - E_A) = \ln \Omega_R(E) - \frac{\partial \ln \Omega(E-E_A)}{\partial E_A} E_A + \mathcal{O}(E_A^2) \approx \ln \Omega_R(E) + \beta E_A
\]

Es importante notar que $\ln \Omega_R(E) = \ln \Omega_{0_R}$ es una constante que corresponde al caso cuando el sistema $A$ tiene energía cero y, por consiguiente, el menor número de estados accesibles $\Omega_A(0)$. Al reemplazar el resultado anterior en la ecuación \eqref{eq:ch3_prob_eq1},

\[
	\ln \Omega(E) \approx \ln \Omega_A(E_A) + \ln \Omega_{0_R} + \beta E_A
\]

Con este resultado, al despejar y simplificar, se obtiene $\Omega_A(E_A)$,

\begin{equation}
	\label{eq:ch3_prob_eq2}
	\Omega_A(E_A) \approx \frac{\Omega(E)}{\Omega_{0_R}} e^{-\beta E_A}
\end{equation} 

El número total de estados accesibles para el sistema $A$ está dado por la suma de la ecuación \eqref{eq:ch3_prob_eq2} para todos los valores de $E_A$,

\begin{equation}
	\label{eq:ch3_prob_eq3}
	\sum_{E_A} \Omega_A(E_A) \approx \frac{\Omega(E)}{\Omega_{0_R}} \sum_{E_A} e^{-\beta E_A}
\end{equation} 

La probabilidad $p(E_A)$ de encontrar al sistema $A$ en un estado con energía $E_A$ está dado por el cociente entre $\Omega_A(E_A)$ y $\sum_{E_A} \Omega_A(E_A)$,

\begin{equation}
	\label{eq:ch3_prob_eq4}
	p(E_A) \approx \frac{e^{-\beta E_A}}{\sum\limits_{E_A} e^{-\beta E_A}}
\end{equation} 

\begin{definition}
	Sea un sistema en contacto con un reservorio de calor a temperatura $T$, donde $\beta = (k_B T)^{-1}$ y $E_r$ son los posibles valores de energía que puede tomar el sistema. Se define la \textbf{función de partición} como la suma, para todas las energías en que se puede encontrar al sistema, de las exponenciales $e^{-\beta E_r}$.\index{Ensamble canónico!función de partición}
	\begin{equation}
		\label{eq:ch3_función_partición}
		Z = \sum_r e^{-\beta E_r}
	\end{equation}
\end{definition}

\begin{definition}
	Sea un sistema en contacto con un reservorio de calor a temperatura $T$. La \textbf{probabilidad} $p(E_r)$ que el sistema se encuentre en un estado con energía $E_r$ está dada por la ecuación \eqref{eq:ch3_probabilidad_canonica}.\index{Ensamble canónico!probabilidad}
	\begin{equation}
		\label{eq:ch3_probabilidad_canonica}
		p(E_r) = Z^{-1} e^{-\beta E_r}
	\end{equation}
\end{definition}

\subsection{Valor esperado de la energía: energía interna}

El valor esperado de la energía $\E{E} = U$ corresponde con la energía interna del sistema y está dado por,

\[
	\E{E} = \sum_r p(E_r) E_r = Z^{-1} \sum_r E_r e^{-\beta E_r}
\]

Como $\partial Z / \partial \beta = - \sum\limits_r E_r e^{-\beta E_r}$, \index{Ensamble canónico!valor esperado de la energía}

\begin{equation}
	\label{eq:ch3_valor_esperado_energía}
	\E{E} = U = - \frac{\partial \ln Z}{\partial \beta}
\end{equation}

\subsection{Valor esperado de la fuerza generalizada conjugada a un parámetro externo $x$}

El valor esperado de la fuerza generalizada $X$ conjugada a un parámetro externo $x$, determinada por la ecuación \eqref{eq:ch2_fuerza_generalizada}, está dado por,

\[
	\E{X} = \sum_r X_r p(E_r) = \sum_r -\frac{\partial E_r}{\partial x} p(E_r) = - Z^{-1} \sum_r \frac{\partial E_r}{\partial x} e^{-\beta E_r} =  Z^{-1} \sum_r X_r e^{-\beta E_r}
\]

Nótese que,

\[
 	\frac{\partial Z}{\partial x} = \sum_r -\beta \frac{\partial E_r}{\partial x} e^{-\beta E_r} = \beta \sum_r X_r e^{-\beta E_r}
\]

Por lo tanto, el valor esperado $\E{X}$ está dado por,\index{Ensamble canónico!valor esperado de la fuerza generalizada}

\begin{equation}
		\label{eq:ch3_valor_esperado_fuerza_conjugada}
		\E{X} = \frac{1}{\beta} \frac{\partial \ln Z}{\partial x}
\end{equation}

\subsection{Entropía}

En el ensamble microcanónico el número de estados accesibles $\Omega(E)$ permite calcular cualquier cantidad termodinámica; de la misma manera, en el ensamble canónico, la función de partición $Z$ permite describir el estado del sistema. Ahora bien, es de resaltar que $Z$ describe el macroestado del sistema, el cual está determinado por la temperatura $T$ y los parámetros externos. Supongamos que se tiene un parámetro externo $x$ cuya fuerza generalizada conjugada a $x$ está dada por $X$ y al variar el parámetro externo una cantidad infinitesimal $dx$ se realiza un trabajo $\dbar W = \E{X} dx$. En tal caso, la función de partición dependerá tanto de la temperatura como del parámetro externo $x$, $Z=Z(T, x)=Z(\beta, x)$. Entonces, una variación infinitesimal de la función de partición estará dada por,

\[
  d \ln Z(\beta, x) = \frac{\partial \ln Z}{\partial \beta} d\beta + \frac{\partial \ln Z}{\partial x} dx
\]

Como la energía interna $U$ está dada por la ecuación \eqref{eq:ch3_valor_esperado_energía}, el valor esperado $\E{X}$ de la fuerza generalizada conjugada a $x$, por la ecuación \eqref{eq:ch3_valor_esperado_fuerza_conjugada} y $\dbar W = \E{X} dx$,

\[
	d \ln Z(\beta, x) = - Ud\beta + \beta \dbar W
\]

Como $d(\beta U) = \beta dU + Ud\beta$,

\[
	d \ln Z(\beta, x) = \beta dU - d(\beta U) + \beta \dbar W = -d(\beta U) + \beta( dU + \dbar W )
\]

De la primera ley de la termodinámica se tiene que $\dbar Q = dU + \dbar W = T dS$, por lo tanto, al reordenar términos se obtiene que,

\[
	d(\ln Z + \beta U) = \beta T dS
\]

Con lo cual se obtiene la entropía expresada en términos de la función de partición, \index{Ensamble canónico!entropía}

\begin{equation}
		\label{eq:ch3_entropía}
		S = k_B \ln Z + \frac{U}{T}
\end{equation}

\subsection{Energía libre de Helmholtz}

De acuerdo a la termodinámica, la energía libre de Helmholtz, $F$, está dada por,

\[
	F = U - TS
\]

Por lo tanto, al comparar esta expresión con la ecuación \eqref{eq:ch3_entropía}, la energía libre de Helmholtz, en términos de la función de partición, está dada por la ecuación \eqref{eq:ch3_energía_libre_Helmholtz}.\index{Ensamble canónico!energía libre de Helmholtz}

\begin{equation}
		\label{eq:ch3_energía_libre_Helmholtz}
		F = - k_B T \ln Z
\end{equation}

\section{Propiedades de la función de partición}

\subsection{Efectos del cambio de referencia de energía}

Sea un sistema con energías $E_r$, tal que su función de partición está dada por,

\[
	Z_0 = \sum_r e^{-\beta E_r} 
\]

En este caso, el valor esperado de la energía, dado por la ecuación \eqref{eq:ch3_valor_esperado_energía}, será,

\[
	\E{E_0} = -\frac{\partial \ln Z_0}{\partial \beta}
\]

Si se cambia la referencia de energía, tal que $E' = E - E_1$, donde $E_1$ es la nueva referencia de energía, entonces, la nueva función de partición $Z_1$ está dada por,

\[
	Z_1 = \sum_r e^{-\beta (E_r - E_1)} = \sum_r e^{-\beta E_r}e^{\beta E_1} = Z_0e^{\beta E_1}
\]

Por lo tanto, el valor esperado de la energía cambia en la misma cantidad,

\[
	\E{E_1} = -\frac{\partial \ln Z_1}{\partial \beta} = -\frac{\partial}{\partial \beta} \left( \ln Z_0e^{\beta E_1} \right) = -\frac{\partial \ln Z_0}{\partial \beta} - E_1 = \E{E_0} - E_1
\]

Como la entropía está dada por la ecuación \eqref{eq:ch3_entropía}, si $S_0$ es la entropía del sistema con energías $E_r$ y $S_1$ cuando se cambia la referencia de energía, entonces,

\[
	S_0 = k_B \ln Z_0 + \E{E_0}/T
\]

\[
	S_1 = k_B \ln Z_1 + \E{E_1}/T = k_B \ln Z_0 + E_1/T + \E{E_0}/T - E_1/T = S_0
\]

\begin{lema}
	Sea $Z_0$ la función de partición de un sistema con valor esperado de la energía $\E{E_0}$. Si la referencia con respecto a la cual se mide la energía del sistema se modifica en un valor $\Delta E$, entonces, la función de partición se escala por un factor $e^{\beta\Delta E}$, el valor esperado de la energía se modifica en el mismo valor, $\E{E} = \E{E_0} + \Delta E$ y la entropía del sistema permanece invariante, $S=S_0$.
\end{lema}

\subsection{Función de partición para la mezcla de sistemas no interactuantes}

Sean dos sistemas $A$ y $B$ con energías $E_r$ y $E'_s$, respectivamente, tal que sus funciones de partición están dadas por $Z_A$ y $Z_B$. 

\[ 
	Z_A = \sum\limits_{r} e^{-\beta E_r}
\]

\[ 
	Z_B = \sum\limits_{s} e^{-\beta E'_s}
\]

Si al mezclar los sistemas estos permanecen como sistemas no interactuantes, tal que la energía del sistema total se puede escribir como $E_{r,s} = E_r + E'_s$, entonces, la función de partición $Z_{AB}$ del sistema mezclado será el producto de las funciones de partición,

\[
	Z_{AB} = \sum_{r, s} e^{-\beta E_{r,s}} = \sum_{r,s} e^{-\beta (E_r + E'_s)} = \sum\limits_{r} e^{-\beta E_r}\sum\limits_{s} e^{-\beta E'_s} = Z_A Z_B
\]

\begin{lema}
	Sea un sistema formado por $N$ subsistemas no interactuantes, cada uno con función de partición $Z_n$. La función de partición $Z$ del sistema completo está dada por el producto de las funciones de partición $Z_n$.
	\begin{equation}
		\label{eq:ch4_funcion_particion_sistemas_no_interactuantes}
		Z = \prod_{n=1}^N Z_n
	\end{equation}	
\end{lema}

\section{Magnetización de un material paramagnético}
\index{Paramagnetismo}

Sea un sistema formado por partículas paramagnéticas con espín 1/2 en contacto con un reservorio de calor a temperatura absoluta $T$ y sometidas a un campo magnético $\vec{B}$. ¿Cuál es la relación entre la energía interna $U=\E{E}$ y la temperatura absoluta $T$? ¿Cómo depende la magnetización $M$ de la temperatura absoluta?

\subsection{Energía interna}

Como el sistema tiene espín $1/2$, la energía de cada partícula puede tomar solo uno de dos posibles valores, $E_1 = -\gls{muB} B$ o $E_2 = \gls{muB}B$, donde $\gls{muB}$ está dado por la ecuación \eqref{eq:ch2_momento_magnetico_paramag}. De esta manera, la función de partición $Z_n$ para la n-ésima partícula está dada por,

\[
	Z_n = e^{\beta \gls{muB} B} + e^{-\beta \gls{muB}B}
\]

Suponiendo que las partículas no interactúan entre si, la función de partición $Z$ está dada por el producto de todas las contribuciones individuales, ecuación \eqref{eq:ch4_funcion_particion_sistemas_no_interactuantes},

\[ 
	Z = \prod\limits_{n=1}^N Z_n = \left( e^{\beta \gls{muB} B} + e^{-\beta \gls{muB}B} \right)^N
\]

Donde, $\ln Z$ está dado por,

\[
	\ln Z = N \ln \left( e^{\beta \gls{muB} B} + e^{-\beta \gls{muB}B} \right)
\]

Por lo tanto, la energía interna se obtiene con la ecuación \eqref{eq:ch3_valor_esperado_energía},

\[
	U = \E{E} = - \frac{\partial \ln Z}{\partial \beta} = -\gls{muB} BN \frac{e^{\beta \gls{muB} B} - e^{-\beta \gls{muB}B}}{e^{\beta \gls{muB} B} + e^{-\beta \gls{muB}B}} = -\gls{muB} B N \tanh\left(\frac{\gls{muB} B}{k_b T} \right)
\]

Lo cual está en concordancia con el resultado obtenido al analizar el mismo sistema a partir de un ensamble microcanónico, ver ecuación \eqref{eq:ch2_valor_esperado_energia_param}.

\subsection{Magnetización}

Como se había estudiando anteriormente, la magnetización $M$ es igual al número de momentos magnéticos por unidad de volumen. Si $V$ es el volumen ocupado por las partículas con momento magnético $\gls{muB}$, la magnetización, para partículas paramagnéticas, se puede expresar como una fuerza generalizada conjugada a la magnitud del campo magnético,

\[
X = V M = - \frac{\partial E}{\partial B} 
\]

Por lo tanto, el valor esperado $\E{X}$ de la fuerza generalizada conjugada a $x$, dado por la ecuación \eqref{eq:ch3_valor_esperado_fuerza_conjugada}, es,

\[
	\E{X} = V\E{M} = \frac{1}{\beta} \frac{\partial \ln Z}{\partial B} = \gls{muB} N \frac{e^{\beta \gls{muB} B} - e^{-\beta \gls{muB}B}}{e^{\beta \gls{muB} B} + e^{-\beta \gls{muB}B}} = \gls{muB} N \tanh\left(\frac{\gls{muB} B}{k_b T} \right)
\]

Que coincide con el resultado obtenido anteriormente con el ensamble microcanónico, ver ecuación \eqref{eq:ch2_magnetizacion_param}. Este resultado, así como el obtenido para la energía interna, nos muestran que la descripción de un sistema físico es razonable que sea independiente del ensamble estadístico utilizado para su descripción.

Con el objetivo que el lector adquiera un mayor nivel de comprensión, se le sugiere resolver las preguntas de autoexplicación del ejemplo trabajado titulado \href{https://colab.research.google.com/github/davidalejandromiranda/StatisticalPhysics/blob/main/notebooks/es_ParticulasConEspinEnCampo_FunParticion.ipynb}{magnetización de un material paramagnético}.

\section{Nanoestructura semiconductora 2D a temperatura absoluta T}
\index{Energía de una nanoestructura a temperatura absoluta T}

Con el objetivo que el lector adquiera un mayor nivel de comprensión, se le sugiere resolver las preguntas de autoexplicación del ejemplo trabajado titulado \href{https://colab.research.google.com/github/davidalejandromiranda/StatisticalPhysics/blob/main/notebooks/es_ElectronEnNanoestructuraTemperatura.ipynb}{nanoestructura semiconductora 2D a temperatura absoluta T}.

\section{Sistemas clásicos a temperatura finita $T$}

La teoría desarrollada para obtener valores esperados a partir de la función de partición se puede aplicar a sistemas clásicos, para ello es necesario escribir dicha función en términos del espacio de fase. Además, para sistemas clásicos, el valor esperado de las contribuciones cuadráticas, tanto de la energía como de momentum, están determinados por el teorema de equipartición, el cual establece que cada grado de libertad cuadrático contribuye con una cantidad $k_BT/2$ al valor esperado de la energía (energía interna).

\subsection{Función de partición}

En un sistema clásico, formado por $N$ partículas, el estado del sistema se especifica en términos del espacio ocupado en el espacio de fase. Para un cierto intervalo de energía entre $E$ y $E + \Delta E$ se tienen $\Omega(E)$ estados accesibles, cada uno de ellos con energía aproximadamente igual a $E$. A cada configuración $(\{q_i\}, \{p_1\})$ en el espacio de fase se le puede asociar una cierta energía $E_r$, en términos de la cual la función de partición está expresada como

\[
	Z = \sum_r e^{-\beta E_r} 
\]

Si se agrupan las configuraciones $(\{q_i\}, \{p_1\})$ en el espacio de fase con energía entre $E$ y $E + \Delta E$, la función de partición se puede expresar en término de la suma de los diferentes valores de energía,

\[
	Z \approx \sum_E \Omega(E) e^{-\beta E} 
\]

Nótese que la única condición impuesta al intervalo $\Delta E$ es que la energía de la configuración $E \leq E_r < E + \Delta E$, por lo tanto, se puede hacer tan pequeño como se quiera el intervalo de energía, incluso el caso límite $\Delta E \to 0$,

\[
	Z = \lim\limits_{\Delta E \to 0} \sum_E \Omega(E) e^{-\beta E}
\]

Como $\Omega(E)$ está dado por la ecuación \eqref{eq:ch3_numero_estados_accesibles_sistemas_clasicos}, para un sistema clásico se tiene, la función de partición toma la forma de una sumatoria de Riemann, dado que al tomar $\Delta E \to 0$, se suman sobre puntos (configuraciones) en el espacio de fase,

\[
	Z = \lim\limits_{\Delta E \to 0} \sum_E \rho e^{-\beta E(q_1, \cdots, q_{3N}, p_1, \cdots, p_{3N})} dq_1 \cdots dq_{3N} dp_1 \cdots dp_{3N}
\]

Por lo tanto, la función de partición, para un sistema clásico, está dada por la ecuación \eqref{eq:ch4_funcion_partición_sistemas_clásicos}.

\begin{definition}
	Sea un sistema formado por $N$ partículas clásicas con configuraciones en el espacio de fase $(\{q_i\}, \{p_i\})$ y densidad de microestados en el espacio de fase $\rho$, en contacto con un reservorio de calor a temperatura absoluta $T$. La \textbf{función de partición} para el sistema está dada por la ecuación \eqref{eq:ch4_funcion_partición_sistemas_clásicos}.
\begin{equation}
	\label{eq:ch4_funcion_partición_sistemas_clásicos}
	Z = \int\limits_{-\infty}^{\infty} \cdots \int\limits_{-\infty}^{\infty} \rho e^{-\beta E(q_1, \cdots, q_{3N}, p_1, \cdots, p_{3N})} dq_1 \cdots dq_{3N} dp_1 \cdots dp_{3N}
\end{equation}
\end{definition}

\subsection{Valor esperado de la energía y el teorema de equipartición}

Supongamos que se tiene un sistema formado por solo una partícula con energía $E=p^2/2m$ y $\rho=1/h$; en tal caso, la función de partición, dada por la ecuación \eqref{eq:ch4_funcion_partición_sistemas_clásicos}, será,

\[
	Z = \frac{1}{h} \int_{-\infty}^{\infty} e^{-\beta \frac{p^2}{2m}} dp = \frac{1}{h}\sqrt{\frac{m}{\pi \beta}}
\]

Como el valor esperado de la energía está dado por la ecuación \eqref{eq:ch3_valor_esperado_energía},

\[
	\E{E} = - \frac{\partial \ln Z}{\partial \beta} = - \frac{\partial }{\partial \beta} \left( \ln h - \frac{1}{2} \ln \beta  + \frac{1}{2} \ln m - \frac{1}{2} \ln \pi \right) = \frac{1}{2\beta} = \frac{1}{2}k_BT
\]

Ahora supongamos que al sistema se le agrega una fuerza recuperadora proporcional a la elongación, entonces, la energía tomará la forma $E=p^2/2m + kx^2/2$, $\rho=h'$ y la función de partición, 

\[
	Z = \frac{1}{h'} \int\limits_{-\infty}^{\infty}\int\limits_{-\infty}^{\infty} e^{-\beta \frac{p^2}{2m}}e^{-\beta \frac{kx^2}{2}} dxdp = \frac{1}{h'}\sqrt{\frac{m}{\pi \beta}}\sqrt{\frac{1}{\pi k \beta}}
\]

Al calcular el valor esperado de la energía se obtiene una contribución $k_BT/2$ para la coordenada espacial y otra contribución igual para el momentum,

\[
	\E{E} = - \frac{\partial \ln Z}{\partial \beta} = \frac{1}{2}k_BT + \frac{1}{2}k_BT = k_B T
\]

Esto nos indica que siempre que se tenga una contribución cuadrática a la energía, bien sea con un término de posición o momentum, dicha contribución aportará $k_B T/2$ al valor esperado de la energía del sistema. Una manera de escribir las contribuciones cuadráticas de manera general es definiendo un grado de libertad $\xi_i$, tal que,

\[
	\varepsilon_i = \xi_i \frac{\partial H}{\partial \xi_i}
\]

Donde $H$ es el hamiltoniano del sistema, con potenciales de interacción entre partículas $f_{i,j}$, dado por,

\[
	H = \sum_i \frac{p_i^2}{2m_i} + \frac{1}{2}\sum_{i,j} f_{i,j} q_i q_j
\]

Nótese que si $\xi_1=q_1$ y $\xi_2=p_1$, debido a las relaciones canónicas de Hamilton,

\[
	\varepsilon_1 = \xi_1 \frac{\partial H}{\partial \xi_1} = \frac{q_1}{2} \frac{\partial H}{\partial q_1} = \frac{q_1}{2} (2f_{1,1}q_1) = f_{1, 1} q_1^2
\]

\[
	\varepsilon_2 = \xi_2 \frac{\partial H}{\partial \xi_2} = p_1 \frac{\partial H}{\partial p_1} = p_1 \frac{p_1}{m} = \frac{p_1^2}{m}
\]

Con lo cual, al utilizar los resultados obtenidos para los valores esperados de la energía, se obtiene $\E{\varepsilon_1} = 2\E{f_{1,1}q_1^2/2} = k_BT$ y $\E{\varepsilon_2} = \E{2p_1^2/2m} = 2 \E{p_1^2/2m} = k_BT$. Esto nos muestra que toda componente cuadrática del hamiltoniano nos aporta $K_BT/2$ al valor medio de la energía, cuando el sistema se encuentra en equilibrio termodinámico.

\begin{lema}
	\textbf{Teorema de equipartición}. Sea un sistema clásico con hamiltoniano $H$ y grados de libertad de la forma $\xi_i$. En equilibrio termodinámico se cumple la ecuación \eqref{eq:ch4_teorema_equiparticion}.
	\begin{equation}
		\label{eq:ch4_teorema_equiparticion}
		\E{\xi_i \frac{\partial H}{\partial \xi_j}} = k_B T\delta_{i,j}
	\end{equation}\index{Teorema de equipartición}
\end{lema}

\section{Valor esperado de la energía obtenidos a partir del teorema de equipartición}

En esta sección se estudiarán unos ejemplos simples de aplicación del teorema de equipartición para obtener el valor esperado de la energía del sistema.

\subsection{Oscilador armónico clásico 1D a temperatura absoluta T}

Sea un oscilador armónico clásico a temperatura absoluta $T$, cuyo movimiento está restringido a una dimensión ¿cuál es el valor esperado de la energía?

La energía de un oscilador armónico clásico está dada por la ecuación \eqref{eq:ch3_oscilador_armonico_clasico_energia_px}, en la cual se observan dos grados de libertad cuadráticos, $p^2/2m$ y $m\omega^2 x^2/2$, por lo tanto, al aplicar el teorema de equipartición se obtiene,

\[
	\E{E} = \E{\frac{p^2}{2m}} + \E{\frac{1}{2}m\omega^2 x^2} = \frac{1}{2}k_BT + \frac{1}{2}k_BT = k_B T
\]

\subsection{Oscilador armónico clásico 2D a temperatura absoluta T}

Sea un oscilador armónico clásico a temperatura absoluta $T$, cuyo movimiento está restringido a dos dimensiones ¿cuál es el valor esperado de la energía?

En este caso, el hamiltoniano del oscilador armónico está dado por,

\[
	H = \frac{p_x^2}{2m} + \frac{p_y^2}{2m} + \frac{1}{2}m\omega_x^2 x^2 + \frac{1}{2}m\omega_y^2 y^2
\]

Como se tienen cuatro componentes cuadráticas en el hamiltoniano, el valor esperado de la energía será,

\[
	\E{E} = 4 \left( \frac{k_B T}{2} \right) = 2 k_B T
\]

\subsection{Oscilador armónico clásico 3D a temperatura absoluta T}

Sea un oscilador armónico clásico en el espacio, a temperatura absoluta $T$ ¿cuál es el valor esperado de la energía?

En este caso, el hamiltoniano del oscilador armónico está dado por,

\[
	H = \frac{p_x^2}{2m} + \frac{p_y^2}{2m} + \frac{p_z^2}{2m} + \frac{1}{2}m\omega_x^2 x^2 + \frac{1}{2}m\omega_y^2 y^2 + \frac{1}{2}m\omega_z^2 z^2
\]

Como se tienen seis componentes cuadráticas en el hamiltoniano, el valor esperado de la energía será,

\[
	\E{E} = 6 \left( \frac{k_B T}{2} \right) = 3 k_B T
\]

\subsection{Gas ideal clásico}
\index{Gas ideal clásico!valor esperado de la energía}
Sea un gas ideal clásico formado por $N$ partículas debilmente interactuantes, confinadas en un volumen $V$. ¿Cuál es el valor esperado de la energía del gas ideal?

En este caso, el hamiltoniano del sistema está dado por,

\[
	H = \sum_{i=1}^{N} \left( \frac{p_{x_i}^2}{2m} + \frac{p_{y_i}^2}{2m} + \frac{p_{z_i}^2}{2m} \right) 
\]

Como se tienen $3N$ componentes cuadráticas en el hamiltoniano, el valor esperado de la energía será,

\begin{equation}
    \label{eq:ch4_valor_esperado_energía_gas_ideal_clásico_teorema_equipartición}
    \E{E} = 3N \left( \frac{k_B T}{2} \right) = \frac{3}{2} N k_B T
\end{equation}

\subsection{Movimiento browniano}
\index{Movimiento browniano}
Sea una partícula con masa $m$, inmersa en un cierto fluido a temperatura T, en el cual describe un movimiento browniano. ¿Cuál es el valor esperado de la magnitud de la velocidad de la partícula?

La magnitud de la velocidad de una partícula que se mueve en el espacio, expresada en coordenadas cartesianas, está dada por,

\[
	v^2 = |\vec{v}|^2 = v_x^2 + v_y^2 + v_z^2
\]

Al aplicar el teorema de equipartición se obtiene,

\[
	\E{\frac{1}{2}mv^2} = \E{\frac{1}{2}mv_x^2} + \E{\frac{1}{2}mv_y^2} + \E{\frac{1}{2}mv_z^2} = \frac{3}{2}k_BT
\]

Entonces, el valor esperado de la magnitud de la velocidad se puede obtener como, 
\[
	\E{|\vec{v}|} \approx \sqrt{\E{v^2}} = \sqrt{\frac{3k_B T}{m}}
\]

Esta expresión para el valor esperado de la magnitud de la velocidad de un partícula que describe un movimiento browniano esta implícita en las predicciones realizadas en 1905 por Albert Einstein \cite{Einstein1905c}. En 1908, Jean Baptiste Perrin demostró experimentalmente la validez de las predicciones de Einstein\footnote{La principal predicción de Einstein para el movimiento browniano corresponde con la expresión para el coeficiente de difusión, el cual se puede determinar experimentalmente analizando el movimiento de las partículas que describen un movimiento browniano.} lo que le mereció el premio Nobel en 1926.

\section{El gas ideal clásico}
\index{Gas ideal clásico!ensamble canónico}

Se conoce como gas ideal clásico a un sistema formado por partículas clásicas no interactuantes, por lo tanto, su energía potencial se considera una constante que no incluye interacciones entre partículas y el hamiltoniano para describir al gas ideal clásico está determinado solo por las contribuciones de la energía cinética de las partículas que lo integran. El hamiltoniano de un gas ideal formado por $N$ partículas con masa $m$ que ocupan un volumen $V$ está dado por la ecuación \eqref{eq:ch4_hamiltoniano_gas_ideal_clásico}.

\subsection{Función de partición de una partícula clásica en 3D}

Supongamos que se tiene una partícula con momentum $\vec{p}=\sum\limits_{i=1}^3 p_i \hat{u}_i$, masa $m$ y hamiltoniano $H=p^2/2m$, que ocupa un volumen $V$, donde $\hat{u}_i$ es el vector unitario en dirección $q_i$. La función de partición $Z_1$ de esta partícula está dada por,

\[
 Z_1 = \frac{1}{h^3} \int_V dq_1 dq_2 dq_3 \int\limits_{-\infty}^\infty e^{-\frac{\beta p_1^2}{2m}} dp_1 \int\limits_{-\infty}^\infty e^{-\frac{\beta p_2^2}{2m}} dp_2  \int\limits_{-\infty}^\infty e^{-\frac{\beta p_3^2}{2m}}  dp_3
\]

Como la integral para cada componente del momentum es evaluada en todos los posibles valores de momentum,

\begin{equation}
    \label{eq:ch4_función_partición_una_partícula_clásica_gas_ideal}
    Z_1 = \frac{V}{h^3} \left( \int\limits_{-\infty}^\infty e^{-\frac{\beta p^2}{2m}} dp \right)^3 = \frac{V}{h^3} \left( 2\pi m / \beta \right)^{3/2}
\end{equation}

\subsection{Función de partición de para $N$ partículas clásicas no interactuantes}

Dado que la función de partición para un sistema formado por partículas no interactuantes se puede escribir como se muestra en la ecuación \eqref{eq:ch4_funcion_particion_sistemas_no_interactuantes}, la función de partición $Z$ para el gas ideal clásico está dada por,

\[
   Z = Z_1^N = \frac{V^N}{h^{3N}} \left( 2\pi m / \beta \right)^{3N/2}
\]

Y su logaritmo natural está dado por,\index{Gas ideal clásico!función de partición}

\begin{equation}
    \label{eq:ch4_ln_Z_gas_ideal_clásico}
    \ln Z = N \ln V - \frac{3}{2} N \ln \beta + \frac{3}{2} N \ln \left( \frac{2\pi m}{h^2} \right)
\end{equation}

\subsection{Valor esperado de la energía}

El valor esperado de la energía $\E{E} = U$ corresponde con la energía interna, el cual se puede obtener a partir de la ecuación \eqref{eq:ch3_valor_esperado_energía},\index{Gas ideal clásico!energía interna}

\begin{equation}
    \label{eq:ch4_valor_esperado_energía_gas_ideal_clásico}
    U = \frac{3N}{2}\beta^{-1} = \frac{3}{2}Nk_BT
\end{equation}

Este resultado concuerda con el obtenido con el teorema de equipartición, ecuación \eqref{eq:ch4_valor_esperado_energía_gas_ideal_clásico_teorema_equipartición}.

\subsection{Valor esperado de la presión}
\index{Gas ideal clásico!ecuación de estado}
La presión $P$ es la fuerza generalizada conjugada al volumen $V$, es decir,

\[
    P = -\frac{\partial E}{\partial V}
\]

Por lo tanto, su valor esperado se puede obtener con la ecuación \eqref{eq:ch3_valor_esperado_fuerza_conjugada},

\[
    \E{P} = \frac{N}{\beta V} = \frac{NkBT}{V}
\]

Este resultado es consistente con el encontrado anteriormente al utilizar un ensamble microcanónico para analizar el gas ideal clásico, ecuación \eqref{eq:ch3_ecuación_estado_gas_ideal_clásico}. Como se espera, la elección del ensamble para analizar una situación física debe ser independiente de la realidad física que se quiere describir.

\subsection{Entropía}

La entropía se puede obtener con la ecuación \eqref{eq:ch3_entropía},\index{Gas ideal clásico!entropía (ensamble canónico)}

\begin{equation}
    \label{eq:ch4_entropía_gas_ideal_clásico}
    S = k_B \left[ N \ln V - \frac{3}{2} N \ln \beta + \frac{3}{2} N \ln \left( \frac{2\pi m}{h^2} \right) + \frac{3}{2}N \right]
\end{equation}

\subsection{Paradoja de Gibbs}
\index{Paradoja de Gibbs}

En 1874 Willard Gibbs descubrió que la entropía predicha por la ecuación \eqref{eq:ch4_entropía_gas_ideal_clásico} plantea una paradoja en cuanto a las propiedades de la entropía, en particular, a su carácter extensivo \cite{Gibbs1874}. Supongamos el siguiente experimento mental: sea un gas ideal formado por $N$ partículas en un contenedor con volumen $V$. Si se divide el contenedor por la mitad de tal manera que en cada lado queden $N/2$ partículas y ambos lados del contenedor estén a igual temperatura y presión, se espera que la suma de la entropía de cada lado del contenedor, $S_{1/2}$ sea igual a la entropía del contenedor sin división, $S$, dada por la ecuación \eqref{eq:ch4_entropía_gas_ideal_clásico}. La entropía de cada lado del contenedor divido está daa por,

\[
S_{1/2} = k_B \left[ \frac{N}{2} \ln \left(  \frac{V}{2} \right) - \left(\frac{3}{2}\right) \left(\frac{N}{2}\right) \ln \beta + \left(\frac{3}{2}\right) \left(\frac{N}{2}\right) \ln \left( \frac{2\pi m}{h^2} \right) + \left(\frac{3}{2}\right) \left(\frac{N}{2}\right) \right]
\]

Se espera que $S - 2S_{1/2}$ sea igual a cero, sin embargo,

\[
    S - 2S_{1/2} = N \ln 2
\]

\begin{figure}[th]
    \centering
    \begin{subfigure}{0.49\textwidth}
        \includegraphics[width=\linewidth]{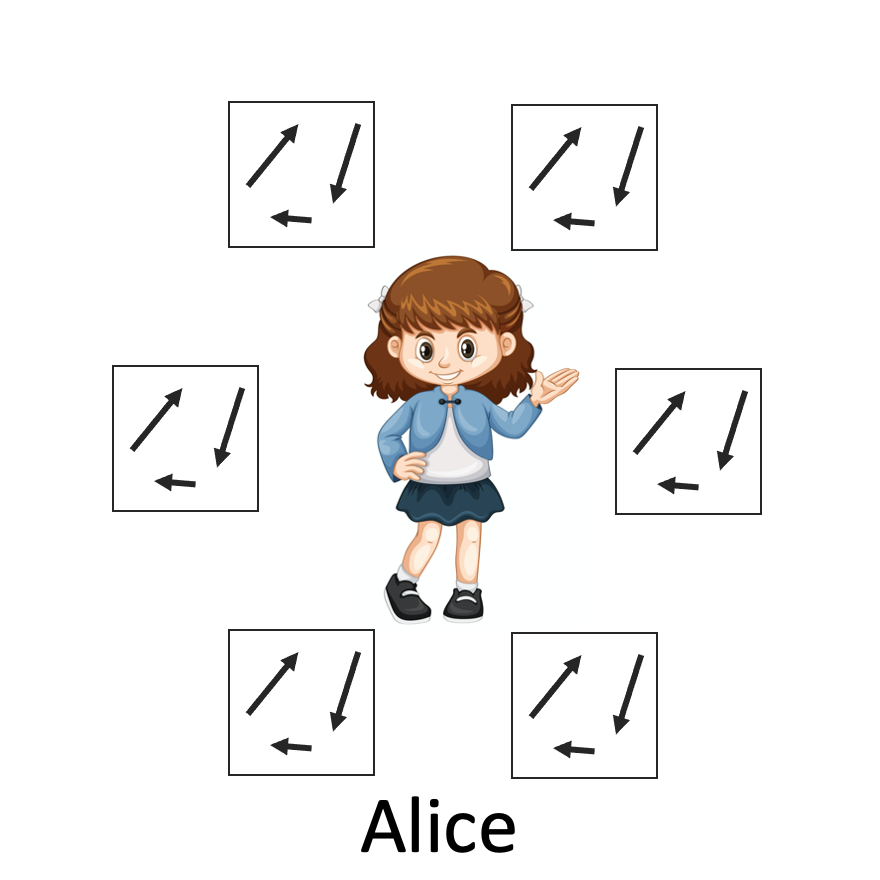}
        \caption{}
        \label{fig:ch4_particulas_indistinguibles}
    \end{subfigure}
    \begin{subfigure}{0.49\textwidth}
        \includegraphics[width=\linewidth]{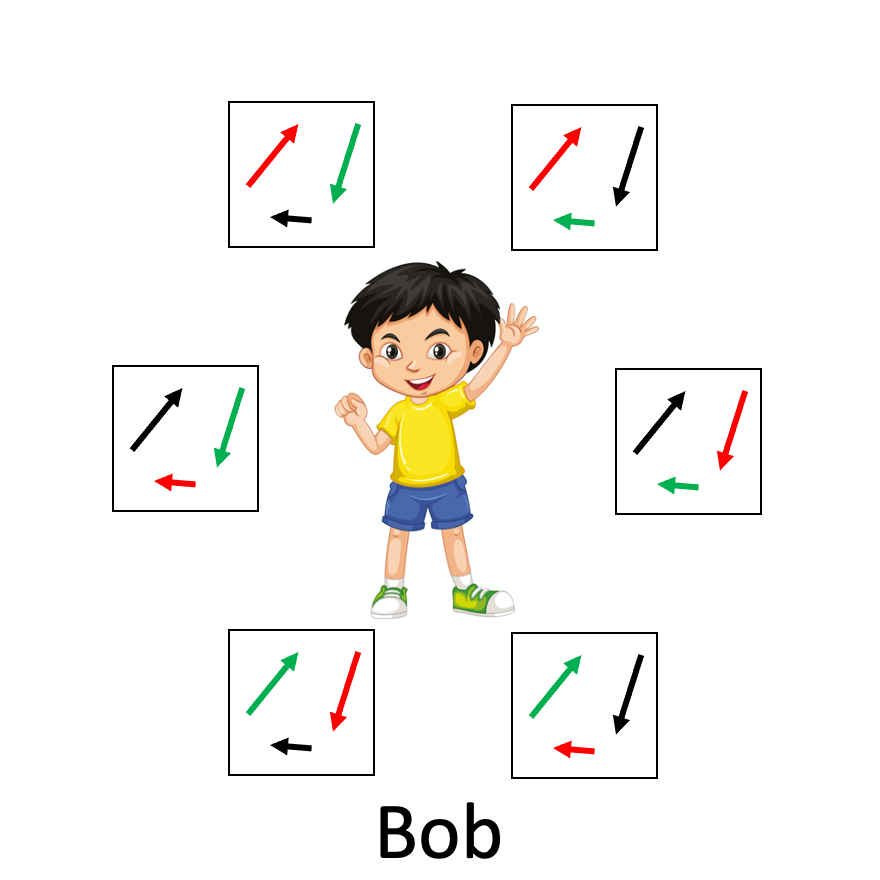}
        \caption{}
        \label{fig:ch4_particulas_distinguibles}
    \end{subfigure}
    \caption{Representación esquemática de un sistema formado por tres partículas, donde se tienen dos observadores, (a) Alice, que es incapaz de distinguir entre una y otra partícula del sistema y (b) Bob, que dice distinguir cada una de las partículas. Alice solo observa un estado indistingible mientras que Bob observa seis estados distinguibles, donde las partículas se distinguen una de la otra por su color. Las flechas representan la velocidad de cada partícula, donde la cola del vector coincide con la posición de la partícula en el instante en que se observa la configuración del sistema.}
    \label{fig:ch4_particulas_gas_ideal_clasico}
\end{figure}

Gibbs interpretó este resultado como la consecuencia de contar estados adicionales cuando se calcula la función de partición $Z$. Para ilustrar este argumento, consideremos un sistema formado por tres partículas no interactuantes, como se muestra en la Figura \ref{fig:ch4_particulas_gas_ideal_clasico}, donde cada flecha representa la velocidad de la partícula y su cola, la posición de la partícula en el espacio. Una configuración, para un tiempo dado, está determinada por seis vectores, tres de posición y tres de momentum, por lo tanto, para una configuración, al tener tres partículas distinguibles, existen seis ($3!$) posibles permutaciones de partículas que dan lugar a la misma configuración del sistema como un todo. Si se consideran $N$ partículas distinguibles, Figura \ref{fig:ch4_particulas_distinguibles}, la cuenta de estados estará multiplicada por un factor $N!$. 

En consecuencia, si se define la función de partición $Z'$ de partículas indistinguibles como,

\begin{equation}
    Z' = \frac{Z}{N!}
\end{equation}

Donde, al tener en cuenta que para $N \gg 1$ se puede utilizar la aproximación de Stirling, $\ln N! \approx N \ln N - N$, el logaritmo natural de la función de partición de partículas indistinguibles estará dado por,

\[
    \ln Z' = \ln Z - N \ln N + N
\]

Y la entropía,

\begin{equation}
    \label{eq:ch4_entropía_gas_ideal_clásico_indistinguible}
    S' = k_B \left[ N \ln \left( \frac{V}{N} \right) - \frac{3}{2} N \ln \beta + \frac{3}{2} N \ln \left( \frac{2\pi m}{h^2} \right) + \frac{5}{2}N \right]
\end{equation}

Al calcular la entropía $S'_{1/2}$ para un compartimento se obtiene que $S' - 2S'_{1/2} = 0$, que coincide con el comportamiento esperado de la entropía. Este resultado sugiere que las partículas que conforman un gas ideal son indistinguibles, por lo tanto, es imposible distinguir entre los seis estados y estos corresponden solo a un estado, Figura \ref{fig:ch4_particulas_indistinguibles}. El primero en notar la indistinguibilidad de partículas clásicas que conforman un gas ideal fue Gibbs en 1874 \cite{Gibbs1874}.

Con el objetivo de complementar el análisis anterior, consideremos dos observadores, Alice y Bob. Alice es incapaz de distinguir una partícula de otra, por lo tanto, al analizar los seis posibles estados obtenidos al permutar las tres partículas, Alice es incapaz de observar ninguna diferencia y, desde su punto de vista, solo existe un estado, Figura \ref{fig:ch4_particulas_indistinguibles}. Por otra parte, Bob dice poder identificar una partícula de la otra, en tal caso, es capaz de identificar cada uno de los seis posibles estados obtenidos al permutar las partículas, Figura \ref{fig:ch4_particulas_distinguibles}. Gibbs fue el primero en caer en la cuenta que la entropía del gas ideal clásico obtenida con la función de partición clásica no es una cantidad extensiva; la explicación del caracter no extensivo de la entropía, según Gibbs, se atribuye a que al realizar los cálculos se consideran estados de partículas distinguibles, a la manera de Bob. Gibbs intuyó que al considerar correcta la observación de Alice, es decir, que es imposible distinguir entre una y otra partícula del gas ideal, se solucionaría la paradoja que surge al analizar la entropía con la función de partición clásica. Para resolver esta paradoja Gibss dividió la entropía obtenida con la función de partición clásica por el número de permutaciones posibles, con lo cual la entropía adquiere el esperado comportamiento como cantidad extensiva y así se resuelve la paradoja.


\subsection{Distribución de Maxwell-Boltzmann}
\index{Distribución de Maxwell-Boltzmann}

Sea $f(\vec{r}, \vec{v})d\vec{r}d\vec{v}$ el número medio de partículas de un gas ideal por unidad de volumen cuyo centro de masas se encuentra entre $\vec{r}$ y $\vec{r} + d\vec{r}$ y su velocidad entre $\vec{v}$ y $\vec{v}+d\vec{v}$. Esta cantidad se puede obtener como,

\[
    f(\vec{r}, \vec{v})d\vec{r}d\vec{v} = C e^{-\beta m v^2/2}d\vec{r}d\vec{v}
\]

Donde $C$ es una constante de normalización tal que al resolver las integrales sobre las posiciones y velocidades se obtiene el número de partículas $N$,

\[
    \int \int f(\vec{r}, \vec{v})d\vec{r}d\vec{v} = N
\]

Al resolver la integral espacial se obtiene el volumen $V$ ocupado por el gas ideal,

\[
    \int \int f(\vec{r}, \vec{v})d\vec{r}d\vec{v} = C \int \int e^{-\beta m v^2/2}d\vec{r}d\vec{v} = C V \int e^{-\beta m v^2/2} d\vec{v} = N
\]

Como, en coordenadas cartesianas, $\vec{v} = v_x \hat{i} + v_y \hat{j} + v_z \hat{k}$,

\[
    \int \int f(\vec{r}, \vec{v})d\vec{r}d\vec{v} = CV \int e^{-\beta m v_x^2/2} dv_x \int e^{-\beta m v_y^2/2} dv_y \int e^{-\beta m v_z^2/2} dv_z
\]

Dado que $\beta^{-1} = k_B T$ y,

\[
    \int e^{-\frac{(x-\mu)^2}{2\sigma^2}} dx = \sigma \sqrt{2 \pi}
\]

Se obtiene,

\[
    \int \int f(\vec{r}, \vec{v})d\vec{r}d\vec{v} = CV \left( \frac{2\pi k_B T}{m} \right)^{3/2} = N
\]

Sea $\eta = N/V$ el número de partículas por unidad de volumen, entonces,

\[
    C = \eta \left( \frac{m}{2\pi k_B T} \right)^{3/2}
\]

Al integrar $f(\vec{r}, \vec{v})d\vec{r}d\vec{v}$ en las posiciones y dividir por el volumen se obtiene la distribución de velocidades de Maxwell-Boltzmann,

\[
    f(\vec{v})d\vec{v} = \frac{1}{V} \int f(\vec{r}, \vec{v})d\vec{r} 
\]

Nótese que para cada componente de la velocidad se obtiene una distribución normal dada por $g_i dv_i$, donde $i$ corresponde con cada una de las componentes del vector velocidad, tal que,

\[
    \frac{f(\vec{v})d\vec{v}}{\eta} = \left[ \frac{g_x(v_x) dv_x}{\eta} \right] \left[ \frac{g_y(v_y) dv_y}{\eta} \right] \left[ \frac{g_z(v_z) dv_z}{\eta} \right]
\]

Además, si $v = |\vec{v}|$, el número medio de partículas por unidad de volumen, $F(v)dv$, con velocidades entre $v$ y $v+dv$ (independiente de su dirección y sentido) está dado por,

\[
    F(v)dv = \int\limits_{\theta = 0}^{\pi} \int\limits_{\varphi = 0}^{2 \pi} f(\vec{v}) v^2 \sen(\varphi) dv d\theta d\varphi  = 4\pi f(v) v^2 dv
\]

\begin{figure}[t]
    \centering
    \includegraphics[width=\linewidth]{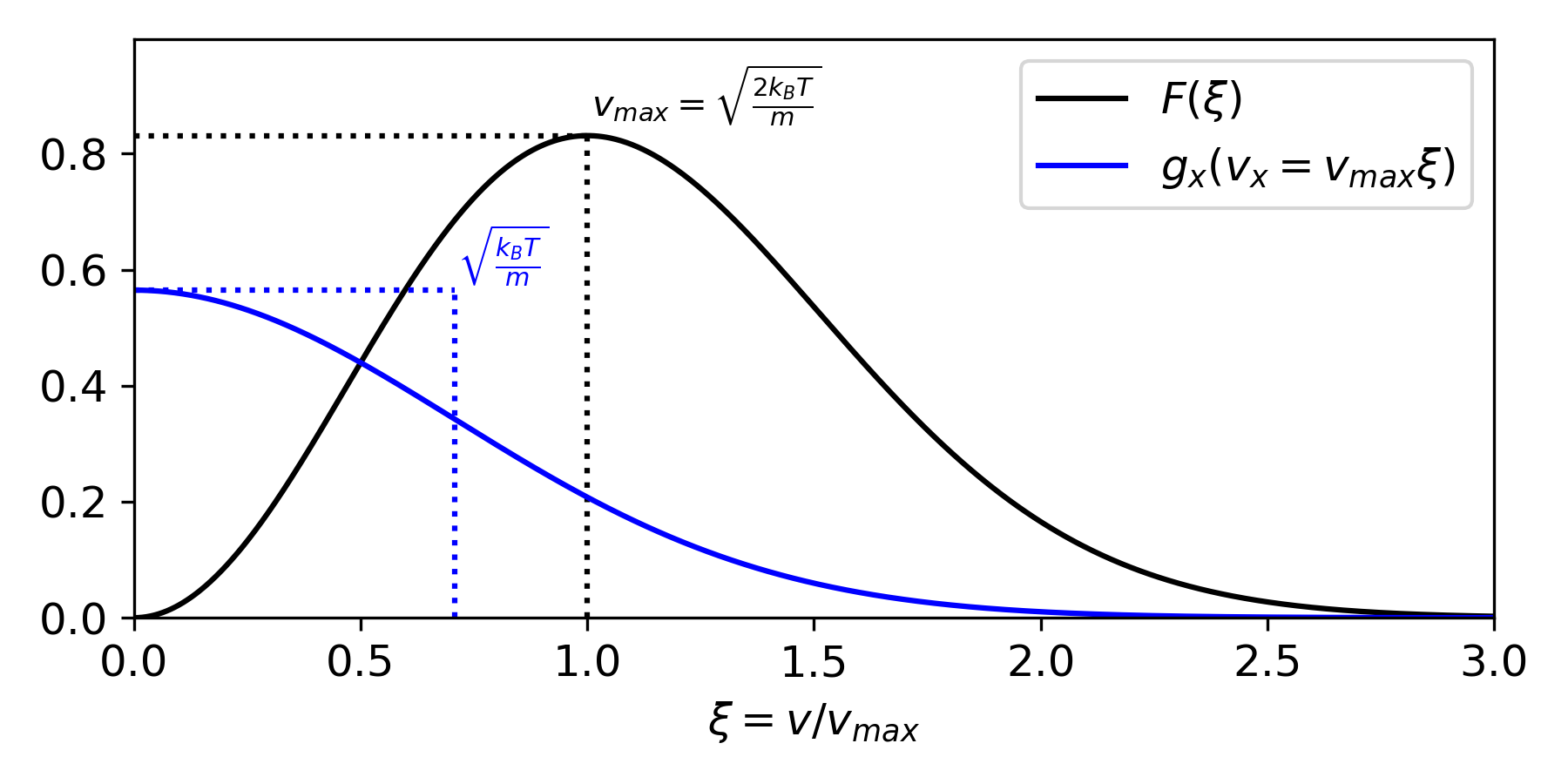}
    \caption{Distribución de velocidades de Maxwell-Botlzmann, donde $g_x(v_x)dv_x$ es el número medio de partículas por unidad de volumen con velocidades entre $v_x$ y $v_x + dv_x$ y $F(v)dv$, el número medio de partículas por unidad de volumen con magnitud de la velocidad entre $v$ y $v+dv$.}
    \label{fig:ch4_Maxwell-Boltzmann_dist}
\end{figure}

En la Figura \ref{fig:ch4_Maxwell-Boltzmann_dist} se compara la el número medio de partículas por unidad de volumen $g_x(v_x)dv_x$ con velocidades entre $v_x$ y $v_x + dv_x$ y el número medio de partículas por unidad de volumen $F(v)dv$ con magnitud de la velocidad entre $v$ y $v+dv$.

Con el objetivo que el lector adquiera un mayor nivel de comprensión, se le sugiere resolver las preguntas de autoexplicación del ejemplo trabajado titulado \href{https://colab.research.google.com/github/davidalejandromiranda/StatisticalPhysics/blob/main/notebooks/es_DistribucionVelocidadesMaxwellBoltzmann.ipynb}{distribución de velocidades de Maxwell-Boltzman}.

\begin{definition}
    \textbf{Distribución de velocidades y posiciones}. El número medio de partículas con posición del centro masas entre $\vec{r}$ y $\vec{r} + d\vec{r}$ y velocidades entre $\vec{v}$ y $\vec{v} + d\vec{v}$, para un gas ideal con $\eta=N/V$ partículas por unidad de volumen está dada por la ecuación \eqref{eq:ch4_distr_velocidades_posición}.\index{Distribución de velocidades y posiciones}
    \begin{equation}
        \label{eq:ch4_distr_velocidades_posición}
        f(\vec{r}, \vec{v})d\vec{r}d\vec{v} = \eta \left( \frac{m}{2\pi k_B T} \right)^{3/2} e^{-\frac{m v^2}{2k_BT}}d\vec{r}d\vec{v}
    \end{equation}
\end{definition}

\begin{definition}
    \textbf{Distribución vectorial de velocidades de Maxwell-Boltzmann}. El número medio de partículas velocidades entre $\vec{v}$ y $\vec{v} + d\vec{v}$, para un gas ideal con $\eta=N/V$ partículas por unidad de volumen está dada por la ecuación \eqref{eq:ch4_distr_velocidades}.\index{Distribución de velocidades de Maxwell-Boltzmann}
    \begin{equation}
        \label{eq:ch4_distr_velocidades}
        f(\vec{v})d\vec{v} = \eta \left( \frac{m}{2\pi k_B T} \right)^{3/2} e^{-\frac{m v^2}{2k_BT}}d\vec{v}
    \end{equation}
\end{definition}

\begin{definition}
    \textbf{Distribución de velocidades de Maxwell-Boltzmann por componentes cartesianas}. El número medio de partículas velocidades entre $v_i$ y $v_i + dv_i$, donde $i={x, y, z}$, para un gas ideal con $\eta=N/V$ partículas por unidad de volumen está dada por la ecuación \eqref{eq:ch4_distr_velocidades_componente}.\index{Distribución de velocidades de Maxwell-Boltzmann por componente cartesiana}
    \begin{equation}
        \label{eq:ch4_distr_velocidades_componente}
        g_i(v_i)dv_i = \eta \left( \frac{m}{2\pi k_B T} \right)^{1/2} e^{-\frac{m v_i^2}{2k_BT}}dv_i
    \end{equation}
\end{definition}

\begin{definition}
    \textbf{Distribución de velocidades de Maxwell-Boltzmann}. El número medio de partículas con magnitud de su velocidad entre $v$ y $v + dv$, para un gas ideal con $\eta=N/V$ partículas por unidad de volumen está dada por la ecuación \eqref{eq:ch4_distr_magnitud_velocidades}.\index{Distribución de velocidades de Maxwell-Boltzmann}
    \begin{equation}
        \label{eq:ch4_distr_magnitud_velocidades}
        F(v)dv = 4\pi \eta \left( \frac{m}{2\pi k_B T} \right)^{3/2} v^2 e^{-\frac{m v^2}{2k_BT}}dv
    \end{equation}
\end{definition}


\newpage

\subsection{Ley de atmósferas}

Sea un gas ideal formado por $N$ partículas, a temperatura $T$ en presencia de un campo gravitatorio en dirección $-\hat{k}$. ¿Cuál es la probabilidad $\prob{z < Z \leq z + dz} = g(z)dz$ de encontrar una molécula del gas ideal a una cierta altura $z$?

Dado que ahora la energía de una partícula del gas ideal depende de la altura, la probabilidad $\prob{z < Z \leq z + dz} = g(z)dz$ que una partícula se encuentre entre $z$ y $z+dz$ está dada por,

\[
    g(z)dz = \left( \frac{1}{h^3 Z'_1} \int dx \int dy \int dp_x e^{-\beta \frac{p_x^2}{2m}} \int dp_y e^{-\beta \frac{p_y^2}{2m}} \int dp_z e^{-\beta \frac{p_x^2}{2m}} \right) e^{-\beta mgz} dz
\]

Donde se asume la referencia de la energía potencial en $z=0$ y $Z'_1$ es la función de partición para una partícula de gas ideal, dada por,

\[
    Z'_1 = \left( \frac{1}{h^3} \int dx \int dy \int dp_x e^{-\beta \frac{p_x^2}{2m}} \int dp_y e^{-\beta \frac{p_y^2}{2m}} \int dp_z e^{-\beta \frac{p_x^2}{2m}} \right) \int_0^\infty e^{-\beta mgz} dz
\]

Como,

\[
    \int_0^\infty e^{-\beta mgz} dz = \frac{k_B T}{mg}
\]

Se obtiene,

\[
    g(z)dz = \frac{mg}{k_B T} e^{-\frac{mgz}{k_B T}} dz
\]

Este resultado nos muestra que la probabilidad de encontrar una partícula de gas ideal decrece exponencialmente con la altura.


\section{Problemas propuestos}

\begin{exercise}
    Calcule la entropía para un gas ideal clásico a temperatura absoluta $T$ y muestre bajo qué consideraciones esta no es una cantidad extensiva.
\end{exercise}

\begin{exercise}
    Calcule el valor esperado de la magnetización para un sistema a temperatura ambiente, $25^\circ C$, formado por partículas paramagnéticas con magnitud del momento angular orbital igual a $3\hbar$ y proyección en el eje $z$ del espín igual a $-\hbar/2$.
\end{exercise}

\begin{exercise}
    Deduzca la función de distribución de Fermi-Dirac y con el resultado obtenido demuestre que cuando la temperatura absoluta tiende a cero todas las partículas están por debajo de la energía de Fermi.
\end{exercise}

\begin{exercise}
    Demuestre que para un gas ideal a temperatura absoluta T su energía interna corresponde con el valor medio de la energía cinética de las partículas del gas y es proporcional a la temperatura absoluta.
\end{exercise}

\begin{exercise}
    Sean dos observadores, Bob y Alice. Bob dice tener la capacidad de distinguir partículas que conforman un sistema, mientras Alice es incapaz de distinguir una partícula de otra.

    \begin{itemize}
        \item[a)] ¿Cómo observaría Bob y Alice un sistema clásico? Argumente su respuesta.
        \item[b)] ¿Son las predicciones de Bob y Alice consistentes con las observaciones experimentales? Argumente su respuesta.
    \end{itemize}

\end{exercise}

\chapter{Capítulo IV. Sistemas en contacto con un reservorio de calor a temperatura T y potencial químico µ: ensamble gran canónico}
Hasta ahora se han estudiado dos formas de preparar un conjunto de sistemas de la misma forma, primero, sistemas aislados con energía en un rango entre $E$ y $E+dE$ (ensamble microcanónico) y segundo, sistemas en contacto con un reservorio de calor a temperatura $T$ con el cual puede intercambiar energía, tal que el sistema con el reservorio se asume aislado del resto del universo (ensamble canónico). Es de notar que el tipo de ensamble utilizado para describir un fenómeno físico debería ser independiente del resultado obtenido, es decir, independientemente de la escogencia del ensamble para describir un fenómeno físico, el resultado obtenido debe ser el mismo. Sin embargo, la descripción de ciertos fenómenos físicos se simplifica seleccionando el ensamble apropiado; por ejemplo, si se quiere estudiar sistemas aislados donde no hay control de su temperatura, un ensamble microcanónico es la mejor escogencia; al describir un sistema en equilibrio termodinámico a una cierta temperatura, con número constante de partículas, la mejor selección es el ensamble canónico. Si se tiene un sistema a una cierta temperatura y cuyo número de partículas puede cambiar, conviene utilizar un ensamble gran canónico (o canónico grande).

En este capítulo se estudiará el ensamble canónico y se utilizará para estudiar sistemas donde se puede variar el número de partículas.

\begin{definition}
    El \textbf{potencial químico} $\mu$ es un parámetro externo que determina el intercambio de partículas entre sistemas. Sea $E$ la energía del sistema y $N$ el número de partículas, entonces, el potencial químico se puede expresar por la ecuación \eqref{ch5_potencial_químico}, donde $v$ indica que la derivada parcial se toma asumiendo los otros potenciales constantes.\index{Potencial químico}
    \begin{equation}
        \label{ch5_potencial_químico}
        \mu = \left( \frac{\partial E}{\partial N}\right)_v
    \end{equation}
\end{definition}

\begin{definition}
    Un \textbf{ensamble gran canónico} consiste en un conjunto de sistemas preparados bajo las mismas condiciones tal que su temperatura $T$ como su potencial químico $\mu$ están definidos y el sistema puede intercambiar tanto partículas como energía con un reservorio con el que se encuentra en contacto, tal que el gran sistema, formado por el sistema y el reservorio, están aislados del resto del universo.\index{Ensamble gran canónico}
\end{definition}

\section{Descripción estadística del sistema y cantidades termodinámicas en un ensamble gran canónico}
\index{Ensamble gran canónico}

\begin{figure}[t]
    \centering
    \includegraphics[width=0.55\linewidth]{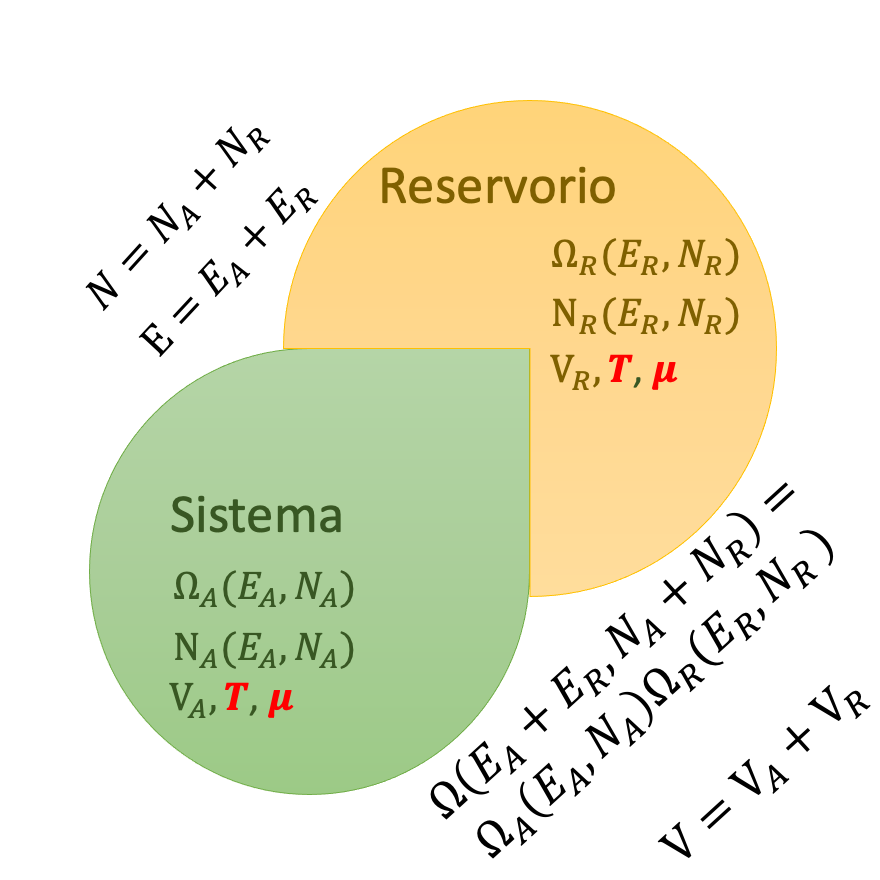}
    \caption{Sistema $A$ en contacto con un reservorio de calor $R$. El sistema $A$ tiene energía $E_A$, número de partículas $N_A$, número de estados accesibles $\Omega_A(E_A, N_A)$, se encuentra a una temperatura $T$ y potencial químico $\mu$, en equilibrio termodinámico con el reservorio de calor $R$. El reservorio $R$ tiene energía $E_R \gg E_A$, número de partículas $N_R\gg N_A$ y número de estados accesibles $\Omega_R(E_R, N_R) \gg \Omega_A(E_A, N_A)$. La energía total $E$ del gran sistema formado por el reservorio y el sistema $A$, es la suma de las energías, $E=E_A+E_R$; de la misma manera, el número total de partículas $N$ está dado por $N=N_A+N_R$. El número de estados accesibles $\Omega(E, N)$, corresponde al producto del número de estados accesibles, $\Omega(E=E_A+E_R, N=N_A+N_R) = \Omega_A(E_A, N_A) \Omega_R(E_R, N_R)$.}
    \label{fig:ch5_sistema_reservorio}
\end{figure}

Sea un sistema $A$ en contacto con un reservorio de calor $R$ con el cual se encuentra en equilibrio termodinámico a una temperatura $T$ y potencial químico $\mu$, como se muestra en la Figura \ref{fig:ch5_sistema_reservorio}. El reservorio de calor es un sistema con la capacidad suficiente para recibir (o entregar) partículas al sistema $A$ sin cambios significativos en su potencial químico y con la suficiente capacidad calorífica para absorber tanto calor del sistema $A$ sin cambios significativos en su temperatura. 

\subsection{Condiciones de equilibrio termodinámico}

Cuando el sistema puede intercambiar energía y partículas con el reservorio, el equilibrio termodinámico se alcanza al cumplir dos condiciones: la condición de equilibrio térmico y la de equilibrio químico. 

\subsubsection{Equilibrio térmico}
\index{Ensamble gran canónico!equilibrio térmico}

Como el gran sistema formado por el reservorio y el sistema $A$ están aislados, el número total de estados accesibles $\Omega(E=E_A+E_R, N=N_A+N_R) = \Omega_A(E_A, N_A) \Omega_R(E_R, N_R)$ es constante y se cumple que,

\[
	\frac{\partial \Omega(E=E_A+E_R, N=N_A+N_R)}{\partial E_A} = 0
\]

Debido a que el número total de estados del gran sistema permanece constante cuando varía la energía del sistema $A$, se tiene la condición de equilibrio termodinámico entre el sistema $A$ y el reservorio de calor,

\[
    \Omega_R(E_A, N_A) \frac{\partial \Omega_A(E_A, N_A)}{\partial E_A} = - \Omega_A(E_A, N_A) \frac{\partial \Omega_R(E_R, N_R)}{\partial E_R} \frac{\partial E_R}{\partial E_A}
\]

Como $E = E_A + E_R$ y $E$ se mantiene constante, entonces, $\partial E_R / \partial E_A = -1$. Al reordenar términos,

\[
    \frac{1}{\Omega_R(E_A, N_A)} \frac{\partial \Omega_A(E_A, N_A)}{\partial E_A} = \frac{1}{\Omega_R(E_R, N_R)} \frac{\partial \Omega_R(E_R, N_R)}{\partial E_R}
\]

Al expresar la anterior ecuación en términos de logaritmo,

\[
    \frac{\partial \ln \Omega_A(E_A, N_A)}{\partial E_A} = \frac{\partial \ln \Omega_R(E_R, N_R)}{\partial E_R}
\]

Como $\beta_i = \partial \ln \Omega_i / \partial E_i$, se encuentra que la expresión anterior corresponde con la condición de equilibrio térmico,

\[
\beta_A = \beta_R \to T_A = T_R
\]

Donde, 

\begin{equation}
	\label{eq:ch5_beta_equilibrio}
	\beta = - \frac{\partial \ln \Omega_R(E-E_A, N-N_A)}{\partial E_A}	
\end{equation}

\subsubsection{Equilibrio químico}
\index{Ensamble gran canónico!equilibrio químico}

Al realizar el mismo análisis anterior con la variación del número de partículas se encuentra que,

\[
	\frac{\partial \Omega(E=E_A+E_R, N=N_A+N_R)}{\partial N_A} = 0
\]
Lo cual da lugar a la siguiente ecuación,

\[
    \frac{\partial \ln \Omega_A(E_A, N_A)}{\partial N_A} = \frac{\partial \ln \Omega_R(E_R, N_R)}{\partial N_R}
\]

El resultado anterior nos permite definir, como se hizo con el parámetro $\beta$, al parámetro externo $\alpha=-\beta \mu$  por medio de la ecuación \eqref{eq:ch5_parametro_alpha}. La condición de equilibrio químico se satisface cuando el potencial químico del sistema $A$ y el reservorio son iguales,

\[
\alpha_A = \alpha_R \to \mu_A = \mu_R
\]
Donde, 

\begin{equation}
	\label{eq:ch5_alpha_equilibrio}
	\alpha = - \frac{\partial \ln \Omega_R(E-E_A, N-N_A)}{\partial N_A}	
\end{equation}

\begin{definition}
        Sea $\Omega(E, N)$ el número de estados accesibles de un sistema con energía $E$ y número de partículas $N$. Se define el parámetro $\alpha$ como, \index{Parámetro $\alpha$}
        \begin{equation}
        \label{eq:ch5_parametro_alpha}
        \alpha = -\beta\mu = \frac{\partial \ln \Omega(E, N)}{\partial N}
    \end{equation}
\end{definition}

\subsection{Número de estados $\Omega_A(E_A, N_A)$ y probabilidad $p(E_A, N_A)$}

Se procederá como se hizo con el ensamble canónico, es decir, se obtendrá la probabilidad que el sistema $A$ se encuentre en un estado con energía $E_A$ y $N_A$ partículas a partir del logaritmo natural del número de estados accesible,s $\ln \Omega(E_A, N_A)$. 

\begin{equation}
	\label{eq:ch5_prob_eq1}
	\ln \Omega(E, N) = \ln \Omega_A(E_A, N_A) + \ln \Omega_R(E - E_A, N - N_R)
\end{equation}

Como $E_R \gg E_A$ y $N_R \gg N_A$, se puede expandir $\ln \Omega_R(E - E_A, N - N_A)$ alrededor de $E_A$ y $N_A$ ,

\[
	\ln \Omega_R(E - E_A, N - N_A) = \ln \Omega_R(E, N) - \frac{\partial \ln \Omega}{\partial E_A} E_A - \frac{\partial \ln \Omega}{\partial N_A} N_A  + \mathcal{O}(E_A^2) + \mathcal{O}(N_A^2)
\]

Al tener en cuenta los parámetros $\alpha$ y $\beta$ dados por las ecuaciones \eqref{eq:ch5_beta_equilibrio} y \eqref{eq:ch5_alpha_equilibrio}, respectivamente,

\[
	\ln \Omega_R(E - E_A, N - N_A) \approx \ln \Omega_R(E, N) + \alpha N_A + \beta E_A
\]

Es importante notar que $\ln \Omega_R(E, N) = \ln \Omega_{0_R}$ es una constante que corresponde al caso cuando el sistema $A$ tiene energía cero y ninguna partícula; por consiguiente, el menor número de estados accesibles $\Omega_A(0, 0)$. Al reemplazar el resultado anterior en la ecuación \eqref{eq:ch5_prob_eq1},

\[
	\ln \Omega(E, N) \approx \ln \Omega_A(E_A, N_A) + \ln \Omega_{0_R} + \alpha N_A + \beta E_A
\]

Con este resultado, al despejar y simplificar, se obtiene $\Omega_A(E_A, N_A)$,

\begin{equation}
	\label{eq:ch5_prob_eq2}
	\Omega_A(E_A, N_A) \approx \frac{\Omega(E, N)}{\Omega_{0_R}} e^{-\alpha N_A - \beta E_A} = \frac{\Omega(E, N)}{\Omega_{0_R}} e^{-\beta (E_A - \mu N_A)}
\end{equation} 

El número total de estados accesibles para el sistema $A$ está dado por la suma de la ecuación \eqref{eq:ch5_prob_eq2} para todos los valores de $E_A$,

\begin{equation}
	\label{eq:ch5_prob_eq3}
	\sum_{E_A, N_A} \Omega_A(E_A, N_A) \approx \frac{\Omega(E, N)}{\Omega_{0_R}} \sum_{E_A, N_A} e^{-\beta (E_A-\mu N_A)}
\end{equation} 

La probabilidad $p(E_A, N_A)$ de encontrar al sistema $A$ en un estado con energía $E_A$ y $N_A$ partículas está dado por el cociente entre $\Omega_A(E_A, N_A)$ y $\sum\limits_{E_A, N_A} \Omega_A(E_A, N_A)$,

\begin{equation}
	\label{eq:ch5_prob_eq4}
	p(E_A, N_A) \approx \frac{e^{-\beta (E_A-\mu N_A)}}{\sum\limits_{E_A, N_A} e^{-\beta (E_A - \mu N_A)}}
\end{equation} 

\begin{definition}
	Sea un sistema en contacto con un reservorio a temperatura $T$ y potencial químico $\mu$, donde $\alpha =-\beta \mu$, $\beta = (k_B T)^{-1}$;  $E_r$ son los posibles valores de energía que puede tomar el sistema y $N_s$, el número de partículas. Se define la \textbf{función de partición $\granZ$ del ensamble gran canónico}, también conocida como gran función de partición, como la suma de las exponenciales $e^{-\beta (E_r-\mu N_s)}$.\index{Ensamble gran canónico!función de partición}
	\begin{equation}
		\label{eq:ch5_función_partición}
		\granZ = \sum_{r, s} e^{-\beta E_r - \alpha N_s} = \sum_{r, s} e^{-\beta (E_r - \mu N_s)}
	\end{equation}
\end{definition}

\begin{definition}
	Sea un sistema en contacto con un reservorio a temperatura $T$ y potencial químico $\mu$. La \textbf{probabilidad} $p(E_r, N_s)$ que el sistema se encuentre en un estado con energía $E_r$ y $N_s$ partículas, está dada por la ecuación \eqref{eq:ch5_probabilidad_canonica}.\index{Ensamble gran canónico!probabilidad}
	\begin{equation}
		\label{eq:ch5_probabilidad_canonica}
		p(E_r, N_s) = \granZ^{-1} e^{-\beta (E_r - \mu N_s)}
	\end{equation}
\end{definition}

\subsection{Valor esperado de la energía: energía interna}

El valor esperado de la energía $\E{E} = U$ corresponde con la energía interna del sistema y está dado por,

\[
	\E{E} = \sum_{r,s} p(E_r, N_s) E_r = \granZ^{-1} \sum_{r, s} E_r e^{-\beta E_r -\alpha N_s }
\]

Como $\partial \granZ / \partial \beta = - \sum\limits_{r,s} E_r e^{-\beta E_r -\alpha N_s}$, \index{Ensamble gran canónico!valor esperado de la energía}

\begin{equation}
	\label{eq:ch5_valor_esperado_energía}
	\E{E} = U = - \frac{\partial \ln \granZ}{\partial \beta}
\end{equation}

\subsection{Valor esperado del número de partículas}

El valor esperado del número de partículas $\E{N}$ está dado por,

\[
	\E{N} = \sum_{r,s} p(E_r, N_s) N_s = \granZ^{-1} \sum_{r, s} N_s e^{-\beta E_r -\alpha N_s }
\]

Como $\partial \granZ / \partial \alpha = - \sum\limits_{r,s} N_s e^{-\beta E_r -\alpha N_s}$, \index{Ensamble gran canónico!valor esperado del número de partículas}

\begin{equation}
	\label{eq:ch5_valor_esperado_número_partículas}
	\E{N} = \mathcal{N} = - \frac{\partial \ln \granZ}{\partial \alpha}
\end{equation}

\subsection{Valor esperado de la fuerza generalizada conjugada a un parámetro externo $x$}

La fuerza generalizada $X$ conjugada a un parámetro externo $x$, suponiendo que el número de partículas permanece constante cuando varía el parámetro externo $x$, está determinada por la ecuación \eqref{eq:ch2_fuerza_generalizada}, por lo tanto, su valor esperado está dado por,

\[
	\E{X} = \sum_{r, s} X_r p(E_r, N_s) = \sum_{r, s} -\frac{\partial E_r}{\partial x} p(E_r, N_s) = - \granZ^{-1} \sum_{r, s} \frac{\partial E_r}{\partial x} e^{-\beta E_r -\alpha N_s}
\]

Como $X_r = \partial E_r/\partial x$,

\[
	\E{X} =  \granZ^{-1} \sum_{r, s} X_r e^{-\beta E_r - \alpha N_s}
\]

Nótese que la variación de la función de partición respecto al parámetro externo $x$ sin que esta variación afecte el número de partículas $N$ está dada por,

\[
 	\left( \frac{\partial \granZ}{\partial x} \right)_N = \sum_{r, s} -\beta \frac{\partial E_r}{\partial x} e^{-\beta E_r - \alpha N_s} = \beta \sum_{r, s} X_r e^{-\beta E_r - \alpha N_s}
\]

Por lo tanto, el valor esperado $\E{X}$ está dado por,\index{Ensamble gran canónico!valor esperado de la fuerza generalizada}

\begin{equation}
		\label{eq:ch5_valor_esperado_fuerza_conjugada}
		\E{X} = \frac{1}{\beta} \left( \frac{\partial \ln \granZ}{\partial x} \right)_N
\end{equation}

Donde el subíndice $N$ en la derivada parcial indica que esta se realiza suponiendo que el número de partículas en los diferentes estados permanece constante cuando se varía el parámetro externo $x$.

\subsection{Entropía}

Para obtener la entropía en términos de la función de partición $\granZ$ se procederá como se hizo con en ensamble canónico,

\[
  d \ln \granZ(\alpha, \beta, x) = \frac{\partial \ln \granZ}{\partial \beta} d\beta + \frac{\partial \ln \granZ}{\partial \alpha} d\alpha + \frac{\partial \ln \granZ}{\partial x} dx
\]

Como la energía interna $U$ está dada por la ecuación \eqref{eq:ch5_valor_esperado_energía}; el valor esperado del número de partículas $\mathcal{N}=\E{N}$, por la ecuación \eqref{eq:ch5_valor_esperado_número_partículas} y el valor esperado $\E{X}$ de la fuerza generalizada conjugada a $x$, por la ecuación \eqref{eq:ch5_valor_esperado_fuerza_conjugada} y $\dbar W = \E{X} dx$,

\[
	d \ln \granZ(\alpha, \beta, x) = - Ud\beta - \mathcal{N}d\alpha + \beta \dbar W
\]

Como $d(\beta U) = \beta dU + Ud\beta$ y $d(\alpha \mathcal{N}) = \alpha d \mathcal{N} + \mathcal{N} d\alpha$,

\[
    d \ln \granZ(\alpha, \beta, x) = \beta dU - d(\beta U) + \alpha d\mathcal{N} - d(\alpha \mathcal{N}) + \beta \dbar W
\]

Reordenando términos,

\[
    d \ln \granZ(\alpha, \beta, x) = -d(\alpha \mathcal{N}) -d(\beta U) + \beta( dU - \mu d\mathcal{N} + \dbar W )
\]

La primera ley de la termodinámica cuando se intercambian $d\mathcal{N}$ partículas a un potencial químico $\mu$ está dada por,

\[
    \dbar Q = dU - \mu d\mathcal{N} + \dbar W = T dS
\]

Por lo tanto, al reordenar términos se obtiene que,

\[
	d(\ln \granZ + \beta U + \alpha \mathcal{N}) = \beta T dS
\]

Con lo cual se obtiene la entropía expresada en términos de la función de partición, \index{Ensamble gran canónico!entropía}

\begin{equation}
		\label{eq:ch5_entropía}
		S = k_B \ln \granZ + \frac{U}{T} - \frac{\mu \mathcal{N}}{T}
\end{equation}

\subsection{Energía libre de Gibbs}

La variación de la energía libre de Gibbs, $dG$, para un sistema en equilibrio termodinámico a potencial químico $\mu$ y con variación de $d\mathcal{N}$ partículas está dada por,

\[
    dG = dF + PdV + VdP + \mu d \mathcal{N}
\]

Como $dU = T \dbar S - PdV$,

\[
    dG = VdP - SdT + \mu d \mathcal{N}
\]

Cuando se tiene presión y temperatura constante,\index{Energía libre de Gibbs}

\begin{equation}
    \label{eq:ch5_energía_libre_Gibbs}
    G = \mu \mathcal{N}
\end{equation}

\subsection{Energía libre de Helmholtz}

De acuerdo a la termodinámica, la energía libre de Helmholtz, $F$, está dada por,

\[
	F = U - TS
\]

Por lo tanto, al comparar esta expresión con la ecuación \eqref{eq:ch5_entropía}, la energía libre de Helmholtz está dada por la ecuación \eqref{eq:ch5_energía_libre_Helmholtz}.\index{Ensamble gran canónico!energía libre de Helmholtz}

\begin{equation}
    \label{eq:ch5_energía_libre_Helmholtz}
    F = \mu \mathcal{N} - k_B T \ln \granZ
\end{equation}

Como  $G = \mu \mathcal{N}$ y $F = G - PV$, se identifica que $PV$ está dada por,

\begin{equation}
    \label{eq:ch5_PV}
	PV = k_B T \ln \granZ	
\end{equation}

En términos de la fugacidad $z$, definida por la ecuación \eqref{eq:ch5_fugacidad},

\begin{equation}
    \label{eq:ch5_energía_libre_Helmholtz_fugacidad}
    F = - k_B T \ln \granZ + \mu \mathcal{N} = -k_B T \ln \left( \frac{\granZ}{z^\mathcal{N}} \right)
\end{equation}

\begin{definition}
    Se define la \textbf{fugacidad} $z$ como,\index{Fugacidad}
    \begin{equation}
        \label{eq:ch5_fugacidad}
        z = e^{-\alpha} = e^{\frac{\mu}{k_B T}}
    \end{equation}
\end{definition}



\section{Sistemas formados por muchas partículas cuánticas}

Las partículas cuánticas se consideran indistinguibles, en el sentido de la imposibilidad para distinguir una de otra, por ejemplo, se asume que un electrón es indistinguible de otro. En la naturaleza se observan sistemas formados por muchas partículas cuánticas, las cuales se pueden dividir en dos tipos, los fermiones y los bosones, de acuerdo a la manera en que estas partículas ocupan los estados accesibles.

Supongamos que se tiene un sistema formado por $N$ partículas cuánticas, descrito por el estado $|\psi\rangle$. Si se tienen $N$ bases continuas espaciales $\{\vec{r}_i\}$ (una para cada partícula) la proyección del estado del sistema sobre dichas bases corresponderá a la función de onda que lo describe,

\[
    \psi(\vec{r}_1, \cdots, \vec{r}_N) = \langle \vec{r}_1, \cdots, \vec{r}_N|\psi\rangle
\]

Esta función de onda tiene la propiedad que permanece invariante ante la permutación de partículas, debido a que se asume que estas son indistinguibles. Ahora definamos el operador permutación de dos partículas $\hat{P}_{i,j}$ tal que si actúa sobre una función de onda $\varphi(\cdots, \vec{r}_i, \cdots, \vec{r}_j, \cdots)$ el resultado será el intercambio de las dos partículas, por ejemplo,

\[
    \hat{P}_{1,2}\varphi(\vec{r}_1, \vec{r}_2) = p_{1,2} \varphi(\vec{r}_2, \vec{r}_1)
\]

Se debe notar que $p_{1,2}$ solo puedo tomar dos posibles valores, $\pm 1$, debido a que al intercambiar las dos partículas se espera que el estado del sistema permanezca invariante (por ser partículas indistinguibles) y su magnitud al cuadrado, $|\varphi|^2$, debe ser uno.

De manera general, si se aplica el operador $\hat{P}_{i,j}$ sobre el estado del sistema formado por $N$ partículas, descrito por la función de onda $\psi(\vec{r}_1, \cdots, \vec{r}_N)$, se tienen dos posibilidades,

\[
    \hat{P}_{i,j} \psi(\cdots, \vec{r}_i, \cdots, \vec{r}_j, \cdots) = \pm \psi(\cdots, \vec{r}_j, \cdots, \vec{r}_i, \cdots)
\]

Hasta donde se conoce (postulado de simetrización), estas son las dos únicas posibilidades que se observan en la naturaleza: estados simétricos ante la permutación de dos partículas conocidas como bosones y estados antisimétricos, para fermiones. Desde un punto de vista teórico, se especula la existencia de otros tipos de simetría que dan origen a las llamadas parapartículas \cite{Messiah1964, Hartle1969}, las cuales no se han observado en ningún experimento y algunos especulan que nunca serán observadas porque tales simetrías son imposibles en nuestro universo.

Pauli estableció que el espín de los fermiones y bosones es muy diferente entre si, siendo medio entero (un número entero dividido por dos) para fermiones y entero para bosones; a este resultado se le conoce como el teorema espín-estadística \cite{Pauli1940}.

\begin{definition}
    \textbf{Bosones}. Partículas que forman sistemas con estados simétricos ante la permutación de dos partículas y cuyo espín es entero. \index{Bosones}
    \[
    \hat{P}_{i,j} \psi(\cdots, \vec{r}_i, \cdots, \vec{r}_j, \cdots) = \psi(\cdots, \vec{r}_j, \cdots, \vec{r}_i, \cdots)
    \]
\end{definition}

\begin{definition}
    \textbf{Fermiones}. Partículas que forman sistemas con estados antisimétricos ante la permutación de dos partículas y cuyo espín es medio entero\footnote{Un espín medio entero, o semi entero, corresponde a un espín descrito por un número entero dividido por dos, por ejemplo: 1/2 y 3/5.} (o semi entero). \index{Fermiones}
    \[
    \hat{P}_{i,j} \psi(\cdots, \vec{r}_i, \cdots, \vec{r}_j, \cdots) = - \psi(\cdots, \vec{r}_j, \cdots, \vec{r}_i, \cdots)
    \]
\end{definition}

\begin{definition}
    El \textbf{postulado de simetrización} establece que los sistemas formados por dos o más partículas indistinguibles son simétricos o antisimétricos ante la permutación de dos partículas. \index{Postulado de simetrización}
\end{definition}

\subsection{Sistemas formados por bosones}

En sistemas formados por bosones se cumple que el estado que lo describe es simétrico ante el intercambio de partículas. Supongamos un sistema formado por dos bosones, descrito por,

\[
    |\psi\rangle = |\varphi_1\rangle|\varphi_2\rangle + |\varphi_2\rangle|\varphi_1\rangle
\]

Al aplicar el operador permutación se obtiene,

\[
    \hat{P}_{1,2}|\psi\rangle = |\varphi_2\rangle|\varphi_1\rangle + |\varphi_1\rangle|\varphi_2\rangle = |\psi\rangle
\]

Si se proyecta en el espacio, con dos bases contínuas espaciales, se puede describir el sistema en términos de funciones de onda, en tal caso, la función de onda que describe el sistema de dos bosones está dada por,

\[
    \langle \vec{r}_1, \vec{r}_2|\psi\rangle = \psi(\vec{r}_1, \vec{r}_2) = \langle \vec{r}_1, \vec{r}_2|\varphi_1\rangle|\varphi_2\rangle + \langle \vec{r}_1, \vec{r}_2|\varphi_2\rangle|\varphi_1\rangle = \varphi_1(\vec{r}_1)\varphi_2(\vec{r}_2) + \varphi_2(\vec{r}_1)\varphi_1(\vec{r}_2)
\]

Al actuar el operador permutación sobre el sistema se obtiene,

\[
    \hat{P}_{1,2}\psi(\vec{r}_1, \vec{r}_2) = \varphi_1(\vec{r}_2)\varphi_2(\vec{r}_1) + \varphi_2(\vec{r}_2)\varphi_1(\vec{r}_1) = \psi(\vec{r}_1, \vec{r}_2)
\]

Esto implica que si $|\varphi_1\rangle = |\varphi_2\rangle$, es decir, que los dos bosones ocupan un mismo estado, existe una probabilidad $|\langle \psi |\psi \rangle|^2 \neq 0$ diferente de cero que esto ocurra.

\begin{lema}
    Uno o más bosones pueden ocupar un cierto estado.\index{Bosones!ocupación de estados}
\end{lema}

\subsection{Sistemas formados por fermiones}

En sistemas formados por fermiones se cumple que el estado que lo describe es antisimétrico ante el intercambio de partículas. Supongamos un sistema formado por dos fermiones, descrito por,

\[
    |\psi\rangle = |\varphi_1\rangle|\varphi_2\rangle - |\varphi_2\rangle|\varphi_1\rangle
\]

Al aplicar el operador permutación se obtiene,

\[
    \hat{P}_{1,2}|\psi\rangle = |\varphi_2\rangle|\varphi_1\rangle - |\varphi_1\rangle|\varphi_2\rangle = -|\psi\rangle
\]

En términos de funciones de onda,

\[
    \langle \vec{r}_1, \vec{r}_2|\psi\rangle = \psi(\vec{r}_1, \vec{r}_2) = \langle \vec{r}_1, \vec{r}_2|\varphi_1\rangle|\varphi_2\rangle - \langle \vec{r}_1, \vec{r}_2|\varphi_2\rangle|\varphi_1\rangle = \varphi_1(\vec{r}_1)\varphi_2(\vec{r}_2) - \varphi_2(\vec{r}_1)\varphi_1(\vec{r}_2)
\]

Al actuar el operador permutación sobre el sistema se obtiene,

\[
    \hat{P}_{1,2}\psi(\vec{r}_1, \vec{r}_2) = \varphi_1(\vec{r}_2)\varphi_2(\vec{r}_1) - \varphi_2(\vec{r}_2)\varphi_1(\vec{r}_1) = -\psi(\vec{r}_1, \vec{r}_2)
\]

Esto implica que si $|\varphi_1\rangle = |\varphi_2\rangle$, es decir, que los dos fermiones ocupan un mismo estado, el estado del sistema antes de aplicar el operador permutación ($|\psi\rangle$) difiere en el signo del estado del sistema al aplicar dicho operador ($-|\psi\rangle$), lo cual es cierto solo en el caso que el estado sea nulo\footnote{Si un sistema permanece en un mismo estado después que un operador actúa sobre él, el vector de estado debe permanecer invariante, lo cual no sucede en este caso, donde $|\psi\rangle \neq -|\psi\rangle$. Note que la igualdad solo se obtiene si $|\psi\rangle=0$.}, es decir, la probabilidad $|\langle \psi |\psi \rangle|^2=0$ que dos fermiones estén en un mismo estado es cero. El primero en darse cuenta de esto fue Pauli y se le conoce como el \textit{principio de exclusión de Pauli}, el cual puede derivarse como una consecuencia del teorema espín-estadística \cite{Pauli1940}.\index{Principio de exclusión de Pauli}

\begin{lema}
    Dos fermiones no pueden ocupar un mismo estado.\index{Fermiones!ocupación de estados}
\end{lema}

\section{Gran función de partición para un sistema con $N$ partículas cuánticas}

\begin{figure}[tb]
    \centering
    \begin{subfigure}{0.49\textwidth}
        \includegraphics[width=\linewidth]{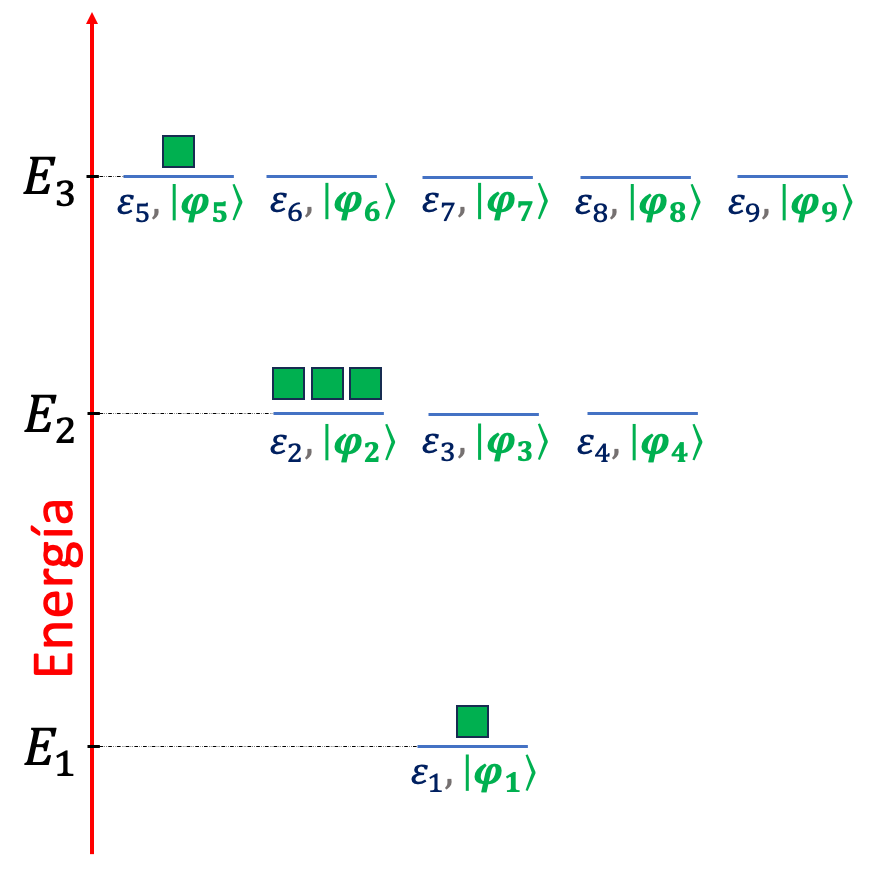}
        \caption{}
        \label{fig:ch5_cinco_bosones}
    \end{subfigure}
    \begin{subfigure}{0.49\textwidth}
        \includegraphics[width=\linewidth]{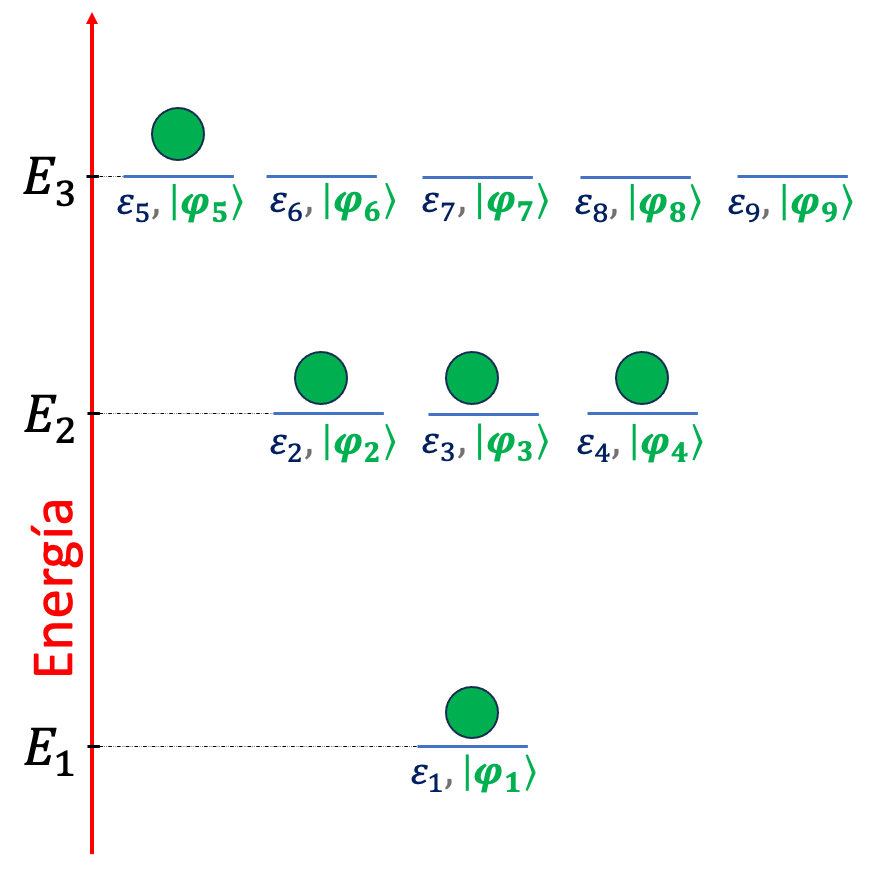}
        \caption{}
        \label{fig:ch5_cinco_fermiones}
    \end{subfigure}
    \caption{Distribución de partículas en los diferentes estados de un sistema formado por (a) cinco bosones y (b) cinco fermiones.  Un estado está determinado por un vector de estado $\ket{\varphi_i}$ al cual se le asocia una cierta energía $\varepsilon_i$. En la figura se observan tres niveles de energía $E_j$ tales que el primer nivel no está degenerado, mientras que los otros dos están degenerados.}
    \label{fig:ch5_cinco_fermiones_cinco_bosones}
\end{figure}

Sea un sistema formado por $N$ partículas, tales que $n_i$ ocupan un estado $\ket{\varphi_i}$ con energía $\varepsilon_i$, como se muestra de manera esquemática en la Figura \ref{fig:ch5_cinco_fermiones_cinco_bosones}. Al preparar un sistema en equilibrio termodinámico con un reservorio a temperatura $T$ y potencial químico $\mu$, se pueden obtener diferentes configuraciones \gls{configuration}, las cuales se identifican con el índice $r$ y, cada configuración, tiene asociada una energía $E_r$ y un número de partículas $N_s = N$, dados por,

\[
    E_r = \sum_{i=1}^I n_i^{(r)} \varepsilon_i
\]

\[
    N_s = \sum_{i=1}^I n_i^{(s)}
\]

La gran función de partición $\granZ$, ecuación \eqref{eq:ch5_función_partición}, se puede escribir, para este caso, como,

\[
    \granZ = \sum_s \sum_r e^{-\beta \left[n_{1}^{(r)}\varepsilon_{1} + n_{2}^{(r)}\varepsilon_{2} + \cdots \right] - \alpha \left[ n_1^{(s)} + n_2^{(s)} + \cdots \right]}
\]

En nuestro estudio de sistemas cuánticos nos interesa estudiar los gases ideales, por lo cual resulta de interés obtener la función de partición correspondiente. Para obtener la expresión de la función de partición para un gas ideal es necesario expresar la gran función de partición en dos términos, uno que solo incluya componentes no interactuante y el otro, los componentes de interacción,

\[
    \granZ = \sum_s e^{-n_{1}^{(s)} ( \beta \varepsilon_1 + \alpha ) - n_{2}^{(s)} ( \beta \varepsilon_2 + \alpha ) + \cdots} + \sum_{s \neq r} e^{-\beta \left[n_{1}^{(r)}\varepsilon_{1} + n_{2}^{(r)}\varepsilon_{2} + \cdots \right] - \alpha \left[ n_1^{(s)} + n_2^{(s)} + \cdots \right]}
\]

Se mostrará que la primera sumatoria representa el término no interactuante, para ello conviene definir $\granZ_i = \sum\limits_{s} e^{-n_{i}^{(s)} ( \beta \varepsilon_i + \alpha )}$, donde la suma en $s$ indica que se deben tomar todos los posibles valores de $n_{i}^{(s)}$, por lo tanto, esta sumatoria se puede re-escribir como,

\begin{equation}
    \label{eq:ch5_granZ_i}
    \granZ_i = \sum_{n_i} e^{-n_{i} ( \beta \varepsilon_i + \alpha )}
\end{equation}

Por otro lado, la segunda sumatoria representa el término interactuante, debido a que involucra varios estados con diferentes números de partículas. En tal sentido, conviene definir $\granZ_{r, s}$ como,

\begin{equation}
    \granZ_{r, s} = \prod_i e^{-\beta n_{i}^{(r)}\varepsilon_{i} - \alpha n_i^{(s)}  }
\end{equation}

Nótese que al escribir la gran función de partición en términos de $\granZ_i$ y $\granZ_{r,s}$, ecuación \eqref{eq:ch5_granZ_niveles_por_partes}, se puede identificar que las contribuciones $\granZ_i$ concuerdan con la descripción de una función de partición de sistemas mezclados no interactuantes, ecuación \eqref{eq:ch4_funcion_particion_sistemas_no_interactuantes}, mientras que la otras contribuciones se puede asociar a la interacción entre partículas.

\begin{equation}
    \label{eq:ch5_granZ_niveles_por_partes}
    \granZ = \prod_i \granZ_i + \sum_{r \neq s} \granZ_{r, s}
\end{equation}

\section{El gas ideal cuántico}
\index{Gas ideal cuántico}

En la anterior sección se obtuvo la expresión para la gran función de partición de un sistema formado por $N$ partículas que ocupan ciertos niveles de energía. Si el sistema en cuestión está formado por partículas cuánticas no interactuantes, estas conforman un gas ideal cuántico, cuya gran función de partición está dada por, \index{Gas ideal cuántico!función de partición}

\begin{equation}
    \label{eq:ch5_granZi_gas_ideal}
    \granZ = \prod_i \granZ_i
\end{equation}

En términos de $\granZ_i$ y utilizando las ecuaciones \eqref{eq:ch5_valor_esperado_número_partículas} y \eqref{eq:ch5_granZ_i}, se obtiene el número de ocupación $\E{n_i}$ para el estado $i$-ésimo,\index{Gas ideal cuántico!número de ocupación}

\[
    \E{n_i} = -\frac{\partial \ln \granZ_i}{\partial \alpha}
\]

\subsection{Gas ideal de bosones: distribución de Bose-Einstein}
\index{Gas ideal cuántico!de bosones}

Debido a que, para un sistema con $N$ bosones, en un mismo estado puede haber desde cero hasta $N$ bosones, la gran función de partición para el estado $i$-ésimo está dada por,

\[
    \granZ_i = \sum_{n_i=0}^N e^{-n_{i} ( \beta \varepsilon_i + \alpha )}
\]

Esta sumatoria corresponde con la suma de términos de una serie geométrica, la cual está dada por,

\[
    S_n = \sum_{n=0}^N a^n = \frac{a^{N+1}-1}{a-1}
\]
Haciendo $a=e^{-\beta\varepsilon_i+\alpha}$ se obtiene,

\[
    \granZ_i = \frac{e^{-(\beta\varepsilon_i+\alpha)(N+1)}-1}{e^{-\beta\varepsilon_i-\alpha}-1}
\]
Si se tienen muchos bosones que conforman el gas ideal cuántico, $e^{-(\beta\varepsilon_i+\alpha)(N+1)} \ll 1$, por lo tanto,

\[
    \granZ_i \approx \frac{1}{1-e^{-\beta\varepsilon_i-\alpha}}
\]
Con esta gran función de partición y la ecuación \eqref{eq:ch5_valor_esperado_número_partículas} se obtiene el número de ocupación para el estado $i$-ésimo de un gas ideal, 

\[
    \E{n_i} = \frac{1}{e^{\beta\varepsilon_i+\alpha}-1}
\]

Si un cierto estado tiene degenerancia $g_i$, como se muestra esquemáticamente en la Figura \ref{fig:ch5_cinco_bosones} para un sistema formado por cinco bosones, se encuentra que debido a la ecuación \eqref{eq:ch5_granZi_gas_ideal}, al obtener el $\ln \granZ_i$ y derivar respecto al parámetro $\alpha$, se obtiene que el número de ocupación está dado por la ecuación \eqref{eq:ch5_dist_bose_einstein}, conocida como función de distribución de Bose-Einstein.

\begin{equation}
    \label{eq:ch5_dist_bose_einstein}
    \E{n_i} = \frac{g_i}{e^{\beta\varepsilon_i+\alpha}-1} = \frac{g_i}{e^{\frac{\varepsilon_i - \mu}{k_B T}}-1}
\end{equation}

\subsection{Gas ideal de fermiones: distribución de Fermi-Dirac}
\index{Gas ideal cuántico!de fermiones}

En el caso de un sistema formados por fermiones, cada estado puede estar poblado máximo por un fermión, como se muestra esquematicamente en la Figura \ref{fig:ch5_cinco_fermiones}. En este caso, 

\[
    \granZ_i = \sum_{n_i=0}^1 e^{-n_{i} ( \beta \varepsilon_i + \alpha )} = 1 + e^{- \beta \varepsilon_i - \alpha}
\]

Con esta gran función de partición y la ecuación \eqref{eq:ch5_valor_esperado_número_partículas} se obtiene el número de ocupación para el estado $i$-ésimo de un gas ideal, 

\[
    \E{n_i} = \frac{1}{e^{\beta\varepsilon_i+\alpha}+1}
\]

Si un cierto estado tiene degenerancia $g_i$, como se muestra esquemáticamente en la Figura \ref{fig:ch5_cinco_fermiones} para un sistema formado por cinco fermiones, se encuentra que debido a la ecuación \eqref{eq:ch5_granZi_gas_ideal}, al obtener el $\ln \granZ_i$ y derivar respecto al parámetro $\alpha$, se obtiene que el número de ocupación está dado por la ecuación \eqref{eq:ch5_dist_bose_einstein}, conocida como función de distribución de Fermi-Dirac.

\begin{equation}
    \label{eq:ch5_dist_fermi_dirac}
    \E{n_i} = \frac{g_i}{e^{\beta\varepsilon_i+\alpha}+1} = \frac{g_i}{e^{\frac{\varepsilon_i - \mu}{k_B T}}+1}
\end{equation}

\subsection{El límite clásico: distribución de Maxwell-Boltzmann}

En el límite clásico $e^{\beta\varepsilon_i+\alpha} \gg 1$, por lo tanto,

\[
    \E{n_i} = g_i e^{-(\varepsilon_i - \mu)/k_B T}
\]

Como $\sum\limits_i g_i e^{-(\varepsilon_i - \mu)/k_B T} = N$,

\[
    e^{\mu} = \frac{N}{\sum\limits_i g_i e^{-\varepsilon_i/k_B T}}
\]

Con lo cual se obtiene la distribución de Maxwell-Boltzmann,

\begin{equation}
    \label{eq:ch5_dist_maxwell_boltzmann}
    \E{n_i} = \frac{N e^{-\varepsilon_i/k_B T}}{\sum\limits_i g_i e^{-\varepsilon_i/k_B T}}
\end{equation}














\chapter{Capítulo V. Estadística Cuántica}
\section{El operador densidad}
\index{El operador Densidad}

La descripción de la naturaleza mediante la teoría cuántica es, de manera intrínseca, probabilística; por ello, se desarrollan métodos estadísticos con el objetivo de obtener información sobre los observables físicos. Esta característica distintiva encuentra sus fundamentos en la interpretación probabilística de la función de onda, en conjunto con el Principio de Incertidumbre. Tal como se discute en los textos estándar, las probabilidades cuánticas encarnan la dualidad de la materia, reproduciendo fenómenos como la interferencia y la difracción. Esta representación estadística, incluso para sistemas de una sola partícula, requiere de un conjunto de sistemas preparados de manera similar, con el fin de generar un conjunto de probabilidades bien definidas para todas las cantidades. Dicho conjunto se conceptualiza como un número 'infinito' de réplicas del mismo sistema que no coexisten en espacio ni tiempo.

Cuando todos los sistemas del conjunto están descritos por la misma función de onda o vector de estado $|\psi\rangle$, se dice que se ha preparado un estado puro. En Mecánica Estadística, $|\psi\rangle$ representa un posible microestado del sistema. Las predicciones físicas de una cantidad observable $\mathbf{A}$ se expresan en términos de promedios o valores medios de la forma:
\begin{equation}
\langle\mathbf{A}\rangle = \langle\psi|\mathbf{A}|\psi\rangle,
\end{equation}
donde el corchete se calcula siguiendo las reglas cuánticas bien establecidas. Por otro lado, las fluctuaciones estadísticas de origen cuántico se obtienen mediante:
\begin{equation}
\langle\mathbf{A}^2\rangle - \langle\mathbf{A}\rangle^2 = \langle\psi|(\mathbf{A} - \langle\mathbf{A}\rangle)^2|\psi\rangle,
\end{equation}
asumiendo que el estado $\psi$ está normalizado, es decir, $\langle\psi|\psi\rangle = 1$. La evolución temporal de este microestado está dictada por la ecuación de Schrödinger, mientras no se realice ninguna medición sobre el sistema. No se profundizará aquí en una discusión exhaustiva sobre los conceptos fundamentales de la Mecánica Cuántica, suponiendo que el lector posee los conocimientos previos necesarios $[1,2]$. Según la Mecánica Cuántica, un estado puro representa la máxima información que puede obtenerse de un sistema cuántico, constituyendo una excepción más que la regla. En la mayoría de las situaciones, la función de onda no se conoce con certeza. Un ejemplo típico lo constituye un sistema de coordenadas $\mathbf{x}$ acoplado a otro sistema de coordenadas $\mathbf{y}$ (este último podría ser un baño térmico que mantiene el sistema a una temperatura constante). Generalmente, no es posible asignar una función de onda a nuestro sistema $\mathbf{x}$, que forma parte del sistema completo $(\mathbf{x}, \mathbf{y})$. En este caso, se afirma que el sistema $\mathbf{x}$ se encuentra en un estado mixto y se debe buscar un objeto matemático más general para describir esta situación. Esta falta de información se asemeja estrechamente al problema estadístico clásico. De hecho, en muchas instancias, este caso se denomina mezcla incoherente, indicando la falta de coherencia cuántica y la ausencia de efectos de interferencia. Como se discutirá más adelante, las fluctuaciones cuánticas y térmicas compiten en este escenario en función de la temperatura. La coherencia cuántica se manifiesta a bajas temperaturas, y, de manera asintótica, a temperatura cero el sistema es dominado exclusivamente por fluctuaciones cuánticas. A medida que la temperatura aumenta, se desarrollan fluctuaciones térmicas que, eventualmente, predominan en la estadística, conduciendo a la decoherencia de los efectos cuánticos (estadísticas clásicas). La descripción del conjunto general, que abarca todos los casos, se logra de manera bastante natural empleando el denominado operador de estado $\boldsymbol{\rho}$ (o operador de densidad), que generaliza el concepto de función de onda. El operador $\rho$ es la cantidad relevante para construir la mecánica estadística cuántica. El término densidad rememora la función clásica $\rho(q, p)$, que indica la densidad de puntos en el espacio de fases para realizar estadísticas clásicas. Un punto singular en el espacio de fases representa uno de los sistemas del conjunto. Un objetivo importante consiste en relacionar el operador de densidad cuántica $\rho$ con la función clásica $\rho(q, p)$. En la sección siguiente, se desarrolla la base del formalismo matemático. Se inicia definiendo los distintos conjuntos que se encuentran al describir un sistema físico.

\begin{definition}
\textbf{Conjunto Puro.} Se define como una colección de sistemas físicos idénticos, tal que todos los miembros del conjunto están caracterizados por el mismo vector de estado o ket \(|\psi\rangle\).
\end{definition}
 Este caso es el habitualmente encontrado en los libros de texto estándar de mecánica cuántica, cuando el estado de un sistema está representado por una única función de onda. Dicho conjunto representa un microestado.

El cálculo de promedios y desviaciones estándar para el conjunto se realiza mediante las ecuaciones (1) y (2). Este estado puro \(|\psi\rangle\) puede ser un estado propio de un observable físico particular, o una superposición lineal de estados propios de un operador arbitrario.

\begin{definition}
\textbf{Conjunto Mixto.} En este caso, la función de onda no se conoce con certeza y existen varias posibilidades que se representan como \(\left\{\left|\psi^{(1)}\right\rangle, \left|\psi^{(2)}\right\rangle, \ldots, \left|\psi^{(i)}\right\rangle, \ldots\right\}\). Esta colección puede ser finita o infinita. A cada ket \(\left|\psi^{(i)}\right\rangle\) se le asigna una fracción de los miembros del conjunto, con población relativa \(w_i\), cumpliendo que
\begin{equation}
\sum_i w_i = 1 .
\end{equation}
Los pesos \(\left\{w_i\right\}\) son números reales positivos (o cero), indicando que este conjunto representa un macroestado del sistema.
\end{definition}

Los vectores de estado \(\left\{\left|\psi^{(1)}\right\rangle, \left|\psi^{(2)}\right\rangle, \ldots, \left|\psi^{(i)}\right\rangle, \ldots\right\}\) están normalizados, pero no necesariamente son ortogonales, y el número de tales estados puede ser mayor que la dimensión del espacio lineal. Los números \(\left\{w_i\right\}\) no deben interpretarse como probabilidades ordinarias, debido a que los estados cuánticos \(\left\{\left|\psi^{(1)}\right\rangle, \left|\psi^{(2)}\right\rangle, \ldots, \left|\psi^{(i)}\right\rangle, \ldots\right\}\) no son 'mutuamente excluyentes'. Esto se entiende en el sentido de que la superposición entre dos estados de la colección no se anula en general, es decir,
\begin{equation}
\left\langle\psi^{(i)} | \psi^{(j)}\right\rangle \neq 0,
\end{equation}
para \(i \neq j\) en el caso general. Se prescribe entonces el cálculo de promedios para el conjunto mixto como sigue:

 Sea \(A\) un observable físico. El promedio para el conjunto, denotado por \([\ldots]\), se distingue de \((\ldots)\) usado para el caso cuántico puro.

\begin{definition}
\textbf{Promedio o Valor Medio para el Conjunto Mixto.}
\begin{equation}
[\mathbf{A}] \equiv \sum_i w_i\left\langle\psi^{(i)}|\mathbf{A}| \psi^{(i)}\right\rangle.
\end{equation}
\end{definition}

En la definición anterior, el promedio cuántico ordinario \(\left\langle\psi^{(i)}|\mathbf{A}| \psi^{(i)}\right\rangle\) para el estado \(\psi^{(i)}\) se pondera por su población relativa \(w_i\), conferiendo al promedio \([\mathbf{A}]\) una naturaleza mixta, tanto cuántica como estadística. Se reescribe la definición usando una base general de estados \(\{|n\rangle\}\), que es ortonormal y completa:

\begin{equation}
\begin{aligned}
{[\mathbf{A}] } & =\sum_i w_i \sum_{n, n^{\prime}}\left\langle\psi^{(i)} \mid n\right\rangle\left\langle n|\mathbf{A}| n^{\prime}\right\rangle\left\langle n^{\prime} \mid \psi^{(i)}\right\rangle \\
& =\sum_{n, n^{\prime}}\left(\sum_i w_i\left\langle n^{\prime} \mid \psi^{(i)}\right\rangle\left\langle\psi^{(i)} \mid n\right\rangle\right)\left\langle n|\mathbf{A}| n^{\prime}\right\rangle .
\end{aligned}
\end{equation}

lo que sugiere la siguiente definición:

\begin{definition}
\textbf{Operador de Estado o Operador de Densidad,} \(\rho\)
\begin{equation}
\rho \equiv \sum_i w_i |\psi^{(i)}\rangle\langle\psi^{(i)}|,
\end{equation}
\end{definition}

Los elementos de matriz de \(\rho\) se dan por

\begin{equation}
\langle n'|\rho| n\rangle = \sum_i w_i \langle n' | \psi^{(i)}\rangle\langle\psi^{(i)} | n\rangle,
\end{equation}

y el valor medio se puede escribir como una traza:

\begin{equation}
[\mathbf{A}]=\sum_{n, n^{\prime}}\left\langle n^{\prime}|\boldsymbol{\rho}| n\right\rangle\left\langle n|\mathbf{A}| n^{\prime}\right\rangle=\operatorname{Tr}(\boldsymbol{\rho} \mathbf{A})
\end{equation}

De la definición, se obtienen algunas propiedades inmediatas del operador \(\rho\), como su hermiticidad, la normalización, y la positividad de sus valores propios, lo que conlleva a que \(\operatorname{Tr}(\rho^2) \leq 1\), y en el caso de un conjunto puro, \(\operatorname{Tr}(\rho^2) = 1\).

\section{Ejemplo: partícula con espín 1/2}
Se trabaja un ejemplo ilustrativo para el espín $1/2$. La dimensión del espacio lineal es 2, y el operador de densidad se representa por una matriz de $(2 \times 2)$. Dada la hermiticidad y normalización, quedan tres parámetros reales independientes para determinar $\rho$. Estos parámetros se identifican con los tres valores medios del promedio del operador de espín, $\left[S_x\right]$, $\left[S_y\right]$, y $\left[S_z\right]$. Esta característica es particular del espín $1/2$. Por conveniencia, se introduce el operador de espín de Pauli $\vec{\sigma}$ como
\begin{equation}
\overrightarrow{\mathrm{S}}=\frac{\hbar}{2} \vec{\sigma},
\end{equation}
con la representación estándar
\begin{equation}
\sigma_x=\left(\begin{array}{cc}
0 & 1 \\
1 & 0
\end{array}\right), \quad \sigma_y=\left(\begin{array}{cc}
0 & -i \\
i & 0
\end{array}\right), \quad \sigma_z=\left(\begin{array}{cc}
1 & 0 \\
0 & -1
\end{array}\right).
\end{equation}
Se observa que las tres matrices de Pauli, junto con la identidad, forman una base del espacio lineal de las matrices complejas de $(2 \times 2)$. Así, en general, se tiene
\begin{equation}
\boldsymbol{\rho}=\frac{1}{2} m_0 \mathbf{1}+\frac{1}{2} \overrightarrow{\mathbf{m}} \cdot \overrightarrow{\boldsymbol{\sigma}},
\end{equation}
con el vector $\overrightarrow{\mathbf{m}}=(m_x, m_y, m_z)$ denominado polarización. La hermiticidad de $\rho$ y las matrices de Pauli implica que todos los coeficientes son reales. Además, dado que las matrices de Pauli son sin traza, la condición de normalización se expresa como
\begin{equation}
\operatorname{Tr}(\boldsymbol{\rho})=1=\frac{1}{2} m_0 \operatorname{Tr}(\mathbf{1}),
\end{equation}
dejando tres parámetros reales independientes dados por $\overrightarrow{\mathbf{m}}=(m_x, m_y, m_z)$, con el operador de densidad escrito de forma general como
\begin{equation}
\rho=\frac{1}{2}(\mathbf{1}+\overrightarrow{\mathbf{m}} \cdot \vec{\sigma}).
\end{equation}

Utilizando el álgebra asociada con las matrices de Pauli, se encuentra que
\begin{equation}
\rho^2=\frac{1}{4}\left[\mathbf{1}+2 \overrightarrow{\mathbf{m}} \cdot \vec{\sigma}+(\overrightarrow{\mathbf{m}} \cdot \vec{\sigma})^2\right]=\frac{1}{2}\left[\left(\frac{1+m^2}{2}\right) \mathbf{1}+\overrightarrow{\mathbf{m}} \cdot \vec{\sigma}\right],
\end{equation}
con $m^2=|\overrightarrow{\mathbf{m}}|^2$. Al tomar la traza, se obtiene
\begin{equation}
\operatorname{Tr}\left(\rho^2\right)=\frac{1+m^2}{2} \leqq 1,
\end{equation}
lo que implica que $m^2 \leqq 1$. Se tiene un conjunto puro si y solo si $m^2=1$. Para el conjunto mixto, $0 \leqq m^2<1$. El caso $m=0$ corresponde a un conjunto no polarizado o aleatorio. Utilizando las propiedades anticonmutativas de las matrices de Pauli:
\begin{equation}
\sigma_x \sigma_y=-\sigma_y \sigma_x=i \sigma_z, \quad \sigma_y \sigma_z=-\sigma_z \sigma_y=i \sigma_x, \quad \sigma_z \sigma_x=-\sigma_x \sigma_z=i \sigma_y,
\end{equation}
se sigue que
\begin{equation}
\left[S_x\right]=\operatorname{Tr}\left(\boldsymbol{\rho} S_x\right)=\frac{\hbar}{2} m_x, \quad \left[S_y\right]=\operatorname{Tr}\left(\boldsymbol{\rho} S_y\right)=\frac{\hbar}{2} m_y, \quad \left[S_z\right]=\operatorname{Tr}\left(\boldsymbol{\rho} S_z\right)=\frac{\hbar}{2} m_z.
\end{equation}

Para el conjunto aleatorio, $\left[S_x\right]=\left[S_y\right]=\left[S_z\right]=0$, es decir, $m=0$ y el operador de densidad se representa como
\begin{equation}
\boldsymbol{\rho}_0=\frac{1}{2} \mathbf{1}.
\end{equation}
Suponiendo que se usa la base de estados que diagonalizan $S_z$, denominados $|\hat{\mathbf{z}} ;+\rangle$ y $|\hat{\mathbf{z}} ;-\rangle$, para los valores propios $\frac{\hbar}{2}$ y $-\frac{\hbar}{2}$ respectivamente, el operador de densidad para el caso aleatorio se puede representar de la forma
\begin{equation}
\begin{aligned}
\boldsymbol{\rho}_0 & =\frac{1}{2}|\hat{\mathbf{z}} ;+><\hat{\mathbf{z}} ;+|+\frac{1}{2}|\hat{\mathbf{z}} ;-><\hat{\mathbf{z}} ;-|= \\
& =w_{+}|\hat{\mathbf{z}} ;+><\hat{\mathbf{z}} ;+|+w_{-}|\hat{\mathbf{z}} ;-><\hat{\mathbf{z}} ;-|,
\end{aligned}
\end{equation}
Con $w_{+}=w_{-}=\frac{1}{2}$, es decir, $\rho_0$ puede considerarse como una mezcla de los estados $\mid \hat{\mathbf{z}}_{;}\rangle+$ y $\mid \hat{\mathbf{z}}_{;}\rangle-$ con pesos iguales. Un problema importante para el conjunto mixto, es que la descomposición en términos de conjuntos puros no es única. En el ejemplo anterior, podemos considerar una base diferente de estados, digamos los estados $\mid \hat{\mathbf{x}}_{;}\rangle+$ y $\mid \hat{\mathbf{x}}_{;}\rangle-$ que diagonalizan la componente $\mathbf{S}_{\mathbf{z}}$ del operador de espín. Tenemos la transformación unitaria:
\begin{equation}
\begin{aligned}
& \left|\hat{\mathbf{z}} ;\rangle+\right.=\frac{1}{\sqrt{2}}\left| \hat{\mathbf{x}} ; \rangle+ \right.+\frac{1}{\sqrt{2}} \left| \hat{\mathbf{x}}_{;}\rangle- \right., \\
& \left|\hat{\mathbf{z}}_{;}\rangle-\right.=\frac{1}{\sqrt{2}}\left| \hat{\mathbf{x}} ; \rangle+ \right.-\frac{1}{\sqrt{2}} \left| \hat{\mathbf{x}}_{;}\rangle- \right.,
\end{aligned}
\end{equation}
que lleva a
\begin{equation}
\rho_0=\frac{1}{2}|\hat{\mathbf{x}} ;\rangle+\langle\hat{\mathbf{x}} ;+|+\frac{1}{2}|\hat{\mathbf{x}} ;\rangle-\langle\hat{\mathbf{x}} ;-|,
\end{equation}
lo que significa que el conjunto aleatorio puede considerarse, al mismo tiempo, como una mezcla de estados $|\hat{\mathbf{x}} ;\rangle+$ y $\mid \hat{\mathbf{x}} ;\rangle-$ con pesos iguales. De hecho, hay un número infinito de posibilidades, diciendo que el conjunto mixto aleatorio puede descomponerse igualmente en términos de negro y blanco, o rojo y verde, o azul y amarillo, etc., al mismo tiempo. Este hecho es una manifestación de la naturaleza cuántica del estado, a pesar de la mezcla máxima. El conjunto puro puede parametrizarse como: $\vec{\mathbf{m}}=(\sin \theta \cos \varphi, \sin \theta \sin \varphi, \cos \theta)$,
con $|\tilde{\mathbf{m}}|^2=1$, con los ángulos $(\theta, \varphi)$ dando la dirección de la polarización. En forma matricial, el operador densidad se lee:
\begin{equation}
\rho=\left(\begin{array}{cc}
\frac{1+\cos \theta}{2} & \frac{e^{-i \varphi} \sin \theta}{2} \\
\frac{e^{i \varphi} \sin \theta}{2} & \frac{1-\cos \theta}{2}
\end{array}\right) .
\end{equation}

Para muchos otros ejemplos, ver la sección de ejercicios.

\section{Sistemas acoplados y no separabilidad}

En el caso separable, discutiremos primero el caso de sistemas no interactivos, es decir, sistemas que son separables.

Un sistema diluido, como por ejemplo un gas ideal, puede aproximarse bien como un sistema de muchas partículas no interactivas (moléculas). Dentro de este límite ideal, las partículas no están correlacionadas, y la Mecánica Estadística puede reducirse a una descripción de una sola partícula. De manera general, consideremos dos sistemas que abarcan dos diferentes espacios de Hilbert de estados, $\mathcal{R}$ y $\mathcal{S}$, con bases $\{\mid N\rangle\}$ y $\{\mid n\rangle\}$, respectivamente. El espacio combinado de los dos sistemas, $\mathcal{R} \times \mathcal{S}$, está abarcado por los estados del producto directo, que en la notación de Dirac, se escriben como:
\begin{equation}
|N n\rangle\equiv| N\rangle\mid n\rangle.
\end{equation}
los productos directos de operadores se representan por productos directos de las matrices correspondientes. Ejemplo. Producto directo de dos matrices $A$ y $B$, donde:
\begin{equation}
A=\left(\begin{array}{cc}
a_{11} & a_{12} \\
a_{21} & a_{22}
\end{array}\right), \quad B=\left(\begin{array}{ccc}
b_{11} & b_{12} & b_{13} \\
b_{21} & b_{22} & b_{23} \\
b_{31} & b_{32} & b_{33}
\end{array}\right) .
\end{equation}

Se define $A \times B$ como la matriz:
\begin{equation}
A \times B \equiv\left(\begin{array}{cc}
a_{11} B & a_{12} B \\
a_{21} B & a_{22} B
\end{array}\right)
\end{equation}

El producto directo resulta en una matriz de $(6 \times 6)$, ya que los términos $a_{ij} B$ se entienden como submatrices de $(3 \times 3)$.
Las siguientes propiedades se demuestran fácilmente:
\begin{enumerate}
    \item Los elementos de la matriz se entienden como:
    \begin{equation}
    \langle N n|\mathbf{A} \times \mathbf{B}| N^{\prime} n^{\prime}\rangle=\langle N|\mathbf{A}| N^{\prime}\rangle\langle n|\mathbf{B}| n^{\prime}\rangle .
    \end{equation}
    
    En particular, los productos escalares se obtienen como:
    \begin{equation}
    \langle N n | N^{\prime} n^{\prime}\rangle=\langle N|N^{\prime}\rangle\langle n| n^{\prime}\rangle=\delta_{NN^{\prime}} \delta_{nn^{\prime}} .
    \end{equation}
    
    \item A partir del punto anterior, obtenemos:
    \begin{equation}
    \operatorname{Tr}(\mathbf{A} \times \mathbf{B})=\operatorname{Tr}(\mathbf{A}) \operatorname{Tr}(\mathbf{B}) .
    \end{equation}
    
    \item Los productos directos de vectores columna (fila) se obtienen utilizando la misma regla definida para matrices. Tomemos por ejemplo el caso a continuación:
    Un estado general del sistema se escribe como
    \begin{equation}
    |\psi\rangle=\sum_{N, n} \mathfrak{C}(N, n)| N\rangle| n\rangle,
    \end{equation}
    pero dado que los sistemas son no correlacionados, se debe tener $\mathfrak{C}(N, n)=C(N) c(n)$ y el estado es separable:
    \begin{equation}
    |\psi\rangle=\sum_{N, n} \mathfrak{C}(N, n)| N\rangle|n\rangle=\left(\sum_N C(N) | N\rangle\right)\left(\sum_n c(n) | n\rangle\right)=| \psi_R\rangle| \psi_S\rangle.
    \end{equation}
\end{enumerate}

Lo mismo es cierto para la matriz de densidad, que en general se escribe como
\begin{equation}
\rho=\sum_{N, M, n, m}|N n\rangle\langle N n| \rho|M m\rangle\langle M m|,
\end{equation}
pero para sistemas no correlacionados, se debería tener
$\langle N n|\rho| M m\rangle=A_{NM} B_{nm}$,
factorizando el operador de densidad como:
\begin{equation}
\boldsymbol{\rho}=\left(\sum_N A_{NM}|N\rangle\langle M|\right)\left(\sum_n B_{nm}|n\rangle\langle m|\right)=\boldsymbol{\rho}_{\mathcal{R}} \times \boldsymbol{\rho}_S,
\end{equation}
en la forma de un producto directo. La separabilidad para el operador de densidad tiene un sentido más amplio en Mecánica Cuántica, pero no profundizaremos esta discusión aquí.

Una situación como la descrita en (5), se llama separabilidad simple. En el caso de un gas ideal, en ausencia de interacciones, todas las partículas son no correlacionadas. Si usamos las coordenadas de las partículas como etiquetas, la separabilidad conduce a:
\begin{equation}
\rho(\mathbf{x}_1, \mathbf{x}_2, \ldots, \mathbf{x}_N)=\rho_1(\mathbf{x}_1) \times \rho_2(\mathbf{x}_2) \times \ldots \times \rho_N(\mathbf{x}_N),
\end{equation}
donde $N$ es el número total de partículas y $\rho_i(\mathbf{x}_i)$ es el operador de densidad para estados de una partícula. Si las partículas son idénticas,
\begin{equation}
\rho_j=\rho_1,
\end{equation}
para todo $j=2,3, \ldots, N$, y la factorización se escribe como
\begin{equation}
\rho(\mathbf{x}_1, \mathbf{x}_2, \ldots, \mathbf{x}_N)=\bigotimes_{i=1}^N \rho_1(\mathbf{x}_i) .
\end{equation}

En particular, si $\rho_1$ está normalizado, obtenemos:
\begin{equation}
\operatorname{Tr}\left[\boldsymbol{\rho}(\mathbf{x}_1, \mathbf{x}_2, \ldots, \mathbf{x}_N)\right]=\prod_{i=1}^N \operatorname{Tr}\left[\boldsymbol{\rho}_1(\mathbf{x}_i)\right]=1 .
\end{equation}

En este caso, el cálculo del operador de densidad se reduce al cálculo del operador de una partícula $\rho_1(\mathbf{x})$.

En el caso de interacción, aparecen correlaciones entre las partículas, y los estados ya no son separables. Supongamos que en el tiempo inicial $(t_0=0)$, los dos subsistemas no están interactuando, y preparamos el estado inicial como separable:
\begin{equation}
|\psi, 0\rangle=|N\rangle| n\rangle .
\end{equation}

Después de eso, la interacción se activa durante un intervalo de tiempo finito, y finalmente se desactiva nuevamente. La interacción hizo que los sistemas estuvieran correlacionados. El estado asintótico $(t \rightarrow \infty)$ es del tipo:
\begin{equation}
|\psi, \infty\rangle=\sum_{M, m} a_{\infty}(M, m ; N, n)| M\rangle| m\rangle,
\end{equation}

donde el coeficiente $a_{\infty}(M, m ; N, n)$ es la amplitud de probabilidad para la transición $|N\rangle| n\rangle\rightarrow|M\rangle| m\rangle$. Dado que los sistemas están correlacionados, tenemos en general que $a_{\infty}(M, m ; N, n) \neq C(M, N) c(m, n)$,
para todos los pares $(M, m)$ y $(N, n)$, y el estado (16) no es separable. En este caso, no podemos asignar una función de onda a un subsistema (ya sea $R$ o $S$). Este resultado se llama principio de no separabilidad.


\subsection{Matriz de densidad de un subsistema}

Hemos visto en la subsección anterior que el lenguaje de estados ket no nos permite, en general, describir un sistema aislado del resto del universo. Pero esto es posible cuando se describe el estado del sistema a través del operador de densidad. Consideremos dos sistemas cuánticos interactuantes, cuyos estados abarcan los espacios $R$ y $S$. Queremos prestar atención al subsistema $S$, dejando $R$ sin detectar ($R$ puede ser un reservorio, y queremos eliminar sus grados de libertad). Los estados del sistema total abarcan $R \times S$, pero en general, los estados físicos no son separables en presencia de interacciones. Asumamos bases $\{\mid N\rangle\}$ y $\{\mid n\rangle\}$ para $R$ y $S$, respectivamente, como en la subsección anterior. Queremos calcular un promedio de un observable que se refiere solo a $S$, que se escribe en la forma
\begin{equation}
\boldsymbol{\Omega}=\mathbf{I}_R \times \boldsymbol{\Omega}_S,
\end{equation}
donde $I_R$ es la identidad en $R$
\begin{equation}
[\boldsymbol{\Omega}]=\operatorname{Tr}\left[\boldsymbol{\rho}\left(\mathbf{I}_R \times \boldsymbol{\Omega}_S\right)\right]=\sum_{N, M, n, m}\langle N n|\rho| M m\rangle\langle M m|\left(\mathbf{I}_R \times \boldsymbol{\Omega}_S\right)| N n\rangle.
\end{equation}

Nótese que $\langle M m|\left(\mathbf{I}_R \times \boldsymbol{\Omega}_S\right)| N n\rangle=\delta_{MN}\langle m|\boldsymbol{\Omega}_S| n\rangle$, y sustituyendo en (18) resulta:
\begin{equation}
[\boldsymbol{\Omega}]=\sum_{n, m}\langle m|\boldsymbol{\Omega}_S| n\rangle\left(\sum_N\langle N n|\boldsymbol{\rho}| N m\rangle\right).
\end{equation}

La cantidad $\sum_N\langle N n|\boldsymbol{\rho}| N m\rangle$ es la traza parcial de $\rho$ en el espacio $R$. Entonces hacemos la siguiente definición:

\begin{definition}
\textbf{ Operador de densidad reducido, $\overline{\boldsymbol{\rho}}_S$, relativo a $S$}.
Sus elementos de matriz en $S$ se dan por:
\begin{equation}
\langle n|\overline{\rho}_S| m\rangle \equiv \sum_N\langle N n|\boldsymbol{\rho}| N m\rangle.
\end{equation}
\end{definition}

Reescribimos la definición anterior (19) de manera formal como:
\begin{equation}
\overline{\rho}_S=\operatorname{Tr}_R(\rho),
\end{equation}
significando que $\overline{\rho}_S$ se obtiene de $\rho$ tomando la traza parcial en $R$. El valor medio (18) ahora se refiere solo al espacio $S$:
\begin{equation}
[\boldsymbol{\Omega}]=\operatorname{Tr}\left[\boldsymbol{\rho}\left(\mathbf{I}_R \times \boldsymbol{\Omega}_S\right)\right]=\operatorname{Tr}_S\left[\overline{\boldsymbol{\rho}}_S \boldsymbol{\Omega}_S\right].
\end{equation}

Al tomar la traza parcial, perdemos la información detallada relativa al subsistema $R$. Queda por demostrar que $\overline{\rho}_S$ es un operador de densidad válido. Esto se logra, si asumimos que $\boldsymbol{\rho}$ es un operador de densidad para todo el universo $R \times S$:
i) hermiticidad,
\begin{equation}
\langle n|\overline{\boldsymbol{\rho}}_S| m\rangle \equiv \sum_N\langle N n|\boldsymbol{\rho}| N m\rangle=\sum_N\langle N m|\boldsymbol{\rho}| N n\rangle^*=\langle m|\overline{\boldsymbol{\rho}}_S| n\rangle^*,
\end{equation}
ya que $\rho$ es Hermitiano;
ii) normalización,
\begin{equation}
\operatorname{Tr}_S(\overline{\rho}_S)=\sum_n\langle n|\overline{\rho}_S| n\rangle \equiv \sum_n \sum_N\langle N n|\boldsymbol{\rho}| N n\rangle=\operatorname{Tr}_S\left[\operatorname{Tr}_R(\boldsymbol{\rho})\right]=\operatorname{Tr}(\boldsymbol{\rho})=1;
\end{equation}
iii) positividad,
\begin{equation}
\langle n|\overline{\rho}_S| n\rangle \equiv \sum_N\langle N n|\rho| N n\rangle \geqslant 0,
\end{equation}
ya que es una suma de términos positivos.
Para medir propiedades del subsistema $S$ no necesitamos el operador de densidad completo $\rho$, sino solo el operador reducido $\overline{\rho}_S$ relativo a $S$. La información detallada del otro subsistema se pierde, pero algunas 'propiedades promedio' de $R$ todavía están contenidas en $\overline{\rho}_S$.

\textbf{Ejemplo 2}
Consideremos dos partículas interactuantes de espín $1/2$, que están acopladas en un estado singlete del espín total:
\begin{equation}
|\psi_0\rangle=\frac{1}{\sqrt{2}}|+\rangle_R|-\rangle_S-\frac{1}{\sqrt{2}}|-\rangle_R|+\rangle_S,
\end{equation}
donde hemos usado las etiquetas $R$ y $S$ para las partículas. El estado anterior se dice que está enredado, y claramente no hay un estado ket para representar a ninguno de los subsistemas $R$ o $S$. El estado (25) es un estado puro, con operador de densidad dado por:
\begin{equation}
\rho= |\psi_0\rangle\langle\psi_0|=\frac{1}{2}\left\{| + \rangle _ { R } | - \rangle _ { S } S \langle - | _ { R } \langle + | + | - \rangle _ { R } | + \rangle _ { S } S \langle + | _ { R } \langle - | - | + \rangle _ { R } | - \rangle _ { S } S \langle + | _ { R } \langle - |\right\}.
\end{equation}

Ahora tomamos la traza parcial relativa a $R$:
\begin{equation}
\overline{\rho}_S = \operatorname{Tr}_R(\rho)=_R\langle+|\rho|+\rangle_R+ _R\langle-|\rho|-\rangle_R= \frac{1}{2}|-\rangle_S S\langle-|+\frac{1}{2}|+\rangle_S S\langle+| \equiv \frac{1}{2}\left(\begin{array}{cc}
1 & 0 \\
0 & 1
\end{array}\right)_S.
\end{equation}

Nótese que $\overline{\rho}_S$ representa un conjunto mixto (mezcla máxima), mientras que el $\rho$ original era un estado puro: al eliminar los grados de libertad de $R$, se pierde información de manera que no se puede recuperar más tarde. Solo para enfatizar el carácter irreversible del proceso, notamos que trazar parcialmente la matriz de densidad del conjunto puro del estado tripleto
\begin{equation}
|\psi_1\rangle=\frac{1}{\sqrt{2}}|+\rangle_R|-\rangle_S+\frac{1}{\sqrt{2}}|-\rangle_R|+\rangle_S
\end{equation}
conduce a la misma matriz de densidad reducida (30).

\subsection{Representando el Operador de Densidad: Matriz de Densidad}
Discutimos en primer lugar el caso de espectro discreto, con una base completa y ortonormal $\{|n\rangle\}$. En muchas instancias, $\{|n\rangle\}$ es la base que hace diagonal un conjunto completo de observables. Expandimos el operador de densidad en términos de esta base:
\begin{equation}
\rho = \sum_i w_i |\psi^{(i)}\rangle\langle\psi^{(i)}| = \sum_{n, m} \sum_i w_i |n\rangle\langle n| \psi^{(i)}\rangle\langle\psi^{(i)}|m\rangle\langle m| = \sum_{n, m}|n\rangle\langle m|\left(\sum_i w_i \langle n|\psi^{(i)}\rangle\langle\psi^{(i)}| m\rangle\right).
\end{equation}

Definiendo los coeficientes lineales como $a_n^{(i)} \equiv \langle n|\psi^{(i)}\rangle$, es decir, $|\psi^{(i)}\rangle=\sum_n a_n^{(i)}|n\rangle$, los elementos de matriz de $\rho$ se escriben como:
\begin{equation}
\langle n|\rho| m\rangle = \sum_i w_i a_n^{(i)} a_m^{(i)*} \equiv \overline{a_n a_m^*},
\end{equation}
donde la barra significa promedio sobre el conjunto mixto y $a^*$ es el conjugado complejo de $a$. Recordamos que $a_n^{(i)} \equiv \langle n | \psi^{(i)}\rangle$ es la amplitud de probabilidad de que el estado $|n\rangle$ esté contenido en $|\psi^{(i)}\rangle$. La probabilidad correspondiente es $P_n^{(i)}=|a_n^{(i)}|^2=|\langle n | \psi^{(i)}\rangle|^2$, que aparece en los elementos de matriz diagonal de $\rho$:
\begin{equation}
\langle n|\rho| n\rangle = \sum_i w_i |a_n^{(i)}|^2 = \sum_i w_i P_n^{(i)} \geqq 0,
\end{equation}
que a su vez, puede interpretarse como una probabilidad, ya que:
\begin{equation}
\sum_n \langle n|\rho| n\rangle = \sum_i w_i \sum_n P_n^{(i)} = \sum_i w_i = 1.
\end{equation}

En otras palabras, el elemento de matriz diagonal $\langle n|\rho| n\rangle$ proporciona la probabilidad de que el estado $|n\rangle$ esté ocupado en el conjunto representado por $\rho$. También podemos referirnos al promedio de una cantidad física $\mathbf{A}$ a la misma base:
\begin{align}
[\mathbf{A}] &= \operatorname{Tr}(\boldsymbol{\rho} \mathbf{A}) = \sum_{n, m}\langle n|\boldsymbol{\rho}| m\rangle\langle m|\mathbf{A}| n\rangle = \sum_{n, m} \sum_i w_i a_n^{(i)} a_m^{(i)*} A_{mn} = \sum_{n, m} \overline{a_n a_m^*} A_{mn}.
\end{align}
Si un operador $\mathbf{B}$ es diagonal en la base dada, es decir, $\langle m|\mathbf{B}| n\rangle=B_n \delta_{mn}$, su promedio se da por:
\begin{equation}
[\mathbf{B}] = \sum_i w_i \sum_n P_n^{(i)} B_n = \sum_n B_n \overline{|a_n|^2}.
\end{equation}

Para el espectro continuo, discutimos los casos importantes de las representaciones de coordenadas y momento. Los elementos de matriz de $\rho$ ahora se convierten en funciones de dos puntos. Para la representación de coordenadas $\{|\vec{\mathbf{x}}\rangle\}$, los coeficientes lineales $\langle\vec{\mathbf{x}}|\psi^{(i)}\rangle$ se denominan 'funciones de onda' y usualmente, se escribe
\begin{equation}
\psi^{(i)}(\vec{\mathbf{x}}) = \langle\vec{\mathbf{x}}|\psi^{(i)}\rangle,
\end{equation}
lo que significa que varían continuamente con $\vec{\mathbf{x}}$. Ahora, representamos el operador de densidad:
\begin{equation}
\langle\vec{\mathbf{x}}|\rho|\vec{\mathbf{x}}'\rangle = \rho(\vec{\mathbf{x}}, \vec{\mathbf{x}}') = \sum_i w_i \langle\vec{\mathbf{x}}|\psi^{(i)}\rangle\langle\psi^{(i)}|\vec{\mathbf{x}}'\rangle = \sum_i w_i \psi^{(i)}(\vec{\mathbf{x}}) \psi^{(i)*}(\vec{\mathbf{x}}') = \overline{\psi(\vec{\mathbf{x}}) \psi^*(\vec{\mathbf{x}}')}.
\end{equation}

Los elementos diagonales se dan por:
\begin{equation}
\langle\vec{\mathbf{x}}|\rho|\vec{\mathbf{x}}\rangle = \rho(\vec{\mathbf{x}}, \vec{\mathbf{x}}) = \sum_i w_i |\psi^{(i)}(\vec{\mathbf{x}})|^2 = \overline{|\psi(\vec{\mathbf{x}})|^2} \geqq 0.
\end{equation}

En Mecánica Cuántica, $|\psi^{(i)}(\vec{\mathbf{x}})|^2$ se interpreta como una densidad de probabilidad, ya que la normalización requiere
\begin{equation}
\int \mathrm{d}\vec{\mathbf{x}}|\psi^{(i)}(\vec{\mathbf{x}})|^2 = 1,
\end{equation}
donde la integral se toma sobre todo el espacio. Se sigue que los elementos diagonales de $\rho$, dados por (22), también son densidades de probabilidad, con
\begin{equation}
\operatorname{Tr}(\rho) = \int \mathrm{d}\vec{\mathbf{x}} \langle\vec{\mathbf{x}}|\rho|\vec{\mathbf{x}}\rangle = \sum_i w_i = 1.
\end{equation}

El promedio de un observable $\mathbf{A}$ se obtiene integrando la función de dos puntos $\rho(\vec{\mathbf{x}}, \vec{\mathbf{x}}')$ de (21) con los elementos de matriz de $\mathbf{A}$:
\begin{equation}
[\mathbf{A}] = \operatorname{Tr}(\rho \mathbf{A}) = \int \mathrm{d}\vec{\mathbf{x}} \int \mathrm{d}\vec{\mathbf{x}}' \rho(\vec{\mathbf{x}}, \vec{\mathbf{x}}') A(\vec{\mathbf{x}}', \vec{\mathbf{x}}).
\end{equation}

Si el observable es local en la representación de coordenadas, $A(\vec{\mathbf{x}}', \vec{\mathbf{x}}) = A(\vec{\mathbf{x}}) \delta(\vec{\mathbf{x}}' - \vec{\mathbf{x}})$, donde $\delta(\vec{\mathbf{x}}' - \vec{\mathbf{x}})$ es la función delta de Dirac, la doble integral (23) se reduce a la integración simple a continuación:
\begin{equation}
[\mathbf{A}] = \int \mathrm{d}\vec{\mathbf{x}} \rho(\vec{\mathbf{x}}, \vec{\mathbf{x}}) A(\vec{\mathbf{x}}),
\end{equation}
es decir, la función $A(\vec{\mathbf{x}})$ se integra con la densidad de probabilidad $\rho(\vec{\mathbf{x}}, \vec{\mathbf{x}})$. Nótese que para un conjunto puro, $\rho(\vec{\mathbf{x}}, \vec{\mathbf{x}})$ se da simplemente por $|\psi(\vec{\mathbf{x}})|^2$, la densidad de probabilidad asociada con la función de onda $\psi(\vec{\mathbf{x}})$.

Pasamos a la representación de momento reemplazando la función de onda en el espacio real por la función de onda en el espacio de momento:
\begin{equation}
\phi^{(i)}(\vec{\mathbf{p}}) = \langle\vec{\mathbf{p}}|\psi^{(i)}\rangle.
\end{equation}

Todas las fórmulas pueden traducirse a la nueva base $\{|\vec{\mathbf{p}}\rangle\}$. Por ejemplo, la densidad de probabilidad asociada con los elementos de matriz diagonal de $\rho$ se obtiene como:
\begin{equation}
\langle\vec{\mathbf{p}}|\rho|\vec{\mathbf{p}}\rangle = \rho(\vec{\mathbf{p}}, \vec{\mathbf{p}}) = \sum_i w_i |\phi^{(i)}(\vec{\mathbf{p}})|^2 = \overline{|\phi(\vec{\mathbf{p}})|^2} \geqq 0.
\end{equation}

Debido al principio de incertidumbre, estamos obligados a usar solo una base, pero podemos pasar de una a otra, digamos de $|\vec{\mathbf{x}}\rangle$ a $|\vec{\mathbf{p}}\rangle$, con la matriz de transformación [1], [2]:
\begin{equation}
\langle\vec{\mathbf{p}}|\vec{\mathbf{x}}\rangle = \frac{1}{(2 \pi \hbar)^{3/2}} \exp \left(-\frac{i \vec{\mathbf{p}} \cdot \vec{\mathbf{x}}}{\hbar}\right).
\end{equation}

La función de densidad clásica que proporciona la densidad de puntos en el espacio de fases, es una función de coordenadas generalizadas y momentos, $\rho(q, p)$. Para hacer conexión con el caso clásico, uno tiene que buscar una representación mixta del operador $\rho$. A primera vista, esto puede parecer prohibido por el principio de incertidumbre, pero se puede seguir un procedimiento debido a Wigner [5] para generar tal función.

Tomemos por ejemplo $\rho(\vec{\mathbf{p}}, \vec{\mathbf{p}})$ y usemos la transformación (24) para pasar a la representación de coordenadas:
\begin{equation}
\rho(\vec{\mathbf{p}}, \vec{\mathbf{p}}) = \int \mathrm{d}\vec{\mathbf{x}} \int \mathrm{d}\vec{\mathbf{x}} \frac{1}{(2 \pi \hbar)^3} \exp \left(\frac{i \vec{\mathbf{p}} \cdot (\vec{\mathbf{x}}' - \vec{\mathbf{x}})}{\hbar}\right) \rho(\vec{\mathbf{x}}, \vec{\mathbf{x}}'),
\end{equation}

luego haga el cambio de variables con jacobiano unitario:
\begin{equation}
\vec{\mathbf{r}} = \vec{\mathbf{x}}' - \vec{\mathbf{x}}, \\
\vec{\mathbf{R}} = \frac{1}{2}(\vec{\mathbf{x}} + \vec{\mathbf{x}}')
\end{equation}

Obtenemos
\begin{equation}
\rho(\vec{\mathbf{p}}, \vec{\mathbf{p}}) = \int \mathrm{d}\vec{\mathbf{R}} \int \mathrm{d}\vec{\mathbf{r}} \frac{1}{(2 \pi \hbar)^3} \exp \left(\frac{i \vec{\mathbf{p}} \cdot \vec{\mathbf{r}}}{\hbar}\right) \rho\left(\vec{\mathbf{R}} - \frac{\vec{\mathbf{r}}}{2}, \vec{\mathbf{R}} + \frac{\vec{\mathbf{r}}}{2}\right).
\end{equation}

El integrando en relación con $\vec{\mathbf{R}}$ define una función que depende de $\vec{\mathbf{R}}$ y $\vec{\mathbf{p}}$, que es de forma mixta y consistente con el principio de incertidumbre.

\textbf{Definición 6} Función de Wigner, $\rho_W$
\begin{equation}
\rho_W(\vec{\mathbf{x}}, \vec{\mathbf{p}}) \equiv \frac{1}{(2 \pi \hbar)^3} \int \mathrm{d}\vec{\mathbf{r}} \exp \left(\frac{i \vec{\mathbf{p}} \cdot \vec{\mathbf{r}}}{\hbar}\right) \rho\left(\vec{\mathbf{x}} - \frac{\vec{\mathbf{r}}}{2}, \vec{\mathbf{x}} + \frac{\vec{\mathbf{r}}}{2}\right).
\end{equation}

En nuestra deducción en (25), hemos demostrado una propiedad importante de la función de Wigner, que se lee:
\begin{equation}
\rho(\vec{\mathbf{p}}, \vec{\mathbf{p}}) = \int \mathrm{d}\vec{\mathbf{x}} \rho_W(\vec{\mathbf{x}}, \vec{\mathbf{p}}),
\end{equation}
es decir, la densidad de probabilidad $\rho(\vec{\mathbf{p}}, \vec{\mathbf{p}})$ se obtiene de la función de Wigner $\rho_W(\vec{\mathbf{x}}, \vec{\mathbf{p}})$ integrando sobre la variable espacial $\vec{\mathbf{x}}$. La relación complementaria también es verdadera, y se deja al lector demostrar su validez:
\begin{equation}
\rho(\vec{\mathbf{x}}, \vec{\mathbf{x}}) = \int \mathrm{d}\vec{\mathbf{p}} \rho_W(\vec{\mathbf{x}}, \vec{\mathbf{p}}).
\end{equation}

Las relaciones (27) y (28) son propiedades deseables para una función candidata a representar la densidad clásica. Pero la definición (26) es completamente cuántica, y uno tiene que tomar el límite de alguna manera no trivial para obtener el caso clásico. En general, se puede mostrar que la propiedad
\begin{equation}
\rho_W(\vec{\mathbf{x}}, \vec{\mathbf{p}}) \geqq 0
\end{equation}
no siempre se satisface en todo el espacio de fases $(\vec{\mathbf{x}}, \vec{\mathbf{p}})$. Las regiones donde $\rho_W(\vec{\mathbf{x}}, \vec{\mathbf{p}}) < 0$ se dice que contienen efectos cuánticos coherentes, cuyo tamaño se reduce con $\hbar \rightarrow 0$. Formalmente, fue mostrado por Wigner [5] que $\rho_W$ satisface la ecuación de Liouville, cuando $\hbar \rightarrow 0$ (ver la siguiente subsección). Para un sistema compuesto por $N$ partículas, la definición (26) se puede generalizar al espacio de fases $\Gamma$ de un sistema de partículas:
\begin{align}
& \rho_W^{(N)}\left(\tilde{\mathbf{x}}_1 + \tilde{\mathbf{x}}_2, \ldots, \overline{\mathbf{x}}_N ; \tilde{\mathbf{p}}_1, \tilde{\mathbf{p}}_2, \ldots, \vec{\mathbf{p}}_N\right) \equiv \\
& \quad \frac{1}{(2 \pi \hbar)^{3N}} \int \ldots \int \mathrm{d}^3 \mathbf{r}_1 \mathrm{d}^3 \mathbf{r}_2 \ldots \mathrm{d}^3 \mathbf{r}_N \exp \left(i \sum_{j=1}^{N} \frac{\tilde{\mathbf{p}}_j \cdot \tilde{\mathbf{r}}_j}{\hbar}\right) \\
& \quad \times \left\langle \tilde{\mathbf{x}}_1 - \frac{\overline{\mathbf{r}}_1}{2}, \ldots, \tilde{\mathbf{x}}_N - \frac{\tilde{\mathbf{r}}_N}{2} \middle| \rho \middle| \tilde{\mathbf{x}}_1 + \frac{\overline{\mathbf{r}}_1}{2}, \ldots, \tilde{\mathbf{x}}_N + \frac{\overline{\mathbf{r}}_N}{2} \right\rangle.
\end{align}

\subsection{Digresión sobre el conjunto de Gibbs}

Aquí discutimos algunos conceptos clave sobre la función de densidad clásica $\rho(q, p)$, que era el objetivo del enfoque de Wigner. En la Mecánica Estadística clásica, un microestado de un sistema de $N$ partículas está representado por un punto en el 'espacio de fases' $\Gamma$. Este es un espacio de $6N$ dimensiones abarcado por las coordenadas generalizadas y los momentos conjugados $\{q_i, p_i\}$, que caracterizan el sistema. Si usamos coordenadas cartesianas,
\begin{equation}
\{q_i\}_{3N} = (x_1, y_1, z_1, \ldots, x_N, y_N, z_N).
\end{equation}

Desde el punto de vista macroscópico, operamos con un conjunto reducido de cantidades que son compatibles con un gran número de microestados, como por ejemplo el caso de un gas ocupando un volumen dado, en condiciones estándar de temperatura y presión. En Mecánica Estadística, no estamos interesados en el movimiento detallado de un sistema de muchas partículas, sino que solo queremos calcular algunas propiedades promedio (cantidades termodinámicas) que computamos usando un conjunto de réplicas ideales del mismo sistema. El conjunto está entonces representado por un enjambre de puntos en el espacio $\Gamma$, siendo los puntos interpretados como diferentes microestados correspondientes a diferentes condiciones iniciales del sistema, todos satisfaciendo las mismas restricciones macroscópicas. Esta idea fue introducida por Gibbs en los fundamentos de la Mecánica Estadística, siendo la cantidad relevante para caracterizar el conjunto la densidad de puntos en el espacio $\Gamma$. Sea $\rho(q_i, p_i, t)$ tal distribución, con la notación significando que $\rho$ depende de todas las coordenadas generalizadas y momentos, y también puede depender explícitamente del tiempo. En otras palabras, $\rho(q_i, p_i, t) \mathrm{d}q^{3N} \mathrm{d}p^{3N}$
es el número de puntos del sistema representativo (microestados) contenidos en el tiempo $t$ en el volumen infinitesimal $\mathrm{d}\Omega = \mathrm{d}q^{3N} \mathrm{d}p^{3N}$, con $\mathrm{d}\Omega$ centrado sobre el punto $\{q_i, p_i\}$

en el espacio de fases. Esos puntos representativos del conjunto evolucionan en el tiempo, trazando una trayectoria en el espacio $\Gamma$ que es cerrada (movimiento periódico) o nunca se autointersecta. Además, las trayectorias de diferentes puntos representativos nunca se intersectan, ya que representan movimientos con diferentes condiciones iniciales (si dos trayectorias se intersectan en un punto dado, ese punto común puede elegirse como una nueva condición inicial, y las dos trayectorias deberían coincidir en todo momento). En la Mecánica Clásica, se puede mostrar que la evolución temporal es una transformación canónica [6], y el volumen del espacio de fases es un invariante canónico (uno de los invariantes integrales de Poincaré). El número de puntos representativos del conjunto contenidos en cualquier volumen infinitesimal $\mathrm{d}\Omega$ también es constante en el tiempo. La forma del elemento infinitesimal cambia, pero su volumen es constante. Ningún punto representativo puede intersectar el límite de $\mathrm{d}\Omega$ en ningún momento (mismo argumento que se dio anteriormente). Entonces, podemos enunciar este resultado como un teorema:

\textbf{Teorema 7 (Liouville)} La densidad $\rho(q, p, t)$ es constante en el tiempo, o
\begin{equation}
\frac{\mathrm{d}\rho}{\mathrm{d}t} = 0.
\end{equation}
Podemos reescribir el teorema (30) en una forma diferente
\begin{equation}
0 = \frac{\mathrm{d}\rho}{\mathrm{d}t} = \frac{\partial \rho}{\partial t} + \sum_i \left(\dot{q}_i \frac{\partial \rho}{\partial q_i} + \dot{p}_i \frac{\partial \rho}{\partial p_i}\right),
\end{equation}
mostrando que la dependencia explícita en el tiempo se cancela con la dependencia implícita a través de las coordenadas y los momentos. Para un sistema Hamiltoniano, se satisfacen las ecuaciones de Hamilton
\begin{equation}
\dot{q}_i = \frac{\partial H}{\partial p_i}, \quad \dot{p}_i = -\frac{\partial H}{\partial q_i},
\end{equation}
las cuales sustituimos en (31), obteniendo:
\begin{equation}
0 = \frac{\mathrm{d}\rho}{\mathrm{d}t} = \frac{\partial \rho}{\partial t} + \sum_i \left(\frac{\partial \rho}{\partial q_i} \frac{\partial H}{\partial p_i} - \frac{\partial \rho}{\partial p_i} \frac{\partial H}{\partial q_i}\right),
\end{equation}
que se escribe a su vez, en términos de un corchete de Poisson [6] como:
\begin{equation}
0 = \frac{\partial \rho}{\partial t} + \{\rho, H\},
\end{equation}
con el corchete de Poisson definido como $\{A, B\} \equiv \sum_i \left(\frac{\partial A}{\partial q_i} \frac{\partial B}{\partial p_i} - \frac{\partial A}{\partial p_i} \frac{\partial B}{\partial q_i}\right)$. La relación (33) es otra forma de enunciar el teorema de Liouville. Puede interpretarse geométricamente [Huang]: el movimiento de los puntos representativos en el espacio $\Gamma$ se asemeja mucho al movimiento de un fluido incompresible. De hecho, la relación (33) tiene la forma de una ecuación de continuidad, si se define una densidad de corriente para el flujo de puntos como:
\begin{equation}
\vec{\mathbf{j}} \equiv \rho \vec{\mathbf{v}},
\end{equation}
con el vector de velocidad escrito como
\begin{equation}
\vec{\mathbf{v}} = (\dot{q}_1, \dot{q}_2, \ldots, \dot{q}_{3N} ; \dot{p}_1, \dot{p}_2, \ldots, \dot{p}_{3N}).
\end{equation}
Debido a las ecuaciones de movimiento de Hamilton (32), obtenemos
\begin{equation}
0 = \frac{\partial \rho}{\partial t} + \{\rho, H\} = \frac{\partial \rho}{\partial t} + \nabla \cdot \vec{\mathbf{j}},
\end{equation}
con el operador 'nabla' definido en el espacio $\Gamma$ como:
\begin{equation}
\boldsymbol{\nabla} \equiv \left(\frac{\partial}{\partial q_1}, \frac{\partial}{\partial q_2}, \ldots, \frac{\partial}{\partial q_{3N}}, \frac{\partial}{\partial p_1}, \frac{\partial}{\partial p_2}, \ldots, \frac{\partial}{\partial p_{3N}}\right).
\end{equation}
Entonces, las variaciones locales de $\rho$ son causadas por el flujo de la corriente de densidad $\vec{\mathbf{j}}$, en cualquier vecindad de puntos representativos del conjunto. Sea $A(q, p)$ una cantidad dinámica del sistema de partículas. A nivel macroscópico, el valor de $A$ que observamos se supone que es el promedio sobre el conjunto, calculado como
\begin{equation}
[A]_C(t) = \frac{\int \mathrm{d}q^{3N} \mathrm{d}p^{3N} \rho(q, p, t) A(q, p)}{\int \mathrm{d}q^{3N} \mathrm{d}p^{3N} \rho(q, p, t)},
\end{equation}
con la distribución $\rho(q, p, t)$ satisfaciendo el teorema de Liouville, y el símbolo $[\ldots]_C$ representando el promedio clásico. En principio, la dependencia temporal de $[A]_C(t)$ debería acercarse a su valor de equilibrio en la situación estacionaria:
\begin{equation}
\frac{\partial \rho}{\partial t} = \{\rho, H\} = 0.
\end{equation}
Una distribución de densidad estacionaria $\rho(q, p)$ dependerá solo de integrales del tiempo independientes de las ecuaciones de movimiento. La suposición más simple es postular que $\rho$ es una función de la energía total $H = E$, que es una cantidad conservada. La distribución de probabilidad a priori igual:
\begin{equation}
\rho = \rho(E) = \begin{cases}
\text{constante,} & \text{si } E-\frac{1}{2}\Delta < H < E+\frac{1}{2}\Delta, \\
0, & \text{de lo contrario}
\end{cases}
\end{equation}
se llama conjunto microcanónico, y representa un sistema aislado. La cantidad $\Delta$ se elige de tal manera que $\Delta \ll E$, y se introduce por conveniencia en el conteo de estados. En el límite termodinámico, las cantidades macroscópicas son independientes de $\Delta$.

La cuestión de cómo el sistema se acerca a tal estado de equilibrio, está en el corazón de la Mecánica Estadística, siendo uno de los problemas centrales en Física desde la época de Boltzmann.

\nextpage

\backmatter

\renewcommand\bibname{Referencias}
\bibliographystyle{plainnat} 
\bibliography{bibliography.bib}

\begin{thebibliography}{34}
\providecommand{\natexlab}[1]{#1}
\providecommand{\url}[1]{\texttt{#1}}
\expandafter\ifx\csname urlstyle\endcsname\relax
  \providecommand{\doi}[1]{doi: #1}\else
  \providecommand{\doi}{doi: \begingroup \urlstyle{rm}\Url}\fi

\bibitem[Abraham and Penrose(2017)]{Abraham2017}
Eitan Abraham and Oliver Penrose.
\newblock {Physics of negative absolute temperatures}.
\newblock \emph{Physical Review E}, 95\penalty0 (1):\penalty0 012125, jan 2017.
\newblock ISSN 24700053.
\newblock \doi{10.1103/PhysRevE.95.012125}.

\bibitem[Boltzmann(1995)]{Boltzmann1995}
L.~Boltzmann.
\newblock \emph{Lectures on Gas Theory}.
\newblock Dover Books on Physics. Dover Publications, 1995.
\newblock ISBN 9780486684550.

\bibitem[Boltzmann(2003)]{Boltzmann1872}
Ludwig Boltzmann.
\newblock Further studies on the thermal equilibrium of gas molecules.
\newblock In \emph{History of Modern Physical Sciences}, pages 262--349.
  {Published} {by} {Imperial} {College} {Press} {and} {distributed} {by}
  {World} {Scientific} {Publishing} {Co}., July 2003.
\newblock \doi{10.1142/9781848161337_0015}.
\newblock URL \url{https://doi.org/10.1142/9781848161337_0015}.

\bibitem[Brown et~al.(2009)Brown, Myrvold, and Uffink]{Brown2009}
Harvey~R. Brown, Wayne Myrvold, and Jos Uffink.
\newblock Boltzmann's h-theorem, its discontents, and the birth of statistical
  mechanics.
\newblock \emph{Studies in History and Philosophy of Science Part B: Studies in
  History and Philosophy of Modern Physics}, 40\penalty0 (2):\penalty0
  174--191, 2009.
\newblock ISSN 1355-2198.
\newblock \doi{https://doi.org/10.1016/j.shpsb.2009.03.003}.
\newblock URL
  \url{https://www.sciencedirect.com/science/article/pii/S1355219809000124}.

\bibitem[Buonsante et~al.(2017)Buonsante, Franzosi, and Smerzi]{Buonsante2017}
P.~Buonsante, R.~Franzosi, and A.~Smerzi.
\newblock {Phase transitions at high energy vindicate negative microcanonical
  temperature}.
\newblock \emph{Physical Review E}, 95\penalty0 (5):\penalty0 052135, 2017.
\newblock ISSN 24700053.
\newblock \doi{10.1103/PhysRevE.95.052135}.

\bibitem[Coleman and Noll(1959)]{Coleman1959}
Bernard~D Coleman and Walter Noll.
\newblock {Conditions for Equilibrium at Negative Absolute Temperatures}.
\newblock \emph{Physical Review}, 115\penalty0 (2):\penalty0 262--265, 1959.

\bibitem[Drossel(2015)]{Drossel2015}
Barbara Drossel.
\newblock \emph{On the Relation Between the Second Law of Thermodynamics and
  Classical and Quantum Mechanics}, pages 41--54.
\newblock Springer Berlin Heidelberg, Berlin, Heidelberg, 2015.
\newblock ISBN 978-3-662-43911-1.
\newblock \doi{10.1007/978-3-662-43911-1_3}.
\newblock URL \url{https://doi.org/10.1007/978-3-662-43911-1_3}.

\bibitem[Dunkel and Hilbert(2013)]{Dunkel2013}
J{\"{o}}rn Dunkel and Stefan Hilbert.
\newblock {Consistent thermostatistics forbids negative absolute temperatures}.
\newblock \emph{Nature Physics}, 10\penalty0 (1):\penalty0 67--72, dec 2013.
\newblock ISSN 17452481.
\newblock \doi{10.1038/nphys2815}.

\bibitem[Einstein(1905)]{Einstein1905c}
A.~Einstein.
\newblock Über die von der molekularkinetischen theorie der wärme geforderte
  bewegung von in ruhenden flüssigkeiten suspendierten teilchen.
\newblock \emph{Annalen der Physik}, 322\penalty0 (8):\penalty0 549--560, 1905.
\newblock \doi{https://doi.org/10.1002/andp.19053220806}.
\newblock URL
  \url{https://onlinelibrary.wiley.com/doi/abs/10.1002/andp.19053220806}.

\bibitem[Frenkel and Warren(2015)]{Frenkel2015}
Daan Frenkel and Patrick~B. Warren.
\newblock {Gibbs, Boltzmann, and negative temperatures}.
\newblock \emph{American Journal of Physics}, 83\penalty0 (2):\penalty0
  163--170, feb 2015.
\newblock ISSN 0002-9505.
\newblock \doi{10.1119/1.4895828}.

\bibitem[Frigerio(1974)]{Frigerio1974}
Norman~A. Frigerio.
\newblock Poisson and non-poisson behavior of radioactive systems.
\newblock \emph{Nuclear Instruments and Methods}, 114\penalty0 (1):\penalty0
  175--177, 1974.
\newblock ISSN 0029-554X.
\newblock \doi{https://doi.org/10.1016/0029-554X(74)90358-9}.
\newblock URL
  \url{https://www.sciencedirect.com/science/article/pii/0029554X74903589}.

\bibitem[Hakonen et~al.(1992)Hakonen, Nummila, and Vuorinen]{Hakonen1992}
P~J Hakonen, K~K Nummila, and R~T Vuorinen.
\newblock {Spin dynamics in highly polarized silver at negative absolute
  temperatures}.
\newblock \emph{Physical Review B}, 45\penalty0 (5):\penalty0 2196--2200, 1992.
\newblock ISSN 01631829.
\newblock \doi{10.1103/PhysRevB.45.2196}.

\bibitem[Hama et~al.(2018)Hama, Munro, and Nemoto]{Hama2018}
Yusuke Hama, William~J. Munro, and Kae Nemoto.
\newblock {Relaxation to Negative Temperatures in Double Domain Systems}.
\newblock \emph{Physical Review Letters}, 120\penalty0 (6):\penalty0 060403,
  feb 2018.
\newblock ISSN 10797114.
\newblock \doi{10.1103/PhysRevLett.120.060403}.

\bibitem[Hartle and Taylor(1969)]{Hartle1969}
James~B. Hartle and John~R. Taylor.
\newblock Quantum mechanics of paraparticles.
\newblock \emph{Phys. Rev.}, 178:\penalty0 2043--2051, Feb 1969.
\newblock \doi{10.1103/PhysRev.178.2043}.
\newblock URL \url{https://link.aps.org/doi/10.1103/PhysRev.178.2043}.

\bibitem[Hecht(1960)]{Hecht1960}
Charles~E Hecht.
\newblock {Thermodynamic potentials for systems at negative absolute
  temperatures}.
\newblock \emph{Physical Review}, 119\penalty0 (5):\penalty0 1443--1444, 1960.
\newblock ISSN 0031899X.
\newblock \doi{10.1103/PhysRev.119.1443}.

\bibitem[Hou(2019)]{Hou2019}
Ji~Xuan Hou.
\newblock {Violation of the temperature-signifies-heat-flow rule in systems
  with long-range interactions}.
\newblock \emph{Physical Review E}, 99\penalty0 (5):\penalty0 052114, may 2019.
\newblock ISSN 24700053.
\newblock \doi{10.1103/PhysRevE.99.052114}.

\bibitem[Klein(1956)]{Klein1956}
Martin~J Klein.
\newblock {Negative Absolute Temperatures}.
\newblock \emph{Physical Review}, 104\penalty0 (3):\penalty0 589--589, 1956.

\bibitem[Mandt et~al.(2013)Mandt, Feiguin, and Manmana]{Mandt2013}
Stephan Mandt, Adrian~E. Feiguin, and Salvatore~R. Manmana.
\newblock {Relaxation towards negative temperatures in bosonic systems:
  Generalized Gibbs ensembles and beyond integrability}.
\newblock \emph{Physical Review A - Atomic, Molecular, and Optical Physics},
  88\penalty0 (4):\penalty0 043643, oct 2013.
\newblock ISSN 10502947.
\newblock \doi{10.1103/PhysRevA.88.043643}.

\bibitem[Messiah and Greenberg(1964)]{Messiah1964}
A.~M.~L. Messiah and O.~W. Greenberg.
\newblock Symmetrization postulate and its experimental foundation.
\newblock \emph{Phys. Rev.}, 136:\penalty0 B248--B267, Oct 1964.
\newblock \doi{10.1103/PhysRev.136.B248}.
\newblock URL \url{https://link.aps.org/doi/10.1103/PhysRev.136.B248}.

\bibitem[Miceli et~al.(2019)Miceli, Baldovin, and Vulpiani]{Miceli2019}
Fabio Miceli, Marco Baldovin, and Angelo Vulpiani.
\newblock {Statistical mechanics of systems with long-range interactions and
  negative absolute temperature}.
\newblock \emph{Physical Review E}, 99\penalty0 (4):\penalty0 042152, apr 2019.
\newblock ISSN 24700053.
\newblock \doi{10.1103/PhysRevE.99.042152}.

\bibitem[of~Arts and Sciences.(1874)]{Gibbs1874}
Connecticut~Academy of~Arts and Sciences.
\newblock \emph{Transactions of the Connecticut Academy of Arts and Sciences},
  volume v.3 (1874-1878).
\newblock New Haven, Published by the Academy, 1866-, 1874.
\newblock URL
  \url{https://www.biodiversitylibrary.org/item/88413#page/128/mode/2up}.

\bibitem[Onsageir(1949)]{Onsageir1949}
L~Onsageir.
\newblock {Statistical Hydrodynamics}.
\newblock \emph{Il Nuovo Cimento (1943-1954)}, 6:\penalty0 279--287, 1949.

\bibitem[Pauli(1940)]{Pauli1940}
W.~Pauli.
\newblock The connection between spin and statistics.
\newblock \emph{Phys. Rev.}, 58:\penalty0 716--722, Oct 1940.
\newblock \doi{10.1103/PhysRev.58.716}.
\newblock URL \url{https://link.aps.org/doi/10.1103/PhysRev.58.716}.

\bibitem[Purcei and Pound(1951)]{Purcei1950}
E~M Purcei and R~V Pound.
\newblock {A Nuclear Spin System at Negative Temperature}.
\newblock \emph{Phys. Rev.}, 81\penalty0 (2):\penalty0 279--280, 1951.
\newblock \doi{10.1103/PhysRev.81.279}.

\bibitem[Ramsey(1956{\natexlab{a}})]{Ramsey1956}
Norman~F Ramsey.
\newblock {Thermodynamics and statistical mechanics at negative absolute
  temperatures}.
\newblock \emph{Physical Review}, 103\penalty0 (1):\penalty0 20--28,
  1956{\natexlab{a}}.
\newblock ISSN 0031899X.
\newblock \doi{10.1103/PhysRev.103.20}.

\bibitem[Ramsey(1956{\natexlab{b}})]{Ramsey1956a}
Norman~F Ramsey.
\newblock {Thermodynamics and Statistical Mechanics at Negative Absolute
  Temperatures}.
\newblock \emph{Phys. Rev.}, 103\penalty0 (1):\penalty0 20--28,
  1956{\natexlab{b}}.
\newblock ISSN 18793169.

\bibitem[Reif(1965)]{Reif1965}
F.~Reif.
\newblock \emph{Fundamentals of Statistical and Thermal Physics}.
\newblock Number v. 10 in Fundamentals of Physics Series. McGraw-Hill, 1965.
\newblock ISBN 9780070856158.

\bibitem[Romero-Roch{\'{i}}n(2013)]{Romero-Rochin2013}
V{\'{i}}ctor Romero-Roch{\'{i}}n.
\newblock {Nonexistence of equilibrium states at absolute negative
  temperatures}.
\newblock \emph{Physical Review E - Statistical, Nonlinear, and Soft Matter
  Physics}, 88\penalty0 (2):\penalty0 022144, aug 2013.
\newblock ISSN 15393755.
\newblock \doi{10.1103/PhysRevE.88.022144}.

\bibitem[Rutherford et~al.(1910)Rutherford, Geiger, and
  Bateman]{Rutherford1910}
E~Rutherford, H~Geiger, and H.~Bateman.
\newblock Lxxvi. the probability variations in the distribution of $\alpha$
  particles.
\newblock \emph{The London, Edinburgh, and Dublin Philosophical Magazine and
  Journal of Science}, 20\penalty0 (118):\penalty0 698--707, 1910.
\newblock \doi{10.1080/14786441008636955}.
\newblock URL \url{https://doi.org/10.1080/14786441008636955}.

\bibitem[Shannon(1948)]{Shannon1948}
C.~E. Shannon.
\newblock A mathematical theory of communication.
\newblock \emph{The Bell System Technical Journal}, 27\penalty0 (3):\penalty0
  379--423, 1948.
\newblock \doi{10.1002/j.1538-7305.1948.tb01338.x}.
\newblock URL \url{https://doi.org/10.1002/j.1538-7305.1948.tb01338.x}.

\bibitem[Swendsen(2015)]{Swendsen2015}
Robert~H. Swendsen.
\newblock {Continuity of the entropy of macroscopic quantum systems}.
\newblock \emph{Physical Review E - Statistical, Nonlinear, and Soft Matter
  Physics}, 92\penalty0 (5):\penalty0 052110, nov 2015.
\newblock ISSN 15502376.
\newblock \doi{10.1103/PhysRevE.92.052110}.

\bibitem[Swendsen and Wang(2015)]{Swendsen2015a}
Robert~H. Swendsen and Jian~Sheng Wang.
\newblock {Gibbs volume entropy is incorrect}.
\newblock \emph{Physical Review E - Statistical, Nonlinear, and Soft Matter
  Physics}, 92\penalty0 (2):\penalty0 020103, aug 2015.
\newblock ISSN 15502376.
\newblock \doi{10.1103/PhysRevE.92.020103}.

\bibitem[Tremblay(1976)]{Tremblay1976}
Andr{\'{e}}-Marie Tremblay.
\newblock {Comment on ''Negative Kelvin temperatures: Some anomalies and a
  speculation''}.
\newblock \emph{American Journal of Physics}, 44\penalty0 (10):\penalty0
  994--995, 1976.
\newblock ISSN 0002-9505.
\newblock \doi{10.1119/1.10248}.

\bibitem[Zermelo(1896)]{Zermelo1896}
E.~Zermelo.
\newblock Ueber einen satz der dynamik und die mechanische wärmetheorie.
\newblock \emph{Annalen der Physik}, 293\penalty0 (3):\penalty0 485--494, 1896.
\newblock \doi{https://doi.org/10.1002/andp.18962930314}.
\newblock URL
  \url{https://onlinelibrary.wiley.com/doi/abs/10.1002/andp.18962930314}.

\end{thebibliography}

\printindex

\end{document}